\PassOptionsToPackage{dvipsnames}{xcolor}
\documentclass[%
    paper=A4,               
    twoside=true,           
    openright,              
    parskip=half,           
    chapterprefix=true,     
    11pt,                   
    headings=normal,        
    bibliography=totoc,     
    listof=totoc,           
    titlepage=on,           
    captions=tableabove,    
    chapterprefix=false,    
    appendixprefix=false,   
]{scrreprt}%

\PassOptionsToPackage{utf8}{inputenc}
\usepackage{inputenc}
\usepackage[author={Auteur1}]{pdfcomment}

\newcommand{\thesisTitle}{Dynamic Scheduling in Fiber and Spaceborne Quantum~Repeater Networks}

\newcommand{\thesisName}{Paolo Fittipaldi}
\newcommand{\thesisSubject}{Quantum Internet}

\newcommand{\thesisDate}{18 Septembre 2025}

\usepackage{morewrites}
\usepackage[pass]{geometry}
\usepackage{scrhack}
\usepackage{tikz}
\usetikzlibrary{shapes.misc, shapes.geometric, positioning, arrows.meta, calc, arrows}
\usepackage{import}
\usepackage[most]{tcolorbox}
\tcbuselibrary{listings}
\usepackage{caption}
\usepackage{subcaption}
\usepackage{adjustbox}
\usepackage{colortbl}
\usepackage{placeins}[section, below]
\usepackage{macros}
\usepackage[defblank]{paralist}  
\usepackage[inline,shortlabels]{enumitem} 
\usepackage{wrapfig}

\usepackage{hyperref}
\hypersetup{
    pdftitle={\thesisTitle},    
    pdfsubject={\thesisSubject},
    pdfauthor={\thesisName},    
    plainpages=false,           
    colorlinks=false,           
    pdfborder={0 0 0},          
    breaklinks=true,            
    bookmarksnumbered=true,     %
    bookmarksopen=true,         %
    hyperfootnotes=false
}
\usepackage[english]{babel} 
\PassOptionsToPackage{
    figuresep=colon,%
    hangfigurecaption=false,%
    hangsection=true,%
    hangsubsection=true,%
    sansserif=false,%
    configurelistings=false,%
    colorize=full,%
    colortheme=bluemagenta,%
    configurebiblatex=true,%
    bibsys=biber,%
    bibfile=main,%
    bibstyle=alphabetic,%
    bibsorting=nty
}{cleanthesis}
\usepackage{cleanthesis}

\definecolor{red}{HTML}{9B0000}
\definecolor{lightred}{HTML}{FF5131}
\definecolor{green}{HTML}{1FAA00}
\definecolor{lightgreen}{HTML}{9CFF57}
\definecolor{purple}{HTML}{7200CA}
\definecolor{blue}{HTML}{0064B7}
\definecolor{verylightgrey}{HTML}{F1F1F1}
\colorlet{ctcolorfootertitlelight}{ctcolorfootertitle!20}

\lstdefinelanguage{diff}{
  morecomment=[f][\color{blue}]{@},
  morecomment=[f][\color{green}]{+},
  morecomment=[f][\color{red}]{-},
  keepspaces=true,
  identifierstyle=\color{black},
}

\lstdefinelanguage{spdiff}{
  morecomment=[s][\color{blue}]{@@}{@@},
  morecomment=[s][\color{blue}]{@rule1@}{@@},
  morecomment=[s][\color{blue}]{@rule2@}{@@},
  morecomment=[f][\color{green}]{+},
  morecomment=[f][\color{red}]{-},
  morekeywords={expression},
  morekeywords={identifier},
  keepspaces=true,
  identifierstyle=\color{black},
}

\newtcolorbox{notice}[1][]{enhanced,
  before skip=3mm,after skip=7mm,
  boxrule=0pt,left=7mm,right=2mm,top=1mm,bottom=1mm,
  colback=ctcolorfootertitlelight,
  sharp corners,
  frame hidden,
  underlay={%
    \path[fill=ctcolorfootertitle,draw=none] (interior.south west) rectangle node[white]{\huge\bfseries !} ([xshift=6mm]interior.north west);
    },
  #1}

\newtcblisting{namedlisting}[2][]{%
  enhanced,
  sharp corners,
  frame hidden,
  boxsep=0pt,
  title={\hspace*{3.5mm}#2},
  titlerule=0pt,
  colbacktitle=ctcolormain!20,
  coltitle=black,
  toptitle=1mm,
  bottomtitle=1mm,
  listing only,
  left=0pt,
  top=0pt,
  bottom=0pt,
  right=0pt,
  toprule=0pt,
  leftrule=0pt,
  bottomrule=0pt,
  rightrule=0pt,
  colframe=ctcolormain,
  listing options={aboveskip=0pt, belowskip=0pt, #1},
}

\tikzset{%
  >=latex,
  edge/.style={
    ->,
    rounded corners=3pt
  },
  node/.style={%
    rectangle,
    rounded corners=1pt,
    draw,
    thick,
    fill=ctcolorgraylighter,
    minimum height=1.5em,
  },
  empty/.style={%
    minimum height=0,
  },
  removed/.style={
    text=red,
    draw=red,
  },
  added/.style={
    text=green,
    draw=green,
  },
  fake/.style={
    densely dotted,
    text=ctcolorgray,
  },
  special/.style={
    trapezium,
    trapezium left angle=60,
    trapezium right angle=120,
    minimum width=2cm,
  },
  entry port/.style={
    alias=this,
    append after command={%
      \pgfextra
        \draw[-(, thick] (this) -- ($(this.north) + (0, 0.25)$);
      \endpgfextra
    },
  },
  exit port/.style={
    alias=this,
    append after command={%
      \pgfextra
        \draw[-o, thick] (this) -- ($(this.south) + (0, -0.35)$);
      \endpgfextra
    },
  },
}

\usepackage{mathtools}
\usepackage[textsize=tiny]{todonotes}

\setuptodonotes{fancyline}

\usepackage{newtxmath}
\usepackage{mathtools}
\usepackage{tabularx}
\usepackage{lscape} 
\usepackage{siunitx}
\usepackage{xfrac} 
\usepackage{braket}
\usepackage{cancel}
\usepackage{array} 
\usepackage{rotating}
\usepackage{pdflscape}
\usepackage{minted} 
\makeatletter
\AddToHook{begindocument/before}{\@ifpackageloaded{minted}{\removefromtoclist[float]{lol}}{}}
\makeatother 

\DeclareMathOperator*{\argmax}{argmax}

\newcommand{\EV}[1]{\vmathbb{E}\left[#1\right]}
\newcommand{\phm}{\phantom-}

\newcommand{\trt}{t_\rightleftharpoons} 

\newcommand{\axiombox}[1]{\begin{center}\fbox{\begin{minipage}{0.9\textwidth}#1\end{minipage}}\end{center}}

\newcommand{\paolofootnote}[1]{\begin{NoHyper}\footnote{#1}\end{NoHyper}}
\newcommand{\paolofootnotemark}{\begin{NoHyper}\footnotemark\end{NoHyper}}
\newcommand{\paolofootnotetext}[1]{\begin{NoHyper}\footnotetext{#1}\end{NoHyper}}

\usepackage{rotfloat}
\usepackage{pdflscape}
\usepackage{eso-pic,zref-user}
\newcounter{cntsideways}

\makeatletter

\AddToShipoutPictureBG{%
  \ifnum\zref@extractdefault{rotate\number\value{page}}{page}{0}=0  
    \PLS@RemoveRotate
  \else 
    \PLS@AddRotate{270}%
  \fi%
}

\newcommand\rotatesidewayslabel{\stepcounter{cntsideways}%
  \zlabel{tmp\thecntsideways}\zlabel{rotate\zref@extractdefault{tmp\thecntsideways}{page}{0}}}

\makeatother

\usepackage{xpatch}
\xapptocmd{\sidewaysfigure}{\rotatesidewayslabel}{}{}
\xapptocmd{\sidewaystable}{\rotatesidewayslabel}{}{}

\begin{document}



\renewcaptionname{english}{\figurename}{Fig.}
\renewcaptionname{english}{\tablename}{Tab.}

\renewcommand*{\lstlistlistingname}{List of Listings}

\pagenumbering{roman}			

%

\newcommand{\jurymember}[5]{\hspace{1.5em}\textbf{{#1} \textsc{{#2}}}, {#3}, {#4} \hfill \textit{{#5}}\\}

\newgeometry{left=2.8cm, right=2.5cm, top=2cm, bottom=2cm}
\begin{titlepage}
	\pdfbookmark[0]{Cover}{Cover}
	\begin{flushleft}
      \includegraphics[height=1.7cm]{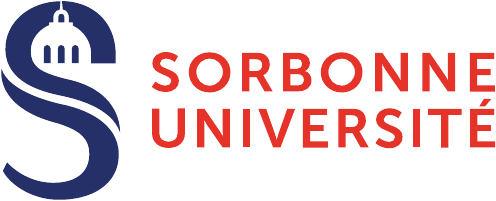}
      \hfill
      \includegraphics[height=1.7cm]{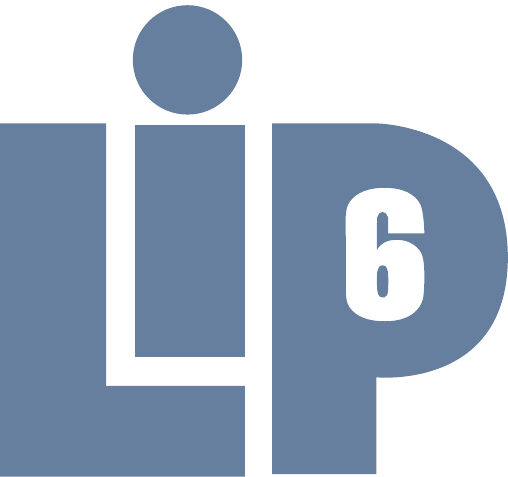}
      \hfill
	\end{flushleft}
	\vfill
	\begin{center}
        \vspace{1em}
        {\large Thèse présentée pour l'obtention du grade de}\\
        \vspace{1em}
        {\LARGE DOCTEUR de SORBONNE UNIVERSITÉ}\\
        \vspace{2em}
		{\small Spécialité}\\
		\vspace{0.5em}
		{\Large Ingénierie / Systèmes Informatiques}\\
		\vspace{2em}
        {\large École doctorale}\\
		\vspace{0.5em}
		{\Large Informatique, Télécommunication et Électronique Paris (ED130)}\\
    \end{center}
	\vfill
	\begin{flushright}
	\hfill
	\vfill
	{\LARGE\thesisTitle \par}
	\rule[5pt]{\textwidth}{.4pt} \par
	{\Large\thesisName}
	\vfill
	\end{flushright}
	Soutenue publiquement le : \textit{\large\thesisDate}
	\begin{flushleft}
		Devant un jury composé de :\\
		  \jurymember{Claudio}{Cicconetti}{Primo Ricercatore}{IIT-CNR}{Rapporteur}
		\jurymember{Daniel}{Oi}{Reader}{University of Strathclyde}{Rapporteur}
		\jurymember{Anne}{Fladenmuller}{Professeure}{Sorbonne Université}{Présidente du Jury}
		\jurymember{Marc}{Kaplan}{CEO}{VeriQloud}{Examinateur}
		\jurymember{Giuseppe}{Vallone}{Professore}{Università di Padova}{Examinateur}
		\jurymember{Rodney}{Van Meter}{Professor}{Keio University}{Examinateur}
		\jurymember{Anastasios}{Giovanidis}{Chargé de Recherche}{CNRS/Sorbonne Université}{Directeur}
		\jurymember{Frédéric}{Grosshans}{Chargé de Recherche}{CNRS/Sorbonne Université}{Co-Encadrant}
	\end{flushleft}
\end{titlepage}
\restoregeometry
\clearpage
\vfill
\begin{flushright}
    \vfill
	\hfill
	\vfill
\end{flushright}
\mbox{}
\vfill
\begin{wrapfigure}{L}{0.25\textwidth}
    \begin{center}
        \vspace{-\baselineskip}
        \includegraphics[width=0.25\textwidth]{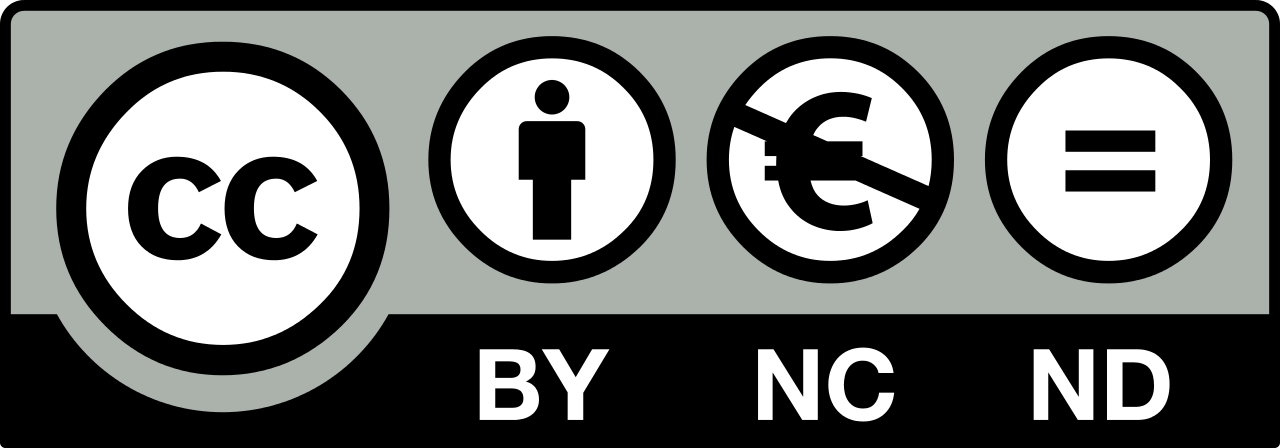}
    \end{center}
\end{wrapfigure}
\noindent
\textbf{Copyright:}\\
Except where otherwise noted,
this work is licensed under\\
\textbf{https://creativecommons.org/licenses/by-nc-nd/4.0/} 
\cleardoublepage

\pagestyle{plain}				
%
%
\pdfbookmark[0]{Acknowledgments}{Acknowledgments}
\addchap*{Acknowledgments}
It is a commonly accepted truth that most people picking up a PhD thesis only read the acknowledgements part. For this reason, while I have your attention, I would like to dedicate this section to emphasizing the importance of proper scheduling in quantum networks and report some interesting results.

While that would be the optimal move, this space is likely better spent acknowledging all the people that helped make this work possible. Network science is the science of distributed problems, and since writing a PhD thesis is very much a distributed problem it is only fair to thank all the nodes that contributed.

I would like to start by thanking my advisors, Dr. Anastasios Giovanidis and Dr. Frédéric Grosshans, for guiding me through my PhD while simultaneously allowing me great freedom to explore the research directions that interested me the most. I'd like to spend a special word for Fred, who took the leading supervision role for most of my thesis, providing support, guidance (including but not limited to science), and an everflowing source of knowledge and ideas. We joked a lot of times about having ideas for fifty more PhD theses: it is a shame that in the end we only got one.

I wish to thank all my colleagues from the QI Team: while experimentalists are forced to show up to the laboratory to advance their project (even when they dislike their colleagues), people that deal with theory and simulations such as myself get the choice to work from home. The fact that I still showed up to the office even in the absence of free coffee must be due to my colleagues. Among them, special thanks go to Santiago, for all the discussions about languages (and cultures, and food, and wine, and anything really), Adriano, fellow Roman and therefore only person in the lab to truly speak my language, Manon, for always asking how I was doing and cheering throughout the thesis writing (Mr Fittipaldi Fittipaldi appreciated that), Verena, for giving me a new enthusiasm about pufferfishes, and Pascal, for the late-night Magic games in the lab. I would also like to thank Yoann, Matteo, Naomi, Kim, Bo, Uta, Mounir, Grégoire, Zoe, Alvaro, Nicolas, Valentina, Laura, Luis, Paul (x2), Dominik, Gizem, Jason, Clément, Tigran, Lucas (x2), Hela, George, and Michael.
I must thank, from the bottom of my heart, Matilde and Alexis for bananas and seagulls (respectively).

I deeply thank Professor Rodney Van Meter and the entirety of the AQUA team from Keio University, who hosted me in Japan during my two research visits. My time in Japan has been a profoundly enriching experience both personally and academically, and working on QuISP was one of the most challenging and rewarding parts of my thesis. From the moment I met the AQUA team, I was warmly welcomed and felt like a member of the team. For that, I wish to also thank Prof. Michal Hajdušek, Prof. Bernard Ousmane Sane, Prof. Shota Nagayama, and my fellow students Naphan, Poramet, Ryosuke, Kentaro, Sara, Monet, Marii, Hyensoo, Kento and Makoto.

I would like to sincerely thank the members of my jury: it is an honor for me to submit my work to such a knowledgeable panel of experts, and I am grateful to them for dedicating a part of their time to my work. I particularly thank Prof. Daniel Oi and Dr. Claudio Cicconetti for their work as referees.

On the personal side, several acknowledgments are in order. As the great Terry Pratchett wrote, ``You haven't really been anywhere until you've got back home'', and all the following people contribute in some sense to my definition of Home.

Firstly, I'd like to thank Ale, my first line of support during these years, always there whether I needed to talk through a stressful time or was looking for someone to celebrate a success with. A good person will offer you support through difficult times, but it takes a special kind of person to make you genuinely believe that you can do it.
This thesis is my greatest achievement so far, and I feel beyond fortunate to get to celebrate it with you.

I also thank Francesco and Giovanni, my housemates during most of my PhD and at this point two of my best friends. Even though I thought of everything, I forgot to thank you two in my thesis. But do you prefer being thanked in a thesis, or knowing that a friend is happy?

I of course thank my friends back in Rome, Ludo, Claudia and Andrea, who have known me since way before I knew what quantum physics was (or how to count to 10, in some cases), and have since then been a constant factor through many changes. Today, we get to see Mr. Plao become Dr. Plao.

Il ringraziamento finale va ai miei genitori. Il lavoro di questa tesi non sarebbe potuto neanche cominciare senza il loro costante supporto e senza la consapevolezza che, comunque fossero andate le cose, sarei sempre potuto tornare a casa. All'inizio del lavoro che riassume tutta la conoscenza che ho accumulato, è giusto dare un riconoscimento speciale alle prime persone da cui ho imparato. 
\cleardoublepage
%
%
\pdfbookmark[0]{Abstract}{Abstract}
\addchap*{Abstract}
\label{chap:abstract}
In this thesis, we analyze the problem of scheduling in the context of quantum networks. Given a quantum network, the scheduling problem amounts to choosing which entanglement swapping operations to perform to better serve user demand. The choice can be carried out following a variety of criteria (e.g.\@ ensuring all users are served equally vs.\@ prioritizing specific critical applications, properly managing load spikes and node failures, adopting heuristic or optimization-based algorithms...), warranting the need for a method to compare different solutions and choose the most appropriate.
We present here a framework to mathematically formulate the scheduling problem over quantum networks and benchmark possible solutions in a variety of environments. Our framework enables the benchmarking of general quantum scheduling policies over arbitrary lossy multicommodity quantum networks.\\
By leveraging the framework, we apply Lyapunov drift minimization (a standard technique in classical network science) to derive a novel class of quadratic optimization based scheduling policies, which we then analyze and compare with a simpler, Max Weight inspired linear class to quantify the performance loss due to the simplification.\\
We start our second chapter with an overview of the pre-existing fiber quantum simulation tools. The rest of the chapter is devoted to the development of numerous extensions to QuISP, an established quantum network simulator focused on scalability and accuracy in modeling the classical communication infrastructure underlying every quantum network. We document the development of our extensions allowing to simulate satellite links and multiple connections in QuISP, with an account of the currently functional extensions (free-space links and connection teardown) and of the ones still under active development (network multiplexing).\\
Since it is likely that a future global-scale quantum network will incorporate satellite interconnections, we devote a chapter to the study of quantum satellite links. We derive an analytical model for the entanglement distribution rates for satellite-to-ground and ground-satellite-ground links and discuss different quantum memory allocation policies for the dual link case. Our findings show that classical communication latency is a major limiting factor for satellite communication, and the effects of physical upper bounds such as the speed of light must be taken into account when designing quantum links, limiting the attainable rates to tens of \unit{\kilo\hertz}. We also investigate the issue of differential latency, a Doppler-like effect caused by the displacement of satellite nodes that changes the timing of incoming photons and adds another upper bound to the generation rate.\\
We conclude the thesis by summarizing our findings and highlighting the challenges that still need to be overcome in order to study the quantum scheduling problem over fiber and satellite large scale quantum networks.\\
\textbf{Keywords:} 
Quantum networks, quantum communication, scheduling, satellites, quantum simulation, QuISP.	
\clearpage

\cleardoublepage
\currentpdfbookmark{\contentsname}{toc}
\setcounter{tocdepth}{2}		
\tableofcontents				
\cleardoublepage

\pagestyle{empty}				
\pagenumbering{arabic}			
\pagestyle{scrheadings}			

%
\chapter{Introduction}
\label{sec:intro}
The promises behind quantum communication at the metropolitan to global scale created an interesting research field that attracts both quantum physicists and network engineers. Due to the field's recent inception, most existing work concerning quantum networks is either at the physical layer (e.g.\@ optical links generating entanglement between two nodes) or at the application layer (e.g.\@ how to exploit a global-scale quantum network to implement a quantum application such as Quantum Key Distribution), whereas a great amount of interesting engineering problems can be formulated and need to be solved in the intermediate layers before quantum networks can truly scale up. In this thesis, we focus on the problem of scheduling.\\
In its simplest form, the scheduling problem boils down to formulating a policy for multiple users to share a common resource. In classical networks, this corresponds to regulating the order in which data packets are sent through a finite-capacity channel. In quantum networks, scheduling means finding the best sequence of entanglement swapping operations to distribute quantum entanglement to multiple user pairs. 
Being such a general problem, there is no unique solution to scheduling, but rather a gamut of policies among which network designers may choose depending on the possibilities, constraints and requirements of a specific realization. The problem of scheduling has been extensively analyzed in classical networking literature \cite[chap.\@ 23, Queuing and Scheduling]{DordalNetworks}, where numerous policies have been developed and their impact on network performance studied in detail. As such, even though the classical solutions are not directly applicable to the quantum case, a key asset in quantum network science and specifically quantum scheduling is an assessment of how much knowledge and methodology can be inherited and/or adapted from the classical domain.

As is widely recognized in classical network science, a well-designed scheduler is a crucial enabler for any realistic network system: a wide array of scheduling policies have been studied and formalized in classical networks \cite{BonaldScheduling} and a few have been explored for quantum ones, but a general language to formalize quantum scheduling policies has not appeared in literature prior to our work.\\
As the first contribution of this thesis (chap.\@ \ref{ch:scheduling}, based on \cite{scheduling_conference} and \cite{scheduling_journal}), we present a framework to formulate the scheduling problem over quantum networks. Through the use of this framework, we demonstrate the relevance of scheduling in quantum network systems and provide some comparisons between several policies in small and medium scale quantum networks in optical fiber. This exemplifies the research work enabled by a scheduling framework and provides a first step towards a fully formalized quantum network scheduling theory. 

After establishing our framework, we give an overview (chap.\@ \ref{ch:simulation}) of the currently available quantum network simulators and a detailed description of QuISP \cite{quisp}, a simulator developed by the AQUA team \cite{aquawebsite} in Keio University (Fujisawa, Japan) which we chose to adopt and extend for the remainder of the thesis. Throughout this chapter, we describe all the software extensions we developed for QuISP, namely the implementation of free-space communication and that of network multiplexing, with the goal of enabling the study of scheduling in fiber and satellite quantum networks.

Taking advantage of said extensions to QuISP, we devote a chapter to investigating the idiosyncrasies of quantum satellite links when compared to a regular optical fiber link (chap.\@ \ref{ch:satellites}, based on \cite{fittipaldisatellites}). Since it is expected that any large-scale quantum network implementation (and ultimately the Quantum Internet) will need to rely on hybrid networks exploiting the strengths of both fiber and free-space optics, it is interesting to study satellite quantum links and get acquainted with the specific problems that they bring to quantum communication. Our findings include bounds on the entanglement distribution rate imposed by physical factors such as the speed of light and the satellite's displacement, and the discussion of various memory allocation policies for an entanglement swapping based satellite link.

In the final chapter (chap.\@ \ref{ch:Conclusion}), we provide a summation of our findings together with an abstract discussion of the scheduling problem in fiber and free-space quantum networks, highlighting the open questions and software development challenges that must be solved before the problem can be approached concretely. These questions are left for future investigation, together with a set of potential new directions for follow-up research that are detailed in the same chapter.
\section{Organization}
The rest of this thesis is organized as follows:
\begin{itemize}
    \item Chapter \ref{chap:IntroQuantumNetworks} provides the required quantum mechanics and network science tools to understand our contributions. Should readers not be interested in the quantum mechanical details of our work, a small shortcut section with no explicit quantum physics is also provided;
    \item Chapter \ref{ch:scheduling} is dedicated to our linear algebraic framework for formulating and solving the scheduling problem in quantum networks. In this chapter, we discuss all the components of our framework and showcase its applicability in analytical discussions and simulation scenarios;
    \item In chapter \ref{ch:simulation}, we provide a concise overview of the current landscape of quantum network simulation and delve into the details of the modifications we made to the \textbf{Qu}antum \textbf{I}nternet \textbf{S}imulation \textbf{P}ackage (QuISP) \cite{quisp,quispsite}, the quantum network simulator on which we based the simulation work for all the remainder of this thesis;
    \item Chapter \ref{ch:satellites} is based on our work in satellite quantum communication: we derive an analytical model for entanglement distribution rates in the satellite-ground and ground-satellite-ground scenarios, and provide simulation results obtained through our extended version of QuISP;
    \item We conclude the thesis with chapter \ref{ch:Conclusion}, where we summarize our results, highlight the open questions before scheduling can be studied in fiber and satellite networks and discuss future directions for our investigation.
\end{itemize}   
\chapter{Introduction to Quantum Networking}
\label{chap:IntroQuantumNetworks}
Small-scale quantum network prototypes have been experimentally realized in the last years \cite{PompiliQN,HermansQNet}. Despite scaling up still being challenging, the endeavor is made interesting by promising applications such as Quantum Key Distribution and Blind Quantum Computing, with more being discovered as the field is explored (more details in sec. \ref{sec:intro_quantumapplications}).\\
Before describing a quantum network in more detail, let us establish some basic quantum concepts with the goal of defining \emph{quantum entanglement}, the resource on which all applications discussed in this thesis will be based. Should the reader not be interested in the quantum physics background motivating this thesis, it is still possible to understand our work: it is advised in this case to only read sec. \ref{sec:intro_noquantum}, where a self-contained description of the required concepts is provided with no quantum physics explicitly involved.
\section{Tracing a Shortcut: A Quantum-Free Introduction}
\label{sec:intro_noquantum}
Despite our work's reliance on quantum physics as a theoretical basis, it is still possible to present the minimal requirements as abstract fundamental concepts, enabling readers with no interest in a quantum physics discussion to understand and appreciate our contributions.
\paragraph{Quantum Entanglement}
In this view, the very first brick on which all the discussion rests is the concept of \emph{quantum entanglement}, which is a physical correlation between two remote parties that cannot be explained through classical physics alone. When two parties are in an entangled state, measurements of certain quantities exhibit correlation even if the two systems carrying entanglement are physically separated. Since a pair of systems can be reliably entangled experimentally, it is interesting to explore the potential applications of such a correlation to information transfer. Being a purely quantum phenomenon, entanglement can only be observed at the quantum scale: examples of quantities that may exhibit quantum entanglement are the degrees of freedom of an electronic or nuclear spin, or the polarization of single flying photons: stationary systems are commonly used for ``storage'' of entanglement, while moving ones can be used to ``transport'' it. Readers need not to understand what exactly entanglement is, only that it is an interesting phenomenon and a universal resource for quantum communication applications (as reviewed in sec.\@ \ref{sec:intro_quantumapplications}): a quantum network's purpose is therefore to enable the distribution of entanglement across arbitrary pairs of nodes.
Each node in a quantum network is equipped with a quantum memory module, comprised of several stationary subsystems called \emph{slots} (such as single atoms, ions or defects in a crystal medium) that can be reliably entangled with incoming and outgoing photons: to establish entanglement between two nodes, we first entangle one of the memory slots with a photon and send it to the receiving node, who will in turn entangle it with one of its local slots and thus create entanglement between the two stationary nodes. Whenever a slot at node $A$ is entangled with a slot at node $B$, we will say the nodes share an entangled pair, or \emph{ebit} (short for entangled bit \cite{wilde}). Depending on the number of available slots, at any given time a node may share multiple ebits with the same or different neighbors. The main problem with quantum entanglement is that it is a fleeting resource, as entanglement between two parties is lost on use: to implement any meaningful application, entanglement between nodes must be continuously created and consumed. Creating entanglement between two parties is a probabilistic process: given two network nodes $A$ and $B$, it is possible to define a rate at which entanglement is established between one of $A$'s slots and one of $B$'s depending on the number of available quantum slots, the physical medium of communication (usually optical fiber or free space) and most importantly the distance between the two nodes. The crucial role of distance as a limiting factor comes from the fact that entanglement distribution requires signals at the single photon level, greatly suffering from attenuation in the physical medium and effectively stunting the distribution rate over long distances.
\paragraph{Entanglement Swapping}
While the attenuation problem cannot be truly solved, there are some ways to work around it: the second fundamental concept we will present is that of \emph{entanglement swapping}, which allows to merge link-level ebits in order to entangle non-neighboring nodes while circumventing the distance limit (see figg.\@ \ref{fig:swapping-example}, \ref{fig:intro_multiswap}). If nodes $A$ and $B$ share an ebit, and node $B$ also shares an ebit with node $C$, then it is possible for $B$ to perform a local operation on its memory slots (called a Bell State Measurement in physical terms) and classically communicate the outcome to either $A$ or $C$. Due to the peculiar nature of measurements in quantum mechanics, this procedure has two effects: it uses up the initial ebits, destroying $B$'s entanglement with both $A$ and $C$, but it also makes it so that the slots at $A$ and $C$, which were previously entangled with $B$, are now entangled with each other, having effectively ``stitched'' the two initial pairs together. This process can be iterated (node $C$ may now further stitch its $A-C$ entanglement with a $C-D$ pair to create $A-D$), enabling distribution of entanglement over long distances, which is the main task of a quantum network as an enabler for quantum applications. It should be stressed that communicating the output of the Bell State Measurement is an integral part of entanglement swapping, which cannot be considered successful until the classical message is received: this means that, while entanglement swapping circumvents distance constraints in terms of attenuation, it is still crucial to consider the latency introduced by classical communication.\\
Other than being a fundamental enabling technology, entanglement swapping also introduces interesting design tradeoffs that are exclusive to quantum networks: entanglement between two given nodes is at the same time a resource for end user applications and for the creation of longer-range entanglement, possibly with multiple competing choices. The quantum scheduling problem, main focus of this thesis, is defined as the task of choosing which ebits should be used for entanglement swapping and which ones should instead be allocated for user consumption.
\section{From Zero to Quantum Internet: Minimal Required Quantum Physics}
Every discussion on quantum communication needs to be firmly grounded in quantum mechanics, which is itself based on four fundamental axioms. In this section, we introduce some useful mathematical notation for quantum mechanics and use it to briefly review its axioms (interested readers will find more details in \cite{NielsenChuang}).
\subsection{The Alphabet of Quantum Physics: The Dirac Notation}
Quantum physics, and especially quantum information, is based on linear algebra for much of its mathematical formalization. A large part of quantum physics can be formalized in terms of vectors and linear maps acting on them, but a trade-specific custom notation called the Dirac Notation \cite{DiracNotation} (or bra-ket notation) is usually employed. Let $\vec{\mathbf{v}}$ be an arbitrary column vector in $\mathbb{C}^N$: in Dirac notation, the symbol $\ket{\mathbf{v}}$ (read ``ket v'') would be used\paolofootnote{Notice that the content of the $\ket{}$ brackets is merely a label, unrelated from the value of the ket: $\ket{p}$, $\ket{\alpha}$, $\ket{7}$ are all valid examples of ket notation.}. Analogously, let $\vec{\mathbf{w}}^T$ be an arbitrary row vector in $\mathbb{C}^N$: it would be notated as $\bra{\mathbf{w}}$ (read ``bra w'') in bra-ket notation.\\
To get a brief glimpse of how the Dirac notation works, picture the space $\mathbb{C}^2$ of 2-vectors of complex components. One of the possible bases of $\mathbb{C}^2$ is $\left\{
\begin{bsmallmatrix}
    1 \\ 0
\end{bsmallmatrix}, 
\begin{bsmallmatrix}
    0 \\ 1
\end{bsmallmatrix}
\right\}$, or $\left\{\ket0,\ket1\right\}$ in bra-ket notation. To notate a scalar product, such as the one between the basis vectors, we write 
$\braket{0|0} = 
\begin{bsmallmatrix}
    1 & 0
\end{bsmallmatrix}
\cdot
\begin{bsmallmatrix}
    1 \\ 0
\end{bsmallmatrix} =
1$, 
$\braket{1|0} = 
\begin{bsmallmatrix}
    0 & 1
\end{bsmallmatrix}
\cdot
\begin{bsmallmatrix}
    1 \\ 0
\end{bsmallmatrix} 
= 0$.
Bras and kets live in a complex Hilbert space\paolofootnote{A Hilbert space is a generalization of Euclidean vector spaces that may have $\mathbb{C}$ as a base field instead of $\mathbb R$ and supports infinite dimensions if needed. While not common in quantum information, general quantum physics routinely works with infinitely-dimensional bras, kets and matrices.} that we call $\mathcal{H}$, and to get from a ket $\ket\psi$ to the corresponding bra $\bra\psi$ one should apply a transposition and take the complex conjugate of all the components. This operation is called the Hermitian Conjugate: $\bra\psi = (\ket\psi^T)^* \eqqcolon \ket\psi^\dagger$.
An interesting feature of bra-ket notation is how natural it is to define linear maps, or operators, that map one ket to another: For instance, the operator $X = \ket{0}\bra{1} + \ket{1}\bra{0}$\paolofootnote{$X = 
\begin{bsmallmatrix}
    1 & 0
\end{bsmallmatrix}
\begin{bsmallmatrix}
    0 \\ 1
\end{bsmallmatrix} 
+
\begin{bsmallmatrix}
    0 & 1
\end{bsmallmatrix}
\begin{bsmallmatrix}
    1 \\ 0
\end{bsmallmatrix}
=
\begin{bsmallmatrix}
    0 & 1 \\ 1 & 0
\end{bsmallmatrix}$.}, when applied to one of the basis states, will flip it:\begin{align*}
    X\ket0 & = \ket{0}\underbrace{\braket{1|0}}_{=0} + \ket{1}\underbrace{\braket{0|0}}_{=1} = \ket{1}\\
    X\ket1 & = \ket{0}\underbrace{\braket{1|1}}_{=1} + \ket{1}\underbrace{\braket{0|1}}_{=0} = \ket{0}.
\end{align*}
Of course, it is also possible to apply $X$ to a generic linear combination of the basis kets $\ket\psi = \alpha\ket0+\beta\ket1 = \begin{bsmallmatrix}
    \alpha \\ \beta
\end{bsmallmatrix}$, and its action will be:
\begin{align*}
    X\ket\psi = \alpha(\ket{0}\underbrace{\braket{1|0}}_{=0} + \ket{1}\underbrace{\braket{0|0}}_{=1}) + \beta(\ket{0}\underbrace{\braket{1|1}}_{=1} + \ket{1}\underbrace{\braket{0|1}}_{=0}) = \alpha\ket1+\beta\ket0 = \begin{bsmallmatrix}
        \beta\\\alpha
    \end{bsmallmatrix}.
\end{align*}
\subsection{The Axioms of Quantum Physics}
Having established a mathematical notation, it is now time to review the basic axioms on which quantum mechanics is based. A small explanation of the key takeaways to understand this thesis follows each axiom.
\axiombox{\paragraph{Axiom I: State Space} The state of a physical system is completely described by a unit vector $\ket\psi$ inside a complex Hilbert space $\mathcal{H}$ of suitable dimension. the vector $\ket{\psi}$ is called the \emph{state vector} and the Hilbert space $\mathcal{H}$ the \emph{state space}.}
This axiom sets the stage for quantum mechanics: whenever a system is examined, its state will be encoded in the complex state vector $\ket{\psi}$ which lives in the Hilbert space $\mathcal{H}$. In order to understand this thesis, there is no need to completely grasp the implications of working in an infinitely dimensional Hilbert space: from here on, $\mathcal{H}$ may be thought of as $\mathbb{C}^{2n}$. Notice that this postulate requires $\ket\psi$ to be a \emph{unit} vector: to account for this, every state vector must always be properly normalized. 
\axiombox{\paragraph{Axiom II: Time Evolution} The time evolution of an \textit{isolated}\paolofootnote{i.e.\@ not interacting with the outside world.} quantum system between $t_1$ and $t_2$ is described by a unitary operator: \begin{align*}
    \ket{\psi_{t_2}} = U\ket{\psi_{t_1}},
\end{align*}
} 
This axiom describes the deterministic evolution of our state vector, essentially laying down the kinematics of quantum mechanics. There is in principle no need to pose any limitation on the shape of $U$: it is a linear operator, i.e.\@ a linear map on the state space, and it is unitary, which means it preserves the norm of $\ket\psi$. It is possible to state the second postulate in terms of Schrödinger's Equation, one of the most important pillars of quantum mechanics:
\begin{align}
    i\hbar\frac{d}{dt}\ket\psi = H\ket\psi,
\end{align}
where $\hbar$ is the reduced Planck Constant and $H$ is an operator called the Hamiltonian\paolofootnote{The Hamiltonian operator of a physical system contains a full description of its dynamics, and its eigenvalues represent the energy associated to each of the system's eigenstates. To give a basic example, the Hamiltonian of an electron in a Coulomb potential $V$ is $\frac{\hat{p}^2}{2m_e} + V(\mathbf{r})$, where $\hat p$ is the momentum operator and $m_e$ the electron's mass. By solving the Schrödinger equation with this Hamiltonian, it is possible to derive the electronic orbitals of an Hydrogen atom and their corresponding energy levels}. Despite its crucial role in general quantum mechanics, the Schrödinger equation is rarely treated in quantum information --- which deals directly with $U$ --- and it has been reported here more for general interest than for direct relevance to the matters discussed in the following.
\axiombox{\paragraph{Axiom III: Measurement}
In quantum mechanics, a measurement is represented by a set of operators $\{M_a,M_b,\ldots\}$, one for each of the possible outcomes of the measurements. Since there needs to be an $M_i$ for each possible measurement outcome, the measurement operators need to satisfy a completeness relation:\begin{align*}
    \sum_i M^\dagger_i M_i = I,
\end{align*}
where $I$ is the identity matrix\paolofootnote{or identity operator, in this context.}. Whenever a measurement is performed on state $\ket\psi$, outcome $i$ is obtained with probability \begin{align*}
    p_i = \bra\psi M_i^\dagger M_i \ket\psi.
\end{align*}
The measurement operation will leave the system in state \begin{align*}
    \frac{M_i\ket\psi}{\sqrt{p_i}},
\end{align*}
according to the obtained outcome.} 
As eloquently put in \cite[Sec. 12.5]{Griffiths}, ``\textit{absent measurement, the wave function\paolofootnote{What is called ``wave function'' in this quote corresponds to our state vector $\ket\psi$.} evolves in a leisurely and deterministic way, according to the Schrödinger equation, and quantum mechanics looks like a rather ordinary field theory}''. What is meant by this quote is that, since measuring a system implies interacting with it from the outside, the linear and deterministic evolution postulated by axiom II is not respected anymore and the system's state evolves according to other rules.\\
The act of measurement in quantum mechanics has consequences that could by themselves support an entire thesis. For our purposes, it is enough to know that measurement in quantum mechanics has an \textit{active} role, in that it irreversibly changes the measured system, and that its outcome is probabilistic: the only predictions that can be made are statistical.\\
To better grasp the mechanism of quantum measurement, let us look at a simple example, namely measuring whether an arbitrary ket $\ket\psi = \alpha\ket0 + \beta\ket1$ is $\ket{0}$ or $\ket{1}$. The operators corresponding to our outcomes are $M_0 = \ket0\bra0$ and $M_1 = \ket1\bra1$. These operators satisfy the completeness relation, and the probabilities to obtain either outcome are
:
\begin{align}
p_0 & = \braket{\psi|0}\braket{0|0}\braket{0|\psi} = \underbrace{\braket{\psi|0}}_{\alpha^*}\underbrace{\braket{0|\psi}}_\alpha = |\alpha|^2 \\
p_1 & = \braket{\psi|1}\braket{1|1}\braket{1|\psi} = \underbrace{\braket{\psi|1}}_{\beta^*}\underbrace{\braket{1|\psi}}_\beta = |\beta|^2.
\end{align}
These expressions also show in practice the physical role of the coefficients in $\psi$ as probability amplitudes, i.e.\@ values that represent probability of measurement when taken in square modulus. This is the physical reason why axiom I requires that state vectors always be normalized, otherwise the probabilities would not sum to 1.\\
Once the measurement is carried out, the system will be either in state $\ket0\braket{0|\psi} = \sqrt{p_0}\ket{0}$ (to be renormalized to $\ket 0$) if the measurement outcome was $0$,  or $\ket1\braket{1|\psi} = \sqrt{p_1}\ket{1}$ (to be renormalized to $\ket 1$) if the outcome was $1$. It should be kept in mind that, as expected, repeated measurements carried out in immediate succession will always yield the same result. This is further evidence that the measurement process in quantum mechanics actively and irreversibly changes the measured state, erasing information about superposition.\\
From this example, one might be led to think that there is something intrinsically special about $\ket0$ and $\ket1$: they are the two outcome states for our measurement, and all states of the type $\alpha\ket0 + \beta\ket1$ collapse into either $\ket0$ or $\ket1$ when measured. This is however not the case, as $\ket0$ and $\ket1$ only play a special role by virtue of their status as basis vectors: all our discussion can be formulated in terms other bases of $\mathcal{H}$, such as $\frac{1}{\sqrt 2}\{\ket0 + \ket 1,\ket 0 - \ket1\}\coloneqq\{\ket +,\ket -\}$
\paolofootnote{$\ket + = \frac{1}{\sqrt 2} 
\begin{bsmallmatrix}
    1\\1
\end{bsmallmatrix}$,
$\ket - = \frac{1}{\sqrt 2} \begin{bsmallmatrix}
    1\\-1
\end{bsmallmatrix}$.}. It is possible to define a measurement $\{M_+,M_-\}$ in this new basis and apply it to a quantum state, and all the discussion presented in the previous paragraphs still applies. The correct way to interpret this information is that measurements, and quantum mechanics in general, are carried out in a specific basis, and bases can be interchanged to make problems more treatable. Once a basis has been chosen, a quantum system can exist in an arbitrary linear combination of the basis vectors called a \emph{superposition state}: a measurement in a given basis will yield an outcome and collapse our system into one of the basis states, but no information can be retrieved about the state prior to the measurement operation. Moreover, there is no notion of value of the measured quantity before the measurement: when a system is in a superposition state, the quantity to measure has no definite value until an outcome is randomly selected by the measurement operation. This property is one of the crucial peculiarities of quantum mechanics and its consequences will be shown to be of paramount importance to communication applications.
\axiombox{\paragraph{Axiom IV: State of Composite Systems} The state space of a composite quantum system is the tensor product of the individual state spaces $\mathcal{H}_{comp} = \mathcal{H}_1\otimes\mathcal{H}_2$. Moreover, if individual quantum systems are prepared in states $\ket{\psi_1},\ket{\psi_2},\ket{\psi_3}\ldots$ the overall quantum state vector of the composite system will be the tensor product of individual states $\ket{\psi_{comp}} = \ket{\psi_1}\otimes\ket{\psi_2}\otimes\ket{\psi_3}\ldots$.}
This final axiom describes how to deal with composite quantum systems, with some interesting consequences. Suppose to have three kets, independently prepared in state $\ket{0}$: we notate such a state as $\ket{\Psi} = \ket0 \otimes \ket0 \otimes \ket0 \coloneqq \ket{000}$. This state lives in the tensor product space $\mathcal{H}^{\otimes3} = \mathbb{C}^2 \otimes \mathbb{C}^2\otimes\mathbb{C}^2$. Analogously, the operators that deal with such a state are defined to be linear maps from $\mathcal{H}^{\otimes3}$ to itself. If for instance one wishes to apply the operator $X$ to the second system in $\ket\Psi$, it is possible to do so by applying the operator $I\otimes X\otimes I$. This yields $(I\otimes X\otimes I)\ket{000} = I\ket0 \otimes X\ket0 \otimes I\ket0 = \ket0 \otimes \ket1 \otimes \ket0 = \ket{010}$. It should be remarked at this point that not all composite system states can be decomposed to the tensor product of individual subsystems: the implications of this property are discussed in sec.\@ \ref{sec:intro_quantumentanglement}.
\subsection{The Quantum Bit}
The quantum bit (also known as \emph{qubit} \cite{qubits}) is the fundamental unit of quantum information. In analogy with classical computing, where the bit may be implemented with any system with two states, a qubit may be realized with any system that features two distinct quantum states\paolofootnote{a simple example of qubit is the electronic spin with its up/down states, represented as $\ket{\uparrow}$/$\ket{\downarrow}$ in the Dirac notation.}. To remain agnostic of the underlying implementation and to reinforce the analogy with classical bits, the states of a qubit are usually mapped to the aforementioned $\ket 0$ and $\ket 1$ in $\mathcal{H}\cong\mathbb{C}^2$.
Unlike the classical bit, whose state may only ever be $0$ or $1$, the state of a quantum system at an arbitrary time may correspond to one of the basis states $\ket 0$, $\ket 1$\paolofootnote{Known as the \emph{computational basis} states \cite{NielsenChuang}} or a quantum superposition $\alpha\ket 0+\beta\ket 1$. Upon the act of measurement, such a superposition state will yield either $0$ or $1$ with a given probability\paolofootnote{Just as measuring the spin of an electron may yield up or down, with varying probabilities depending on the applied magnetic field, even though the spin itself is a quantum variable and generally in a superposition state.}.
There are several physical candidates as platforms for qubit implementation (including but not limited to trapped ions \cite{trappedionreview}, superconducting circuits \cite{SCqubitsreview}, nitrogen-vacancy centers in diamond \cite{NVqubitsreview}, single photon polarization \cite{SlussarenkoPhotonicReview}\ldots), a complete review of which is outside the scope of this thesis: for our purposes, it is sufficient to understand how a qubit works in the abstract sense and how it interacts with other qubits.
\subsection{Multiple Qubits and Quantum Entanglement}
\label{sec:intro_quantumentanglement}
A pair of classical bits can take the states $00$, $01$, $10$ and $11$. By applying axiom IV, the state of a pair of qubits may be described as a linear combination of four basis states\paolofootnote{Since qubits are represented as vectors $\ket q$ in a complex $2$-dimensional Hilbert space $\mathcal{H}$, by axiom IV the state of a two-qubit system must exist in the $4$-dimensional Hilbert space $\mathcal{H} \otimes \mathcal{H}$ and its four eigenstates must be $\ket {0} \otimes \ket {0}$, $\ket {0} \otimes \ket {1}$, $\ket {1} \otimes \ket {0}$, $\ket {1} \otimes \ket {1}$, which are commonly notated as $\ket{00}, \ket{01}, \ket{10}$ or $\ket{11}$.} $\ket \Psi = \alpha\ket{00} + \beta\ket{01} + \gamma\ket{10} + \delta\ket{11}$. If one measures both qubits, there is a probability corresponding to the square modulus of the corresponding coefficient to obtain $00$, $01$, $10$ or $11$. However, it is also possible to measure only one qubit and leave the other unperturbed: to visualize this, let us start from a pair of qubits in an even superposition state $\ket {\Psi} = \frac{1}{2}(\ket{00} + \ket{01} + \ket{10} + \ket{11})$. If we only want to measure the first qubit in the computational basis, our measurement operators are $M_0 = \ket 0\bra 0\otimes I$ and $M_1 = \ket 1\bra 1\otimes I$. Running the measurement will yield $0$ (or $1$) with probability $(\sfrac{1}{2})^2 + (\sfrac{1}{2})^2 = \sfrac{1}{2}$ and project the system in the state $\ket {\Psi_0} = \frac{1}{\sqrt 2}\ket0(\ket 0 + \ket 1)$ (or $\ket {\Psi_1} = \frac{1}{\sqrt 2}\ket1(\ket 0 + \ket 1)$), where the coefficients have been renormalized.

Let us now apply the same reasoning to an interesting set of two-qubit states:
\begin{align*}
\ket{\Phi^+} = \frac{1}{\sqrt 2}(\ket{00} + \ket{11})\quad & \quad
\ket{\Psi^+} = \frac{1}{\sqrt 2}(\ket{01} + \ket{10}) \\
\ket{\Phi^-} = \frac{1}{\sqrt 2}(\ket{00} - \ket{11})\quad & \quad
\ket{\Psi^-} = \frac{1}{\sqrt 2}(\ket{01} - \ket{10}).
\end{align*}
These states constitute another basis for the space of two-qubit states and they are known as the Bell states \cite[Section 1.3.6]{NielsenChuang}. These states exhibit a peculiar property: they cannot be decomposed to tensor products of single-qubit computational basis states. Physically, this implies that when measuring one qubit in the computational basis, the state of both qubits is collapsed. To observe this, let us repeat our single-qubit measurement experiment. Suppose a pair of qubits is in $\ket{\Phi^+}$: measuring the first qubit alone has once again equal probability to yield $0$ or $1$. However, applying $\ket{0}\bra{0}\otimes I$ (or $\ket{1}\bra{1}\otimes I$) to $\ket{\Phi^+}$ collapses it to $\ket{00}$ (or $\ket{11}$), effectively collapsing the second qubit as well. It can be readily verified that this is the case for all the Bell states\paolofootnote{Mathematically speaking, running this experiment on $\ket{\Psi^-}$ or $\ket{\Phi^-}$ collapses the second qubit in $-\ket1$ or $-\ket0$. However, quantum physical states differing by a global phase shift are indistinguishable, and are therefore considered the same state: $-\ket{\mathbf{u}} = \ket{\mathbf{u}}$.}. The collapse happens instantly and independently of the distance between the two qubits: measuring the second qubit immediately after the collapse of the first will deterministically output the same value. This correlation between the two qubits is called \emph{quantum entanglement} and the qubits are said to have been an \emph{entangled pair} (performing the measurement irreversibly breaks entanglement). Such an intrinsic correlation is exclusive to quantum mechanics, sparking interest for a variety of technological applications.

At first, it may seem like entanglement enables information transfer at superluminal speed, since acting on a local qubit instantly affects a remote one: this is however not the case, because the result of the local measurement is not predetermined nor controllable in any way, meaning that a parallel classical information channel is needed to implement any kind of information transfer through entanglement. To better visualize the physical intuition behind quantum entanglement, one should not think of a measurement on one system affecting another, but rather view the two qubits as a single \emph{non-local} physical system that is affected as a whole by the measurement operation. Despite not enabling superluminal communication, quantum entanglement is still the fundamental enabling phenomenon sustaining numerous quantum applications, a brief review of which will be given in sec. \ref{sec:intro_quantumapplications}.

As a final note, we remark that entanglement is not necessarily only bipartite: entangled states of three or more qubits (such as the GHZ state $\frac{1}{\sqrt 2}(\ket{000} + \ket{111})$ and W state $\frac{1}{\sqrt 3}(\ket{100} + \ket{010} + \ket{001})$ for three parties \cite{GHZandW} or more) exist and are commonly studied in quantum information, but they are outside the scope of this document.
\subsection{Entanglement Swapping}
\label{sec:intro_swapping}
\begin{figure}
    \centering
    \subfloat[Before Swapping;]{
\includegraphics[width=.5\textwidth]{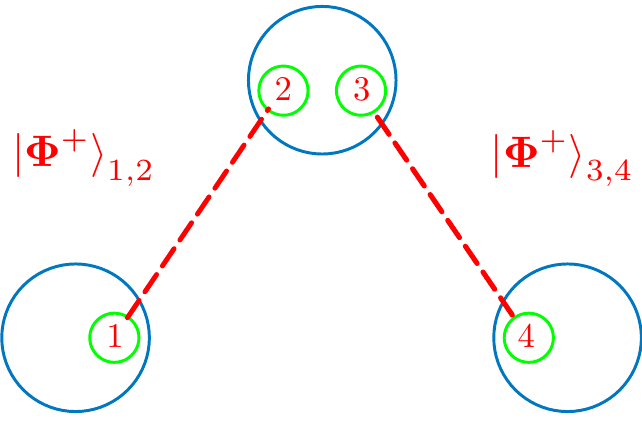}
\label{fig:intro_swapping_example_before}
}
\subfloat[After Swapping;]{
\includegraphics[width=.5\textwidth]{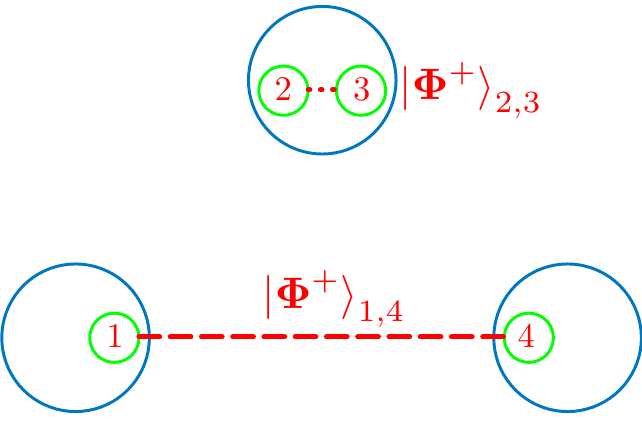}
\label{fig:intro_swapping_example_after}
}
\caption[Graphical example of entanglement swapping.]{Example of entanglement swapping: a local measurement at the middle node splices states $\ket{\Phi^+}_{12}$ and $\ket{\Phi^+}_{34}$ into $\ket{\Phi^+}_{14}$. $\ket{\Phi^+}_{23}$ is created as a side effect and is not relevant to the procedure.}
\label{fig:swapping-example}
\end{figure}

As introduced before, most quantum applications hinge on the possibility of sharing entangled states over long distances. However, this requires sending single photons across long segments of optical fiber. Both the probability of reception and the quality of the received quantum state decrease exponentially with the distance across which the photon must travel, making direct transmission across fiber links an unsuitable solution as distances increase. 
In classical networks, the problem is alleviated by amplification and/or retransmission of the signal at intermediate stations, which is impossible in quantum systems because both operations require creating a copy of the transmitted photons: one of the elementary results of quantum physics is the No-Cloning Theorem \cite{WoottersNoCloning}, which states that it is impossible to copy the quantum state of a system without perturbing it, because preparing the copy requires knowledge of the state in the first place\paolofootnote{A measurement of a state yields either $0$ or $1$, with no information about the coefficients of $\ket0$ and $\ket1$, making it impossible to characterize a quantum state with a single measurement. Characterization of quantum states is usually done by building statistics over many different measurements of an ensemble of qubits in the same quantum state, a process called \emph{quantum tomography} \cite{AltepeterTomo}. We remark that it is indeed possible to prepare an ensemble of qubits in an arbitrary quantum state $\ket\psi$: what is ruled out by quantum physics is the existence of a black box that takes one copy of an unknown state and yields two or more copies of the same state. As a very specific case, it is technically possible to copy the quantum state of a system that is restricted to orthogonal states. However, a qubit restricted to orthogonal states would amount to a system that can only ever be in one of two states with no superposition, i.e.\@ functionally equivalent to a classical bit.}: an alternative solution must be found to distribute entanglement over long distances.\\
One of the peculiar features of quantum entanglement is that it can be \emph{swapped} \cite{SwappingZukowski}: if a node $B$ shares two entangled pairs, one with partner $A$ and one with partner $C$, it is possible to perform a joint measurement on $B$'s qubits with the goal of projecting them in one of the four Bell states. This operation, called a Bell State Measurement (BSM), has two simultaneous effects: it erases the $A-B$ and $B-C$ entanglement, but most importantly it projects the qubits at $A$ and $C$, who have never interacted before, into an entangled state.\\
To visualize this phenomenon, let us examine a simplified example: node $B$ has two qubits, each entangled to a qubit at nodes $A$ and $C$ respectively. Both entangled pairs are in state $\ket{\Phi^+}$, as shown in fig. \ref{fig:intro_swapping_example_before}.
The overall state of our system is $\ket{S} = \ket{\Phi^+}\otimes\ket{\Phi^+} = \frac{1}{2}(\ket{0000} + \ket{0011} + \ket{1100} + \ket{1111})$ (Notice how the first two qubits and the second two qubits are entangled).
Let us now measure the two middle qubits in the Bell basis, with measurement operators $M_{\Phi+} = I\otimes\ket{\Phi^+}\bra{\Phi^+}\otimes I$, $M_{\Psi+} = I\otimes\ket{\Psi^+}\bra{\Psi^+}\otimes I$, $M_{\Phi-} = I\otimes\ket{\Phi^-}\bra{\Phi^-}\otimes I$, and $M_{\Psi-} = I\otimes\ket{\Psi^-}\bra{\Psi^-}\otimes I$.\\
The possible outcomes to our measurement are, with probability $\sfrac{1}{4}$ each:
\begin{itemize}
    \item $M_{\Phi+}\ket{S} = \frac{1}{2}(\ket{0000} + \ket{0110} + \ket{1001} + \ket{1111})$;
    \item $M_{\Phi-}\ket{S} = \frac{1}{2}(\ket{0000} - \ket{0110} - \ket{1001} + \ket{1111})$;
    \item $M_{\Psi+}\ket{S} = \frac{1}{2}(\ket{0011} + \ket{0101} + \ket{1010} + \ket{1100})$;
    \item $M_{\Psi-}\ket{S} = \frac{1}{2}(\ket{0011} - \ket{0101} - \ket{1010} + \ket{1100})$. 
\end{itemize}
In all these cases, the two initial $AB$, $BC$ pairs are disentangled and pairwise entanglement is created between the two outermost qubits ($AC$ entanglement) and the two qubits at $B$, as shown in fig. \ref{fig:intro_swapping_example_after} for the $\ket{\Phi^+}$ case.\\
Even though the two outermost qubits are now entangled, the nodes that hold them have no knowledge about which of the four Bell states they share: measuring $A$'s qubit as $0$ or $1$ yields no information about $C$'s qubit unless the BSM's outcome is known. This imposes a significant constraint on entanglement swapping: for the procedure to be meaningful and useful, the BSM's outcome must be classically communicated to the nodes. Once the nodes are made aware of the outcome of the BSM, they can apply local correction operations to their qubits\paolofootnote{Manipulating a state does not disentangle it: one of the two qubits of an entangled state can undergo transformations such as rotations and bit-flips without affecting its entangled state, as long as no measurement is carried out.} and recover the $\ket{\Phi^+}$ (or, equivalently, one of the other three) state $100\%$ of the times.\\
In summary, given two entangled $AB$ and $B'C$, it is always possible to jointly measure $B$ and $B'$ in a way that creates $AC$ entanglement, provided a classical channel is present to communicate the outcome of the Bell measurement. This might at first sound like a simple experimental nonideality, but it is actually a crucial physical constraint: swapped entanglement is unusable until the correction message has been received, meaning that the latency introduced by classical communication (very high in satellite links) will place upper bounds on the quantum communication rates. This problem is explored in detail in chap. \ref{ch:satellites}.
\section{The Building Blocks of Quantum Networking}
\subsection{Quantum Repeaters}
\label{sec:IntroRepeaters}
Swapping is the key enabling phenomenon behind the idea of \emph{quantum repeaters} \cite{BriegelRepeaters}, which in their simplest iteration are quantum processing nodes capable of performing BSMs. Instead of distributing entanglement over a long fiber link, repeaters are placed as intermediate stations to divide the link in several sub-links, across which entanglement distribution has a much higher probability of success. Despite at first seeming beneficial for increasing entanglement rates, simple swapping stations do not change the exponential scaling of the success probability, as all sub-links are required to successfully distribute entanglement in a synchronous way. To meaningfully affect the rate of entanglement distribution, individual sub-links need to be able to ``lock'' successful entanglement attempts while they wait for the others to be established: once all sub-links are entangled, swapping operations are performed. 

``Locking'' successfully entangled qubits can be achieved through \emph{quantum memory}, which analogously to classical memory is able to store quantum information for some amount of time. One peculiarity of quantum memory is that it preserves the entanglement of the stored qubits, making it a cornerstone in the realization of quantum repeaters. To be suitable for quantum repeater applications, memories must have storage times at least comparable with the expected time it takes to establish entanglement across a sub-link, they must not degrade the stored qubit's quality too much and, depending on the specific repeater implementation, they might need to allow on-demand retrieval of the stored quantum information. In sec.\@ \ref{sec:IntroQM}, we give a brief review of the state of the art in quantum memory technology and relevant figures of merit.

In the presence of quantum memory, the scaling on distance of the probability of entanglement distribution improves by a square-root factor when a single repeater is added. In general, the distribution rate improves exponentially with the number of quantum repeaters \cite{AzumaRepeaters} if imperfections such as memory errors and non-unitary probability of success for BSMs are neglected, leading eventually to a polynomial scaling instead of exponential.
\subsubsection{Generations of Quantum Repeaters}
Depending on entanglement generation and error suppression mechanisms, quantum repeaters may be divided into three consecutive generations \cite{1g2g3g}:
\begin{itemize}
    \item \textbf{1G Quantum Repeaters} use heralded entanglement generation (HEG) \cite{LagoRiveraLink} to minimize loss errors (the generation probability is conditioned on the heralding event) and perform standard heralded entanglement purification \cite{BennettPurification} --- a technique that consists in consuming multiple lower-quality entangled pairs to probabilistically obtain a higher-quality one --- after each individual swapping operation. They are the repeaters with the lowest technological requirements, making them interesting for short-term quantum networks, but due to the need for two-way signaling in both the generation and purification phases they can attain at most polynomial scaling on the entanglement generation rate. The results presented in this thesis are mostly geared toward 1G repeaters, with possibility of extension to 2G.
    \item \textbf{2G Quantum Repeaters} still adopt HEG to establish entanglement, but they aggregate the generated physical qubits into logical blocks on which to perform full-fledged quantum error correction \cite{RoffeQEC} (QEC) to correct operation errors. The generation rates are much higher than with 1G repeaters because QEC only requires one-way classical signaling between adjacent stations (as opposed to HEP at each round, that requires two-way signaling between nonadjacent stations) but the hardware requirements also increase because QEC needs a block of physical qubits for each logical qubit (upping memory requirements) and fault-tolerant quantum gates.
    \item \textbf{3G Quantum Repeaters} are radically different from other generations, in that they are much closer to classical networks: information is directly encoded in a block of qubits and relayed like a classical package from a source repeater to a destination repeater. The repeaters exploit QEC to correct both loss and operation errors, achieving high communication rates and overcoming both loss and operation errors with theoretically moderate block size ($\sim 200$ physical qubits per logical qubit). Since 3G quantum repeaters correct errors along the way, they eliminate two-way signaling, a major source of latency for 1G and 2G, and are compatible with a more classical-like network model. However, their technological requirements are as high as 2G and they need closer spacing because QEC can only correct a limited amount of losses. 3G repeaters will therefore not be discussed in this thesis.
\end{itemize}

\subsection{Quantum Memory}
\label{sec:IntroQM}
Quantum memory represents a fundamental ingredient of quantum communication because it either solves or greatly alleviates several implementation issues. To start approaching the topic of quantum memory, one could say that any physical system capable of taking a photon as input and giving it back at a later time while preserving its quantum state may be called a quantum memory. By this definition, both a fiber delay line and a reflecting cavity can be taken as rudimentary examples of quantum memory. In this section, we review some examples of quantum memory and provide a breakdown of the main figures of merit in order to justify the approximations that will be taken in the next chapters.

The quantum memories that are used in experimental realizations are usually more complex than the simple examples presented above, and they work by reversibly mapping the state of the input photonic qubit to that of a stationary qubit. The key figures of merit to examine when comparing memory implementations are:
\begin{itemize}
    \item The Fidelity \cite[Section 9.2.2]{NielsenChuang} $F$ of the retrieved state to the input state, which is a measure of how ``close'' a quantum state is to another and is mostly affected by the quantum decoherence dynamics inside the medium;
    \item The Storage-and-Retrieval efficiency $\eta(\Delta t)$, which is the probability to correctly retrieve a photon from the memory after a time interval $\Delta t$ has elapsed;
    \item The Lifetime $\tau$, i.e.\@ the expected lifetime of a stored qubit before it is lost to decoherence; 
    \item The multiplexing capabilities of the memory, i.e.\@ how many independent spatial, temporal or spectral modes of light it can store. In our work, this figure corresponds to the number of qubits a memory can simultaneously store;
    \item The ability (or lack thereof) to read the memory on demand: some implementations of quantum memory simply store the quantum state of a photon and automatically return it after a fixed time, while other schemes allow for on-demand readout triggered by an external event.
\end{itemize}

\begin{figure}
    \centering
   
    \includegraphics[width=\linewidth]{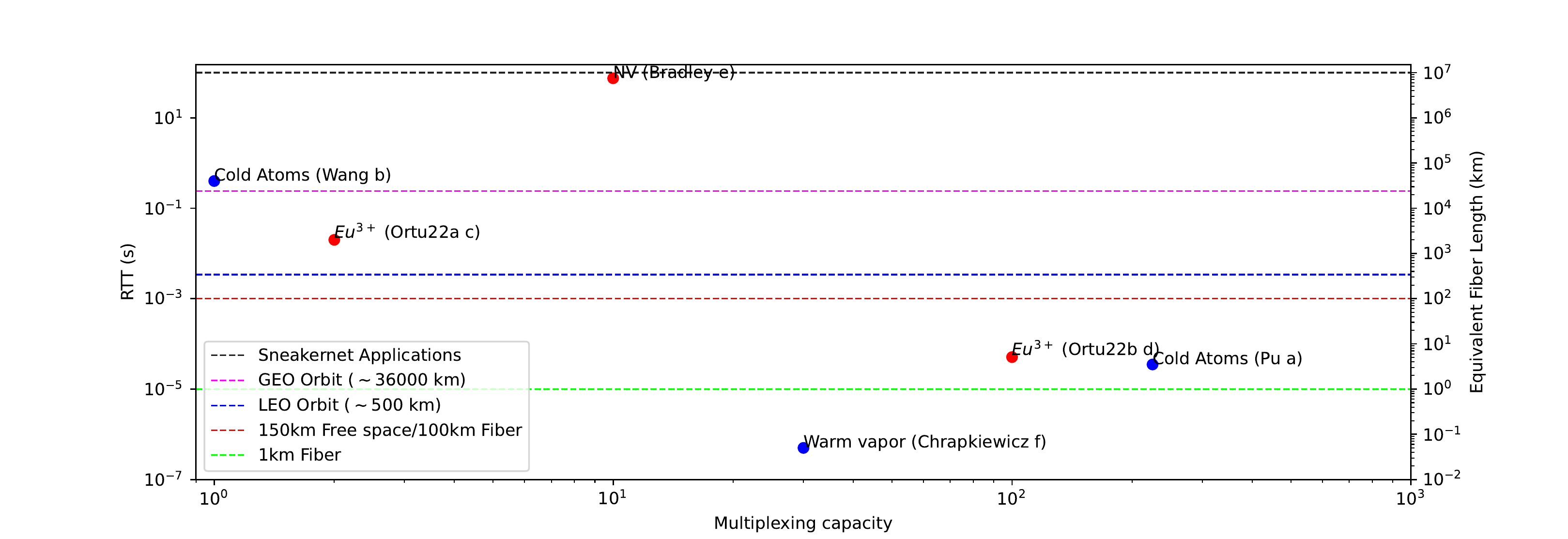}
    \caption[Comparison plot of different quantum memory implementations in terms of their multiplexing capabilities and storage lifetime.]{Comparison of different quantum memory implementations in terms of their multiplexing capabilities and storage lifetime. In this picture, the tradeoff between multimodality and quality storage is made apparent. Horizontal lines represent one round-trip time across the corresponding length of fiber and/or free-space. Red points represent solid-state memories, while blue ones correspond to gaseous systems. References cited in the picture: \begin{enumerate*}[label=(\alph*)]
            \item \cite{PuMemory}
            \item \cite{WangMemory}
            \item \cite{OrtuMemoryA}
            \item \cite{OrtuMemoryB}
            \item \cite{BradleyMemory}
            \item \cite{ChrapkiewiczMemory}
        \end{enumerate*}}
     \label{fig:MemoryRecap}
\end{figure}

\paragraph{Gaseous Atomic Ensembles}
A first example of quantum memory platform is that of atomic ensembles, which are clouds of gases with specific quantum states arrangements that allow mapping the state of a photon to a long-lived collective excitation of the gaseous ensemble. Historically, the first realizations of an ensemble memory were demonstrated with room-temperature ensembles (commonly called warm vapor memories \cite{Hosseini2011WarmVapor}) obtaining results that are acceptable for quantum applications while retaining the relatively low experimental complexity of a room-temperature system. Ensemble memories are usually operated through Raman interactions, an example of which is Electromagnetically Induced Transparency (EIT) \cite{HarrisEIT}, a technique in which the photon pulse is adiabatically slowed down and trapped inside the atom ensemble by a control laser that interacts with its propagation velocity.
Since these memories work by mapping the photon to a collective cloud excitation, thermal noise poses severe limitations on the accuracy of the retrieved state. To circumvent this problem, laser-cooled cold-atom ensembles were engineered and demonstrated to attain much higher performance \cite{Cao2020EIT}, at the cost of more complex experimental arrangements.\\
Ensemble memories are usually high-quality memories in terms of SaR efficiency and fidelity. Recent efforts show that they can support hundreds of modes \cite{PuMemory}, but high lifetimes often require cavity-enhanced setups \cite{WangMemory}, which directly compete with multiplexing and bandwidth. 
\paragraph{Solid State Memories}
Quantum memories can also be realized on solid state systems. Solid state quantum memories are usually realized either with arrays of stationary qubits (such as mapping photonic states to spin excitations in nitrogen-vacancy centers) or by engineering the absorption profile of rare-earth doped crystals through techniques such as the Atomic Frequency Comb (AFC) \cite{AfzeliusAFC}. Solid state memories often have good multiplexing capabilities and varying fidelities, and their lifetime is high although often fixed, since it is usually difficult to implement on-demand readout\paolofootnote{feasible by writing the photon on an AFC, then mapping the excitation to a spin state. At readout, the spin state is re-mapped to an excitation of the AFC which is then naturally read out when it dephases sufficiently}. Another key advantage of solid state memories is their high integrability, crucial when realizing on-chip solutions.\\
As demonstrated in \cite{BradleyMemory}, memories realized through single defects in solids (one NV center and nine nuclear spins in diamond in this case) deserve a special mention as the only memory implementation that provides high lifetime with satisfactory multiplexing and control. As shown in our recapitulation plot (fig. \ref{fig:MemoryRecap}), NV memories are the most promising for satellite quantum communication scenarios with their $
\unit{\milli\second}$ lifetime. Furthermore, they are close to suitability for sneakernet\paolofootnote{In network science, the term \emph{sneakernet} is a humorous term that refers to data transfer by physically carrying a medium such as a USB stick from one machine to another. Sneakernet ``protocols'' have high latency, but they are sometimes convenient due to their high bandwidth allowing transfer of hundreds of TBs of data in seconds. In quantum communication, sneakernet techniques correspond to distributing entangled pairs and then physically separating the entangled parties. In \cite{QNetShip}, such a quantum scheme is proposed on the transatlantic scale with a ship.}-like scenarios where the satellite establishes entanglement with one ground station and carries the entangled qubit until a second ground station becomes visible.

Despite quantum memory being an extremely active research field, no platform and/or protocol has yet been found that offers high fidelity, efficiency and lifetime with good multiplexing capabilities and on-demand readout. However, the discussed parameters pose as a useful interface to develop higher layers of the network stack and investigate the impact of various memory parameters in an hardware-agnostic way.
We conclude this section with tab.\@ \ref{tab:intro_memorysummary}, where we report some relevant parameters for different implementations of quantum memory and fig.\@ \ref{fig:MemoryRecap}, a graphical comparison of the lifetimes and multiplexing capabilities of different quantum memory platforms. More details about the current state of the art in quantum memory technology can be found in \cite{LeiMemoryReview}.
\begin{table}[]
     \caption[Summary of quantum memory figures of merit.]{Summary of quantum memory figures of merit to compare the different platforms and highlight their tradeoffs. References cited in the table: \begin{enumerate*}[label=(\alph*)]
             \item \cite{PuMemory}
             \item \cite{WangMemory}
             \item \cite{OrtuMemoryA}
             \item \cite{OrtuMemoryB}
             \item \cite{BradleyMemory}
             \item \cite{ChrapkiewiczMemory}
             \item \cite{HosseiniMemory}
        \end{enumerate*}}
    \centering
    \begin{tabularx}{1.05\linewidth}{|>{\centering\arraybackslash}p{.2\linewidth}|>{\centering\arraybackslash}p{.1\linewidth}|>{\centering\arraybackslash}X|>{\centering\arraybackslash}p{.16\linewidth}|>{\centering\arraybackslash}p{.1\linewidth}|>{\centering\arraybackslash}p{.17\linewidth}|}
    \hline
         \textbf{Platform} & \textbf{Fidelity} & \textbf{SaR Efficiency} & \textbf{Lifetime} & \textbf{Modes} & \textbf{On Demand Readout} \\ \hline
         Cold Atoms & $90\%^a$, $92\%^b$ & $38\%^b$ & $35\mu s^a$, $400\mu s^ b$ & $225^a$, $1^b$ & yes\\ \hline
        Warm Vapor & $98\%^g$ & $78\%^g$ & $5\mu s^f$ & $30^f$ & yes \\ \hline
        Europium Crystals & $85\%^c$ & $7\%^c$, $18\%^d$ & $20 ms^c$, $51\mu s^d$ & $2^c$, $100^d$ & no \\ \hline
        NV Centers & $97\%^e$ & n.c. & $75s^e$ & $10^e$ & yes \\ \hline
        Fiber Loop ($15$km) & $\sim 1$ & $50\%$ & $75\mu s$ & many & no \\ \hline
        Fiber Loop ($45$km) & $\sim 1$ & $12.5\%$ & $225\mu s$ & many & no \\ \hline
    \end{tabularx}
    \label{tab:intro_memorysummary}
\end{table}

\section{Quantum vs. Classical Networking}
In this section, we motivate the interest in quantum networks through a brief review of some potential applications and highlight the differences between classical and quantum network systems to identify the quantum-specific challenges and better frame our quantum scheduling problem.
\subsection{Quantum Network Use Cases}
\label{sec:intro_quantumapplications}
Due to the absence of a medium-to-large scale realization of a quantum network, the use case space has not been completely explored yet and new applications are often discovered during research. In the following, we present a brief review of three potential applications for quantum networks, selected in terms of their relative experimental maturity and conceptual accessibility.
\paragraph{Quantum Key Distribution}
The most developed application of quantum communication is by far Quantum Key Distribution, a technology that allows establishment of a cryptographic key whose security relies not on computational assumptions, as is the case for classical cryptography, but on the very laws of quantum physics \cite{BB84}. The cryptographic applications of quantum communication directly stem from the elementary properties of quantum mechanics: as an example, the No-Cloning theorem was presented in sec.\@ \ref{sec:intro_swapping} as a physical limitation that rules out the amplification of quantum signals, but from the point of view of cryptography is is an invaluable asset in that it forbids the creation of copies of a message if it is encoded in qubits. If a malicious party manages to read a quantum channel they will perturb the original message, thereby alerting the intended recipient of their presence and/or making the transmitted information meaningless. \\
Quantum Key Distribution is a fairly mature technology: certified commercial QKD systems have already been deployed and a large amount of research has been conducted regarding its various specializations \cite{QKDReview} (discrete/continuous variable, device independent, prepare-and-measure vs.\@ entanglement based\ldots) and the search for their potential vulnerabilities.

One interesting fact about QKD is that it technically does not require repeaters, since it can be implemented in a prepare-and-measure \cite{WehnerQInternetVision} way. This makes it the most realistic application for very short-term quantum network research.
\paragraph{Distributed Quantum Computing}
Another interesting application for quantum networking is Distributed Quantum Computing (DQC) \cite{CaleffiDQC}: when picturing a network, the first thing that comes into mind is a large-scale, heavily distributed communication system with the purpose of transferring information between end users. However, there are cases in which it is useful to have a small-scale network localized inside a single device. One such case is DQC, a proposal that aims to scale up the power of quantum computers not by simply adding qubits (which is not a sustainable long-term strategy as every error-corrected logical qubit requires to be encoded in blocks of physical qubits, introducing scaling issues), but rather by creating interconnections between several smaller quantum processing units (QPUs). DQC is a promising solution to the very relevant issue of upscaling quantum processors, but to properly work it requires a correctly designed quantum networking infrastructure: in the DQC paradigm, every QPU has a number of communication qubits whose task is to communicate with other QPUs. At any given time, the communication qubits of two or more QPUs will need to be entangled to exchange data: to avoid the requirement of a fully connected QPU array, entanglement is distributed to the communication qubits by means of swapping through a small scale quantum network.
\paragraph{Blind Quantum Computing}
Given the price and difficulty of realization of quantum computing systems, it is not expected for end-users to have a private quantum computer in the near future. However, this does not necessarily hinder public access to quantum technologies thanks to Blind Quantum Computing (BQC) \cite{Fitzsimons2017BQC}, a broad set of protocols that enable access to a QPU on a remote quantum server from a less capable local quantum client or, ideally, a completely classical client. BQC is extremely interesting because it grants access to sophisticated quantum computation routines, and it is possible to leverage quantum physics and encryption to keep the server unaware of the calculation the client is running and its result through careful design of the networking protocols that connect the servers to the clients.
\subsection{Classical Networks}
\label{sec:IntroClassicalNetworks}
Quantum networks are not expected to provide a replacement to classical ones, but rather an additional layer of enhancement and new functionality. As such, quantum systems must be designed to work in close cooperation with the preexisting classical infrastructure, making it instrumental for a quantum network engineer to have some degree of understanding of classical networking.

In practice, communication in a network system is realized through a set of \textit{algorithms} and \textit{protocols}: an algorithm is a sequence of instructions that starts from a well-defined input to obtain a precise output and thus implement the solution to a specific problem, while a protocol is a suite of algorithms and specifications that answer a macroscopic necessity such as the general purpose operation of a network\paolofootnote{In physics literature, the word ``protocol'' is often employed when discussing operational sequences of steps: physics papers often discuss QKD protocols or entanglement generation protocols meaning ``instructions to follow to achieve the goal in question''. This interpretation of the word, closer to the concept of an algorithm, is different from the network science definition of the word, which is better thought as a common set of abstract rules to format and share data between heterogeneous network nodes.}. An example of algorithm is the Dijkstra algorithm \cite{dijkstra}, which allows to find the shortest path between two nodes in a graph, while an example of protocol is the TCP protocol \cite{rfc9293TCP}, a set of specifications describing exactly what kinds of communication packets must be used to communicate, what information is inside each of them and according to which algorithms they are to be created, relayed and read by recipients. 

\begin{figure}
    \centering
    \includegraphics[height=7cm]{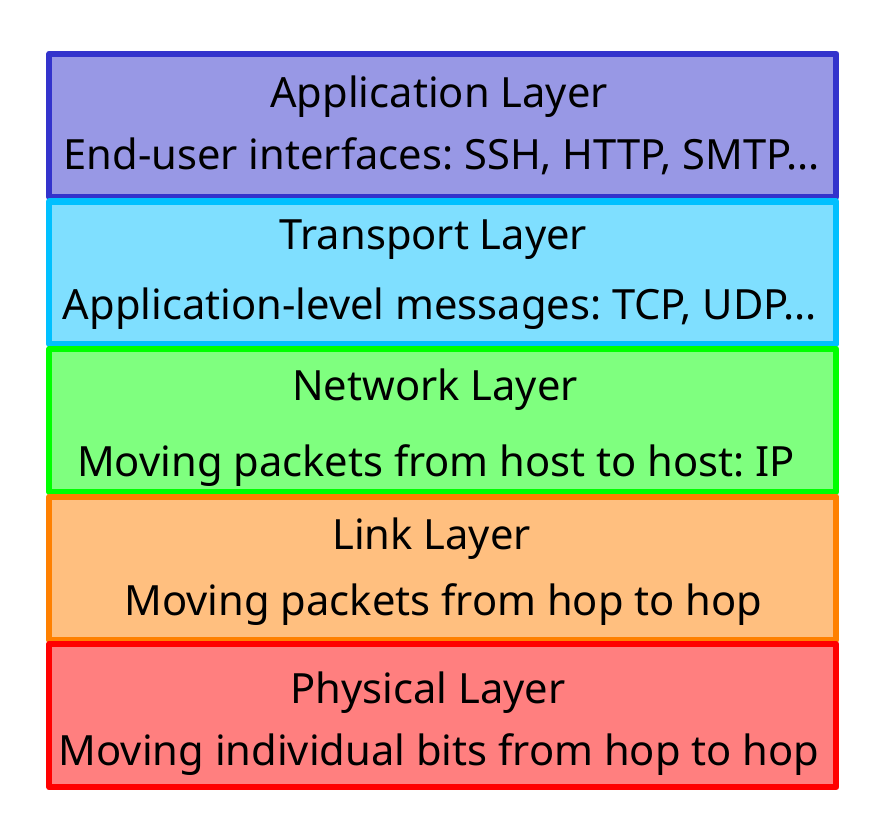}
    \caption[The Internet protocol stack.]{The Internet protocol stack, as defined in \cite[Secion 1.5]{KuroseRossBook}. For each layer, we provide an elementary description of its tasks and, where applicable, some examples of protocols employed in the current Internet stack.}
    \label{fig:Intro_InternetStack}
\end{figure}

Due to their complexity and need for interoperability, a layered model known as the Internet protocol stack \cite[Secion 1.5]{KuroseRossBook} has been standardized to describe networks. The Internet stack starts from the physical layer, which is comprised of raw bit streams propagating through a physical medium (wires, optical fibers, free space), and builds increasingly more abstract layers up to the application layer, which interfaces directly with the end users. 
Each layer has its own regulating protocols, and the way layers communicate with each other is standardized. A summary of the Internet stack is shown in fig.\@ \ref{fig:Intro_InternetStack}. 
The Internet stack and the more general idea of \emph{separation of concerns} enabled the design and implementation of numerous protocols that have lead the classical Internet to its current worldwide scale. On the Internet stack's trail, several stacks have been proposed for an hypotetical quantum internet. \cite{IllianoQStack} provides an interesting review of all the currently active different proposals for a quantum internet stack.

Before discussing the main focus of this thesis, the problem of scheduling, let us take a short deviation and examine the issues of network routing and multiplexing, two interesting prerequisites to the scheduling problem.
\paragraph{Routing}
Routing is one of the fundamental challenges to tackle when designing a network system. In the context of the third layer of the Internet stack (the Network Layer), routing intuitively amounts to finding a path between two network nodes. Although the problem is easily solved in small-scale networks, complications arise when the number of nodes grows, because it becomes unsustainable for every node to hold information about the path to every other node in the network. Instead, nodes usually possess what is known as a \emph{routing table}: for every possible destination in the network, the routing table tells the node which neighbor it should forward packets to. In a chain of nodes $ABCD$, node $A$ does not necessarily know how to reach node $D$, but it knows that packets addressed to it should be handed to node $B$. Node $B$ will then hand them to node $C$, who knows how to finally reach its neighbor $D$. Each handover between nodes is called a hop.
Since every node needs to know the next hop to each destination, routing in large-scale networks amounts to populating each node's routing table. The details of this procedure depend on the specific routing protocol, but abstract high-level insight may be gained by thinking of repeated updates: every node continuously sends (receives) routing table updates to (from) its neighbors and thus ``learns'' about all reachable destinations. The continuous nature of this process allows routes to change, reacting in real-time to changes in the network topology (e.g.\@ due to component failure).
We direct the interested reader willing to deepen their understanding of classical routing protocols to the \textit{RIP} protocol \cite{rfc1058RIP}, one of the first and simplest routing protocols to be adopted on the Internet, and to the \textit{OSPF} protocol \cite{rfc2328OSPF}, which is currently employed for Internet routing.\\
On the quantum side, the problem of routing is an under active research: given the short retention time of quantum memories, finding the best route along which to distribute entanglement is critical for performance, especially as quantum networks scale up. Since routing is a protocol issue, it is likely that multiple solutions to the problem exist, depending on factors such as the underlying hardware and the chosen performance metric to analyze routing protocols. While a complete survey of quantum routing literature would be outside the scope of this thesis, we mention some interesting examples of quantum routing research: \cite{ChakrabortyDistributedRouting} formulates and benchmarks three routing algorithms on various network topologies, using the latency of service as a performance metric and comparing attainable performance in the case of continuous and on-demand entanglement generation along the routes. \cite{PantRouting} focuses on the creation of a framework for quantum routing and highlights the deep impact that early quantum memories can have on network performance, especially when considering multi-path routing (the same two users generating entanglement along multiple cooperating paths) and/or simultaneous routing of multiple user pairs. Finally, \cite{CaleffiRouting} derives a stochastic model of the entanglement distribution rate along a path as a function of physical parameters such as entanglement generation mechanisms, imperfect BSMs and swapping imperfections and formulates a routing protocol that leverages the model to calculate routes. The protocol is proven to be optimal if the entanglement generation rate is chosen as a performance metric.
\paragraph{Multiplexing}
\label{subsec:Muxing_classical}
When deploying a network system, it is unfeasible to follow a fully connected model, where every node would be directly connected to every other node: the complexity of network problems scales much more leniently if some links are shared among users, so that every node is connected to all other nodes but not necessarily as a first neighbor. Building a network this way brings great improvement in terms of network complexity, but it introduces another interesting engineering problem: most of the links in the network will be shared by multiple pairs of users, requiring some scheme to subdivide and assign the limited available resources. The problem is usually subdivided in two subproblems, namely \emph{multiplexing} and \emph{scheduling.}
As a first step, multiplexing consists in subdividing the resource in several units, called \emph{resource blocks} (RB). One or multiple quantities may be chosen to form the RBs: As an example, one might divide the operation of the link in time slots and impose that during time slot $t_i$, only messages corresponding to connection $i$ may pass through the links. This is known as Time Division Multiplexing (TDM). Other examples create resource blocks in a ``parallel'' way, partitioning variables such as the transmission bandwidth or the available transmission and reception antennas. Of course, resource blocks may be formed by aggregating multiple degrees of freedom in a single unit. In quantum networks, one of the natural quantities to exploit when constructing RBs is the number of entangled pairs. The task of allocating entangled pairs to competing sets of end users is known as a \emph{quantum scheduling} problem. 
\subsubsection{Scheduling} 
\label{sec:ClassicalScheduling}
Once roads have been built (i.e.\@ the routing protocol has traced routes connecting all the communicating nodes), traffic lights must be placed at each intersection. In this analogy, traffic lights correspond to a scheduling policy. In realistic network systems, numerous messages need to traverse the network simultaneously, but the channels between the nodes have finite capacity, constraining the number of packets that can go through at the same time. As such, just like vehicles queue up at intersections, it is common for each node to possess a backlog of incoming messages that need to be relayed to neighboring nodes to proceed to their destination. The problem of scheduling amounts to determining the order in which backlogged packets are processed and relayed. There is no single solution to the scheduling problem: depending on the specific requirements of each system, nodes could relay first the packets that arrived earlier, they could alternate service of requests between several users or they could have a more complex priority system regulating request service \cite[Section 4.2.5]{KuroseRossBook}. Each node could also serve the requests in a random order, or implement more refined selection schemes based on additional status information such as the traffic conditions in other parts of the network. 
All these potential solutions are examples of \emph{scheduling policies}: a scheduling policy is defined as the set of rules a node follows to determine the order in which packets are relayed through it. A scheduling policy may be static or dynamic: in the static case, the policy's decisions and priorities are predetermined and never changed, while in the dynamic case scheduling decisions are calculated in real-time as a function of the network's state variables. Despite static policies being generally simpler to understand and implement, due to the short time scales typical of quantum networking this thesis will focus exclusively on dynamic scheduling policies, which are quicker to react to changes in the network and more adaptable to different demand and performance regimes.\\
Since no policy is universally best for all network scenarios, comparing different policies is crucial and can be achieved through a variety of figures of merit (e.g.\@ average waiting time for a request to be served, responsiveness under network changes, communication overhead\ldots). To obtain an easily readable comparison while retaining insight on the performance of different policies, the main figure of merit throughout this thesis will be the \textit{stability region}, i.e.\@ a mathematical representation of the set of all loads that can be properly served under the scheduling policy in question. The formal definition of this crucial concept will be given in sec.\@ \ref{sec:StabilityDef}.
\paragraph{The Max-Weight Policy}
\label{sec:intro_maxweight}
One interesting example of classical scheduling policy is the Max Weight (MW) policy\paolofootnote{A quantum version of this policy will be examined in detail in chap.\@ \ref{ch:scheduling}.} \cite{TassiulasMaxWeight}, an optimization-based policy that is ubiquitous in classical networks literature: under the MW policy, every node $i$ stores the incoming packages in a set of queues (one queue per destination node $j$). The length of each queue $w_{i\rightarrow j}$ is called the \textit{weight} of destination $j$ at node $i$. Taking a discrete time axis, we can define a \textit{scheduling decision} $\mathbf{r}(t)$ as a vector with as many entries as there are weights: at every time step, each entry of $\mathbf{r}(t)$ specifies how many packets associated to each destination are to be let through each node. For example, $r_{A\rightarrow C}(4) = 3$ means that at time step $4$ node $A$ should relay $3$ packets from the queue of packets addressed to node $C$.
At every time step, the scheduling decision $\mathbf{r}(t)$ is obtained by solving the optimization problem
\begin{align}
    \mathbf{r}(t) = \argmax_{\mathbf{r}(t)} \mathbf{w}^T(t)\mathbf{r}(t) = \argmax_{\{r_{i\rightarrow j}(t)\}_{i,j}}\sum w_{i\rightarrow j} r_{i\rightarrow j}(t)
    \label{eq:classicalMW_objective}
\end{align}
with constraints coming from the specific network implementation, commonly placing limits on the channel capacity (i.e.\@ the number of packets that can simultaneously pass through the same link), the ability of some nodes to transmit/receive, or imposing that the scheduler never asks for service of more requests than the ones currently in the queues. Solving the MW problem amounts to prioritizing service of longer queues over shorter ones in order to reduce queue accumulation and balance out the service of different kinds of requests.\\
The MW policy is commonly taken as the exemplary scheduling policy in classical networks. It is a \textit{global} policy, meaning that the scheduling decision is taken by a central authority and sent to the network nodes, and it was proven to be throughput-optimal (the meaning of this term will be clarified in sec. \ref{sec:StabilityDef}) in \cite{TassiulasMaxWeight}.
\subsection{Key Figure of Merit: Stability of a Network System}
\label{sec:StabilityDef}
As established before, there is no single, universal solution to the scheduling problem. Designing a network architecture requires comparing several policies to make a selection, and it is therefore instructive to formalize the key performance indicator of a scheduling policy. Such a performance metric is called \emph{stability}, which is a direct measure of how congested with traffic a network is under a given scheduling policy. In this section, we define stability and an important graphical aid for comparing scheduling policies called the \emph{stability region}.

As a first step, it is important to remember that a scheduler interfaces with its underlying network through requests: in the classical domain, packets that are entering a node want to be relayed to their next hop while in the quantum one, entangled pairs need to be swapped in order to serve user demand. Abstractly speaking, the scheduling problem is about deciding which requests to handle at which time. 
For each kind of request (i.e.\@ for each pair $(i\rightarrow j)$ of local node and destination node), a queue is defined as a container to which requests are added (enqueued) when they are issued and removed (dequeued) when they are handled or no longer valid.

Intuitively, for a system to be performing well all the queues need to retain finite length: a system that effectively serves all incoming requests will have empty queues most of the time, while one that does not will have its queues accumulate demands. In the worst cases, the length of the queues will diverge.
Let us assume, for simplicity, that each kind of request comes with a fixed average rate $\alpha_{ij} \geq 0$. Therefore, the load on the network can be represented by the \textit{load vector} $\mathbf{a}$. Every scheduling policy is stable under null load ($\mathbf{a}$ = 0), and for every network there always exists a load vector for which a given scheduling policy is unstable. This concept can be formalized through the idea of stability of a queueing system, for which numerous formal definitions exist \cite{MeynTweedieBook}. 
In the context of this thesis, we define a system to be stable if and only if the time it takes for the cumulative queue length to return to zero is finite on average, i.e.\@ the underlying Markov process is positive recurrent, as per \cite{BramsonStabilityLecNotes}: no matter how much the queues accumulate, a stable system guarantees that all its queues will eventually return to zero. Informally, this is equivalent to saying that, on average, more requests are served than arrive.
Having established the extremes, we can define the stability region of a given scheduling policy as the region in the space of load vectors for which a given scheduling policy is stable. The stability region's boundary is a Pareto front, meaning that if a given load is unstable, then all loads strictly higher than it are unstable. This makes the stability region compact in the space of load vectors. In formal terms, if load vector $\mathbf{a} = (a_1,a_2,\ldots a_N)$ is unstable, then $\mathbf{a'} = \{(a'_i | a'_i = a_i \text{ for } i \notin \mathcal{K}, a'_i > a_i \text{ for } i \in \mathcal{K}\}$ is unstable for all subsets $\mathcal K$ of the index set of $\mathbf a$. Given a network system, different scheduling policies may have different stability regions: the ideal solution to the scheduling problem over a network is the policy whose stability region is \textit{maximal}, i.e.\@ a superset of all other stability regions. Such a policy is deemed throughput-optimal.

\begin{figure}
    \centering
    \subfloat[Dumbbell Network Topology;]{
\includegraphics[width=.5\textwidth]{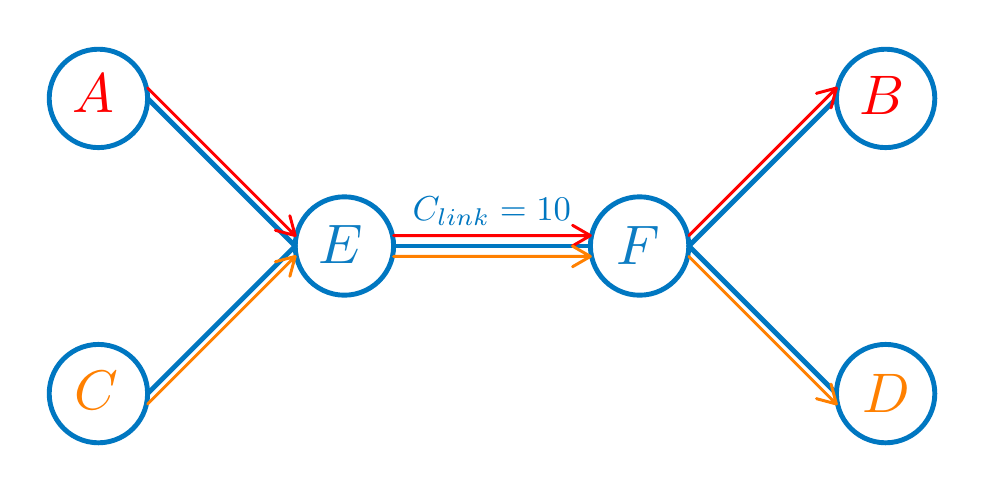}
\label{fig:example_stab_topo}
}
    \subfloat[Example of Maximal and Non-Maximal Stability Regions;]{
\includegraphics[width=.5\textwidth]{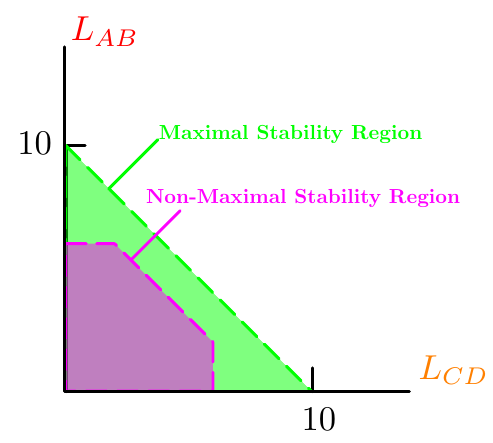}}
\caption[Example of maximal and non-maximal stability regions for a simple dumbbell network.]{Example of simple dumbbell network topology with two communication flows $(A\rightarrow B)$ and $(C\rightarrow D)$. In the ideal case, the limiting factor is the capacity of the $EF$ bottleneck link, leading to the maximal stability region shown in the right figure. However, other nonidealities could limit communication (e.g.\@ high losses, traffic from other users\ldots). In that case, the shape of the stability region is usually more similar to the purple nonmaximal one.}
\label{fig:example_stabregion}
\end{figure}

Even though it is not always possible or straightforward to prove optimality of a policy, the stability region is still a useful aid for comparing scheduling policies: a policy with a larger stability region is usually better performing. However, mathematical optimality does not necessarily imply the policy is a good choice from the engineering point of view: policies with larger stability regions might require more information about the network system, which in turn demands a more complex communication infrastructure, creating interesting engineering tradeoffs to explore during the network design phase.
An example of maximal and non-maximal stability regions is provided in fig.\@ \ref{fig:example_stabregion}.\\
Stability regions are an interesting metric of comparison, and we rely on them extensively to compare quantum scheduling policies in chap.\@ \ref{ch:scheduling}.
\subsection{Quantum-specific Networking Challenges}
Having described the elementary ingredients of quantum networks and before moving to the problem of quantum scheduling, let us highlight the challenges that differentiate quantum networks from classical ones and make them a new field to which classical solutions are not directly applicable. 

In light of the previous foundational definitions, a quantum network is defined as an array of distributed and interconnected quantum repeaters. The ultimate task of a quantum network is to establish quantum entanglement between arbitrary sets of $n$ nodes by overcoming an array of quantum-specific networking challenges, detailed in the following. 
\paragraph{Nondirectional Communication} Formulating network problems over quantum networks requires some paradigm shifts from the classical case. The first one is related to the \emph{directionality} of communication: in a classical network we have a source node that sends a message, and this message will eventually reach a destination node\paolofootnote{This is also true for networks based on 3G quantum repeaters, which are outside the scope of this thesis.}. There is a clear direction in which information propagates. 
If we now shift to quantum networks based on 1G and 2G quantum repeaters, we notice that communication is no longer directional: in a quantum network, the ultimate goal is to distribute entangled pairs to end users, and it is achieved by distributing link-level Bell pairs and then performing entanglement swapping operations until the two end nodes that issued the request share one or more pairs. Links are therefore established in a more ``growing and merging'' fashion through the creation of Virtual Quantum Links \cite{RoutingMemoriesWehnerKerenidis, ChakrabortyDistributedRouting} (VQLs), i.e.\@ intermediate entangled pairs that are longer than the link-level but not yet end-to-end. Distributing entanglement by creating VQLs contrasts with the idea of source and destination in favor of several smaller links that are spliced together to form a longer one. This is a crucial difference with classical networks that prevents application of classical routing and scheduling protocols without nontrivial adaptation.
\paragraph{Multipartite Entanglement} In the bipartite case ($n=2$), users request pairwise entanglement. However, this is not necessarily the case: several quantum applications such as Multiparty Computation \cite{CrepeauMPComp}, Electronic Voting \cite{CentroneEVoting} or Conference Key Agreement \cite{MurtaCKA} require multipartite entangled states. In this thesis, we will focus on the bipartite case because the diversity and complexity of multipartite entanglement generation protocols \cite{GHZandW,WalterMultipartite,FischerMultipartite} would require significantly more discussion. Moreover, it is always possible to suboptimally create arbitrary multipartite states starting from a number of entangled pairs \cite{MeignantMultipartite, PirkerMultiParty}, implying that the $n=2$ case may be taken as universal when optimal resource usage is not required.
\paragraph{Performance of Quantum Memory} Another crucial difference is the performance of memory: if left unperturbed, modern classical memory systems have retention times spanning from a few years for solid state drives up to multiple decades for hard disk drives, which is essentially infinite compared to typical networking time scales. However, it is quite rare for quantum memories to have retention times above the millisecond scale. In classical computing, when short-retention memories are employed (such as capacitive DRAM cells, which naturally lose the stored charge over time), the practice of \emph{memory refresh} is adopted, which consists of periodically reading and re-writing the contents of a cell. This strategy is not viable with quantum memories because reading the memory's content implies measuring it, which destroys its quantum nature. This makes memory and in particular its imperfections a fundamental limitation of quantum network science, and protocols must be devised accordingly.
\paragraph{Ebits as Channel Capacity} Even though it would seem natural to draw an analogy between classical communication packets and ebits, as both travel through their respective network, there are radical differences in the role that those two actors play: while a well-performing classical network is one with no packets queued up at the nodes, the best case for a quantum network is to have as many entangled qubits as possible along relevant links. In this sense, entangled qubits are closer to the classical concept of channel capacity: users issue communication requests which can be satisfied only if the channel connecting them has enough capacity, i.e.\@ only if enough Bell pairs are generated between the users. The other side of this analogy is that all quantum information protocols must account for high variability of link capacities.
\paragraph{Nomenclature: Quantum Switches and End Nodes} The literature often traces a distinction between \emph{repeater nodes}, which are only tasked with entanglement distribution, and \emph{end nodes}, which are the ones to which entanglement is distributed and consume it for applications. We will not enforce this distinction in the following, so that any repeater in our networks will also be a potential end node: this is equivalent to having end nodes interconnected to the corresponding repeaters, without actually restricting the set of nodes that may request entanglement. Moreover, some contributions adopt the term \emph{quantum switch} when referring to a quantum repeater with $>2$ neighbors: we will not trace this distinction and treat the terms ``quantum repeater'' and ``quantum switch'' as synonyms.
\subsection{Quantum Scheduling}
The scheduling problem as introduced for classical networks cannot directly carry over to 1G and 2G quantum networks, which are the main focus of this thesis. This incompatibility arises from the previously mentioned nondirectionality of 1G/2G quantum communication: ebits start at the link level and they are ``stitched'' together until end-to-end entanglement is established. It must be noted that, since we consider memory-endowed quantum networks, the creation of an end-to-end pair usually happens in multiple intermediate steps, as shown in fig.\@ \ref{fig:intro_multiswap}:
\begin{figure}
    \centering
    \subfloat[Link-level entanglement is established along links $AB$ and $BC$;]{\includegraphics[width=.45\linewidth]{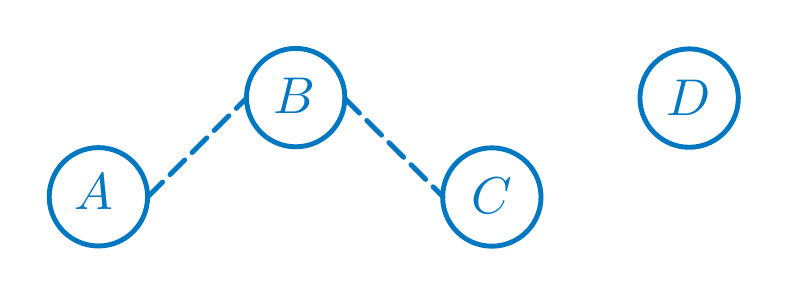}}\hfill
    \subfloat[Entanglement is swapped from $AB$-$BC$ to $AC$;]{\includegraphics[width=.45\linewidth]{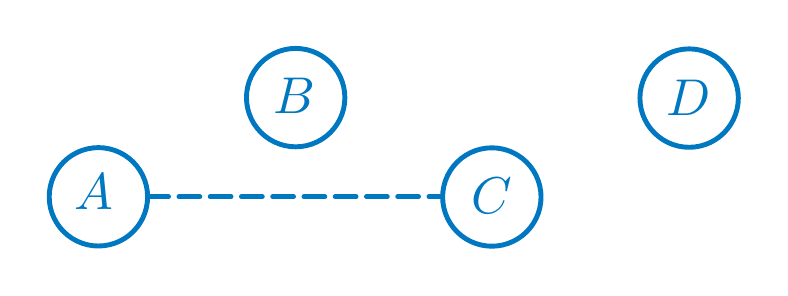}}
    
    \subfloat[Link-level entanglement is established along link $CD$ while $AC$ waits in memory;]{\includegraphics[width=.45\linewidth]{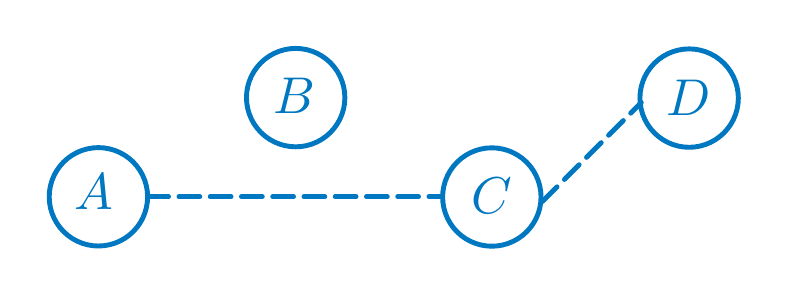}}\hfill
    \subfloat[The final $AC$-$CD$ swap is performed to entangle $AD$.]{\includegraphics[width=.45\linewidth]{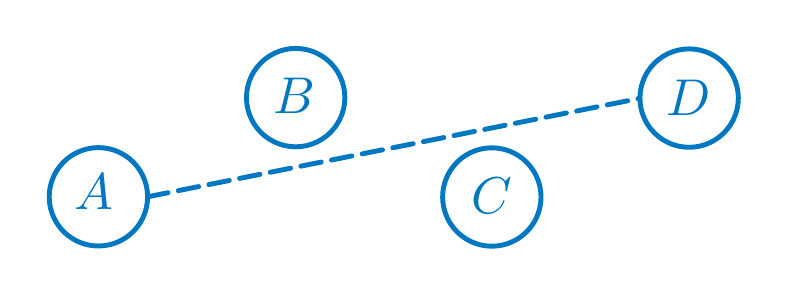}}
    \caption[Creation of an end-to-end entangled pair in multiple steps.]{Creation of an end-to-end entangled pair in multiple steps (dashed lines represent one entangled pair of qubits shared between two nodes): quantum memories allow storage of intermediate ebits until all sublinks are ready.}
    \label{fig:intro_multiswap}
\end{figure}
this idea is referred to in literature as a Virtual Quantum Link \cite{RoutingMemoriesWehnerKerenidis} because it creates shared ebits, which we recall are analogous to information capacity, between two non-connected nodes. VQLs are a direct consequence of the presence of memory at the nodes and they greatly improve the performance of a quantum network. However, creating a VQL is not always advantageous, and deciding which VQLs to create is a nontrivial problem, as nodes are faced with the decision of whether to swap (creating longer ebits and freeing up memory slots at intermediate stations) or accumulate a buffer supply of shorter ebits to better respond to future requests, which may concern routes outside of the pre-created VQLs.

In light of this information, the quantum scheduling problem may be intuitively formulated as finding the best sequence of entanglement swapping operations to perform in order to establish pairwise entanglement between the requesting users. Unlike classical networks, there are currently no established scheduling policies for quantum networks, although considerable effort is being devoted to solving the scheduling problem in specific quantum circumstances with a variety of methods.

Starting from a single repeater serving multiple users, policy-related work has been carried out by applying stochastic analysis \cite{VardoyanSwitchStochastic, DaiSwitchProtocols, ChandraScheduling}, convex analysis \cite{TillmanSwitch} and Lyapunov stability methods (defined in sec.\@ \ref{sec:IntroLDM}) \cite{VasantamSwitch,PanigrahySwitch}.  
Moving to several repeaters arranged in a linear topology serving entanglement to the extremes, an optimal theoretical bound for entanglement distribution is derived in \cite{DaiScheduling} and expanded upon in \cite{DaiTowsleyScheduling}, while \cite{KaminRepeaterChain} and \cite{DeAndradeRepeaterChain} propose performance benchmarks in terms of low-level physical parameters.

Finally, the scheduling problem has been tackled in a full-network scenario in \cite{SkrzypczykArchitecture} with a focus on architecture design and heuristic scheduling policies and in satellite quantum networks in \cite{WilliamsSatScheduling}.

It should be noted that, given all the previously presented differences with the classical case, it must be ascertained whether there is any advantage in devising scheduling policies for quantum networks, given that the more elaborate a policy is, the more additional operative requirements it poses upon a network infrastructure. Moreover, since there is no universal scheduling policy that can guarantee the best performance over all networks, large part of the interest resides in the comparison between different policies over real network topologies. The framework we build in this thesis is, to our knowledge, the first general toolbox allowing the formalization and benchmark of arbitrary scheduling policies over general quantum network topologies.
\section{Satellite Networking}
To conclude our introductive section, we add the final building block to our hybrid network puzzle, i.e.\@ a description of the peculiarities of free-space links with spaceborne nodes, and how they can be integrated in quantum networks with the goal of interconnecting several fiber subnetworks.\\
There are three main categories of satellite orbits, with different fitness for quantum networking:
\begin{itemize}
    \item \emph{Low Earth Orbit} (LEO) includes altitudes that range from $\sim200\unit{\kilo\meter}$ (to minimize atmospheric drag) to $\sim2000\unit{\kilo\meter}$ (due to the Van Allen radiation belt and the high beam divergence). It is the lowest category, meaning it suffers the least from atmospheric attenuation, but it also features the fastest satellites (revolution period in the hours range) and shortest visibility windows. LEO is commonly used for high-resolution imaging and detection. Moreover, it is the most used orbit region for satellite quantum communication \cite{Micius,dFdP}.
    \item \emph{Medium Earth Orbit} (MEO) is the orbit band that goes from $\sim2000$ km to $35786$ km. This kind of orbit has longer visibility windows (with revolution periods up to $12$ h) and requires less frequent handovers\paolofootnote{i.e.\@ one satellite going out of view and handing over service of a ground station to another satellite.} but noticeably increases losses. To see any benefit from using MEO satellites for quantum communication, it is necessary to equip the ground stations with SNSPDs\paolofootnote{Superconducting Nanowire Single Photon Detectors.} \cite[Sec. D2]{dFdP}, costly detectors that require cryogenic equipment. Outside quantum, the main application of MEO satellites is navigation.
    \item \emph{Geostationary Orbit} (GEO) is a high orbit ($35786$ km) with an interesting quality: satellites on a GEO orbit look stationary to an observer on Earth, removing the need for tracking equipment. Although it is unfeasible to directly use GEO satellites for quantum communication due to their altitude and the  $\sim 240$ ms round-trip time, it could be interesting to envision the usage of such satellites as quantum repeaters interconnecting MEO and/or LEO nodes. GEO satellites are commonly used for applications such as meteorology, long-distance communication and error correction in navigation.
    \end{itemize}
The first problem introduced by satellite communication is that of latency: even at the lowest orbits ($\sim200$ km, Low Earth Orbit range), long distance satellite links introduce latencies that are in the order of milliseconds and highly variable due to the fast displacement of the satellite. These latencies are usually tolerable in classical applications but can prove fatal if not accounted for in quantum scenarios, due to the short retention time of quantum memories, which as seen in sec.\@ \ref{sec:IntroQM} is often in the milliseconds range, and to the fact that quantum protocols commonly have synchronization requirements well below the ms threshold, as discussed in sec.\@ \ref{sec:satellites_difflatency}. The problem of latency is even more accentuated in 1G and 2G networks because their nondirectional nature requires more satellite-ground roundtrips, ultimately placing a physical upper bound on entanglement rate (more details in sec.\@ \ref{sec:satellite_TheorySingleLink}).

Another pre-existent issue that is amplified by the fiber to satellite passage is that of losses: long links passing through the atmosphere are bound to suffer from high losses, which combined with the extremely low intensities required by quantum information applications means usually more than $99\%$ of the signals are lost. To mitigate this issue, one can ensure to always work in the \emph{downlink} configuration (photons going from satellite to ground): most of the photons' path is outside the atmosphere, and in a downlink configuration atmospheric refraction happens in the final section of the path. This means that downlink beams diverge less than in the uplink scenario, with signal strength differences as high as $30$ dB \cite{SatLosses}. Due to the high losses of long-distance links, LEO (and sometimes MEO) satellites are most commonly used for quantum applications, entailing visibility windows (and therefore link uptime) in the order of a few minutes to a couple hours per day.

While at first it might seem that the limited visibility of satellites should be an additional challenge for scheduling (having to prepare resources in the best possible way to make maximum use of the visibility window), the problem does not arise because satellite movement and entanglement generation/swapping happen on radically different timelines. It is however an interesting research question to examine what happens when a network has a satellite path and an alternative low-throughput fiber-and-repeaters path: the scheduling policy will need to exploit the two links appropriately and balance user load in terms of requests and demanded quality.

The short visibility window issue of LEO and MEO satellites --- a problem also present in classical communications --- is commonly solved through the use of \emph{satellite constellations} \cite{LoSatConstellations}, i.e.\@ a set of multiple satellites spaced along the same orbit so that one of them is always visible to each relevant ground station. The constellation method introduces the issue of \emph{handover}, meaning that protocols must specifically account for the moment in which the ``active'' satellite goes out of sight and the new satellite arrives by specifying how the relevant data and control information are passed between members of the constellation. Handover may be carried out with the ground station as a middle man or through the broader domain of satellite-to-satellite communication, creating interesting quantum-specific questions as to which one is the better way and how exactly quantum information and entanglement should be handed over across satellites.\\
Readers interested in more detail concerning satellite quantum communication may find in \cite{dFdP} a complete review describing problems, use cases and expected performance of quantum satellite networks, and in \cite{Micius} a full account of the design of an experimental satellite system (the Chinese Academy of Science's Micius mission).
\chapter{A Linear Algebraic Framework for Quantum Internet Dynamic Scheduling}
\label{ch:scheduling}
In this chapter, we discuss our linear algebraic framework to formulate and solve the problem of scheduling in quantum networks. A preliminary version of the framework was presented at the \textit{IEEE International Conference for Quantum Computing and Engineering 2022} \cite{scheduling_conference} and was later extended to a journal article published in \textit{IEEE Transactions on Quantum Engineering} \cite{scheduling_journal}.
\section{Introduction}
As experimental demonstrations of quantum repeater links and small-scale quantum networks \cite{Bernien3Metres,LagoRiveraLink,HermansQNet} start to surface, the vision of a future Quantum Internet moves closer to reality \cite{rfc9340,WehnerQInternetVision,CacciapuotiInternet,VanMeterArchitecture}.

Despite it still being a long-term goal, the road is partially paved by the development of the classical internet, that identified and solved all the problems intrinsic to scaling a network up and operating it in a distributed way. The solutions to such problems are not directly translatable to quantum networks in general because quantum hardware is fundamentally different, creating the need for a new branch of network science with its own set of specialized tools. We propose a novel framework to formulate and solve the problem of scheduling entanglement swapping operations in quantum networks, and showcase its potential through application examples.
\section{Network Description}
The abstract network model on which our scheduling framework is defined is an arbitrary connected graph $\mathcal{G} = (\mathcal{V},\mathcal{E})$. Each vertex of $\mathcal{G}$ is taken to represent a quantum repeater and each edge a lossy fiber link. In order to build a model that translates well to Discrete Time Simulation, we choose to work with discrete time and call $\Delta t$ the time step length. Along each edge, entangled pairs are continuously created through a generic link-level entanglement distribution scheme (such as a $\chi^{(2)}$ crystal, a Bell State Analyzer in the middle \cite{DLCZ} or at one of the stations \cite[sec. V-C]{AzumaRepeaters}, NV center-based systems \cite{LiuNV}\ldots) with a constant average rate, which might be heterogeneous across links.
Inside each node, there are quantum memories in which qubits may be stored and quantum processing hardware such that it is possible to perform Bell State Measurements between any two of those qubits. This enables nodes to perform entanglement swapping operations. To simplify the analysis, we will assume the quantum memories to have an infinite number of slots.\\
A subset of $n$ pairs of nodes $\{(\mathit{Alice}_1, \mathit{Bob}_1),\ldots,(\mathit{Alice}_n, \mathit{Bob}_n)\}$ is taken to represent final users, requesting entangled pairs of qubits to implement a generic application. Each $\mathit{Alice}_n$ node is assumed to be connected to its corresponding $\mathit{Bob}_n$ by $m$ fixed routes, along which link-level entanglement must be converted into end-to-end by means of entanglement swapping operations. We stress that the routes are not necessarily disjoint, creating competition along some of the network's links that must be handled by a suitable scheduling policy.\\
Time is discretized, and at each time step qubits may be lost to decoherence. For each qubit, we compute the probability $\eta$ to still be present in memory at the end of a time step, and run a random check against it at each step. If qubit $A$ of an entangled pair is lost, then qubit $B$ is assumed to be instantly lost as well. All nodes are assumed to also be connected through a classical communication network, for them to exchange control information at runtime.
\section{The Queuing Model}
\label{sec:scheduling_queuingmodel}
In our case, a scheduling policy may be seen as a black box that takes as input some information about the current network state and outputs a scheduling decision consisting of a set of entanglement swapping operations. Therefore, before discussing scheduling and benchmarking policies we need a clear mathematical language to describe the network state. In classical networks, this is commonly done through queuing models: despite requiring some quantum-specific adaptations due to how entanglement grows through virtual quantum links, queuing methods are a suitable tool for quantum networks as well.\\
In particular, we will formulate two interdependent queuing models so that, for each edge $ij$ in the extended edge set $\tilde{\mathcal{E}}=\mathcal{V}\times\mathcal{V}$, we will derive equations describing:
\begin{itemize}
    \item an \emph{ebit queue} $q_{ij}(t)$, keeping track of all the entangled pairs currently available between nodes $i$ and $j$;
    \item a \emph{demand queue} $d_{ij}(t)$, tracking the amount of entanglement requests between $i$ and $j$.
\end{itemize}
In our discrete time scheme, the scheduling decision will be elaborated and applied at the end of each time step.
\subsection{Ebit Queues}
In our framework, ebit queues are the mathematical object responsible for keeping track of how many entangled pairs are currently shared between any two nodes. Each possible pair of nodes has a corresponding ebit queue. 
As we will describe in the following, ebit queues are affected by link-level pair generation, losses, and entanglement swapping operations. 
Ebit queues will be further divided in \emph{physical} queues, which are the ones corresponding to a fiber link in the network, and \emph{virtual} queues, which do not correspond to a fiber link. This distinction will be made clearer in the following, while describing the individual terms that compose an ebit queue.
The complete time evolution of an ebit queue can be written as: 
\begin{align}
q_{ij}(t+1) = \underbrace{q_{ij}(t)}_{\mathclap{\text{Previous State}}} + \overbrace{a_{ij}(t)}^{\substack{\text{Link-Level}\\\text{Generation}}} - \underbrace{\ell_{ij}(t)}_{\text{Losses}} \pm \overbrace{s(t)}^{\mathclap{\text{Swapping}}}
\label{eq:scalarevolution_ebitqueue}
\end{align}
\subsubsection{Generation}
The generation term is the one modeling the link-level generation of entangled pairs: it is a non-negative integer term, and it represents the number of pairs generated across link $ij$ during the current time step. It should be stressed that this term only concerns generated pairs: for all virtual queues, i.e.\@ queues that do not map to a physical fiber link, this term is constant and equal to $0$. In the following, we simplify the analysis by taking $a_{ij}(t)$ to be a Poissonian random process with constant average rate $\alpha_{ij}$. However, this is not the only choice possible, as this term is an open interface to the specific entanglement generation mechanism adopted.
\subsubsection{Losses}
This term models quantum memory losses, based on the discussions of sec. \ref{sec:IntroQM}: $\ell_{ij}(t)$ is a non-negative, random integer that is binomially distributed with $q_{ij}(t)$ as the number of trials (meaning that newly-arrived ebits are immune to loss) and success probability $(1-\eta)$, with $\eta = \exp{\left(-\frac{\Delta t}{\tau}\right)}$, $\tau$ being the expected lifetime of a qubit in the memory and $\Delta t$ the duration of a time step.

We remark that the statistical distribution of ebit survival times follows the geometric distribution defined by $\eta$, whose mean value $\tfrac{1}{1-\eta}$ tends to the expected $\tfrac{\tau}{\Delta t}$ for small $\frac{\Delta t}{\tau}$, $\tau$ being the expected lifetime of ebits in the memories.

Finally, the dependence of the loss probability on the duration of a time step makes the framework relevant as a tool to select the correct $\Delta t$ in real implementations.

\subsubsection{Swapping}
\label{sec:scheduling_sysdesc_swapping}
The effect of swapping on ebit queues is more complex than arrivals and losses because it is a multi-queue operation. As such, we need to introduce our formalism for entanglement swapping and interqueue operations before describing its impact on the queues. 

As introduced in sec.\@ \ref{sec:intro_swapping}, entanglement swapping is a process that allows to splice two shorter entangled pairs into a longer one. If one has an entangled pair between nodes $(A,B)$ and one across $(B,C)$ it is possible to perform a Bell State Measurement on the two qubits at $B$ and consume the two original entangled pairs to obtain one entangled pair across $(A,C)$.
Since an entanglement swapping operation amounts to consuming one pair each from two parent queues to create a new pair in a child queue, our framework models a single entanglement swapping operation as one dequeue operation from each of the parent queues and one enqueue into the child queue. As a notation example, we discuss the $A[B]C$ swap, which corresponds to a swap at node $B$, using $AB$ and $BC$ as parent queues and $AC$ as child queue. Applying this transition consumes one pair from $AB$ and one from $BC$ to create one pair in $AC$. In terms of queues, this entails a $-1$ on both $q_{AB}$ and $q_{BC}$, and a $+1$ on $q_{AC}$. As an additional example, $F[S]T$ would refer to a swapping operation happening at node $S$, with parent queues $FS$ and $ST$ and child queue $FT$. In queue terms, dequeue $1$ from $q_{FS}$ and $q_{ST}$, enqueue $1$ in $q_{FT}$. 

\begin{figure}[t]
    \centering
    \includegraphics[width=0.8\linewidth]{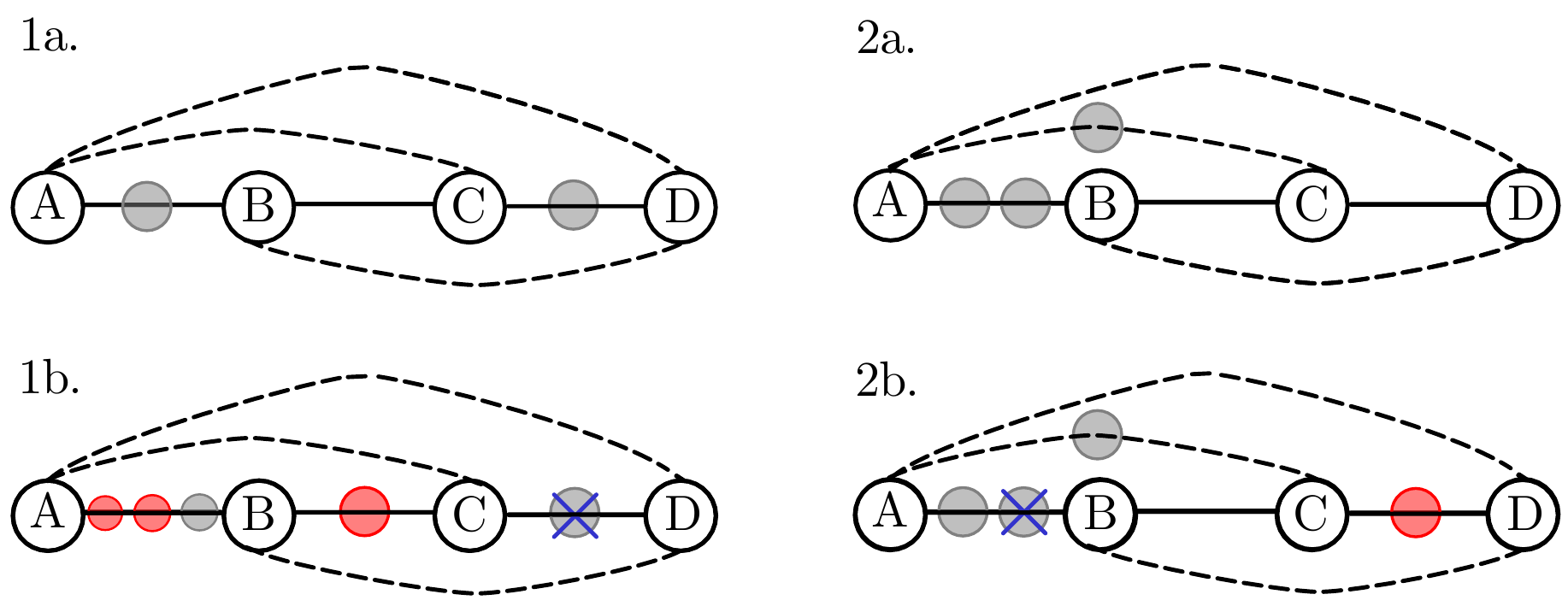}
    \caption[Example of two time steps in our scheduling framework.]{Explicit example of two successive time steps over a simple chain topology. Continuous lines represent physical links, and therefore correspond to physical queues in our model, whereas dashed lines symbolize virtual queues, i.e. pairs of nodes that are not directly connected by fiber but may share ebits after some entanglement swapping operations. 
   Grey circles represent ebits in the queue at the beginning of the current time step. 
   Their number is encoded in $\mathbf{q}(t)$ in our model.
   Red circles represents ebits arrived during that time step ($\mathbf{a}(t)$). 
   Blue crosses represent loss of an ebit ($\boldsymbol{\ell}(t)$). 
   Upper figures (a) represent the state at the beginning of the corresponding time step, lower figures (b) at the end of it. Passing from time step $1$ to $2$, the scheduling decision $r_{A[B]C}(t=1) = 1$ has been applied, removing one ebit each from queues $AB$ and $BC$ and adding one to $AC$.}
    \label{fig:scheduling_simexample}
\end{figure}

\begin{figure}[t]
    \centering
    \includegraphics[width=0.8\linewidth]{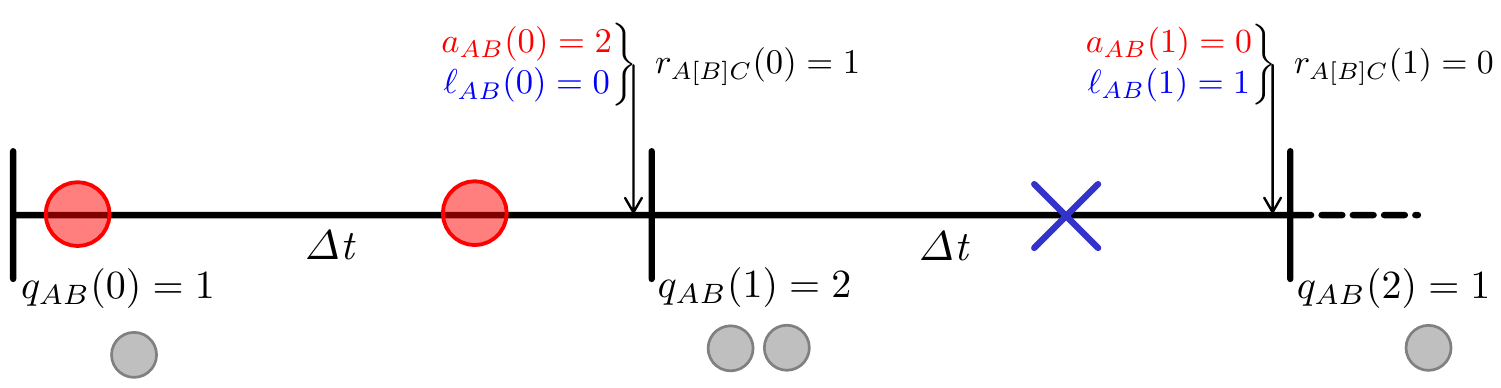}
    \caption[Timing scheme of the two time steps depicted in fig.\@ \ref{fig:scheduling_simexample}.]{Same example as fig. \ref{fig:scheduling_simexample}, as seen internally by queue $AB$ to highlight the timing of the various phenomena at play. Queue snapshots $q_{ij}(t)$ are taken at the very beginning of a time step, whereas arrivals and losses happen stochastically. At the end of each time step, arrivals and losses are counted and the scheduling decision is taken. Note that ebits arriving during the current time step are not subject to losses in this model.}
    \label{fig:scheduling_simtiming}
\end{figure}

Transitions are a way to mathematically formalize the scheduler's task: at the end of each time step, the scheduler is expected to provide a list of which entanglement swapping operations to apply given the current network state. In figg.\@ \ref{fig:scheduling_simexample} and \ref{fig:scheduling_simtiming} we provide a walkthrough of two consecutive time steps in order to cement our notation and clarify the timing of all the actors in play. Fig.\@ \ref{fig:scheduling_simexample} shows the complete network, while fig.\@ \ref{fig:scheduling_simtiming} is focused on queue $AB$ to highlight the timing of all the queue events.\\

$\bullet$ During time step 1:
\begin{enumerate}
    \item At the beginning of the time step (fig. \ref{fig:scheduling_simexample}, $1a$), the queue states are: $q_{AB}(1) = q_{CD}(1) = 1$, $q_{BC}(1) = 0$;
    \item At the end of the time step (fig. \ref{fig:scheduling_simexample}, $1b$), new ebits have been generated across $AB$ and $BC$ ($a_{AB}(1) = 2$, $a_{BC}(1) = 1$, red circles) and one has been lost across $CD$ ($\ell_{CD}(1) = 1$, crossed-out grey circle). The scheduling decision is taken from this configuration as $r_{A[B]C}(t=1) = 1$: one swap at node $B$ with queues $AB$ and $BC$ as parents and $AC$ as child. The time step is concluded by the application of the scheduling decision.
\end{enumerate}

$\bullet$ During time step 2:
\begin{enumerate}
    \item The initial configuration (fig.\ref{fig:scheduling_simexample}, $2a$) sees two stored pairs in $AB$ which were not employed in the last time step ($q_{AB}(2) = 2$) and the freshly swapped one in $AC$ ($q_{AC}(2) = 1$);
    \item At the end of the time step (fig.\ref{fig:scheduling_simexample}, $2b$), one pair was lost across $AB$ ($\ell_{AB}(2) = 1$) and one generated across $CD$. The scheduler may now decide $r_{A[C]D}(2) = 1$ to move to $AD$ or store the pairs for future use.
\end{enumerate}

To categorize transitions in terms of their net effect on queues, we say that a given transition $i[j]k$ is
\textit{incoming} for queue $(i,k)$, because it adds pairs to it, and \textit{outgoing} for queues $(i,j)$ and $(j,k)$, because it takes pairs from them. A queue's evolution can therefore be summarized as follows, i.t.\@ and o.t.\@ being shortcuts for incoming and outgoing transitions:
\begin{align}
    q_{ij}(t+1) = q_{ij}(t) + a_{ij}(t) - \ell_{ij}(t) - \smashoperator{\sum_{o \in \text{o.t.}}}r_o(t) + \smashoperator{\sum_{k \in \text{i.t.}}}r_k(t). 
    \label{eq:scheduling_scalarevolQ}
\end{align}
For clarity, we reiterate that while all terms of (\ref{eq:scheduling_scalarevolQ}) are calculated for every queue, $a_{ij}(t)$ across a virtual queue will always be zero, because virtual queues do not generate ebits. Moreover, it is quite unusual for a physical pair to have incoming transitions, but not impossible: it may happen in some topologies such as the $ABC$ triangle with $AB$ as an Alice-Bob pair and $ACB$ as service route. In this edge case, transition $A[C]B$ is incoming for a physical queue. 

Conversely, it should be stressed that the loss term $\ell_{ij}(t)$ is calculated in the same way for all queues, because ebit storage is always handled by memories at the network nodes.

\subsubsection{Vector Formulation}
A description of the whole system requires $|\tilde{\mathcal{E}}|$ equations like (\ref{eq:scheduling_scalarevolQ}), ushering a natural transition to a model built with matrices and vectors. 

The first vector terms are $\mathbf{q}(t)$, $\mathbf{a}(t)$ and $\boldsymbol{\ell}(t)$, whose $N_{\text{queues}}$ entries correspond to the individual $q_{ij}(t)$, $a_{ij}(t)$ and $\ell_{ij}(t)$ values (the ordering is irrelevant as long as it is consistent). Moreover, since the effect of swapping on the queues is linear, it is possible to describe it by introducing the vector $\mathbf{r}(t)$, which has $N_{\text{transitions}}$ elements --- and a matrix $\mathbf{M}$ with $N_\text{queues}$ rows and $N_\text{transitions}$ columns to translate the transition rates into their net effect on queues. 

The $\mathbf{r}(t)$ vector embodies the scheduling decision and it is a mere list of all the $r_{i[j]k}$ terms, while the $\mathbf{M}$ matrix introduces an efficient encoding of the network topology and routes.
For each of its columns, associated to transition $i[j]k$, the $\mathbf{M}$ matrix has $-1$ on the rows associated to queues $(i,j)$ and $(j,k)$, and $+1$ on the row associated to queue $(i,k)$. All other terms are zero. The $\mathbf{M}$ matrix of a small network is given in tab. \ref{tab:scheduling_Mexample} as an example in order to provide the reader with intuition on how it is built. It should be remarked that in practice, the $\mathbf{M}$ matrix is easily and automatically generated from the network routes.
Having defined all the components, system-wide queue evolution can be restated as the following simple linear equation:
\begin{align}
\mathbf{q}(t+1)  & = \mathbf{q}(t) - \boldsymbol\ell(t) + \mathbf{a}(t) + \mathbf{Mr}(t).
\label{eq:VectorialEvolutionOfq}
\end{align}

\begin{table}[tb]
    \centering
    \caption[$\mathbf{M}$ matrix for the $4$-nodes chain network.]{$\mathbf{M}$ matrix for the linear $ABCD$ network.}
    \begin{tabular}{l | c c c c}
    & ${A[B]C}$ & ${B[C]D}$ & ${A[B]D}$ & ${A[C]D}$\\\hline
    $AB$ & $-1$ & $\phm0$ & $-1$ & $\phm0$ \\
    $BC$ & $-1$ & $-1$ & $\phm0$ & $\phm0$ \\ 
    $CD$ & $\phm0$ & $-1$ & $\phm0$ & $-1$ \\
    $AC$ & $+1$ & $\phm0$ & $\phm0$ & $-1$ \\
    $BD$ & $\phm0$ & $+1$ & $-1$ & $\phm0$ \\
    $AD$ & $\phm0$ & $\phm0$ & $+1$ & $+1$ \\
    \end{tabular}
    \label{tab:scheduling_Mexample}
    \vspace{-0.5cm}
\end{table}
Looking at table \ref{tab:scheduling_Mexample}, notice that, as this work only involves bipartite entanglement, all columns of $M$ have two $-1$ terms and one $1$. 
The fact that only bipartite entanglement is considered entails that every column of $M$ sums to -1, i.e. every swap operation has the net effect of removing one pair from the system.

\subsubsection{Ebit Consumption}
Up to now, the scheduler can freely swap pairs in the network but there is no mechanism for users to employ the received pairs. The missing piece of the puzzle for ebit queues is consumption: whenever there is availability of entangled pairs across one of the final $(\mathit{Alice}_n,\mathit{Bob}_n)$ pairs, the scheduler must be able to use the available pairs to serve requests, i.e. consume the distributed resource. This is implemented in the model by extending the matrix $\mathbf{M}$ through concatenation of a negative identity block to obtain $\tilde{\mathbf{M}} = \left[ \mathbf{M} \middle|-\vmathbb{I}_{N_{\text{queues}}}  \right]$, and the $\mathbf{r}(t)$  vector to have $N_{\text{transitions}} + N_{\text{queues}}$ components, as shown in tab. \ref{tab:scheduling_Mtildeexample}. 
\begin{table}[h]
    \centering
    \caption[$\tilde{\mathbf{M}}$ matrix for the $4$-nodes chain network.]{$\tilde{\mathbf{M}}$ matrix for the linear $ABCD$ network.}
    \begin{tabular}{l | c c c c c c c c c c}
    & ${A[B]C}$ & ${B[C]D}$ & ${A[B]D}$ & ${A[C]D}$ & $AB$ & $BC$ & $CD$ & $AC$ & $BD$ & $AD$ \\\hline
    $AB$ & $-1$ & $\phm0$ & $-1$ & $\phm0$ & $-1$ & $\phm0$ & $\phm0$ & $\phm0$ & $\phm0$ & $\phm0$ \\
    $BC$ & $-1$ & $-1$ & $\phm0$ & $\phm0$ & $\phm0$ & $-1$ & $\phm0$ & $\phm0$ & $\phm0$ & $\phm0$ \\ 
    $CD$ & $\phm0$ & $-1$ & $\phm0$ & $-1$ & $\phm0$ & $\phm0$ & $-1$ & $\phm0$ & $\phm0$ & $\phm0$ \\
    $AC$ & $+1$ & $\phm0$ & $\phm0$ & $-1$ & $\phm0$ & $\phm0$ & $\phm0$ & $-1$ & $\phm0$ & $\phm0$ \\
    $BD$ & $\phm0$ & $+1$ & $-1$ & $\phm0$ & $\phm0$ & $\phm0$ & $\phm0$ & $\phm0$ & $-1$ & $\phm0$ \\
    $AD$ & $\phm0$ & $\phm0$ & $+1$ & $+1$ & $\phm0$ & $\phm0$ & $\phm0$ & $\phm0$ & $\phm0$ & $-1$ \\
    \end{tabular}
    \label{tab:scheduling_Mtildeexample}
\end{table}

What this extension achieves is to have a set of new transitions that only remove one pair from a given queue, modeling actual consumption of the distributed pair by the users. Extending $\mathbf{M}$ to $\tilde{\mathbf{M}}$ empowers the scheduler but also adds a new facet to the decision problem: if a given queue has $n$ pairs inside, the scheduler not only needs to balance swapping and storage for future use, it might also have to account for direct consumption of some of the available ebits. 

Putting all the terms together, the vector of ebit queues evolves as:
\begin{align}
\mathbf{q}(t+1)  & = \mathbf{q}(t) - \boldsymbol\ell(t) + \mathbf{a}(t) + \mathbf{\tilde{M}r}(t).
\label{vectorialevolQwithMtilde}
\end{align}

\subsection{Demand Queues}
\label{sec_sub_DemandQueues}
The ultimate purpose of a communication network is to serve the requests that users issue. Therefore, we need to include in our discussion a mechanism that allows to keep track of user demand: at any given time, every $(\mathit{Alice}_n, \mathit{Bob}_n)$ pair will issue a given number of demands $b_{(\mathit{Alice}_n, \mathit{Bob}_n)}$ and store them in a backlog called the \emph{demand queue}. Every time a direct consumption operation is scheduled and a pair is consumed along link $ij$, a demand is contextually removed from the demand queue of link $ij$. This physically corresponds to the users measuring their qubits and consuming one ebit to realize the specific application they are implementing.

Thus, it becomes natural to introduce another set of queues to describe the evolution of demands.  
Similarly to ebits, demands arriving to the system and being held for future service are modeled through queues: alongside every ebit queue, there exists a demand queue $d_{ij}(t)$ that keeps track of the number of user-issued requests (as introduced in \cite{DaiSwitchProtocols} for a single switch and generalized here for an arbitrary topology). At each time step, every demand queue $d_{ij}(t)$ receives $b_{ij}(t)$ demands, which for simplicity and generality are again modeled as a Poisson process with constant average value $\beta_{ij}$ (as in the case of ebit generation, this term may be interpreted as an open interface to more refined traffic patterns). To maintain the model's uniformity, all edges belonging to $\tilde{\mathcal{E}}$ have a demand queue, but only the ones that are associated to an $(\mathit{Alice}_n, \mathit{Bob}_n)$ pair have nonzero arrivals. For all the other links, $b_{ij}(t) = 0\,\forall\,t$.

Demand queues have a simpler evolution than ebit queues, since a demand is only a request for one ebit to be distributed across a given $(\mathit{Alice}, \mathit{Bob})$ pair: demands enter their queues when they are received and exit when they are served. Demand service can be naturally controlled by the $ij$ terms of the $\mathbf{r}(t)$ vector, i.e.\@ the same terms that control ebit consumption. We therefore introduce the matrix $\tilde{\mathbf{N}} =  \left[\vmathbb{0}_{N_{\text{queues}}\times N_{\text{transitions}}} \middle| -\vmathbb{I}_{N_{\text{queues}}}\right]$ as a mean of interfacing with the consumption part of the $\mathbf{r}(t)$ vector without being affected by the scheduling one, which is irrelevant to demand queues. An example of the $\tilde{\mathbf{N}}$ matrix is shown for clarity in tab. \ref{tab:scheduling_Ntildeexample}.
\begin{table}[h]
    \centering
    \caption[The $\tilde{\mathbf{N}}$ matrix for the $4$-nodes chain network.]{$\tilde{\mathbf{N}}$ matrix for the linear $ABCD$ network.}
    \begin{tabular}{l | c c c c c c c c c c}
    & ${A[B]C}$ & ${B[C]D}$ & ${A[B]D}$ & ${A[C]D}$ & $AB$ & $BC$ & $CD$ & $AC$ & $BD$ & $AD$ \\\hline
    $AB$ & $\phm0$ & $\phm0$ & $\phm0$ & $\phm0$ & $-1$ & $\phm0$ & $\phm0$ & $\phm0$ & $\phm0$ & $\phm0$ \\
    $BC$ & $\phm0$ & $\phm0$ & $\phm0$ & $\phm0$ & $\phm0$ & $-1$ & $\phm0$ & $\phm0$ & $\phm0$ & $\phm0$ \\ 
    $CD$ & $\phm0$ & $\phm0$ & $\phm0$ & $\phm0$ & $\phm0$ & $\phm0$ & $-1$ & $\phm0$ & $\phm0$ & $\phm0$ \\
    $AC$ & $\phm0$ & $\phm0$ & $\phm0$ & $\phm0$ & $\phm0$ & $\phm0$ & $\phm0$ & $-1$ & $\phm0$ & $\phm0$ \\
    $BD$ & $\phm0$ & $\phm0$ & $\phm0$ & $\phm0$ & $\phm0$ & $\phm0$ & $\phm0$ & $\phm0$ & $-1$ & $\phm0$ \\
    $AD$ & $\phm0$ & $\phm0$ & $\phm0$ & $\phm0$ & $\phm0$ & $\phm0$ & $\phm0$ & $\phm0$ & $\phm0$ & $-1$ \\
    \end{tabular}
    \label{tab:scheduling_Ntildeexample}
\end{table}

Demand evolution may therefore be stated as:
\begin{align}
\mathbf{d}(t+1) &= \mathbf{d}(t) + \mathbf{b}(t) + \mathbf{\tilde{N}r}(t).
\label{vectorialevolD}
\end{align}
By construction, the last $N_\text{queues}$ components of the $\mathbf{r}(t)$ vector regulate both demand and ebit consumption: one demand always consumes one ebit.

\section{Scheduling Policies}
\label{sec:policies}
\subsection{General Overview}
\label{sec:sched_intro}
After introducing all the components of the model, we move to describing scheduling policies and how they can be tested through our tools. 
We first outline what a scheduling policy is in the context of our work and follow up with subsections dedicated to three categories of scheduling policies:
\begin{itemize}
    \item A greedy scheduler, i.e.\@ a policy that performs random swaps whenever pairs are available and is used as a lower bound for performance;
    \item Quadratic schedulers, mathematically derived and expected to be the best performing category;
    \item Max Weight schedulers, obtained by slightly modifying the quadratic schedulers to obtain comparable performance at lower computational cost.
\end{itemize}
We define a \textit{Scheduling Policy} as any arbitrary set of rules that at every time step $t$ takes as its input some degree of information about the network state and returns a scheduling decision $\mathbf{r}(t)$, i.e.\@ a scheduling vector as defined in the previous section. 

We first subdivide policies according to their localization degree: in distributed policies, the nodes themselves determine the operations to perform; in global ones, the system features a physical scheduler to which all the nodes communicate their status information and receive orders from. It is moreover possible to categorize policies in terms of information availability: we remark that in all policies that we analyze in the following we work on the assumption that $(\mathbf{q}(t),\mathbf{d}(t))$, i.e.\@ the exact state of the system at the beginning of time step $t$, is known to all parties. However, since networks are distributed systems, it may happen that some other crucial information (such as the realizations of the random processes $a_{ij}(t)$ and $\ell_{ij}(t)$ for faraway queues) is not available or outdated when the scheduling decision is taken. To describe this phenomenon, we define three information availability levels, through which our schedulers will further be categorized:\begin{itemize}
    \item \textbf{Full Information (FI)}: the scheduler has access to all information about the network. It instantly knows the outcome of every random process in the network and can take the most informed scheduling decisions without having to rely on expectations;
    \item \textbf{Partial Information (PI)}: the scheduler only has access to our $(\mathbf{q}(t),\mathbf{d}(t))$ assumption, but no information about the generation and loss processes. Decisions are calculated by estimating the queues' state through average values (very low requirements in terms of additional communication infrastructure);
    \item \textbf{Local Information (LI)}: a blend between FI and PI exclusive to distributed scheduling policies. In LI conditions, each node blends together partial information about remote nodes and up-to-date information about its first neighbors.
\end{itemize}
The concept of imperfect information raises the question of \emph{feasibility} of a scheduling decision, which is detailed in the following section.
\subsection{Feasibility of a Scheduling Decision}
\label{sec:scheduling_infeasibility}
To start discussing feasibility, let us assume a centralized, FI scheduler. 
Let $\mathcal{I}(t)$ be the set of information accessible to the scheduler at time $t$. 
As previously discussed, an FI scheduler will have access to the information set $\mathcal{I}^\text{FI}(t) = \left\{\mathbf{q}(t),\mathbf{d}(t),\mathbf{a}(t),\boldsymbol\ell(t),\mathbf{b}(t)\right\}$, i.e. the state $(\mathbf{q}(t),\mathbf{d}(t))$ of the system at time $t$ plus the realizations of all the random quantities at play, making it so that the scheduler perfectly knows the state of the system at the end of the time step.
Other, more realistic schedulers, will only have access to a subset of this information, as we will see later.

As shown in sec.\@ \ref{sec:scheduling_sysdesc_swapping}, the net effect of a scheduling decision $\mathbf{r}(t)$ on the ebit and demand queues is given respectively by $\mathbf{\tilde{M}r}(t)$ and $\mathbf{\tilde{N}r}(t)$. We can set two bounds on the decision:
\begin{enumerate}
    \item The net number of outgoing ebits from any given queue can never exceed what is physically available:
    \begin{align}
        -\mathbf{\tilde{M}r}(t)\leq 
\mathbf{q}(t) - \boldsymbol{\ell}(t) + \mathbf{a}(t).
\label{constr_feasibilityEBIT}
    \end{align}
    \item Along a queue, the number of consumed ebits should never be higher than the demands:
    \begin{align}
    -\mathbf{\tilde{N}r}(t)\leq 
\mathbf{d}(t) + \mathbf{b}(t).
\label{constr_feasibilityDEMAND}
    \end{align}    
\end{enumerate}
We refer to those bounds as the \emph{feasibility} bounds. 

If we now suppose (as will be the case for most of the scheduling policies presented hereafter) to have incomplete access to information, one or more of the random processes' realizations become inaccessible, making it impossible to exactly formulate the feasibility bounds. Despite it still being possible to design scheduling policies that perform well while only using educated guesses based on averages, it is not possible to guarantee that their decisions at each time instant will respect (\ref{constr_feasibilityEBIT}) and (\ref{constr_feasibilityDEMAND}). 

Infeasibilities in general arise when $n$ ebits are available in a queue and $n'>n$ are scheduled out of it; they may be caused by a central scheduler relying on outdated information and scheduling more pairs than available, or by conflicts between two node-local schedulers (see sec. \ref{sec:locquad}) that try to draw from the same queue. 

Infeasibile decisions themselves do not prevent a network from operating (performing more measurements than there are available ebits simply results in failure of the excess measurements), but infeasibility that is not properly managed may entail sensible degradation of performance. Therefore, a working quantum network stack also needs a specific discipline to manage infeasible orders. 
In the context of this work, conflicting requests are managed in a random order, to mimic a real network adopting a first-come-first-serve (FCFS) discipline. 

As an example, suppose to have one ebit in queue $BC$. It may happen that a scheduling policy requests $r_{A[B]C} = 1$ and $r_{B[C]D} = 1$, two operations that feature $BC$ as a parent and therefore compete for the single available ebit. At this point, one needs to choose the discipline according to which priority is assigned: whereas this basic example may be solved by simple random selection, we illustrate in the following a more in-depth example to show the full complexity of this problem and the solution we adopt in our work. 

It could happen that the scheduler ordered to feed $q_{AC}$ through $r_{A[B]C} = 1$, exploit the new $AC$ pair in $r_{A[C]D} = 1$ and finally serve one request with $r_{AD} = 1$. Each of these operations depends on the one before it, and if the execution sequence is not respected the system will serve one less $AD$ request, possibly also wasting the intermediate links in the process and ultimately degrading performance. 

Therefore, to ensure proper priority is respected, we introduce a ranking system for swapping and consumption operations to preserve execution order. Swapping transitions and consumption orders are grouped by ranks, and the ranks are executed sequentially. All conflicts inside a rank are managed through random selection. 

\begin{figure}
\includegraphics[width=.9\linewidth]{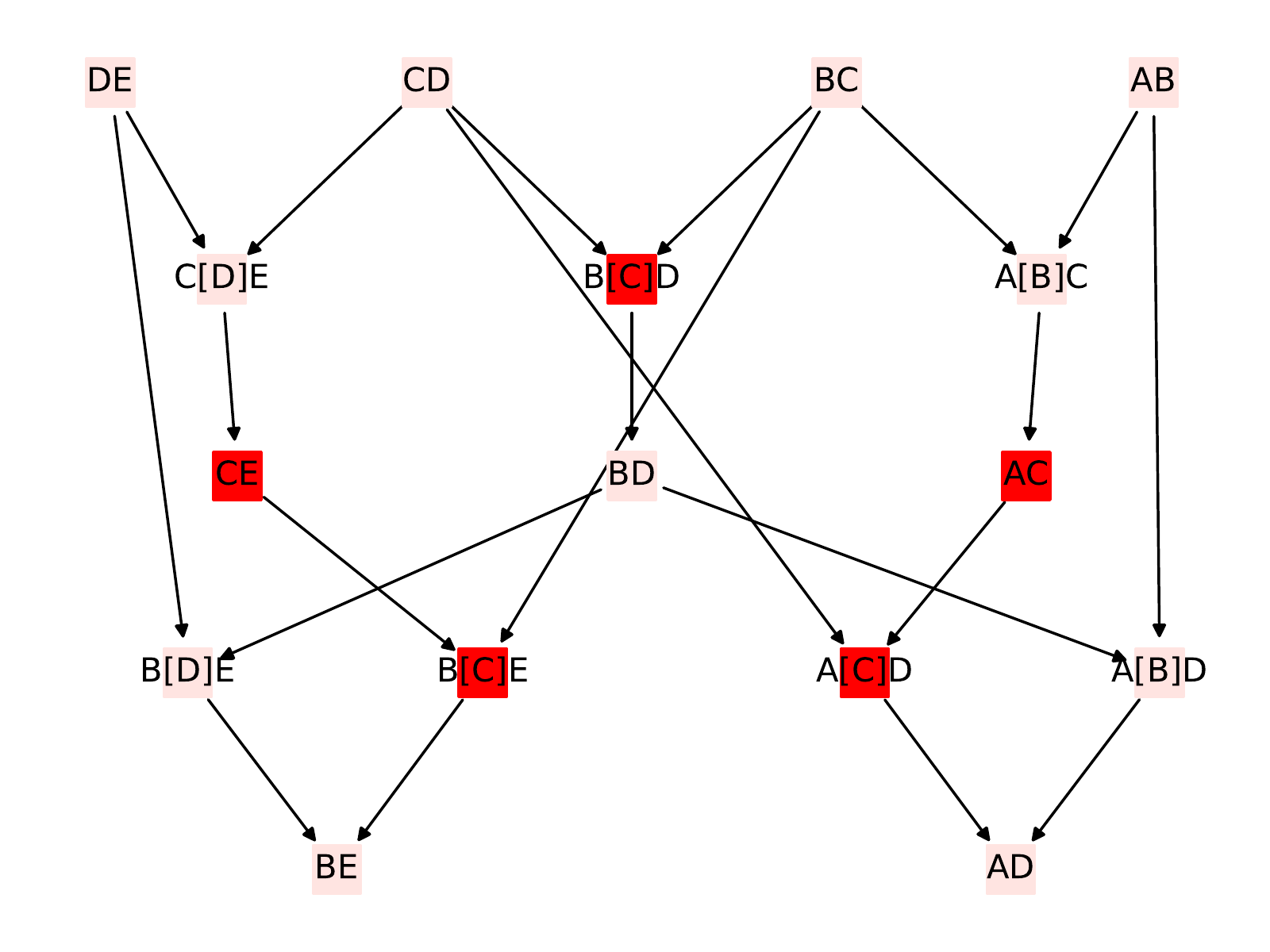}
\caption[Scheme of the rank-based conflict management system.]{The scheme of our rank system for an $ABCDE$ chain topology. Every square with only two letters inside (e.g.\@ $DE$) represents a consumption operation along a given link, while three-letter squares (e.g.\@ $C[D]E$) represent swapping transitions. A set of squares at the same height are grouped in one rank, starting from zero at the top (direct consumption from physical queues) and increasing going down. Arrows represent the ''paths`` to follow to obtain one of the final, user-requested pairs.
Focusing on the bright red squares in this scheme, which all involve node $C$ in some way, we can provide an example of how the conflict-management system works. Whenever it needs to apply a scheduling order, node $C$ will sequentially: 
\begin{enumerate*}
\item Perform transition $B[C]D$ (rank $1$) as many times as requested;
\item Satisfy consumption orders along $CE$ and $AC$ in a random order, since they are competing rank $2$ operations;
\item Perform transitions $B[C]E$ and $A[C]D$ in a random order, since they are competing rank $3$ operations.
\end{enumerate*}

As discussed in the main text, $CE$ and $AC$'s consumption orders are satisfied in a random order, but always after the upstream $B[C]D$ transition and always before the downstream $B[C]E$ and $A[C]D$ transistions.
\label{fig:RankSystem}}
\end{figure}

To form the ranks, we start by assigning rank $0$ to consumption orders from physical queues: these orders will be executed first. Secondly, swapping operations whose parents are physical queues are assigned rank $1$, so that they are executed second. After that, ranks are assigned in iterative order: for $n=0,1,2\ldots$, even ranks $2n$ are assigned to consumption orders along queues that are fed by transitions of rank at most $2n - 1$, and odd ranks $2n+1$ to transitions whose parent queues have rank at most $2n$, as depicted in fig. \ref{fig:RankSystem}. Such a system allows for a good balance between fair management of conflicting requests while preserving the sequentiality of ``ascending'' entanglement distribution operations and is also easily implemented in practice: we assume that every node knows the rank of the queues and transitions that involve it, and we envision to send, instead of a single ``apply decision'' control signal, a series of ``apply rank $n$ operations'' signals. To help intuition, an example of how the ranking system works is provided and commented in fig. \ref{fig:RankSystem}.

In the following sections we propose some examples of scheduling policies and provide details on their degree of localization and information availability.
\subsection{Greedy Scheduling}
\label{sec_sub_Greedy}
The Greedy Scheduler is a nontrivial, distributed scheduling policy that works with minimal communication between the nodes. It is a natural and immediate solution to the scheduling problem, and it is commonly found in classical network literature as a test case. Under a greedy scheduling policy, all nodes perform swapping operations as soon as they are available, regardless of user demand. When several competing operations are available, the node selects randomly. It should be noted that, although it disregards user demand, the greedy scheduler we examine is still \emph{routing-aware}: if the route $ABCD$ is to be served, the scheduler will never attempt ``downward'' transitions like $A[D]C$. 

The greedy scheduler's advantage lies in the fact that it requires no additional communication infrastructure on top of the one already employed by ebit generation and swapping, since the policy works on strictly local information. The downside to such simplicity is found in the low performance of this policy, that is only interesting in extremely simple networks or as a lower bound for other policies to beat in order to justify the additional communication overhead required. Simulation data for the greedy policy, as well as a comparison with more refined schedulers, is provided in sec. \ref{sec:scheduling_results}.

\subsection{Quadratic Scheduling}
In this section, we will derive a family of quadratic scheduling policies through the technique of Lyapunov Drift Minimization, which we briefly present before delving into its application to our case.
\subsubsection{Lyapunov Drift Minimization}
\label{sec:IntroLDM}
Lyapunov Drift Minimization (LDM) is a standard technique that is often used in classical network science to stabilize queuing systems \cite[sec.\@ 8.4]{MeynTweedieBook}. In short, choosing an arbitrary Lyapunov function and showing it satisfies certain conditions allows to infer that a queuing s system is stable. We provide in this section a demonstration of how and why LDM works, and follow up with its application to quantum networks. 
As a first step, let the Lyapunov Function $V(\mathbf{q}(t),\mathbf{d}(t)) := V(t)$ be an arbitrary, non-negative, convex $\vmathbb{N}^n\xrightarrow{}\vmathbb{R}$ function of the current state of the queues. A common choice \cite{GeorgiadisResAllocation} in network science is to use the square norm of the queue backlog vector as $V(t)$. This method entails great simplification of the analysis of highly multivariate systems, because it reduces the problem to a scalar one: when $V(t)$ is small, all the queues are small, and when it is big, at least one queue is accumulating.

After choosing a suitable Lyapunov function, the next step is to define its \textit{drift} $\Delta V(t)$ as: 
\begin{equation}
   \Delta V(t) = \EV{V(t+1) - V(t)|{\mathcal{I}(t)}}.
\end{equation}
where $\mathcal{I}(t)$ is defined as the set of available information at time t (as per \ref{sec:scheduling_infeasibility}), under the assumption that all policies presented in this work have at least knowledge of the state of the queues at time $t$ $(\mathbf{q}(t),\mathbf{d}(t))$.
Some intuition about the drift formulation can be gained by thinking of the Lyapunov function as a potential, akin to the electrical one in physics: the drift is positive if from $t$ to $t+1$ the system evolves into an higher-potential, less stable state, and negative otherwise. 
It is possible to prove \cite[sec.\@ 8.4.2]{MeynTweedieBook} that if $\Delta V(t)$ is negative on the entire state space of the system, except possibly for a compact subset of $(\mathbf{q}(t),\mathbf{d}(t))$ values, then the Markov chain describing the system is positive recurrent, i.e. the network is stable and user requests will not accumulate boundlessly. Such property is known as the Foster-Lyapunov criterion. Intuitively, the drift being positive only on a compact set means that there is a region of the state space in which the system evolves away from stability: since the drift is negative everywhere outside said region the system is always pushed back inside it, so that the Lyapunov function is never allowed to diverge. To visualize this, one may think of a charged particle in a potential well: even if it manages to exit in some way, it is eventually pushed back by the higher potential region.
In its most general form, the Foster-Lyapunov criterion can be phrased as:
\begin{equation}
    \Delta V(t) \leq -f(t) + g(t),
    \label{eq:fosterlyap}
\end{equation}
where $f$ and $g$ are two non-negative functions and the right-hand side is positive on a compact region of the state space of our system. 
Therefore, the practical goal is to find a bound for the drift and minimize it, in order to satisfy the Foster-Lyapunov criterion:
\begin{align}
    \min_{R(t)\in\mathcal{R}}{\Delta V(t)} & \leq -f(t) + g(t) 
    \label{myopicoptimization}
\end{align}
where $\mathcal{R}$ is the set of all feasible scheduling policies.

Notice that everything in our equation is defined only in terms of $t$ and $t+1$: the optimization must be repeated at every time step because of the $t$ dependence, and since the system only sees up to $t+1$ we call this process a \textit{myopic} optimization. Solving the myopic problem at every time step can be proven \cite[appendix]{GiovanidisRACH} to be a suboptimal solution to the infinite horizon Markov Decision Problem of stabilizing the network at steady state.  

\subsubsection{Application to Our Framework}
\label{subsec:QuantumLDM}
We start applying LDM to our problem by remarking that, while our model has two queuing systems, we only seek to stabilize demand queues because ebit queues play the role of a resource, and their accumulation is not an indicator of the ability of the network to serve user demand. Accumulating ebit queues merely amount to more ebits being available and more freedom to the scheduler, especially under unlimited memory assumptions. Additionally, we remark that experimental quantum networks will have a finite number of quantum memory slots at every node, enforcing a hard upper bound on $\mathbf{q}(t)$.

Since LDM will involve working with arbitrary vectors $\mathbb{N}^n$ unconstrained by the available network resources, we generalize our definition of $\mathbf{d}(t)$ to avoid unphysical results:
\begin{align}
\mathbf{d}(t+1) &= (\mathbf{d}(t) + \mathbf{b}(t) + \mathbf{\tilde{N}r}(t))^+,
\end{align}
$(\cdot)^+$ being a shorthand for $\max{(\cdot,0)}$. This is a failsafe measure that prevents the queues in our mathematical model from going negative even if an hypothetical scheduling policy prescribes more service than there are requests. 

To apply drift minimization to our case, the first step is to choose a Lyapunov function. As is commonly done in classical networks, we opt for the square norm of the queue backlog:
\begin{align}
    V(t) = \frac{1}{2}\mathbf{d}^T(t)\mathbf{d}(t).
\end{align}
From there, we obtain the drift:
\begin{align}
    \Delta V = \frac{1}{2}\EV{\mathbf{d}^T(t+1)\mathbf{d}(t+1) - \mathbf{d}^T(t)\mathbf{d}(t)\big|\mathcal{I}(t)}
    \label{eq:drift_I}
\end{align}
If we let $\mathbf{d}(t)+\mathbf{b}(t)=\tilde{\mathbf{d}}(t)$ and note that $\left[\max{(x,0)}\right]^2 \leq x^2$ we can bound the drift as: 

\begin{align}
{
    \frac{1}{2}\EV{\mathbf{d}^T(t+1)\mathbf{d}(t+1) - \mathbf{d}^T(t)\mathbf{d}(t)\big\lvert\mathcal{I}(t)} \leq }\\
    {
    \leq \frac{1}{2}\EV{(\mathbf{\tilde d}(t)+\mathbf{\tilde{N}r}(t))^T(\mathbf{\tilde d}(t)+\mathbf{\tilde{N}r}(t)) - \mathbf{d}^T(t)\mathbf{d}(t)\big\lvert\mathcal{I}(t)} = }\\
    {
    = \frac{1}{2}\left[\EV{\mathbf{\tilde d}^T(t)\tilde{\mathbf{d}}(t)\big\lvert\mathcal{I}(t)} - \mathbf{d}^T(t)\mathbf{d}(t) +
    U(\mathbf{r}(\mathcal{I}(t)),t)\right],
  }
    \label{eq:objderivation}
\end{align}
where
\begin{align}
{
  U(\mathbf{r}(\mathcal{I}(t)),t):=}&\ {2\EV{\mathbf{\tilde d}(t)\big\lvert\mathcal{I}(t)}^T\mathbf{\tilde{N}r}(\mathcal{I}(t))\ +} \nonumber \\ & \hspace{1cm} {+ \mathbf{r}^T(\mathcal{I}(t))\tilde{\mathbf{N}}^T\mathbf{\tilde N r}(\mathcal{I}(t)).
}
\end{align}
We could pull $\mathbf{d}(t)$ out of the expectation because it is fully determined by $\mathcal{I}(t)$ ($\mathbf{d}(t) \in \mathcal{I}(t)$ for all schedulers by assumption). Furthermore, we have chosen to enforce that the scheduling policies we consider are deterministic and their decisions are completely determined by $\mathcal{I}(t)$, allowing us to also pull $\mathbf{r}(t)\equiv\mathbf{r}(\mathcal{I}(t))$ out of the expectation.\\
We chose to make the dependence of $U$ on both $t$ and $\mathbf{r}(\mathcal{I}(t))$ explicit to highlight the role of the scheduling decision: whereas stochastic quantities directly depend on $t$, $\mathbf{r}(\mathcal{I}(t))$ behaves as a control parameter: starting from $\mathcal{I}(t)$, the scheduler must tune $\mathbf{r}(\mathcal{I}(t))$ in order to make the controllable part of the drift as negative as possible.
Notice that choosing $\mathbf{r}(\mathcal{I}(t))=0$ leads to $U(\mathbf{r}(\mathcal{I}(t)),t) = 0$: therefore, either the optimal scheduling decision is to take no action, or there is an optimal decision that makes $U$ negative, playing the role of $-f(t)$ in the Foster-Lyapunov criterion (see (\ref{eq:fosterlyap})).
\subsubsection{Quadratic Policies}
\paragraph{Full Information Quadratic Scheduler}
The derivation presented in the previous section yielded an expression that has a direct effect on stability: the more negative $U(\mathbf{r}(\mathcal{I}(t)),t)$ is, the stabler the network. In other words, the task of a scheduler in this context is to choose at every time step a decision $\mathbf{r}(t)$ such that $U(\mathbf{r}(\mathcal{I}(t)),t)$ is as negative as possible.\\
The natural tool to solve this problem is optimization. Let us assume, as an initial ideal case, to be in Full Information conditions. All information about the network state is available, and it is therefore possible to drop the expectation from $U(\mathbf{r}(\mathcal{I}(t)),t)$ since $\mathcal{I}^\text{FI}(t) = \left\{\mathbf{q}(t),\mathbf{d}(t),\mathbf{a}(t),\boldsymbol\ell(t),\mathbf{b}(t)\right\}$, i.e. the realizations of all random variables are exactly known. Using $U(\mathbf{r}(\mathcal{I}(t)),t)$ as an objective function, we formulate a central scheduling policy that at each time step solves the following quadratic integer program:
\begin{align}
\begin{cases}
\min{\mathbf{w}^{\text{FI}}}(t) \cdot \mathbf{r}(t) + \frac{1}{2}\mathbf{r}(t)^T\mathbf{\tilde N}^T\mathbf{\tilde Nr}(t)\\
\text{s.t. } \mathbf{r}(t)\in{\mathcal{R}^{\text{FI}}(t)}
\end{cases}
\label{quadschedprob}
\end{align}
with weights
\begin{align}
{
\mathbf{w}^{\text{FI}}(t)} = (\mathbf{d}(t) + \mathbf{b}(t))^T\tilde{\mathbf{N}}).
\label{w_fullinfo}
\end{align}

Since we assumed complete information availability, we can use as constraints the feasibility conditions mentioned in sec. \ref{sec:scheduling_infeasibility} ($d$ being a shorthand for the dimension of $\mathbf{r}(t)$):
\begin{align} 
    {\mathcal{R}^{\text{FI}}(t)} = \left\{\mathbf{r}(t)\in\mathbb{N}^d
| -\mathbf{\tilde{M}r}(t)\leq 
\mathbf{q}(t) - \boldsymbol{\ell}(t) + \mathbf{a}(t) \land -\mathbf{\tilde{N}r}(t)\leq 
\mathbf{d}(t) + \mathbf{b}(t)\right\}
\label{quadconstrFULL}
\end{align}
This constraint set binds the system so that, along every queue:
\begin{itemize}
    \item No more outgoing transitions are scheduled than there are stored ebits;
    \item No more ebits are consumed than there is demand.
\end{itemize}
Solving this problem at every time step will yield a scheduling decision $\mathbf{r}(t)$ that relies on the complete information set, even though such a policy carries a crucial flaw that hinders its experimental realizability: since this is a centralized policy, there must be a physical scheduling block that acts as an authority; all the nodes in the network submit local status information and receive a scheduling decision to apply. In the time it takes for the information to reach the scheduling agent and for the decision to be relayed back to the nodes and applied, the physical layer of the network has continued stochastically generating and losing ebits, so that when the decision finally arrives it is based on outdated information. Two possible solutions to this problem are addressed in the following, in the form of two policies that rely on less information being available. 
\paragraph{Partial Information Quadratic Scheduler}
One solution to the stale information problem detailed in the previous section could be to replace all unavailable information with sensible expectation values and thus implement a partially informed quadratic scheduler. We assume to be in Partial Information conditions, meaning that for each queue the scheduler has access to: \begin{itemize}
    \item The average arrival rate $\alpha$;
    \item The loss parameter $\eta$;
    \item The average demand rate $\beta$;
    \item The system state ($\mathbf{q}(t),\mathbf{d}(t)$) \emph{at the beginning} of each time step.
\end{itemize}
Therefore, in PI conditions, $\mathcal{I}^\text{PI}(t)=\{\mathbf{q}(t),\mathbf{d}(t),\alpha,\beta,\eta\}$.
This information set relaxes the requirements because the network can take a snapshot of its state at the beginning of each time step and exploit the leftover time to communicate it to the scheduler. The scheduler will in turn use average parameters to build an expectation for the queues' state at the end of the time step and take its decision based on that. Note that if these requirements are still too tight, it is always possible to formulate a policy that knows the exact state of the system with $n$ time steps of delay, or even hybrid localized policies where every node knows the state of the surrounding queues with a delay that depends on their physical distance. 

To formulate our partially informed policy, we re-use the (\ref{quadschedprob}) problem, replacing every quantity for which information is not available with an expected value. Of course, such a heuristic modification degrades the performance of the scheduling policy.
To rely only on information contained in $\mathcal{I}^\text{PI}(t)$, we change the weights of the problem to

\begin{align}
{\mathbf{w}^\text{PI}(t)} = \EV{(\mathbf{d}(t) + \mathbf{b}(t))|{\mathcal{I}^\text{PI}(t)}}^T\mathbf{\tilde N} = (\mathbf{d}(t) + \beta\mathbf{1}_{\dim{(\mathbf{d})}})^T\mathbf{\tilde N},
\label{w_partinfo}
\end{align}

where $\mathbf{1}_{\dim{(\mathbf{d})}}$ is the vector of all ones with appropriate dimension, and the constraints to

\begin{align}
{\mathcal{R}^\text{PI}(t)} = \left\{\mathbf{r}(t)\in\mathbb{N}^d
\ \middle|\right. & \left.-\tilde{\mathbf{M}}\mathbf{r}(t)\leq 
\EV{\mathbf{q}(t) - \boldsymbol\ell(t) + \mathbf{a}(t)| {\mathcal{I}^\text{PI}(t)}},\ \right. \nonumber\\ 
&\left. -\tilde{\mathbf{N}}\mathbf{r}(t)\leq 
   \EV{\mathbf{d}(t) + \mathbf{b}(t)|{\mathcal{I}^\text{PI}(t)}}\right\},
\end{align}
which in practice reads:
\begin{align}
{\mathcal{R}^\text{PI}(t)} = \left\{\mathbf{r}(t)\in\mathbb{N}^d
\ \middle|\ \right.&\left.-\tilde{\mathbf{M}}\mathbf{r}(t)\leq \eta\mathbf{q}(t) + \alpha\mathbf{1}_{\dim{(\mathbf{d})}}\right., \nonumber\\ 
&\left.-\tilde{\mathbf{N}}\mathbf{r}(t)\leq 
   \mathbf{d}(t) + \beta\mathbf{1}_{\dim{(\mathbf{d})}}\right\}.
\label{quadconstrPART}
\end{align}
This class of partially informed policies still outperforms greedy ones but removes the stale information problem.

It should be stressed that, since this policy relies on a heuristic guess made using averages, it is not guaranteed that its decisions will satisfy the feasibility conditions (conflicts are managed as shown in sec. \ref{sec:scheduling_infeasibility}).

The performance of this policy is reviewed in sec. \ref{sec:scheduling_results}.
\paragraph{Local Information Quadratic Scheduler}
\label{sec:locquad}
As mentioned before, information availability is one of the main points to consider when choosing a scheduling policy: a well-designed policy must be able to take sensible decisions while leveraging the available information to the best extent possible.

Following this idea, we propose a distributed, optimization-based original policy and subsequently benchmark it to assess its expected performance.

Since we are describing a distributed policy, we shift our point of view to that of a node in the network (distributed policy): we assume that every node $i$ in the network has access to all relevant average values, which can be communicated before the network is booted or measured in a rolling average fashion. 

Additionally, let node $i$ have access to the queue state of the full network at the start of each time step ($\mathbf{q}(t),\mathbf{d}(t)$), where the same remarks we gave in the previous section apply. 

Finally, due to how entanglement generation and swapping are implemented, node $i$ should have access to how many qubits are stored in its memory slots and with whom they are entangled, which means that node $i$ also knows exact arrivals and exact losses for all the queues connected to it, both physical and virtual, and can exploit this additional information when taking a scheduling decision. This set of available information has been defined in sec. \ref{sec:sched_intro} as Local Information.

To formalize this, let $\mathcal{C}_i$ be the set of queues connected to node $i$, i.e. the set of edges $e$ in the extended set $\tilde{\mathcal{E}}$ such that $e$ is connected to node $i$. Using this concept, we can define a node-local version of the information set $\mathcal{I}^\text{LI}_i(t)$ which contains the entirety of the information available to node $i$:
\begin{equation}
{\mathcal{I}^\text{LI}_i(t)} = \left\{\mathbf{q}(t),\mathbf{d}(t),\eta,\beta,\alpha,a_{e}(t),\ell_{e}(t),b_{e}(t),\ \forall e\in\mathcal{C}_i\right\}\nonumber,
\label{infosetLOC}
\end{equation}

where $a_{e}(t),\ell_{e}(t)$ and $b_{e}(t)$ correspond to the additional local exact information that is unique to each node.

Instead of phrasing a global optimization problem, node $i$ may now formulate an individual problem and solve it to obtain a strictly local scheduling decision to apply directly, without waiting for a discrete scheduler to send back a decision. To do so, the node builds all the relevant quantities (backlogs, arrivals, losses) with exact information from the queues it is connected to and expectation values from the other queues. The $i$-localized quadratic integer program can thus be written as:

\begin{align}
\begin{cases}
\min {\mathbf{w}^{\text{LI}}_{i}(t)} \cdot \mathbf{r}(t) + \frac{1}{2}\mathbf{r}^T(t)\mathbf{\tilde N}^T\mathbf{\tilde N}\mathbf{r}(t) \\
s.t. \ \mathbf{r}(t)\in{\mathcal{R}^\text{LI}_{i}(t)}
\end{cases}
\end{align}
where the weights are given by
\begin{equation}
{\mathbf{w}^{\text{LI}}_{i}(t)} = \EV{\mathbf{d}(t) + \mathbf{b}(t)|{\mathcal{I}^\text{LI}_i(t)}}^T\mathbf{\tilde N},
\label{quadweightsLOC}
\end{equation}

In accordance with its previous definition, the set $\mathcal{R}^\text{LI}_{i}(t)$ of all possible scheduling decisions $\mathbf{r}(t)$ at time slot $t$ localised at node $i$ is defined as:
\begin{align}
{\mathcal{R}^\text{LI}_{i}(t)} = & \left\{\mathbf{r}(t)\in\mathbb{N}^d
\ \middle|\right.\nonumber\\
&\left.-\mathbf{\tilde M}\mathbf{r}(t)\leq 
\EV{\mathbf{q}(t) - \boldsymbol\ell(t) + \mathbf{a}(t)| {\mathcal{I}^\text{LI}_i(t)}},\ \right. \nonumber\\ 
&\left. -\mathbf{\tilde N}\mathbf{r}(t)\leq 
   \EV{\mathbf{d}(t) + \mathbf{b}(t)|{\mathcal{I}^\text{LI}_i(t)}}\right\}.
\label{quadconstrLOC}
\end{align}
In practice, each individual expected value in the weights and constraints' expressions will locally resolve to a form similar to (\ref{quadconstrFULL})  and (\ref{w_fullinfo}) (i.e.\@ all exact values) for queues that are connected to node $i$ and to (\ref{quadconstrPART}) and (\ref{quadweightsLOC}) (all averages) for queues that are not.
As an example, node $A$ will be able to formulate a problem that includes the constraint $-\mathbf{\tilde M}_{AB,} \cdot \mathbf{r}(t) \leq q_{AB}(t) - \ell_{AB}(t) + a_{AB}(t)$ (where $\mathbf{\tilde M}_{AB,}$ is row $AB$ of $\mathbf{\tilde M}$) because queue $AB$ is directly connected to it, but will have to resort to $-\mathbf{\tilde M}_{CD,} \cdot \mathbf{r}(t) \leq \eta q_{CD}(t) + \alpha$ for queue $CD$, because it has no up-to-date information about it. In the simple $ABCD$ chain topology, node $B$ (connected to $A$ and $C$ but not $D$) will formulate its constraints as (substituting the relevant entries from (\ref{quadconstrFULL}) and (\ref{quadconstrPART}) into (\ref{quadconstrLOC})):
\begin{multline}
\mathcal{R}^\text{LI}_{i}(t) = \left\{ \vphantom{ \begin{bmatrix}
        q_{AB}(t)\\q_{BC}(t)\\ \eta q_{CD}(t)\\ \eta q_{AC}(t)\\ q_{BD}(t)\\ \eta q_{AD}(t)\\
    \end{bmatrix}} \mathbf{r}(t)\in\mathbb{N}^d \middle|
    -\mathbf{\tilde M}\mathbf{r}(t)  \leq \right. 
    \begin{bmatrix}
        q_{AB}(t)\\q_{BC}(t)\\ \eta q_{CD}(t)\\ \eta q_{AC}(t)\\ q_{BD}(t)\\ \eta q_{AD}(t)\\
    \end{bmatrix} + 
    \begin{bmatrix}
        a_{AB}(t)\\a_{BC}(t)\\ \alpha\\ \alpha\\ a_{BD}(t)\\ \alpha\\
    \end{bmatrix} - 
    \begin{bmatrix}
        \ell_{AB}(t)\\\ell_{BC}(t)\\ 0\\ 0\\ \ell_{BD}(t)\\ 0\\
    \end{bmatrix},  \\
    -\mathbf{\tilde N}\mathbf{r}(t) \leq 
    \left.
    \begin{bmatrix}
        d_{AB}(t)\\d_{BC}(t)\\ d_{CD}(t)\\ d_{AC}(t)\\ d_{BD}(t)\\ d_{AD}(t)\\
    \end{bmatrix} +
    \begin{bmatrix}
        b_{AB}(t)\\b_{BC}(t)\\ \beta\\ \beta\\ b_{BD}(t)\\ \beta\\
    \end{bmatrix}
    \right\}.
\end{multline}
The locally informed quadratic scheduler provides a practically implementable alternative to the globally informed policy while still retaining good enough performance. We remark once again that, whereas the centralized fully informed method came from abstract mathematical arguments, this scheduler was modified and is thus partially heuristic. Therefore, it is reasonable to expect some degree of performance degradation: one of the tasks of our analysis is to characterize this margin of degradation in order to gauge how close a distributed scheduler can get to its centralized, idealistic variant. 

\subsection{Max Weight Scheduling}
\label{sec_sub_mwsched}
The quadratic policies that have been detailed in the previous section are valid solutions to the scheduling problem in quantum networks. However, situations might arise in which computational complexity is a stricter constraint than network performance. To accommodate such cases, we present in this section a class of policies that perform almost as well as the quadratic ones, for a fraction of the computational cost.

Looking at the policies presented until now, we notice two interesting points: 
\begin{itemize}
    \item The objective function features a linear term that depends on queue length plus a quadratic penalty that does not;
    \item The linear terms are reminiscent of the objective function for the Max Weight \cite{TassiulasMaxWeight} policy, an extremely well-established result of classical network theory that we discussed in sec.\@ \ref{sec:intro_maxweight} (p.\@ \pageref{sec:intro_maxweight}). 
\end{itemize}
It is therefore natural to propose a class of semi-heuristic scheduling policies derived by taking our quadratic objectives and suppressing the quadratic penalty, which does not depend on the queue backlog. For brevity, we explicitly formulate only the fully informed variant of the Max Weight scheduler. The partial and local information quadratic schedulers can be turned to their linear variants following the same steps. 
The fully informed Max Weight problem is obtained by simply suppressing the quadratic term from (\ref{quadschedprob}):
\begin{align}
\begin{cases}
\min {\mathbf{w}^{\text{FI}}(t)} \cdot \mathbf{r}(t)\\
s.t. \ \mathbf{r}(t)\in{\mathcal{R}^{\text{FI}}}(t), 
\end{cases}
\end{align}
and solving it with the same weights and constraints as (\ref{quadconstrFULL}). The partial and local information policies may be similarly constructed by suppressing the quadratic term from the respective quadratic policy. 
The performance analysis for the globally, partially and locally informed linear schedulers is provided in section \ref{sec:scheduling_results}.
\subsection{Quadratic vs. Max Weight Scheduling}
\begin{figure}[h]
\centering
\subfloat[Fully informed Max-Weight;]{\includegraphics[width=.4\textwidth]{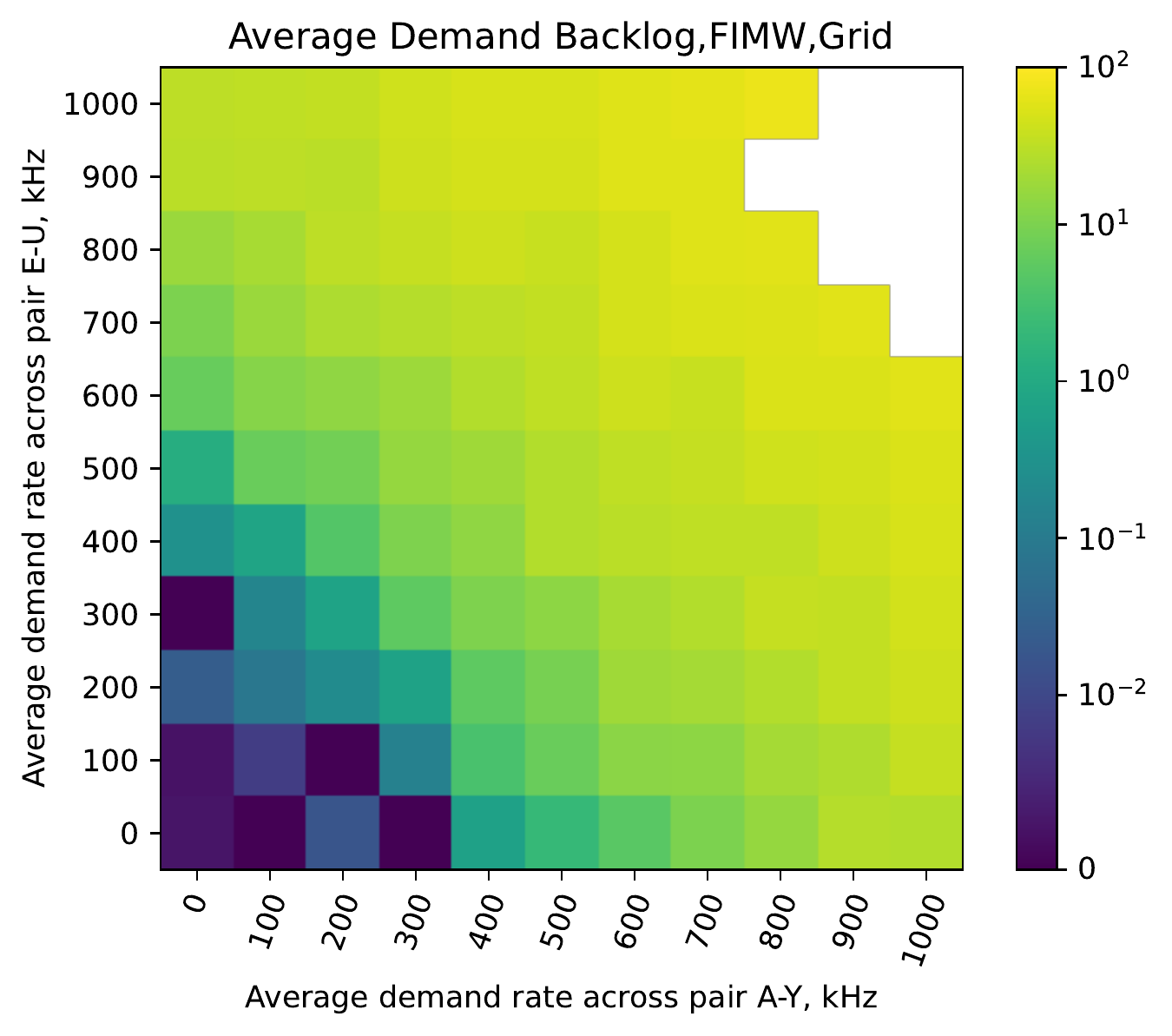}}
\subfloat[Fully informed Quadratic;]{\includegraphics[width=.4\textwidth]{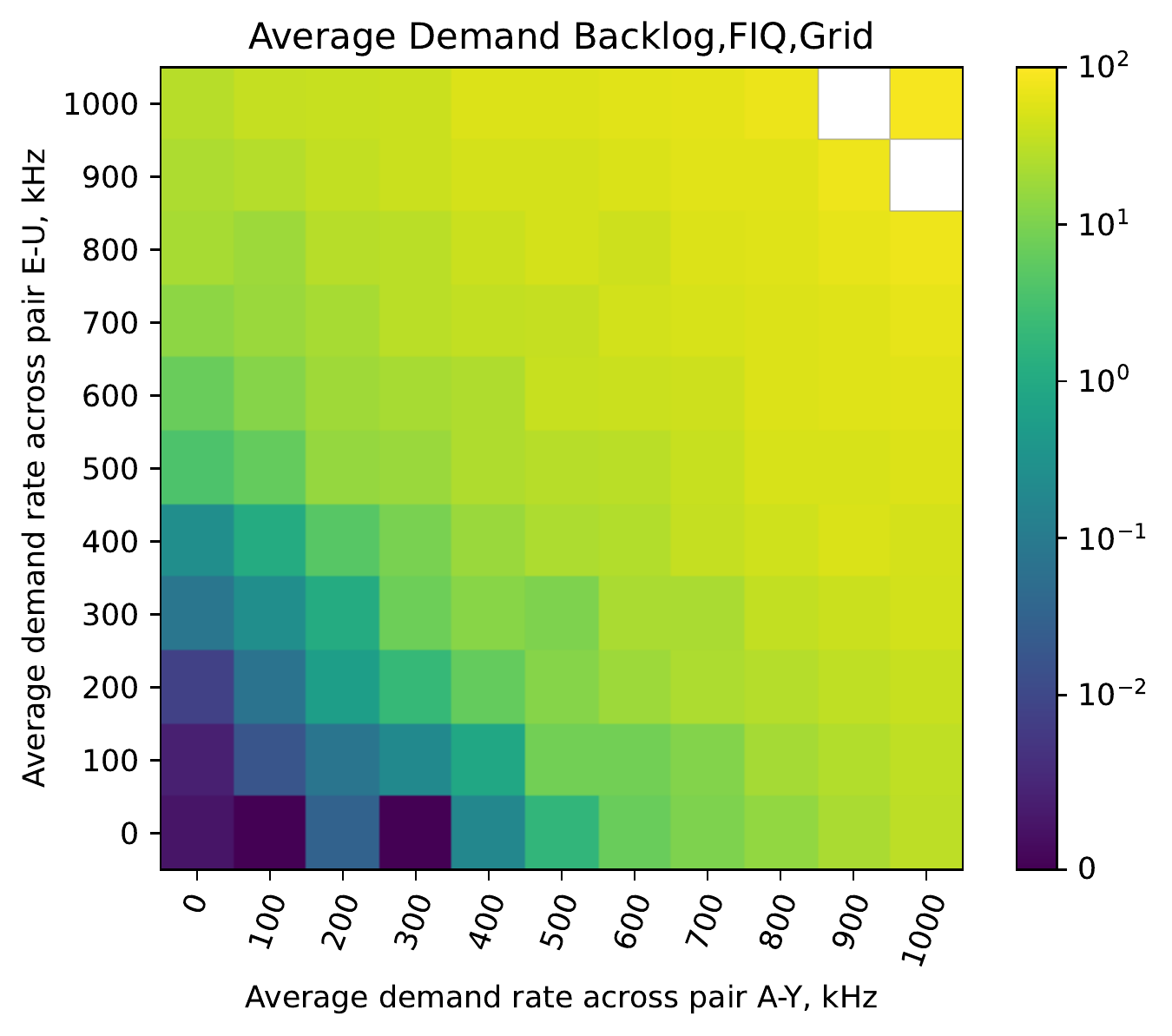}}

\subfloat[Locally informed Max-Weight;]{\includegraphics[width=.4\textwidth]{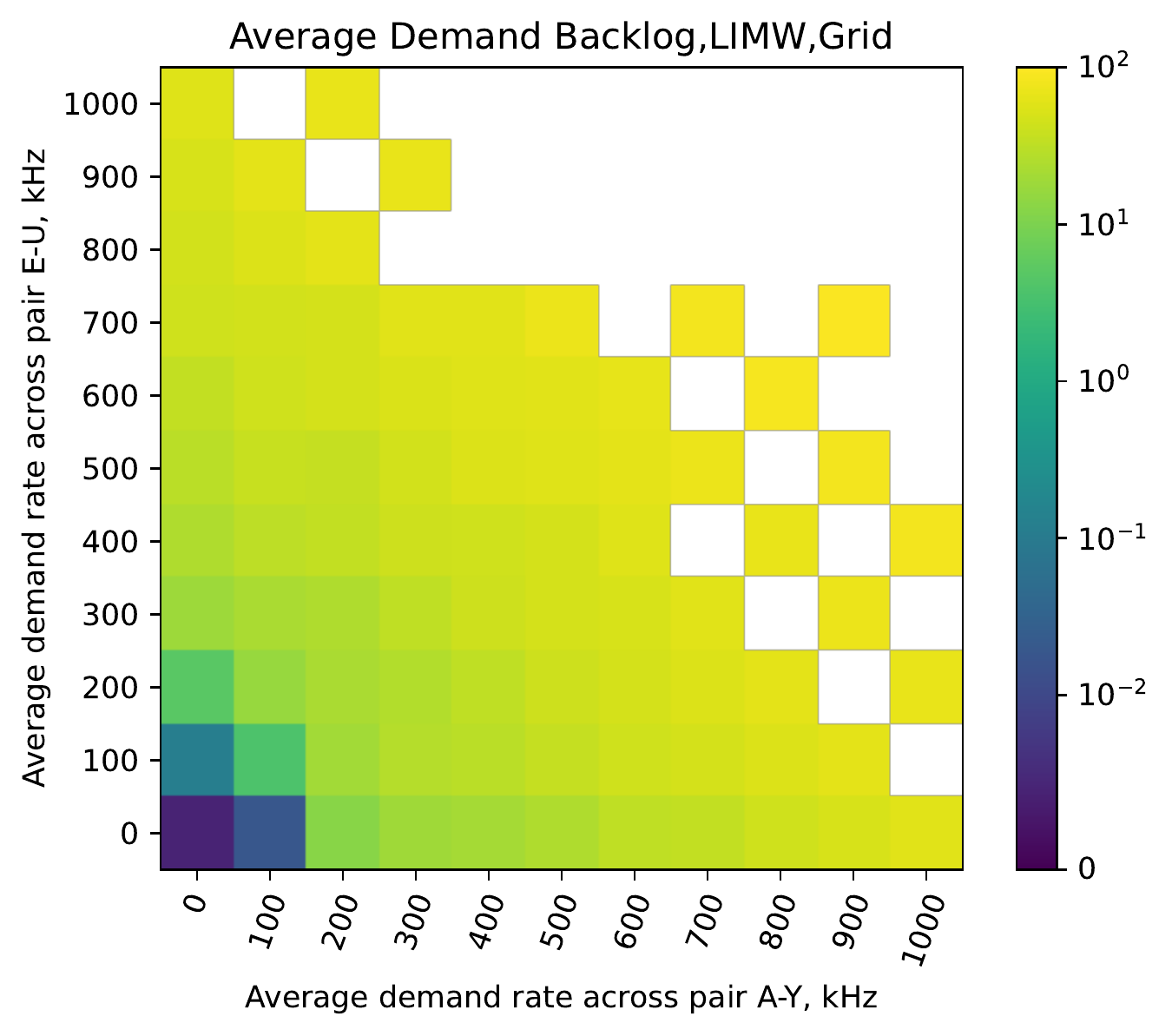}}
\subfloat[Locally informed Quadratic;]{\includegraphics[width=.4\textwidth]{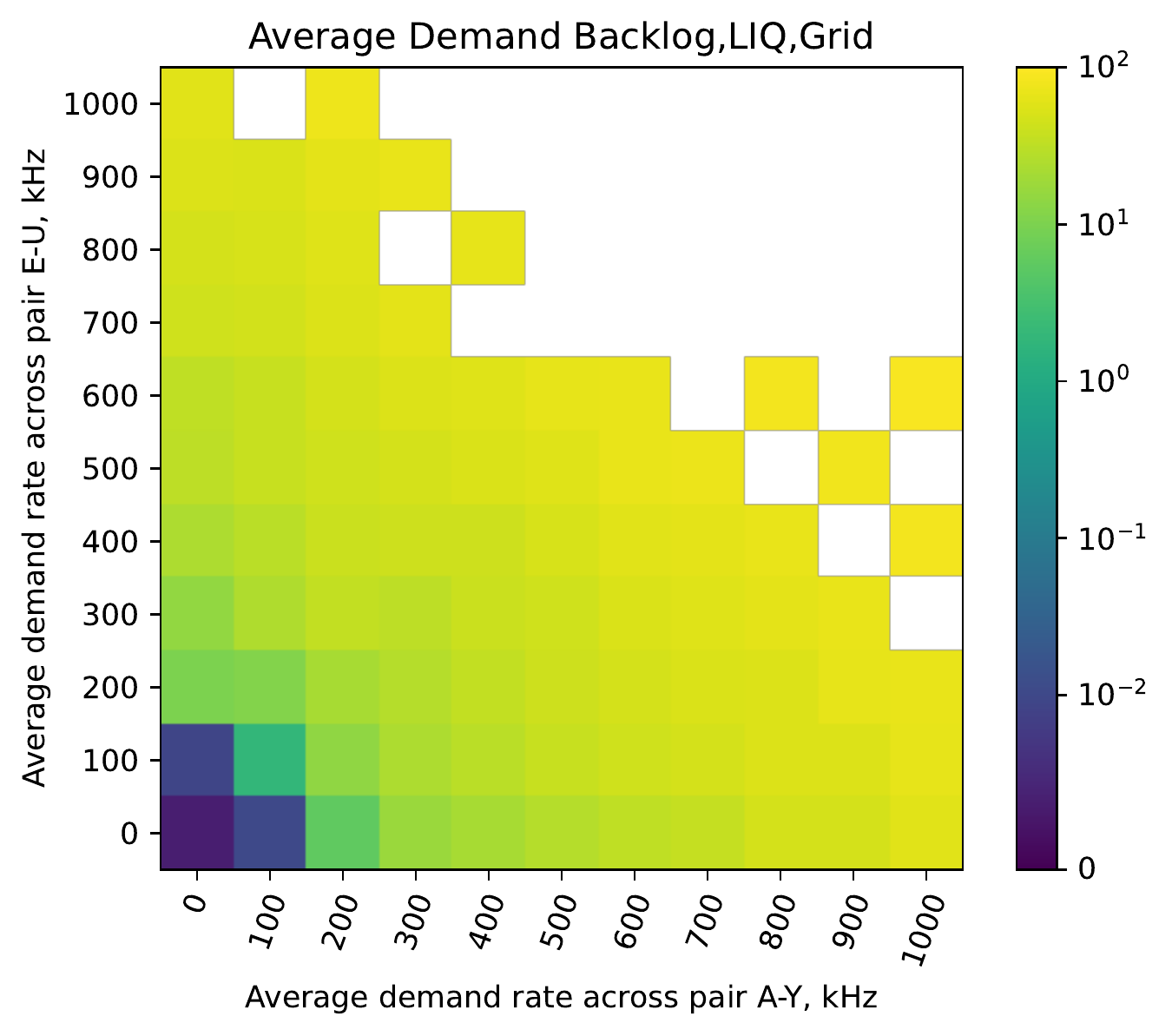}}

\subfloat[Average-only Max-Weight;]{\includegraphics[width=.4\textwidth]{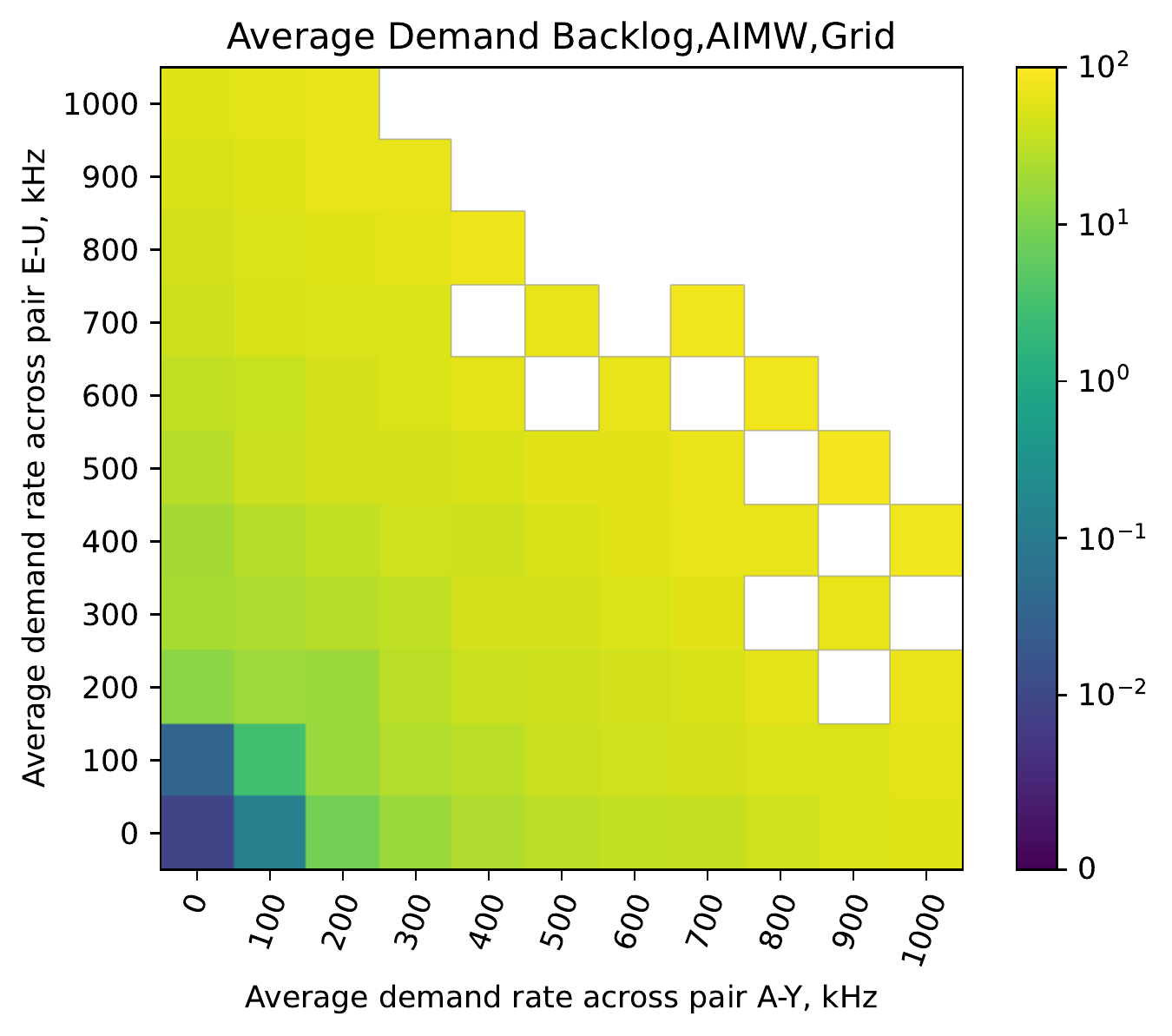}}
\subfloat[Average-only Quadratic.]{\includegraphics[width=.4\textwidth]{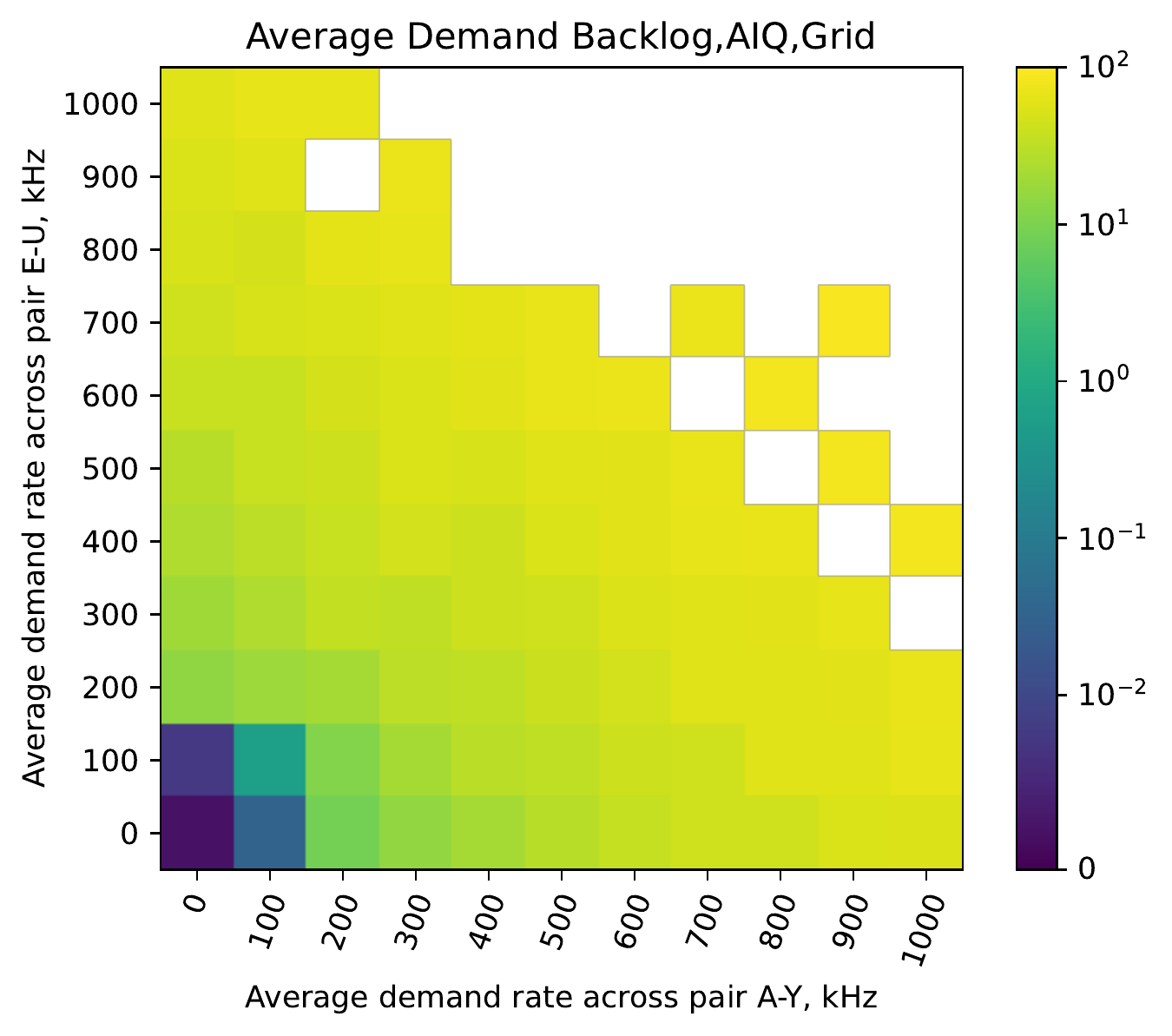}}
\caption[Comparison of the performance of the linear scheduling policies and their quadratic counterparts.]{Comparison of the performance of the linear scheduling policies and their quadratic counterparts. For brevity, we only report results for the Grid topology, while stating that the same phenomenon is observed for all topologies: the margin of performance between Max-Weight and Quadratic schedulers is almost imperceptible in our tests. This figure was calculated with eight parasitic pairs, whose average load was fixed at $100$ kHz. Analogously to the main results in fig. \ref{fig:scheduling_journal_stabgrids}, simulations were run for $1000$ time steps of $1 \mu s$, discarding the first $100$ observations to reduce the impact of transients. 
The white points were skipped by the simulator and directly deemed unstable, since one or more strictly lower-load points were found to be unstable. More details on this computational economy technique can be found in sec.\@ \ref{sec:scheduling_simulator}.}
\label{fig:scheduling_comparison_quadratic}
\end{figure}

The computational requirements of quadratic optimization are dramatically higher than those of linear optimization. Thus, it is useful before diving into a full quadratic simulation campaign to run a few pilot scenarios to assess whether the performance difference between these two scheduler classes is worth the additional computational load. In all the cases presented in this chapter, we were unable to observe any tangible performance difference justifying the additional computational costs and decided therefore to only include Greedy and Max Weight policies in all the following results. We provide in fig. \ref{fig:scheduling_comparison_quadratic} one example of such Quadratic vs.\@ Max Weight comparison as evidence to our claim.
\section{Results}
\label{sec:scheduling_results}
\subsection{Simple Example: Max Weight Scheduling over a small network}
\begin{figure}[h]
    \centering
    \includegraphics[width=0.8\linewidth]{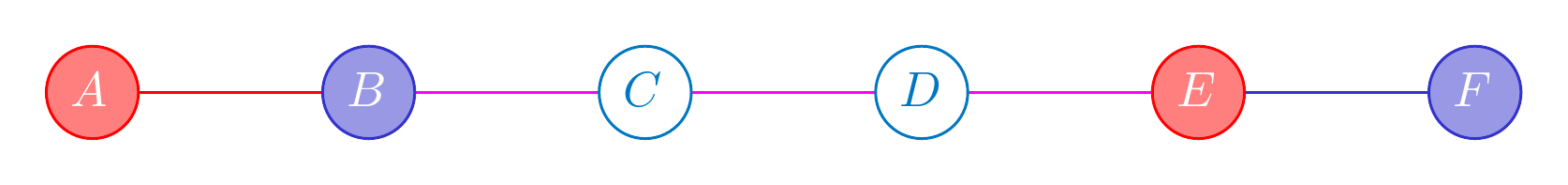}
    \caption[A simple chain topology.]{Simple chain topology. The user pairs are $AE$ (in red) and $BF$ (in blue) in order to create congestion. The purple links are relevant to both service pairs, and are therefore the resource that needs to be balanced by the scheduler.}
    \label{fig:simple6chaintopology}
\end{figure}

\begin{figure}[h]
\centering
\subfloat[]{\includegraphics[height=6.3cm]{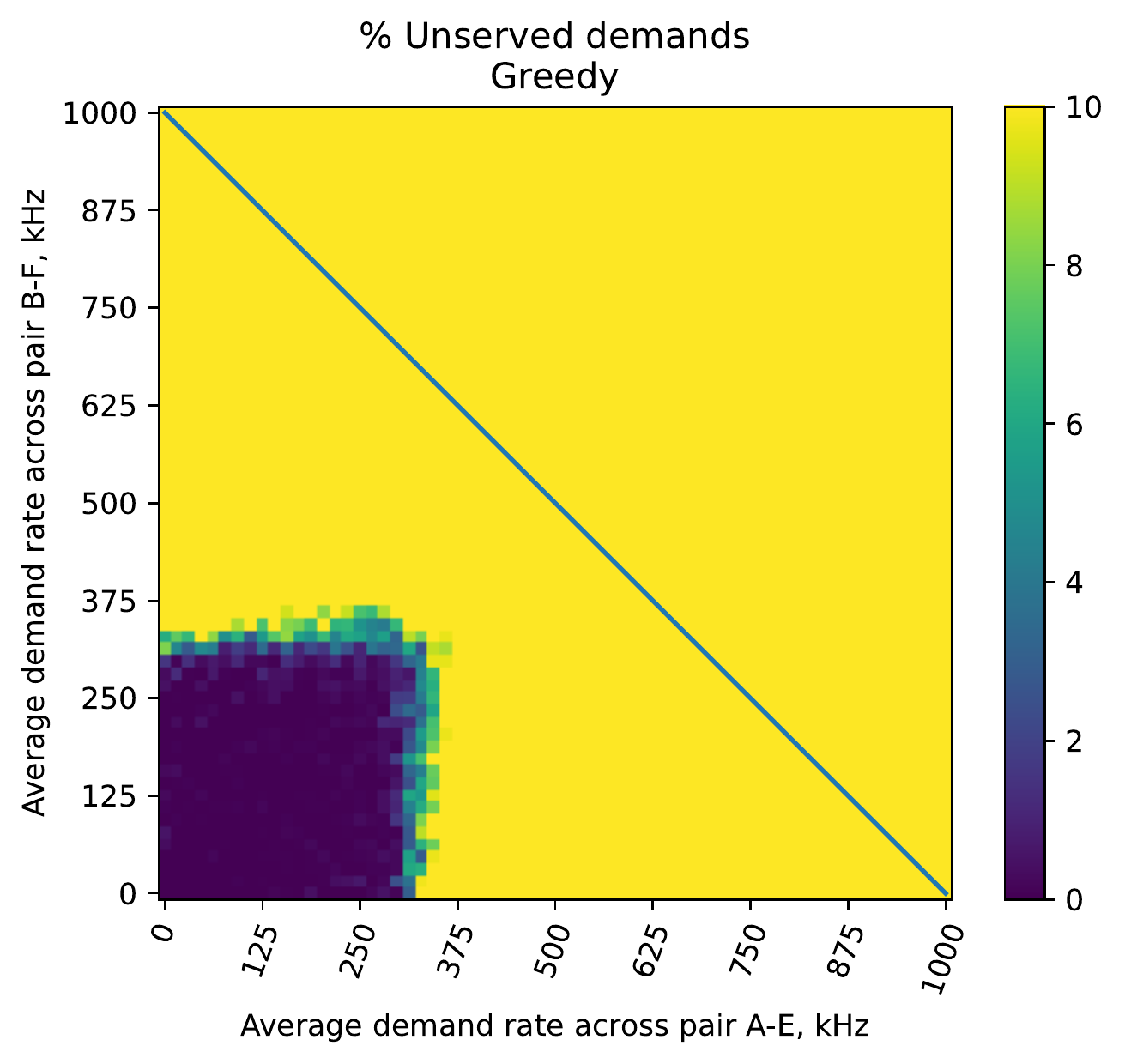}
\label{fig:scheduling_example_resGS}}
\subfloat[]{\includegraphics[height=6.3cm]{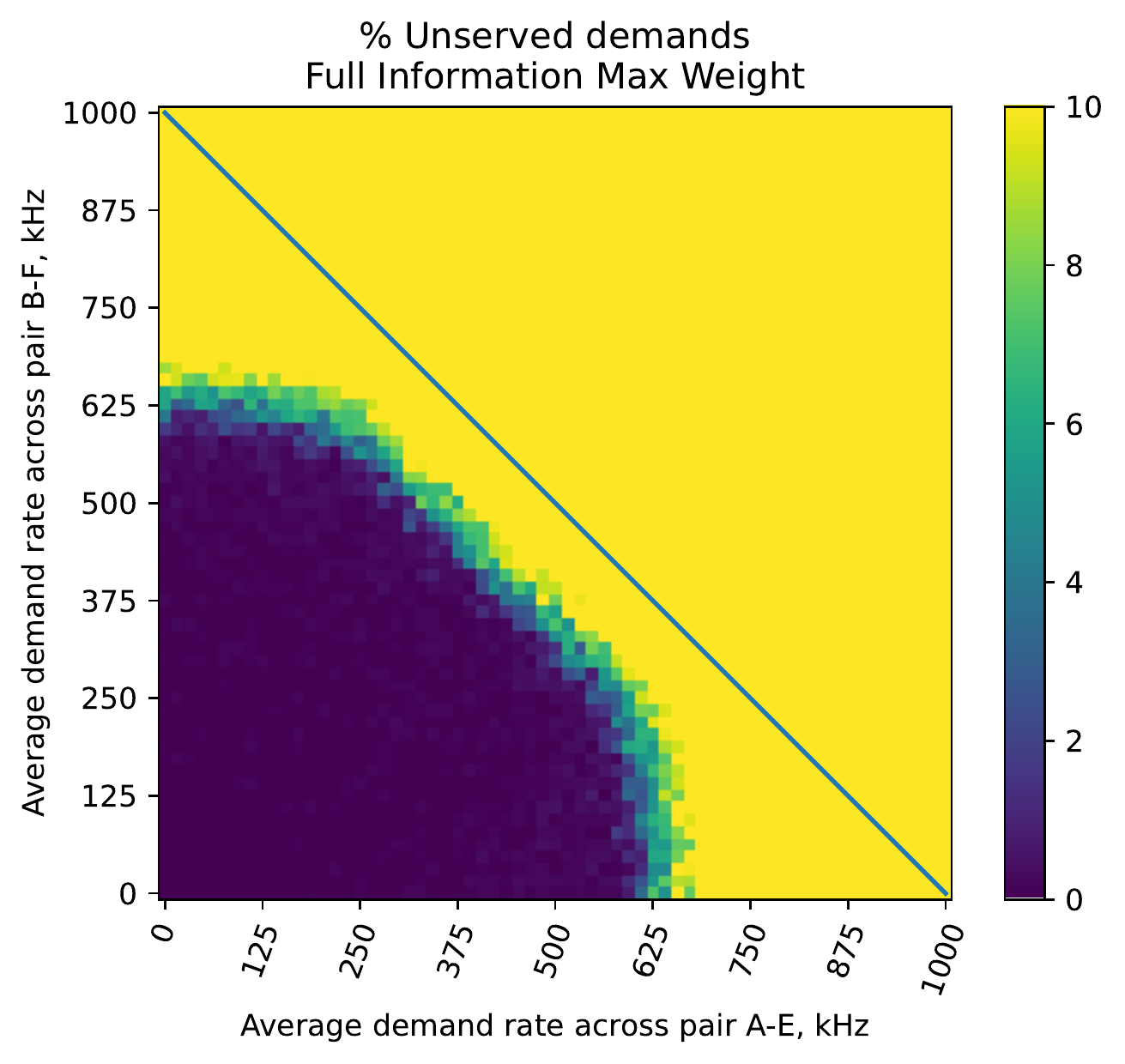}
\label{fig:scheduling_example_resFI}}

\subfloat[]{\includegraphics[height=6.3cm]{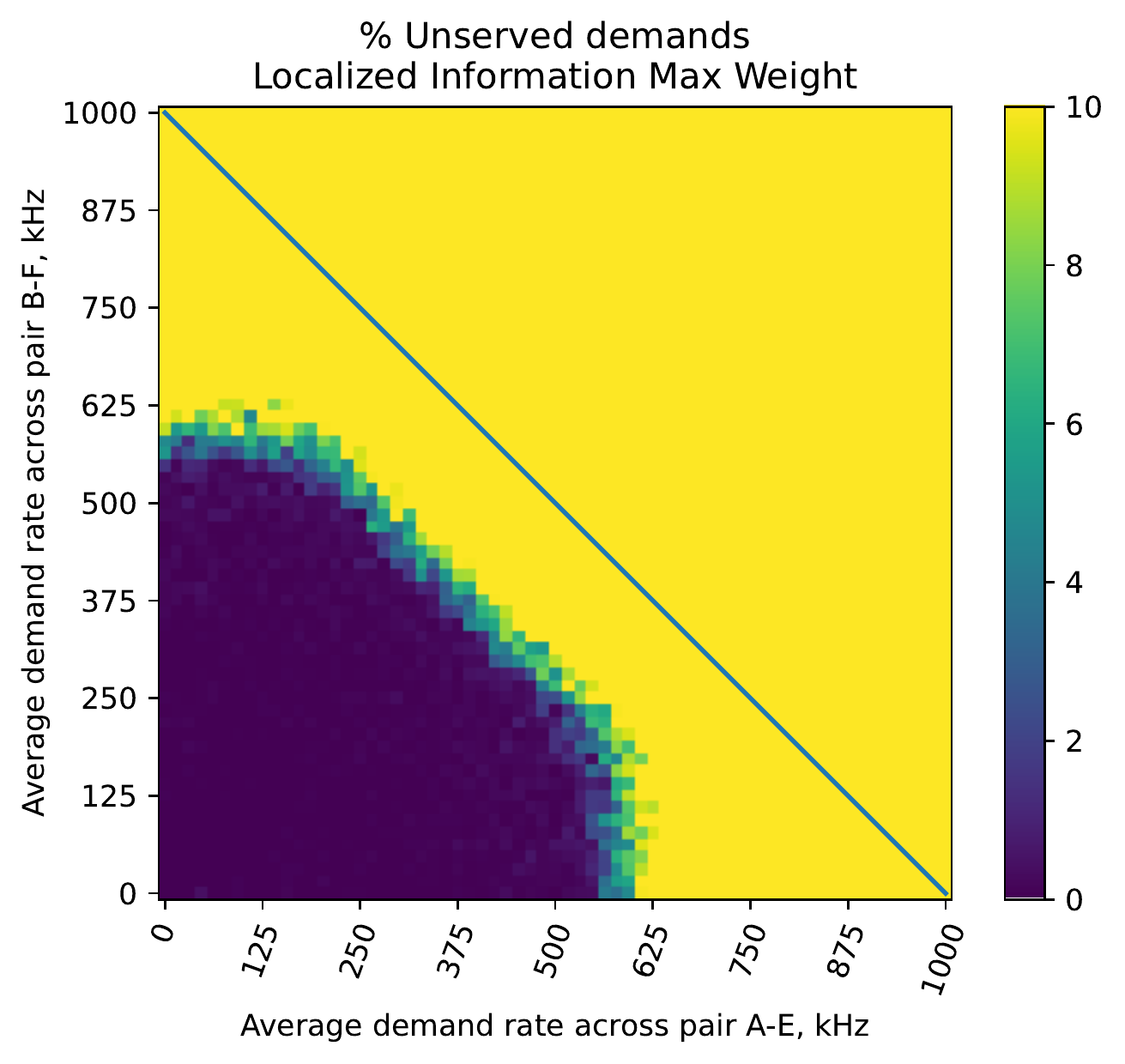}
\label{fig:scheduling_example_resLI}}
\caption[Simulation results for various quantum scheduling policies over a simple chain.]{Simulation results for our simple chain network example: percentage of unserved demands by the three schedulers. The greedy scheduler (a) exhibits a square-shaped rate region whose sides go from $0$ kHz up to $\sim 300$ kHz: this means that increasing demand across one commodity does not impair service of the other (with the current values of loss and infinite memory). The fully informed (b) and locally informed's (c) regions have a similar shape with a cut corner: the diagonal segment in the upper right corresponds to a limitation of cumulative demand, that is of the sum of the two rates. The full information can serve individual demand up to $\sim 600$ kHz, twice the performance of the greedy scheduler, and a cumulative demand of $\sim800$ kHz, while the local information can serve $\sim550$ kHz and a cumulative $\sim700$ kHz. The diagonal segment is parallel to the optimal $(\alpha,0)-(0,\alpha)$ bound (pictured), and in the lossless case it rests on it.}
\label{fig:scheduling_conferenceStabRegions}
\vspace{-.5cm}
\end{figure}

In this section, we isolate an elementary example of quantum scheduling problem and provide some simple results to demonstrate the application of our framework. This section serves both as a first result of our quantum scheduling investigation and as a template analysis before diving into more complex examples. The topology we analyze is a chain of six quantum nodes (depicted in fig. \ref{fig:simple6chaintopology}).
We choose $(A,E)$ and $(B,F)$ as the end-user pairs, which yields the service routes $ABCDE$ and $BCDEF$. The goal of the scheduling policies we analyze is to correctly manage the shared $BCDE$ bottleneck in order to maximize the stability region of this very simple network. 
\begin{table}[tb]
    \centering
    \caption{Simulation parameters for fig. \ref{fig:scheduling_conferenceStabRegions}.}
    \begin{tabularx}{0.6\textwidth}{| X | c |}
    \hline
    Number of time steps      & $100000$ \\
    \hline
    Time step duration        & $1\mu s$\\
    \hline
    Memory efficiency ($\eta$)& $0.9$ \\
    \hline
    Link-level entanglement generation rate ($\alpha$) & $1 MHz$ for all links \\
    \hline
    Ebit queues weight in scheduling calculation\paolofootnotemark\,$\gamma$ & 1\\
    \hline
    \end{tabularx}
    \label{tab:conferenceSimParams}
    \vspace{-0.5cm}
\end{table}\paolofootnotetext{At the time these results were derived, we were experimenting with including ebit queue in the calculation of the scheduling weights to make resource usage more uniform and force preventive swapping. However, we have since observed that removing ebit queues from the calculation always increases scheduling performance. The factor has thus been removed.}
The first step to our analysis will be to generate the $\tilde{\mathbf{M}}$ matrix. The process is automatically carried out by our simulator, and we report the obtained $\mathbf{M}$ (without the consumption part) for instructional purposes in tab. \ref{tab:M6Chain}. In particular, we remark the absence of the $AF$ queue and related transitions: since it is not included in any of the service routes, the corresponding entries of $\tilde{\mathbf{M}}$ are not generated\paolofootnote{One could argue that it could be useful to include $AF$ in order to serve the user pairs through transitions such as $A[F]E$ or $B[A]F$. However, since experimentally entanglement swapping is a probabilistic process and this inclusion would entail two additional swaps at minimum, we choose to leave it out for the sake of simplicity.}.
The scheduling policies that we analyze over this simple example are Greedy, Full Information Max Weight and Local Information Max Weight. Our results are presented in fig. \ref{fig:scheduling_conferenceStabRegions}, and the simulation parameters used to calculate them are summarized in tab. \ref{tab:conferenceSimParams}.
Concerning the shape of the regions reported in fig. \ref{fig:scheduling_conferenceStabRegions}, we remark that the ideal shape for a stability region in this context would be the $(0,0)-(\alpha,0)-(0,\alpha)$ triangle, meaning that all pairs coming to the $BCDE$ bottleneck are effectively employed in demand service. The maximal region can be observed by employing a Full Information Max Weight Scheduler with no losses and $\gamma=0$.\\
The results in fig. \ref{fig:scheduling_conferenceStabRegions} already showcase the difference in performance between a greedy policy and an optimization-based one. In particular, it is possible to observe a diagonal segment in the optimized scheduler's stability region which corresponds to the commodities directly competing for the available resources. This is also evidenced by the square shape of the greedy scheduler's stability region: since the greedy scheduler randomly performs swaps regardless of demand, service across one user pair is not affected by increasing demand across the other, while demand-aware schedulers try to balance service across all users. \\
Moreover, other than the noticeable improvement that comes with optimization-based policies, it is apparent how the greedy scheduler seems to be an acceptable competitor for low loads (below $\sim300$ kHz). The following sections will demonstrate how the greedy scheduler's performance rapidly becomes inadequate in more complex scenarios. 

The definition of network stability discussed in sec. \ref{sec:StabilityDef} is abstract and statistical, meaning that, strictly speaking, it is only appropriate for systems that run for an infinite amount of time. As the results presented here arise from simulations running for a finite time, it is not rigorously correct to extract stability conclusions from them and it is therefore not possible to draw a clear-cut stability boundary. 
However, some reasonable guesses can be proposed: the $(0,0)$ point, for instance, is bound to be stable because no demand is ever produced. Likewise, all points above $(0,\alpha)$ and to the right of $(\alpha,0)$ are surely unstable because there is physically not enough resource to serve them, no matter the scheduling policy. Having fixed the bounds, we select a performance metric (the percentage of unserved demands in this case, in order to observe a clearer stability boundary) and an instability threshold ($10\%$ here) and plot our chosen performance metric as a colormap to provide an estimate of the stability region. To calibrate the threshold, we verified that in the lossless case the Full Information scheduler's stability bound falls on the diagonal line between $(0,\alpha)$ and $(\alpha,0)$, i.e.\@ the highest possible stability bound. 

\subsection{Moving to Complex Results}
Although our simple example already provides interesting insight into the benefits of a well-designed scheduling policy and the corresponding benchmarking toolbox, the real interest of network analysis surfaces when moving to more complex examples featuring nontrivial topologies with multiple concurring commodities.
We propose here the analysis of our scheduling policies over four complex network topologies:
\begin{description}
    \item[-] \textbf{A $5\times5$ grid};
    \item[-] \textbf{A $6\times6$ grid} from which $\sim25\%$ of nodes have been randomly removed;
    \item[-] \textbf{An Erdős–Rényi random graph} \cite{ErdosRandom}, i.e.\@ a random graph generated by taking $N_{E-R}$ points and connecting each pair of them with a given probability $p_{E-R}$;
    \item[-] \textbf{A Watts-Strogatz random graph} \cite{WattsRandom}, i.e.\@ a random graph that starts from a ring of $N_{W-S}$ nodes, each connnected to its $K_{W-S}$ first neighbors ($K_{W-S}$ needs to be even) and then rewires each edge to a random node with probability $p_{W-S}$. Watts-Strogatz graphs are particularly relevant in network science because, unlike Erdős–Rényi graphs, they exhibit the small-world property, which means that their edges are highly clustered: a Watts-Strogatz graph is therefore more similar to a set of subnetworks connected by a backbone, which is a typical configuration for information networks.
\end{description}

Since the $\tilde{\mathbf{M}}$ generation process starts from the set of service routes, making the topology more complex alone does not guarantee an interesting network scenario. Suppose a grid is chosen as a nontrivial topology and the two user pairs placed on the four corners, each pair connected by one service route: since most of the grid is not included in the service routes, the generated $\tilde{\mathbf{M}}$ will not contain any information outside the two service routes, yielding a problem that is not much more complex than our preliminary example.

To make the problem more interesting and realistic, one can increase the number of user pairs and/or the number of service routes serving each user pair. In our examples, we adopt both solutions: we increase the number of users to ten Alice-Bob pairs and we increase the number of service routes per user pair to two (creating another degree of freedom for the scheduler, which now needs to exploit the less congested routes while easing the load on the more congested ones). This yields a problem in which ten commodities compete for resources, which implies the stability region requires ten axes to be correctly represented. To avoid this issue, we divide our ten user pairs in two groups:
\begin{itemize}
    \item $2$ \emph{main pairs}, over which a full stability sweep as detailed in the previous example will be performed. These pairs are selected by hand: \begin{itemize}
        \item the first is chosen as the pair of nodes connected by the longest shortest path $\lambda$ in the topology;
        \item The second one is selected by removing each edge in $\lambda$ with a tunable probability and taking the new longest shortest path.
    \end{itemize}
    \item $8$ \emph{parasitic pairs}, which are selected by randomly placing $16$ users and connecting them pairwise.
\end{itemize}
We set the demand across all parasitic pairs to be constant and equal to a given value $L$, while sweeping demand across the main pairs and plotting our performance metrics as was done for $AE$ and $BF$ in the simple chain example. We provide several $L$ values to better assess how the scheduling policies perform under increasing network load. 

\begin{figure}[h]
\subfloat[A complete $5 \times 5$ grid;]{
\includegraphics[width=.5\textwidth, trim =53 0 53 20, clip]{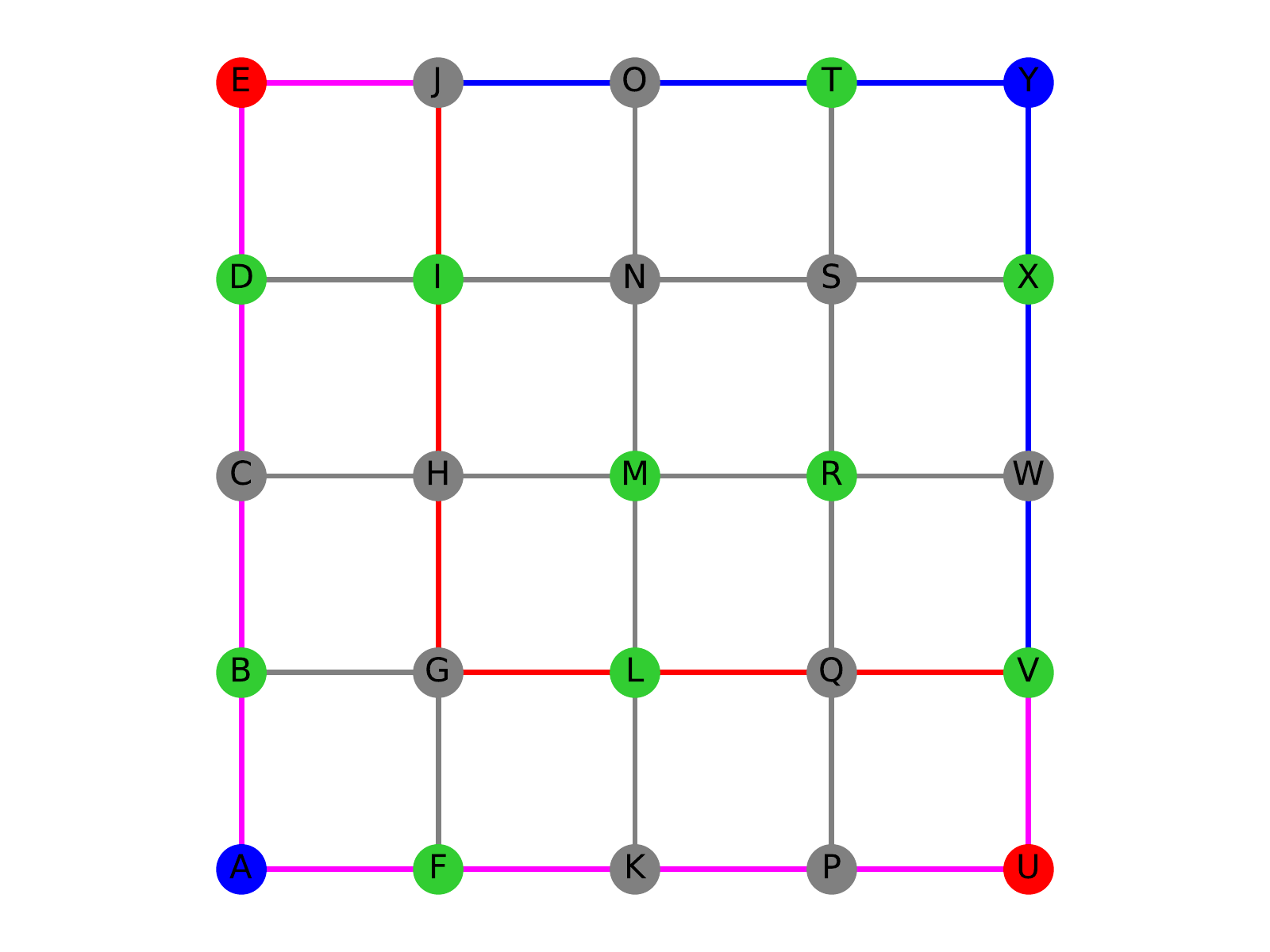}
\label{fig:gridtopo}
}
\hfill
\subfloat[A 6x6 grid whose nodes had a probability $p=0.25$ of being removed;]{\includegraphics[width=.5\textwidth]{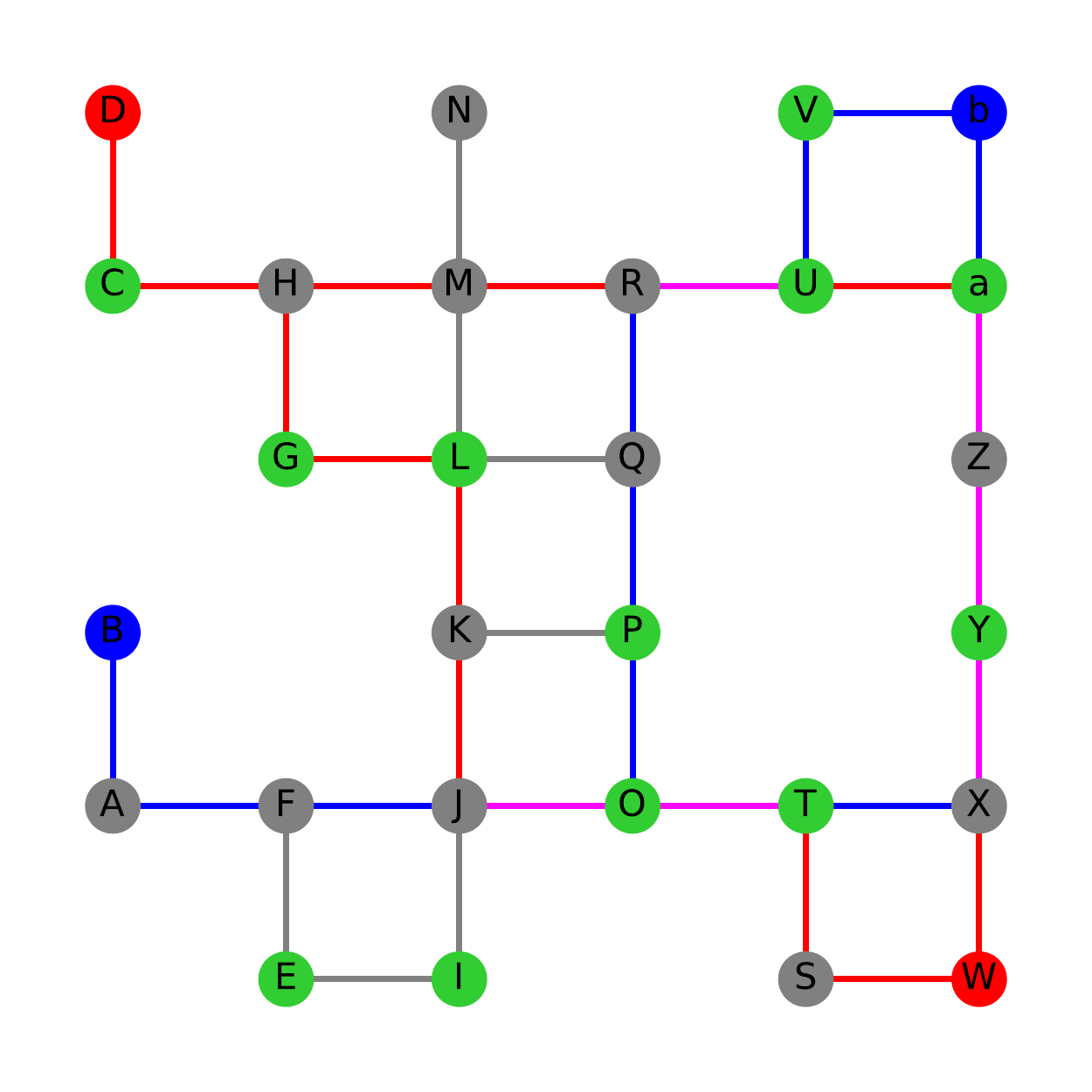}}\label{fig:pgridtopo}
\hfill
\subfloat[An Erdős–Rényi random graph, with $n=25$ and $p=0.125$;]{\includegraphics[width=.5\textwidth, trim =145 145 145 145, clip]{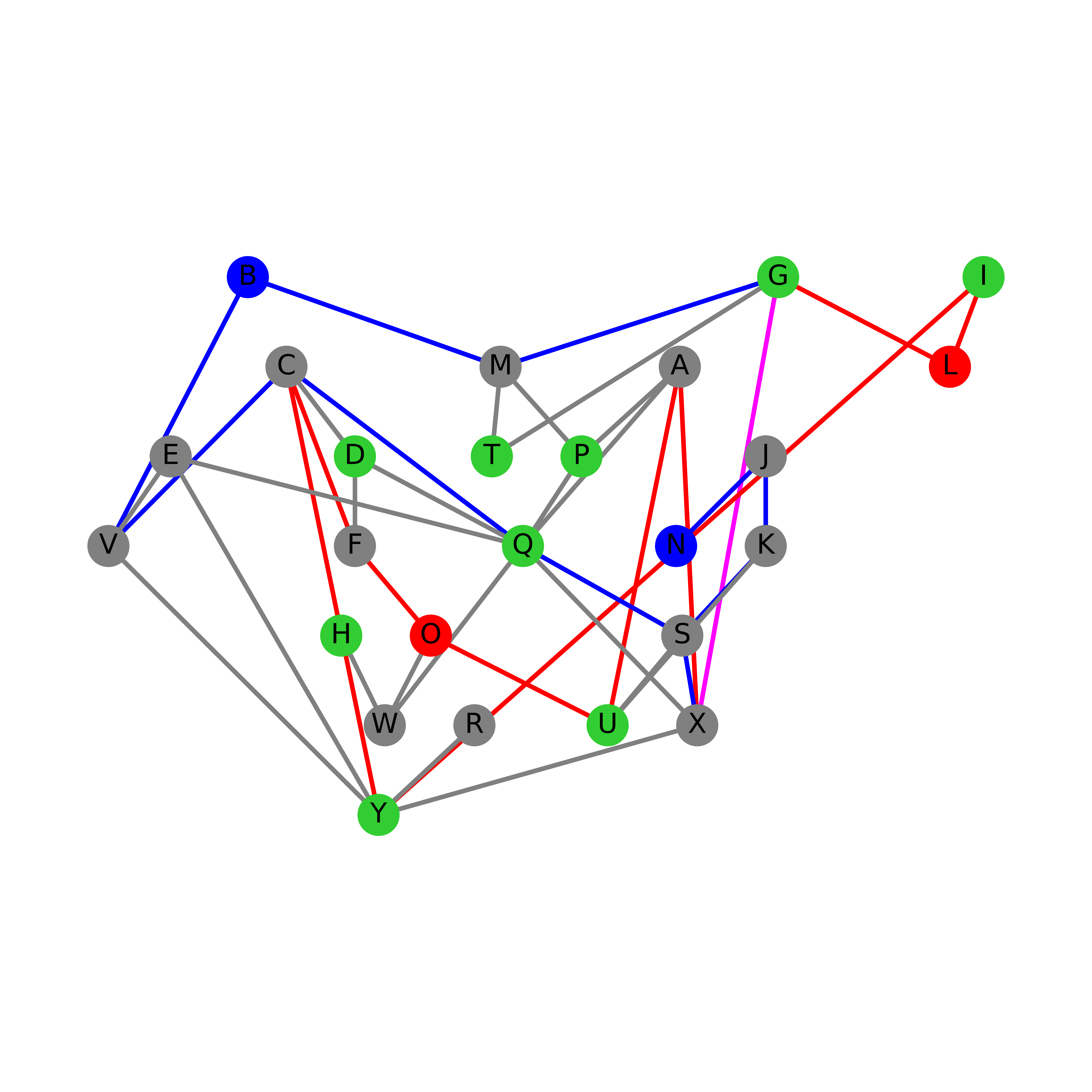}}\label{fig:ERtopo}
\hfill
\subfloat[A Watts--Strogatz random graph, with $n=25$, $n_{\text{neighbors}} = 4$ and $p=0.2$.]{\includegraphics[width=.5\textwidth]{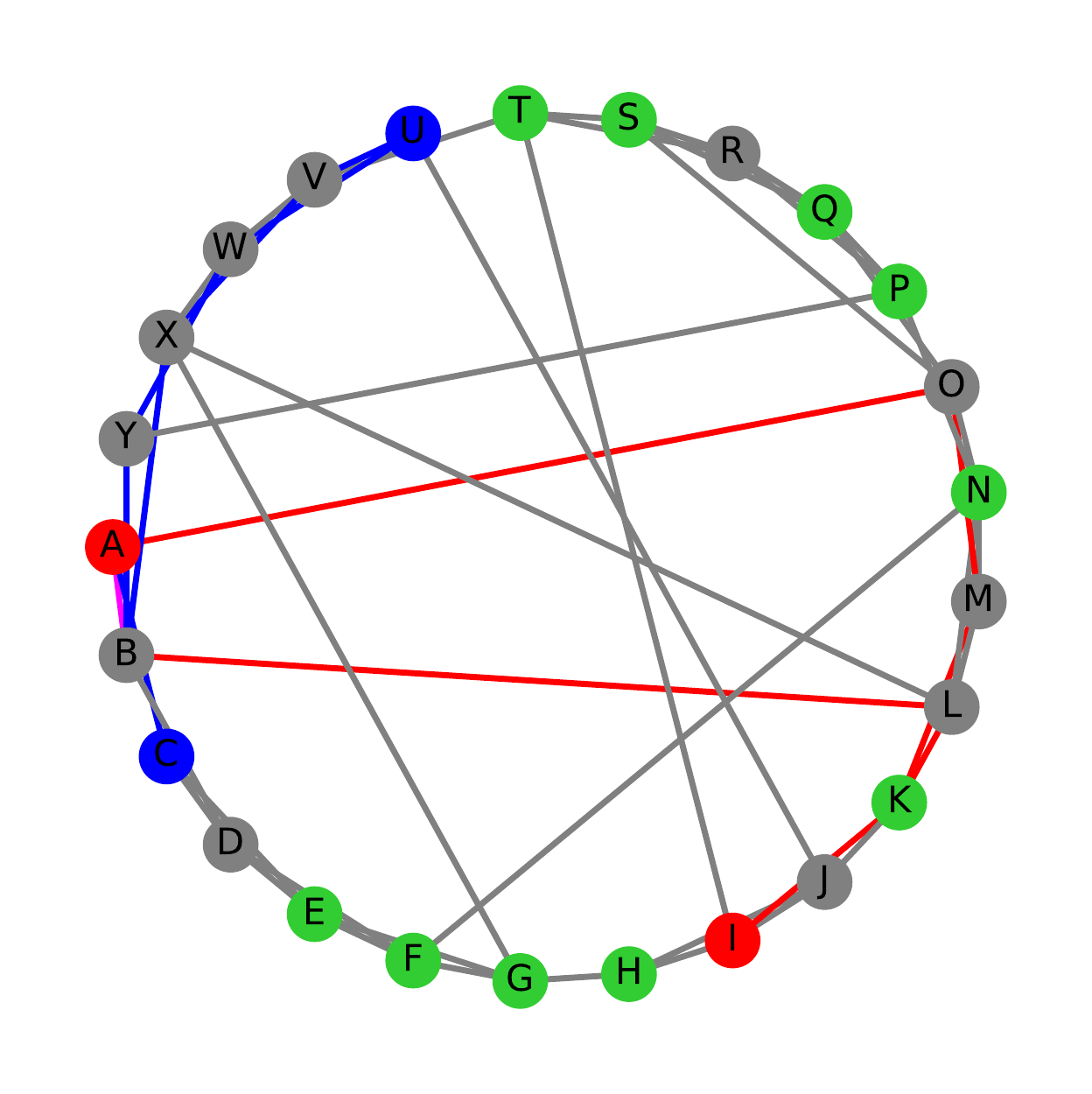}}\label{fig:WSTopo}
\caption[Quantum network topologies: grid, probabilistic grid, Erdős–Rényi and Watts--Strogatz.]{The four topologies analyzed in this section. The main service pairs and the routes connecting them have been highlighted in red and blue, with purple representing shared edges, i.e. edges that appear in both pairs' service routes. In green, we provide a visual example of the random parasitic pairs: every green node is paired with another colored node and requests entanglement with a fixed rate. At every run of the simulator, we redraw the green pairs to study the effect of traffic without bias towards a specific configuration.}
\label{fig:scheduling_topologies}
\end{figure}

In order to even out any effects due to the random placement of the parasitic users, each plot we will present comes from an average of $10$ simulation runs, each with the same main pairs and a freshly randomized set of parasitic pairs.
The topologies we examine in the following, together with an example of parasitic users placement, are presented in fig. \ref{fig:scheduling_topologies}.
\subsection{Advanced Results: Multicommodity Scheduling over Complex Topologies}
\subsubsection{Performance Metrics}
\begin{figure}[h]
    \centering
    \includegraphics[width=\linewidth]{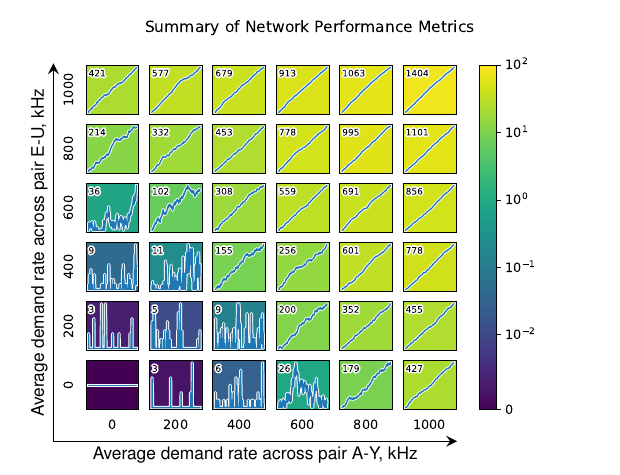}
    \caption[Global summary of the network performance metrics supported by our simulator.]{Global summary of all the network performance metrics that can be analyzed through our simulator, when running the full information Max Weight scheduler over the grid topology from fig. \ref{fig:gridtopo}. 

Inside each cell:
\begin{enumerate*}
    \item The plot shows the temporal evolution of total demand from start to finish; 
      it allows to easily distinguish stable regime (with finite excursion) from the unstable one 
      (with a linear trend);
    \item The background color represents the average demand backlog throughout a simulation run; 
    \item The top-left number is the maximum excursion of the total demand in the network;
      in the stable regime, it can be seen as a rough upper bound on the amount of quantum memory
      required at each node to achieve this performance level.
\end{enumerate*}
}
    \label{fig:scheduling_perfmetrics}
\end{figure}

In this section, we present the results of our full simulation campaign. First of all, we showcase in fig.\@ \ref{fig:scheduling_perfmetrics} the performance metrics that can be analyzed through our simulator by presenting a stability grid calculated for a $5 \times 5$ grid with the main pairs in opposite corners, controlled by a full information Max Weight scheduler. The metrics in object are the average cumulative length of the demand queues (represented in each cell's background color), the temporal trend throughout the simulation of the total demand (line plot) and the maximal excursion of total demand, to provide a sense of scale to the results. In the following, as the main interest is the comparison of stability regions, we will use the average cumulative demand queue length as our performance metric.
\subsubsection{Stability Grids}
\begin{figure}[h]
\hspace{-3cm}
\subfloat[Complete grid;]{
\includegraphics[width = 9cm]{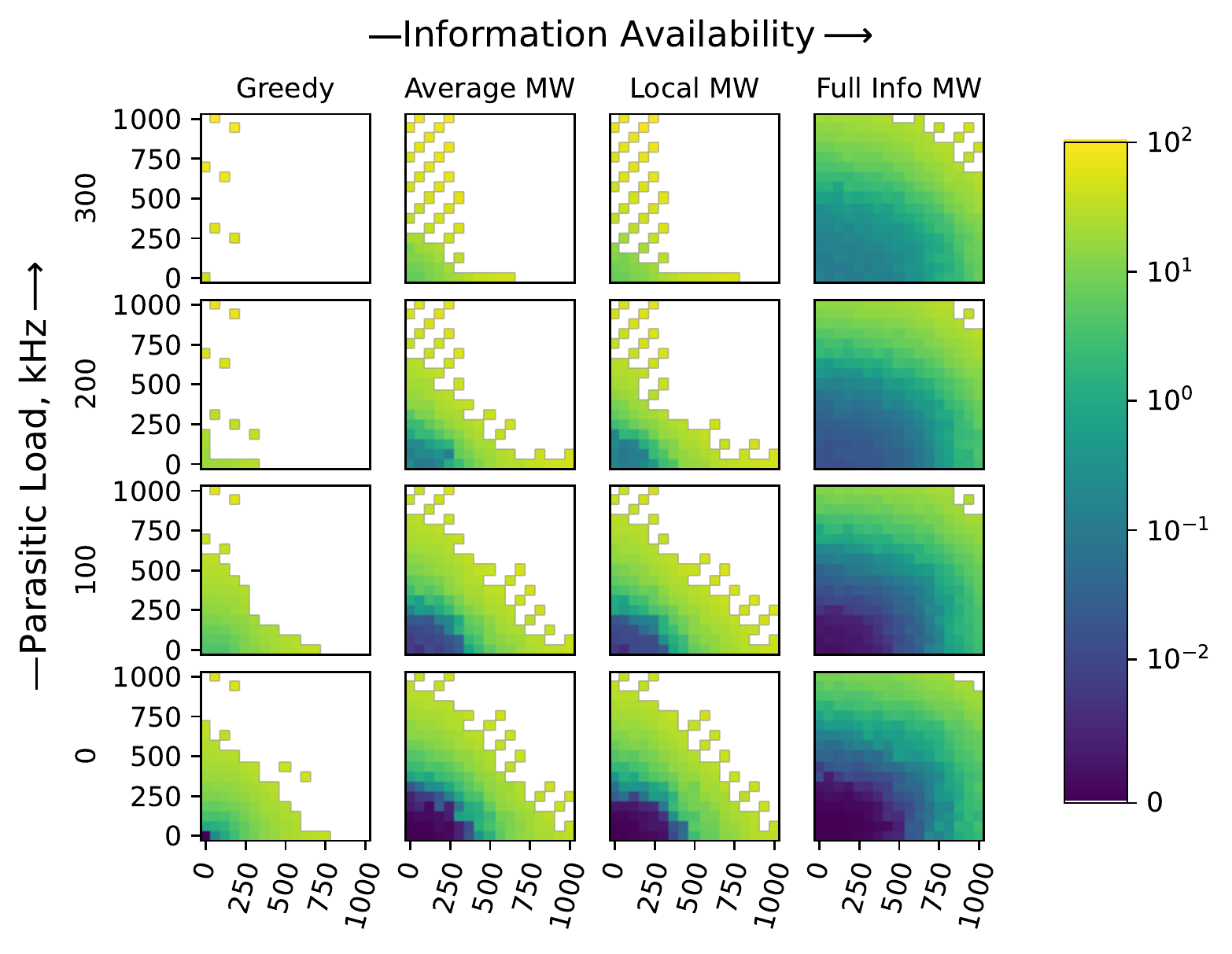}
\label{fig:finalgrid}}
\subfloat[Grid with probabilistically removed nodes;]{
\includegraphics[width = 9cm]{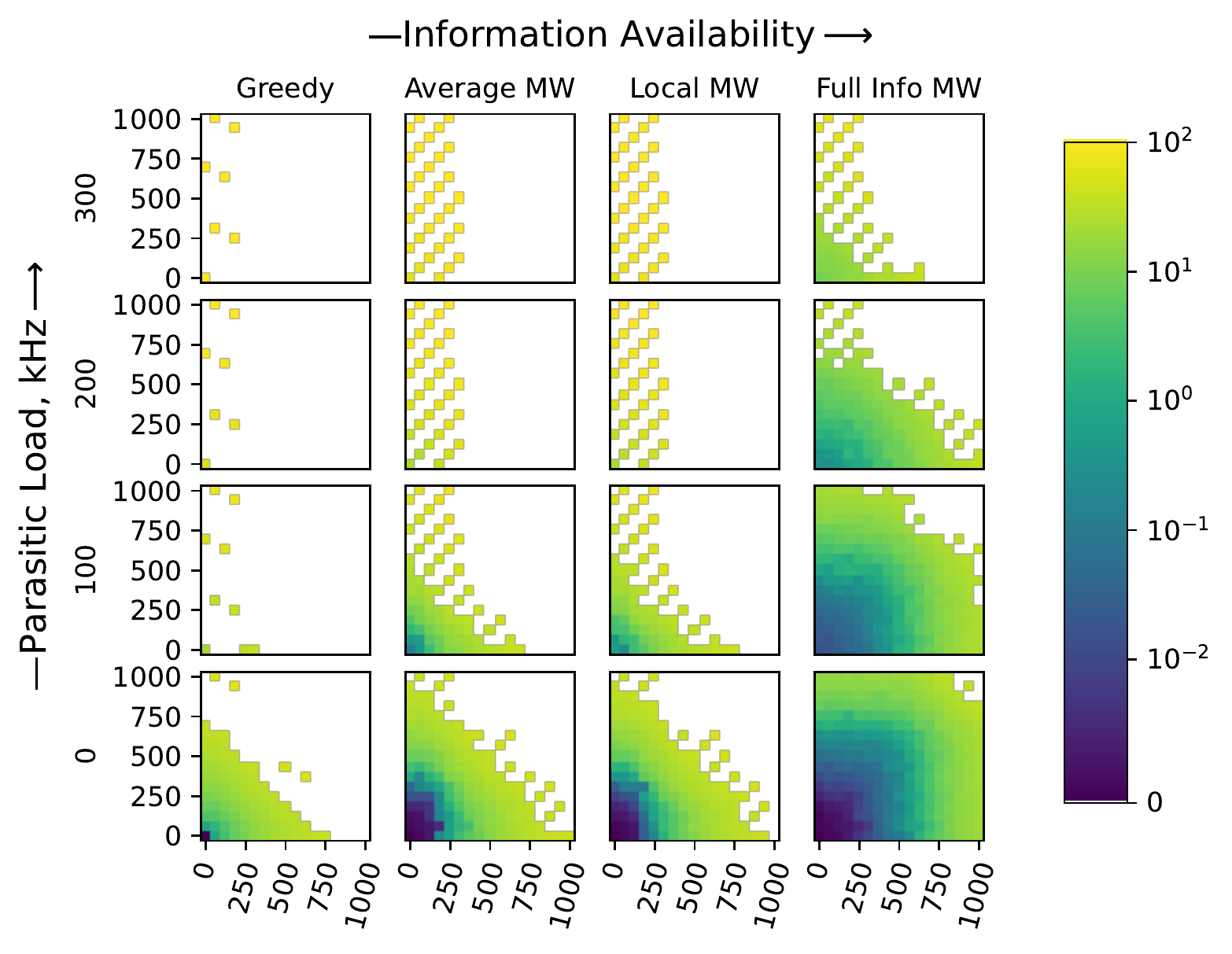}
\label{fig:finalholedgrid}}

\hspace{-3cm}
\subfloat[Erdős–Rényi random graph]{
\includegraphics[width = 9cm]{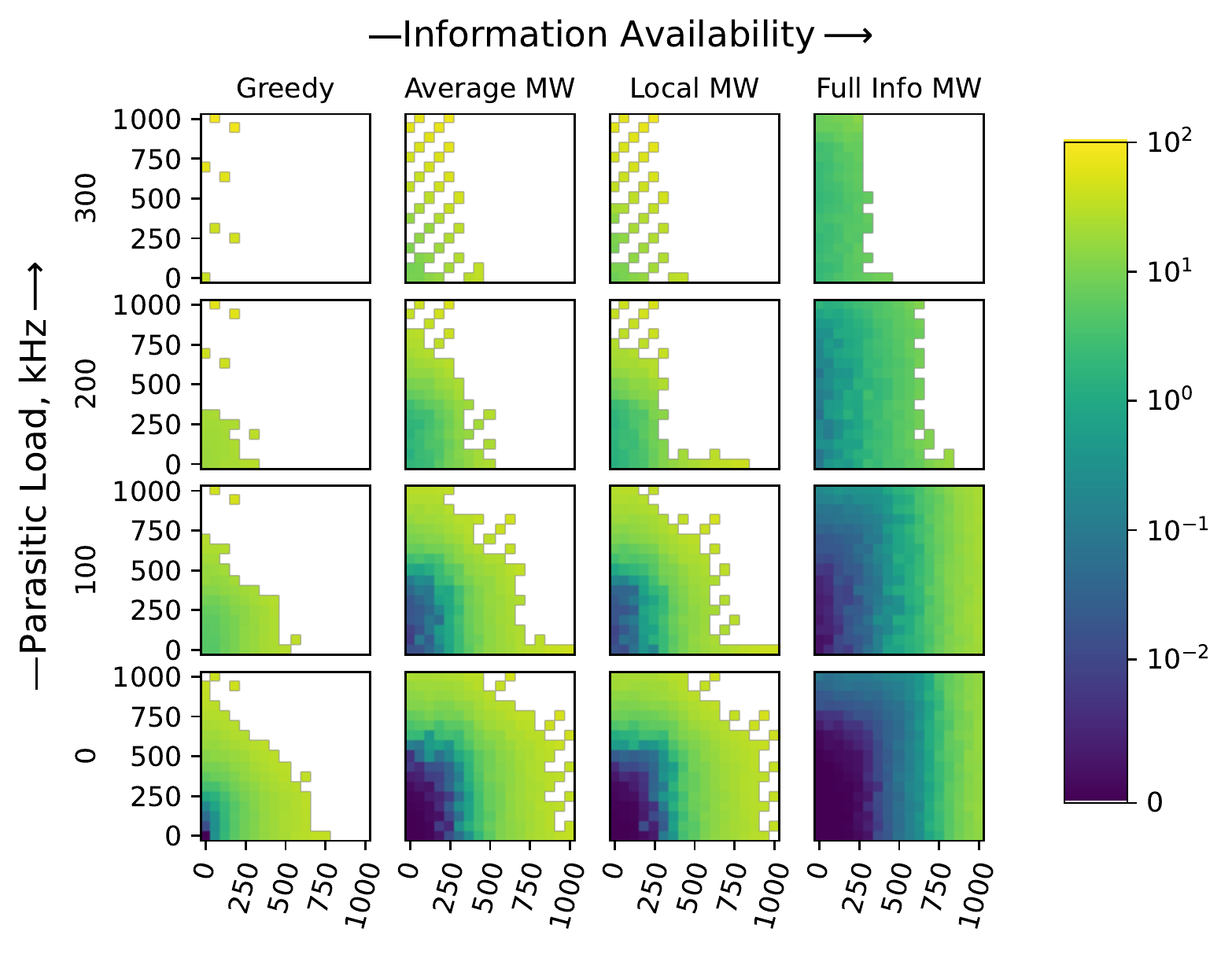}
\label{fig:finalER}}
\subfloat[Watts--Strogatz random graph.]{
\includegraphics[width = 9cm]{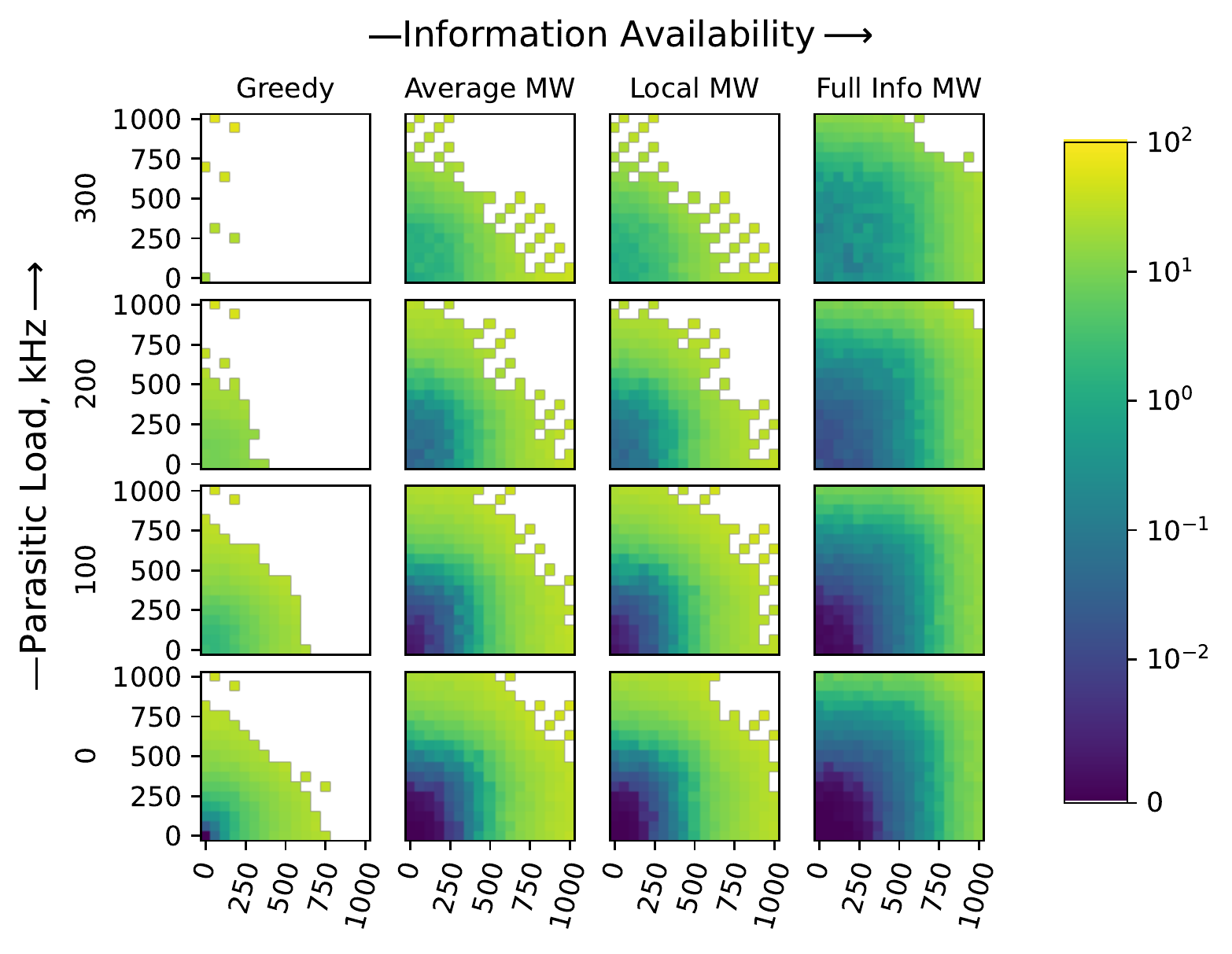}
\label{fig:finalWS}}
\caption[Stability results for several users over the topologies of fig.\@ \ref{fig:scheduling_topologies}.]{For each of the four topologies, we provide a grid of plots obtained by simulating different operating points. As mentioned in the main text, there are ten pairs of users, of which two are fixed and eight randomized. Each cell of the grids is a plot of the average demand backlog vs.\@ the load across the two main pairs (reported in $kHz$ on the small axes of the individual cells) under certain operating conditions. The conditions in which every plot was calculated are fixed by the Information Availability and Parasitic Load meta-axes, the former indicating which scheduler was employed to control the network (Greedy to Full Information, in increasing order of available information), the latter the load placed upon the randomized parasitic pairs in $kHz$. As discussed in the main text, a dark blue point is deemed stable and a yellow one unstable, while the middle grounds are somewhat ambiguous due to the finite-time nature of the simulation. 
The white points have not been calculated by the simulator to save time (see sec. \ref{sec:scheduling_simulator} for details). We recall that every cell in the grids comes from averaging ten different traffic configurations, where a configuration consists of the same two main pairs and a fresh set of eight parasitic ones. The shape of each stability region may be seen as a measure of competition between user pairs: the more diagonal the boundary of the dark blue region, the higher the direct competition between the main pairs. The difference in area of regions along one given column is a direct measurement of how the main and parasitic pairs compete (and therefore how the network serves requests under increasing stress), while the differences along one row show how well the scheduler leverages additional information.}
\label{fig:scheduling_journal_stabgrids}
\end{figure}

Running our analysis over all topologies and schedulers and plotting the average demand backlog, we obtain four arrays of plots that show the performance of our network as a function of the information granted to the scheduling policy (Greedy to Fully Informed Global) and the load on the parasitic pairs, shown in fig. \ref{fig:scheduling_journal_stabgrids}. From these arrays of plots, insight on several levels may be obtained. Firstly, by comparing fig. \ref{fig:scheduling_journal_stabgrids} to the simple result presented in fig. \ref{fig:scheduling_conferenceStabRegions} one can appreciate how the Greedy Scheduler becomes a much worse contender in larger network scenarios. This is due to the much larger number of decisions that can be taken on the scaled-up problem, which render the Greedy Scheduler an inadequate solution in any realistic network scenario.

Moreover, looking at all the plots for any given topology, we observe that changing the scheduler entails noticeable change on the capacity region of a quantum network, providing proof that not only the scheduling problem is an interesting one to formulate in the context of quantum networking, but its solution brings non-negligible performance margins to the operation of a quantum network. Another piece of information that may be gathered resides in the shapes of the stability margin: when the deep blue region is not shaped like a rectangle it means that the two plotted pairs are in direct competition, as increasing demand along one of the axes reduces the amount of demand that can be served along the other one. To an end user employing our tool for network design, this would mean that the network is bottlenecked by routing, since there is a set of routes across which the scheduler must balance service to two or more competing commodities. In fact, by once again comparing with fig. \ref{fig:scheduling_conferenceStabRegions}, it is apparent that in this complex scenario there is much less direct competition for entangled resources. This is due to using two routes per user pair, so that it is easier for the scheduler to avoid congestion by exploiting alternate routes whenever possible in order to ease congestion and lessen competition. We stress that, even in the presence of dual routes, the absence of a diagonal bound does not imply the network is not congested: traffic from the parasitic pairs is still competing with the main pairs and stressing the network (as demonstrated by the reduction in size of the stability region when going up along the parasitic load axis), requiring careful scheduling decisions.

Another point that can be made from the results in fig. \ref{fig:scheduling_journal_stabgrids} comes from looking at the difference between the fully informed global scheduler and the local ones: as mentioned before, the fully informed Max Weight scheduler can be interpreted as a performance upper bound for a Max Weight policy. Therefore, when designing an original scheduler, one may gauge its performance by comparing stability regions with the fully informed scheduler. There is a noticeable difference between FI and LI but it may be deemed acceptable because of the information trade-off: the region still has the same shape and, although smaller, is still comparable to the upper bound, meaning that the locally informed policy we are proposing performs very well in this scenario.
\section{The Simulator}
\label{sec:scheduling_simulator}
All the results shown in this work were obtained through an ad-hoc simulator implemented in Python, relying on the \texttt{gurobi} \cite{gurobi} solver for the optimization calculations and the \texttt{networkx} \cite{networkx} library as a graph backend. In the following, we provide a quick breakdown of how our simulator works, from the point of view of a user that is not necessarily experienced with writing code. Interested readers may find more information on the simulator's GitHub repository \cite{gitrepo}. 
From a black-box perspective, the focus of the code design phase of our work was on an object-oriented model of the network system that is as modular and layered as possible. The motivation driving this approach was that an ideal version of the controlling code should be abstract enough not to be aware whether it is driving our model, another more refined simulator or even a real network. In the following, we give a brief rundown of the kind of parameters that a user of our framework and simulator may expect to tune. 
The simulator's input files are composed of two sets of ingredients for the user to provide: the first set of parameters is devoted to the generation of the network topology, the choice of service pairs and demand rates. Users are free to choose one of the topologies we propose in this work (with tunable parameters) or provide an entirely custom network graph. 
After selecting the topology, the user selects the set of scheduling policies that the simulator will analyze. As before, it is possible to select one of the policies we analyzed here or provide a custom one. The code provides seamless access to all the information we used in our policies through simple specification of an ``information availability'' parameter. 
The second set of input values is related to physics and low-level simulation parameters, enabling fine-tuning of generation rates across physical links and losses at nodes, but also number and duration of the time steps. 
A set of parameters related to the optimization of the simulator's performance concludes the user inputs for our code. A complete breakdown of the parameters that users may edit is provided, together with typical values whenever possible, in tab. \ref{tab:scheduling_simparams}.
\begin{table}[]
    \centering
    \caption[A selection of parameters that can be tuned in our simulator.]{A selection of parameters that can be tuned in our simulator, together with the values that were adopted for the results of this section. In this table, ``p.\@ pairs'' is a shorthand for parasitic pairs.}
    \begin{tabular}{|>{\centering\arraybackslash}m{.2\linewidth}|>{\centering\arraybackslash}m{.21\linewidth}||>{\centering\arraybackslash}m{.2\linewidth}|>{\centering\arraybackslash}m{.21\linewidth}|}
    \hline
         \textbf{Memory lifetime} & $10 \mu s$& \textbf{Time step duration} & $1\mu s$\\
    \hline
    \textbf{Number of time steps} & 1000 & \textbf{Load points for service pairs} & min, max, number of points \\ \hline
         \textbf{Ebit generation rate} & $1MHz$, can be specified per-link & \textbf{Topology} & Chosen among predefined or custom graph  \\ \hline
         \textbf{Number of p.\@ pairs}& $8$ & \textbf{Number of routes per pair} & $2$\\ \hline
         \textbf{Service pairs} & Can be provided or autoselected & \textbf{Load points for p.\@ pairs} & min, max, number of points\\ \hline
         \textbf{Number of sets of p.\@ pairs to average} & $10$  &\textbf{List of scheduling policies} & Each has name, localization, info access, type (linear, quadratic, greedy or custom)\\ \hline 
    \end{tabular}
    \label{tab:scheduling_simparams}
\end{table}
\paragraph{Code Optimization Schemes}
The results presented in this chapter required an extensive simulation campaign which was run on a node of LIP6's HPC \texttt{tall} cluster (Intel Xeon Silver 4216, 64 threads @ $2.1 \unit{\giga\hertz}$). Therefore, a number of methods were adopted to limit the computational impact of our simulator.

The computational bottleneck in our discussion is by far the solution of optimization problems: as much as modern algorithms have sped this process up (to the point that it is feasible to conceive a real-time scheduler based on it), solving optimization problems remains a costly procedure, especially when run billions of times as in our examples. Therefore, the goal of all the proposed schemes is to try and skip as many optimizer calls as possible throughout the code. 

The first and simplest way this is done is through \textit{memoization}, i.e.\@ the practice of logging the input and output of a costly function whenever it is called. As long as the function's output is only determined by its input parameters, memoization enables the code to look the input up before calling and, if the function has already been called once with the same input, skip the calculation because the result is already known. Memoization is more impactful when the function is often called with the same parameters: in our case, it is more useful for stable points since their queues are generally fluctuating close to zero, thus more likely to repeat the same states.

The second scheme we implemented is \textit{parallel programming}: since each load configuration in our simulations is independent from the other, we observe sizable speedups (almost linear in the number of computational cores) when parallelizing our code. Instead of running one configuration at a time, the simulator is simultaneously running batches of as many points as there are computational cores. Parallelizing the code is extremely useful in cases like ours because it allows the simulator to fully exploit the multiple cores provided by HPC architectures.

Finally, a third measure we adopted was to exploit the fact that the stability region is a Pareto front to skip the simulation of some pixels: as mentioned before, finding an unstable point means that all the points with strictly greater load are unstable as well.
Since it is impossible to predict the exact shape of the output stability region, our optimization scheme was implemented by randomizing the order in which load points are tested for stability. Although this is not necessarily the best ordering to achieve maximum speedup, this measure still provided noticeable benefits.

\section{Limitations of the Framework and Future Outlook}
In this section, we discuss the main limitations and open questions in our model, and propose some seed ideas for future directions. The first limitation to talk about is the modelization of strictly quantum imperfections such as decoherence, that degrade the quality of a quantum state without necessarily meaning the state is lost. Despite being well aware of the paramount importance of noise in quantum modeling, the history of the classical Internet shows that a successful large-scale network infrastructure is best thought of in terms of separate functional layers, and a layered architecture has already been proposed for a prospective future quantum internet \cite{QStack} that effectively separates the Link Layer, where quantum error correction should be implemented, from the Network Layer, which is the scope of our work. While we are aware that in real implementations, especially initial ones, theoretically separate layers leak and blend with each other, the Quantum Internet should eventually converge to a well defined network stack, making it redundant to treat noise in the same layer as scheduling. Thus, while we remain interested in an expansion of our work that treats quantum imperfections, the lack of explicit state quality modeling does not make our work irrelevant.

A similar concern could be raised for the memory at the network nodes: despite this being another issue that is very close to hardware, its integration with scheduling policies would seem crucial because it could intrinsically change how a scheduling decision is taken: if a node only has a finite number of memory slots, the scheduler would have the additional constraint of free space (or lack thereof, in some cases having to ``waste'' ebits in order to free up memory). As a matter of fact, a similar problem has been analysed over a single switch in \cite{VardoyanSwitchStochastic} and \cite{NainSwitch}, showing that the memory requirements of an isolated quantum switch are quite low (on the order of 5 slots) to achieve performance comparable to that of a switch with unlimited memory slots, making the memory problem not as concerning. Moreover, \cite{PromponasMemory} formulates the problem of exploiting limited memory slots and develops a Max-Weight memory allocation policy for quantum nodes that could be adapted to our scenario. 

Furthermore, it is possible to look at the memory problem from a different direction: while a solution inside our framework could in principle be to add compound constraints to the optimization problems, we stress that results such as fig. \ref{fig:scheduling_perfmetrics} (maximal excursion numbers) gauge the accumulation of total demand in a stable network, effectively providing an upper bound for memory requirements in the design of a real quantum network system.

The third limitation of our work is how the framework scales: The fact that the number of queues we need to account for grows quadratically with the number of nodes in the network entails quick growth of the $\mathbf{M}$ matrix, which makes the integer programs required by several policies presented here increasingly complex. While this is not as much of a problem currently as it was in the past decades, it is still an issue that is worth closely investigating, perhaps to find scheduling strategies that require only a subset of the extended edge set (akin to an overlay network, as demonstrated in \cite{Pouryousef2022QON}). We note here that easing scaling concerns would also enable a future extension of our framework to multipartite entanglement: as mentioned in the beginning, an extension in this direction would require the definition of new multipartite virtual queues, together with ad-hoc transitions that interface them with the bipartite ones, greatly increasing the overall number of queues and therefore the problem's complexity.

Moreover, we do not provide analytical proofs of Lyapunov stability or optimality of the proposed families of policies, which are of great interest in network science and could be promising directions for future work. To provide a starting point, we direct the interested reader to the well-known optimality results of Max Weight on classical networks \cite{TassiulasMaxWeight} and to \cite{VasantamSwitch}, where Lyapunov stability and throughput optimality are analytically proven for a Max Weight policy in the case of a switch without quantum memory serving three users in a star topology: it is possible to show, by translating the referenced paper's model into our framework, that the Max Weight policy presented in the cited work is equivalent to the Fully Informed Max Weight analyzed in ours. 
Interestingly, in the case proposed by \cite{VasantamSwitch} our class of Quadratic policies reduces to Max Weight: since \cite{VasantamSwitch} employs Bernoullian ebit arrivals and no quantum memory, the components of the scheduling decision $\mathbf{r}(t)$ can be at most $1$. Coupled with the structure of $\mathbf{\tilde{N}}$, this entails that at all time steps the quadratic penalty has the same value across all possible scheduling decisions, reducing the more general Quadratic objective function to a Max Weight one and showing that in this special case our policies are optimal, which offers potential for investigation as to how generalizable the optimality claim is. We note that, since we included additional information in the Lyapunov drift definition, a formal proof of stability would also require averaging over all the additional environmental factors that were included in the conditioning of (\ref{eq:drift_I}) other than the queue state ($\mathbf{q}(t),\mathbf{d}(t)$).

Finally, it would be interesting to delve into other physical imperfections, such as finite speed of communication between nodes, which entail a stricter definition of what information is local and accessible to a node at a given time. One interesting implication of such analysis would be the case in which only one of the qubits in an ebit is lost, and what happens if the loss is not communicated before other swapping operations are undertaken, i.e.\@ the error propagates along the swapping route. 
All these considerations would require a more refined physical model both in terms of mathematics and of simulation, which is the reason why in the remainder of this thesis we employ more refined simulation software.
\begin{sidewaystable}
\small
    \centering
    \caption[Generated $\mathbf{M}$ for the $6$-nodes chain network.]{Generated $\mathbf{M}$ for the $ABCDEF$ chain.}
    \begin{tabular}{l | c c c c c c c c c c c c c c c c}
         & ${B[D]F}$ & ${A[B]E}$ & ${C[E]F}$ & ${A[C]D}$ & ${B[E]F}$ & ${B[D]E}$ & ${B[C]F}$ & ${B[C]D}$ & ${D[E]F}$ & ${C[D]F}$ & ${B[C]E}$ & ${A[C]E}$ & ${C[D]E}$ & ${A[B]D}$ & ${A[D]E}$ & ${A[B]C}$ \\\hline
$DE$ & $\phm0$  & $\phm0$  & $\phm0$  & $\phm0$  & $\phm0$ & $-1$  & $\phm0$  & $\phm0$ & $-1$  & $\phm0$  & $\phm0$  & $\phm0$ & $-1$  & $\phm0$ & $-1$  & $\phm0$ \\
$AB$ & $\phm0$ & $-1$  & $\phm0$  & $\phm0$  & $\phm0$  & $\phm0$  & $\phm0$  & $\phm0$  & $\phm0$  & $\phm0$  & $\phm0$  & $\phm0$  & $\phm0$ & $-1$  & $\phm0$ & $-1$ \\
$BC$ & $\phm0$  & $\phm0$  & $\phm0$  & $\phm0$  & $\phm0$  & $\phm0$ & $-1$ & $-1$  & $\phm0$  & $\phm0$ & $-1$  & $\phm0$  & $\phm0$  & $\phm0$  & $\phm0$ & $-1$ \\
$AC$ & $\phm0$  & $\phm0$  & $\phm0$ & $-1$  & $\phm0$  & $\phm0$  & $\phm0$  & $\phm0$  & $\phm0$  & $\phm0$  & $\phm0$ & $-1$  & $\phm0$  & $\phm0$  & $\phm0$  & $+1$ \\
$BD$ & $-1$  & $\phm0$  & $\phm0$  & $\phm0$  & $\phm0$ & $-1$  & $\phm0$  & $+1$  & $\phm0$  & $\phm0$  & $\phm0$  & $\phm0$  & $\phm0$ & $-1$  & $\phm0$  & $\phm0$ \\
$AD$ & $\phm0$  & $\phm0$  & $\phm0$  & $+1$  & $\phm0$  & $\phm0$  & $\phm0$  & $\phm0$  & $\phm0$  & $\phm0$  & $\phm0$  & $\phm0$  & $\phm0$  & $+1$ & $-1$  & $\phm0$ \\
$BE$ & $\phm0$ & $-1$  & $\phm0$  & $\phm0$ & $-1$  & $+1$  & $\phm0$  & $\phm0$  & $\phm0$  & $\phm0$  & $+1$  & $\phm0$  & $\phm0$  & $\phm0$  & $\phm0$  & $\phm0$ \\
$AE$ & $\phm0$ & $+1$ & $\phm0$ & $\phm0$ & $\phm0$ & $\phm0$ & $\phm0$ & $\phm0$ & $\phm0$ & $\phm0$ & $\phm0$ & $+1$ & $\phm0$ & $\phm0$ & $+1$ & $\phm0$ \\
$CD$ & $\phm0$  & $\phm0$  & $\phm0$ & $-1$  & $\phm0$  & $\phm0$  & $\phm0$ & $-1$  & $\phm0$ & $-1$  & $\phm0$  & $\phm0$ & $-1$  & $\phm0$  & $\phm0$  & $\phm0$ \\
$CE$ & $\phm0$  & $\phm0$ & $-1$  & $\phm0$  & $\phm0$  & $\phm0$  & $\phm0$  & $\phm0$  & $\phm0$  & $\phm0$ & $-1$ & $-1$  & $+1$  & $\phm0$  & $\phm0$  & $\phm0$ \\
$EF$ & $\phm0$  & $\phm0$ & $-1$  & $\phm0$ & $-1$  & $\phm0$  & $\phm0$  & $\phm0$ & $-1$  & $\phm0$  & $\phm0$  & $\phm0$  & $\phm0$  & $\phm0$  & $\phm0$  & $\phm0$ \\
$CF$ & $\phm0$  & $\phm0$  & $+1$  & $\phm0$  & $\phm0$  & $\phm0$ & $-1$  & $\phm0$  & $\phm0$  & $+1$  & $\phm0$  & $\phm0$  & $\phm0$  & $\phm0$  & $\phm0$  & $\phm0$ \\
$BF$ & $+1$ & $\phm0$ & $\phm0$ & $\phm0$ & $+1$ & $\phm0$ & $+1$ & $\phm0$ & $\phm0$ & $\phm0$ & $\phm0$ & $\phm0$ & $\phm0$ & $\phm0$ & $\phm0$ & $\phm0$ \\
$DF$ & $-1$  & $\phm0$  & $\phm0$  & $\phm0$  & $\phm0$  & $\phm0$  & $\phm0$  & $\phm0$  & $+1$ & $-1$  & $\phm0$  & $\phm0$  & $\phm0$  & $\phm0$  & $\phm0$  & $\phm0$
    \end{tabular}
    \label{tab:M6Chain}
\end{sidewaystable}
\chapter{Simulating Quantum Networks}
\label{ch:simulation}
As the quantum networks field begins to develop and evolve, so does the part of it devoted to quantum network simulation. A simulator's goal is to reproduce the behavior of a physical system to an arbitrary level of detail, and several tools are available that focus on different aspects, from modeling the deep physical intricacies of a single quantum link to abstracting some detail away in order to simulate large internet-like recursive networks. We start the chapter with sec.\@ \ref{sec:DTSvsDES}, a description of the differences between Discrete Events and Discrete Time simulators. We then provide in sec.\@ \ref{sec:simulators_review} a brief overview of the main quantum network simulators that are currently available. Sec.\@ \ref{sec:intro_opp_quisp} delves in more detail about OMNeT++ and QuISP, the software tools we use for the remainder of this thesis. Sec.\@ \ref{sec:adapting_quisp_sat} marks the beginning of our original contributions, as we describe our implementation of satellite links in the simulator QuISP (which enable the results presented in chap.\@ \ref{ch:satellites}). Sec.\@ \ref{sec:adapting_quisp_muxing} provides an account of our implementation of multiplexing in QuISP (a crucial prerequisite for scheduling problems). In sec.\@ \ref{sec:DSSandQuISPSched}, we formulate a simple scheduling problem and compare results between QuISP and the simple simulator from chap.\@ \ref{ch:scheduling}. We wrap the chapter up with sec.\@ \ref{sec:quisp_currentstate}, a summary of the current status of our contributions to QuISP.
\section{Simulation Paradigms: Discrete Time vs\@ Discrete Events}
\label{sec:DTSvsDES}
The evolution of a system in time may be modeled either through \emph{discrete time simulation} (DTS) or \emph{discrete events simulation} (DES) \cite{DiscreteTime}, two distinct simulation paradigms each with its advantages and disadvantages. 
Due to the fundamental differences between these two approaches, which we outline in the following, it is not possible to unify them: simulators conform to one or the other depending on their specific requirements.
\paragraph{Discrete Time Simulators} are the most straightforward way to program a simulation of a natural process. Given a time interval from $t_\mathrm{start}$ to $t_\mathrm{end}$, we subdivide it into time steps of length $\Delta t$. Starting from $t_\mathrm{start}$, the simulator will increase the current time by $\Delta t$, recalculate the system's state variables, record data and repeat the loop until $t_\mathrm{end}$ is reached.\\
DTSs are simple to understand and can be built to be accurate simulation tools, but there are a series of details that must be taken into careful consideration, mostly coming from the choice of $\Delta t$: since the simulation discretely steps from one time tick to the next, the behaviour of state variables is technically undefined between steps. Moreover, since the modeled system is continuous but the simulator only evolves at fixed steps, multiple events happening during the same step must be properly sequenced by the simulator to avoid modeling them in the wrong order and polluting the results. We show an example of this particular problem in sec.\@ \ref{sec:scheduling_infeasibility}.\\
Another typical trait of DTSs is that they require care in the choice of $\Delta t$: the shorter the time interval, the closer the discrete system is to the continuous one, which in theory can arbitrarily increase accuracy. However, shorter time steps mean more updates over the same time interval, so that the simulator becomes more computationally demanding and slower to run. On the other hand, a longer $\Delta t$ will greatly improve performance, but it will invalidate the results if care is not taken in its selection.
The ad-hoc simulator \cite{gitrepo} we developed in chap.\@ \ref{ch:scheduling} is an example of DTS.
\paragraph{Discrete Events Simulators} trade some of the performance and intuitiveness of DTSs for increased fidelity to the modeled phenomenon. The inner workings of a DES are slightly more complex, but the end result is a more flexible simulator that is less prone to user error and eliminates the difficulty of properly choosing $\Delta t$. The key feature of a DES is an event timeline: components of the simulation are allowed to schedule events on the timeline, and the simulator's time cursor skips from one event to the next. Events can be scheduled before the simulation starts, or they can be added as it runs. For example, one event could be ``sender sends message'', scheduled at $t_\mathrm{send}$: when the time cursor arrives at this event, the simulation kernel will schedule the event ``recipient receives message'' at $t_\mathrm{receive} = t_\mathrm{send} + t_\mathrm{transmission}$. If no other events are scheduled between $t_\mathrm{send}$ and $t_\mathrm{receive}$, the time cursor then skips immediately at $t_\mathrm{receive}$.\\
The immediate advantage of a DES is that, since the cursor skips from event to event, periods of time in which everything is idle are not simulated. This is a desirable feature for systems that spend the majority of the time idling, such as a satellite link when the satellite node is not in sight.\\
The six leading simulators analyzed in sec.\@ \ref{sec:simulators_review} are all DES, including QuISP \cite{quisp}, the simulator used in this thesis from sec.\@ \ref{sec:intro_opp_quisp} onwards.

A more in-depth comparison of DTSs and DESs may be found in \cite{DiscreteTime}. In general, DTSs are better for prototyping, simplicity of development and gaining high-level insight about high-rate systems, which would be expensive to simulate with a DES due to the high volume of events. DESs, on the other hand, tend to be more flexible, powerful and adapted to systems with relatively long idle times (such as quantum networks). 
\section{Main Existing Quantum Network Simulators}
\label{sec:simulators_review}
In this section, we discuss some of the leading simulation efforts for quantum networks. The simulators were chosen considering the public availability of their source code, their recent activity and their scope (many quantum simulators exist that are either closed-source commercial solutions, academic projects that have not been updated in a long time or developed with a focus on other domains such as quantum computing, which is outside our scope). We provide a summary of the key differences between the simulators in tab.\@ \ref{tab:sim_simssummary}.
\paragraph{NetSquid} \cite{netsquid} (\textbf{Net}work \textbf{S}imulator for \textbf{Qu}antum \textbf{I}nformation using \textbf{D}iscrete events) is a discrete-events quantum network simulator developed in Python at QuTech (Delft, Netherlands). As a physics-based simulator, NetSquid allows its users to model quantum networks implementing several applications (such as Quantum Key Distribution \cite{BB84}) over a variety of physical platforms (trapped ions, NV centers, atomic ensembles\ldots). Given the accuracy of its physical models, NetSquid offers a large amount of configurable parameters at all layers of the network stack. Should the user not be interested in fine-grained configuration, NetSquid also features a repository of \emph{snippets}, which are pre-configured building blocks modeling hardware and protocols provided by the NetSquid community. A user developing a simulation in NetSquid may opt for full customization of the network stack or employ snippets as high-level building blocks to assemble into larger scenarios. At the time of writing, there are plans to make the source code of NetSquid publicly available. In \cite{netsquid}, the usage of NetSquid is demonstrated at various levels of abstraction for the study of a single multiparty switch, a three-node simple swapping network with entanglement distillation and a long repeater chain.
\paragraph{SimulaQron} \cite{SimulaQron} is an application-oriented quantum simulator, also developed in Python at QuTech (Delft, Netherlands). Its source code is publicly available \cite{SimulaQronRepo} under the BSD 3-Clause license. SimulaQron differs from NetSquid in that it specifically seeks to enable application development before the hardware for experimental quantum networks is available. To this end, SimulaQron abstracts its physical quantum backend (which can be delegated to another simulator, possibly NetSquid itself) and chooses to focus on the classical communication involved in the creation and manipulation of entangled links. The peculiarity of SimulaQron is that, to accurately gauge the impact of classical communications on quantum networking, it supports being run on a distributed array of interconnected computers, each simulating a single quantum node\paolofootnote{Running the simulator on a single machine is still supported.}. The nodes can exchange messages to create entanglement and instruct each other to run quantum manipulations on the backend (such as correcting swapping operations) through the Classical-Quantum Combiner \cite{SimulaQron} (CQC) interface, a low-level instruction set that implements the operations typical of a quantum processor (such as quantum gates) and specifies a message format for applications to communicate through IP packets. Users can program in CQC through Python \cite{CQCDocsPython} and C \cite{CQCDocsC} libraries, with plans by the team to develop a full-fledged compiler to replace the libraries.
\paragraph{QuNetSim} \cite{QuNetSim} (\textbf{Qu}antum \textbf{Net}work \textbf{Sim}ulator) is an application-oriented quantum network discrete-events simulator written in Python developed by the TQSD Team \cite{tqsdwebsite} at the University of Munich, Germany. QuNetSim places its emphasis on assembling pre-developed modules that represent experimental devices and simulating their interaction. Being a simulator oriented to protocol design and testing, QuNetSim focuses on the layered structure of the network system, explicitly simulating the Application, Transport and Network layers. The simulation of quantum hardware is abstracted to pre-existing qubit backends such as Qiskit \cite{qiskit}, or even real hardware. Despite the ``building blocks'' nature of the simulator, users may design their own custom network modules and integrate them in QuNetSim simulations. In \cite{QuNetSim}, QuNetSim is employed to study issues such as anonymous distribution of an entangled pair through a GHZ state and testing of a quantum routing protocol. QuNetSim's source code is publicly available on GitHub \cite{QuNetSimRepo} under the MIT license.
\paragraph{ReQuSim} \cite{ReQuSim} (\textbf{Sim}ulation platform for \textbf{Qu}antum \textbf{Re}peaters) is an event-based Monte Carlo simulator centered around quantum repeater strategies developed at the Dahlem Center for Complex Quantum Systems at the Free University of Berlin, Germany. Its source code is publicly available \cite{ReQuSimRepo} under the MIT license. ReQuSim's focus is placed on repeater network design, specifically on the tradeoffs between low-level physical parameters and high-level performance metrics such as the achievable entanglement distribution rate. ReQuSim supports several detailed quantum error and time-dependent quantum noise models, allowing the benchmark of repeater strategies for both fiber and free-space links with a high level of sensitivity to experimental imperfections. Classical communication latency is taken into account by introducing delays in quantum processing (with no explicit simulation of classical messages). Given its depth in terms of imperfection modeling, the simulator keeps runtimes reasonable by probabilistically drawing the number of transmitted photons through channels (instead of keeping track of each sent photonic packet individually) and by updating time-dependent error models only at relevant times instead of continuously\paolofootnote{This particular approach is one of the main advantages of developing a DES.}. In \cite{ReQuSim}, Wallnöfer et al.\@ employ ReQuSim to correlate the effect of parameters such as memory noise, number of purification rounds and initial fidelity of the distributed entangled states to the achievable QKD key rate and wait times for entanglement distribution. 
\paragraph{SeQUeNCe} \cite{sequence} (\textbf{S}imulator of \textbf{QU}antum \textbf{Ne}twork \textbf{C}ommunication) is a Python modular discrete-events simulator developed at Argonne National Laboratory in Lemont, USA. SeQUeNCe is a modular full-stack simulator (the layers being Hardware, Entanglement Management, Resource Management, Network Management and Application) that allows benchmarking network performance in terms of quantum parameters, protocol choice at the different layers and latency of classical communication. In \cite{sequence}, Wu et al.\@ use SeQUeNCe to conduct a study of end-to-end throughput of various quantum flows in a quantum network in the metropolitan Chicago area. The presented results directly relate the throughput of entanglement distribution to low-level hardware parameters such as the efficiency of the quantum memories, the delay of classical communication channels or the allocation of memory slots to the different connections. SeQUeNCe's source code is available on GitHub \cite{SequenceRepo} under the BSD 3-Clause license.
\paragraph{QuISP} \cite{quisp} (\textbf{Qu}antum \textbf{I}nternet \textbf{S}imulation \textbf{P}ackage) is a large-scale quantum network simulator written in C++, developed by the AQUA Team at Keio University in Fujisawa, Japan \cite{aquawebsite}. Similarly to NetSquid and SeQUeNCe, QuISP poses as a full-stack toolbox for quantum network simulation. Of the presented tools, QuISP is the more network-oriented one because it was built on top of OMNeT++ \cite{omnetpp}, a well-established simulation platform in classical networking. The main peculiarity of QuISP is the concept of RuleSet, a program that is generated at runtime and describes all actions a quantum node needs to perform to implement a given application. Since QuISP is the simulator that was adopted for the remainder of this thesis, more details on it are available in the following section. The source code of QuISP is available on Github \cite{quisprepo} under the BSD 3-Clause license, while OMNeT++'s source code is available on GitHub \cite{oppGithub} under the Academic Public License, which allows free academic use but requires a separate license for commercial applications.
\begin{table}[]
     \caption[Comparison of six quantum network simulators.]{Summary of the differences between the analyzed quantum network simulators.}
    \centering
    \begin{tabularx}{\linewidth}{|>{\raggedright\arraybackslash}X|>{\raggedright\arraybackslash}X|>{\raggedright\arraybackslash}X|>{\raggedright\arraybackslash}X|>{\raggedright\arraybackslash}X|}
    \hline
         \textbf{Simulator} & \textbf{Main Focus} & \textbf{Developed By} & \textbf{Language} & \textbf{License}\\ \hline
         NetSquid & General purpose network simulation & QuTech, Delft (NL) & Python & Free download and use, currently closed source\\ \hline
        SimulaQron & Application development on distributed simulation cluster & QuTech, Delft (NL) & Python/C & BSD 3-Clause\\ \hline
        QuNetSim & Application prototyping by combining pre-developed and custom blocks & TQSD Team, University of Munich (DE) & Python & MIT License\\ \hline
        ReQuSim & Benchmarking quantum repeater strategies & Dahlem Center for Complex Quantum Systems, Free University of Berlin (DE) & Python & MIT License\\ \hline
        SeQUeNCe & General purpose network simulation & Argonne National Laboratory, Lemont (US) & Python & BSD 3-Clause \\ \hline
        QuISP & General purpose network simulation & AQUA Team, Keio University, Fujisawa (JP) & C++ & BSD 3-Clause, OMNeT++ under Academic Public License \\ \hline
    \end{tabularx}
    \label{tab:sim_simssummary}
\end{table}
\section{OMNeT++ and QuISP}
\label{sec:intro_opp_quisp}
\begin{figure}
    \centering
    \includegraphics[width=\linewidth]{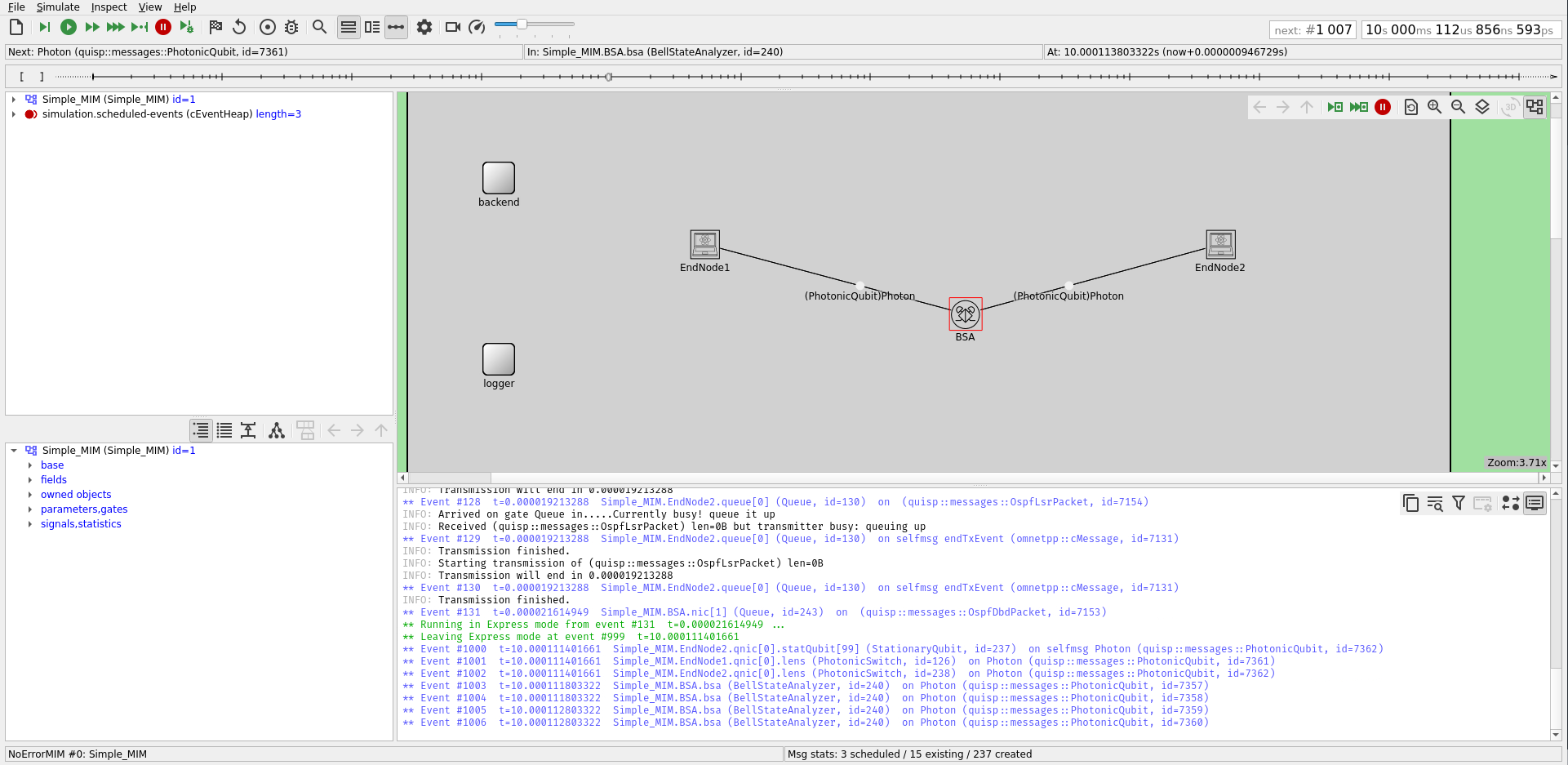}
    \caption[OMNeT++'s graphical interface.]{OMNeT++'s graphical interface displays a fully animated model of all the messages exchanged by the network nodes. Events are logged and can be inspected by the user. A headless mode is also available for better performance.}
    \label{fig:opp_screenshot}
\end{figure}

Before delving into the details of QuISP, it is useful to discuss the software upon which it is based, the \textbf{O}bjective \textbf{M}odular \textbf{Ne}twork \textbf{T}estbed in C$\mathbf{++}$ (OMNeT++) \cite{omnetpp, omnetppsite}. OMNeT++ is not a simulator in the strict sense: it is better defined as a discrete events \emph{simulation framework}, meaning it is an abstract set of building blocks that can be combined and specialized to develop network simulators for different architectures. OMNeT++ is a well-known and vastly employed tool in classical network science, and various models have been developed by the community to simulate network scenarios of high relevance such as 5G communication or vehicular networks\paolofootnote{More models can be found at \cite{oppExamples}.}. OMNeT++ is a stable, open and feature-complete platform that allows development of simulation models through a graphical IDE based on Eclipse \cite{Eclipse}. The developed simulations can be executed in a faster headless mode or through an animated GUI that allows on-the-fly inspection of all the components of a network (as shown in fig.\@ \ref{fig:opp_screenshot}).
The main drawbacks of OMNeT++ are its somewhat steep learning curve --- formulating custom simulation models must be done in C++ --- and its installation process, which for the academic version requires compilation from source.

Every simulation using OMNeT++ is defined through abstract \textit{modules}, 
i.e.\@ the components of the network that need to communicate, \textit{channels} that interconnect said modules and \textit{messages} for the modules to exchange through the channels.
Modules may be \textit{simple} (basic functional units for the network executing arbitrary C++ code) or \textit{compound} (built by combining multiple simple modules called submodules). In practice, a module is represented by a NED (\textbf{NE}twork \textbf{D}escription) file, the purpose of which is to describe it from a structural perspective (i.e.\@ I/O ports, submodules and how they are interconnected...) and, in the case of simple modules, a C++ source file describing the internal working of the module. Since compound modules are obtained from aggregating simple modules, they require no C++ code. In the following, we report a commented version of the compound \texttt{.ned} file for a standalone BSA node from QuISP.
\begin{minted}{cpp}
// Package directives: the .ned syntax supports importing definitions from 
// other .ned files to define modules.
package modules;
import ned.DatarateChannel;
import ned.DelayChannel;
import ned.IdealChannel;
import modules.PhysicalConnection.BSA.*;
import modules.QNIC.PhotonicSwitch.*;
import modules.Common.Router;
import modules.Common.Queue;
module BSANode //name of the module.
{   
    // In this section, the module's parameters are defined.
    parameters: 
        int address = default(0);
        string node_type = default("BSA");
    gates: // These are the module's I/O ports.
        inout quantum_port[];
        inout port[];
    // In this section, the module's submodules (defined in separate 
    // .ned files) are instantiated. Notice that it is possible to have 
    // vectors of submodules.
    submodules: 
        bsa: BellStateAnalyzer {
         }
        router: Router {
            gates:
                fromQueue[sizeof(parent.port)];
                toQueue[sizeof(parent.port)];
        }
        //sizeof(port) is determined dynamically from the connections section.
        nic[sizeof(port)]: Queue { 
            parameters:
                address = 0;
         }
        bsa_controller: BSAController {
       }
    //here, submodules are connected to each other and to the module's I/O ports. 
    // "port++" is a shorthand for "first free port".
    connections: 
        for i=0..sizeof(port)-1 {
            router.toQueue[i] --> nic[i].in;
            router.fromQueue[i] <-- nic[i].out;
            nic[i].line <--> port[i];
        }
        for i=0..sizeof(quantum_port)-1 {
            quantum_port[i] <--> bsa.quantum_port++;
        }
        bsa_controller.to_router --> router.fromApp;
        router.toApp --> bsa_controller.from_router;
        bsa.to_bsa_controller --> bsa_controller.from_bsa;
}
\end{minted}
Every simple module's C++ code has at least an \texttt{initialize()} method, called at simulation startup, and a \texttt{handleMessage()} method, called whenever a message is received by the module. It is up to the simulation developer to re-implement these methods to emulate the functionality of the desired network architecture.
Concerning channels, each of them has a \texttt{processMessage()} method. By providing an implementation for this method, users may model channel behavior ranging from simple message loss to more complex noise and latency models by acting directly on the message object.
Once all the building blocks have been properly described and implemented, an additional NED file is created to assemble compound modules and channels into a network: this will be the top-level network architecture on which simulations will run. A NED file may contain several networks among which the final simulation user may choose.
Every object in OMNeT++ has parameters, which are user-tuned via \texttt{.ini} files and are visible to the underlying C++ code.

Such a simple basic structure has the advantage of high scalability and fidelity to the simulated network system: since everything happens through messages, the simulation mirrors the internal workings of a network system on a message-by-message basis. Moreover, since the channels that interconnect modules are themselves abstract and expected to be re-implemented by the simulation developer, it is possible to implement and study simulated networks that closely reproduce the physical transfer of signals on an arbitrary level of detail. 

Built on top of OMNeT++, QuISP provides a suite of modules and channels designed for the simulation of optical fiber quantum networks. Quantum hosts are modeled in QuISP as \emph{quantum nodes} (QNodes for short), compound OMNeT++ modules containing:
\begin{description}
    \item[Quantum Network Interface Cards (QNICs)\protect{\paolofootnotemark}]\paolofootnotetext{QNICs should be considered analogous to classical network interfaces: for each neighbor, the node has one independent QNIC that is exclusively connected to that neighbor. The QNIC in question will hold link-level pairs entangled to the neighbor's QNIC and pairs entangled with remote nodes' QNICs created through swapping.}, which comprise the quantum part of the node hardware, i.e.\@ a register of quantum memory slots and a Bell State Analyzer to interfere the photons emitted by memory slots with incoming remote photons;
    \item[Quantum Repeater Software Architecture (QRSA)] module, which encases all the control logic of the node and is itself divided in:
    \begin{itemize}
        \item ConnectionManager (CM), mapping end-to-end connection identifiers to the quantum hardware allocated to them;
        \item HardwareMonitor (HM), in charge of tracking the quality of physical links through link-level tomography;
        \item RealTimeController (RT), the interface between QRSA and QNIC through which quantum hardware can be directly manipulated;
        \item RoutingDaemon (RD), a submodule whose task is discovering the network topology through the OSPF \cite{rfc2328OSPF} protocol and maintaining a routing table;
        \item RuleEngine (RE), the heart of the QRSA, tasked with executing the \emph{RuleSets}, generic programs generated by the CM under request by an application that implement quantum networking algorithms. More detail on the RuleSets will be provided in the following.
        \end{itemize}
    \item[Application] module, which is only present if the node is an end node\paolofootnote{as opposed to a repeater node, that has no application module and therefore waits passively for a RuleSet to execute (usually one for entanglement swapping and/or purification).}, tasked with generating traffic in the form of requests of $N$ entangled pairs with a given partner (possibly with specific requirements on entanglement quality);
    \item[Router] module, whose purpose is relaying internal messages between the submodules and dispatching incoming classical message from the external world to the appropriate submodule.
\end{description}
To gain insight into how QuISP works, let us look at how connections are set up. As mentioned before, the core concept of QuISP's network architecture is the RuleSet \cite{MatsuoRuleset,CocoriMaster,QInternetArchitecture}, an arbitrary program that quantum nodes can run that implements a given quantum application. To establish a connection and distribute entanglement between two nodes, RuleSets need to be created and distributed to each of the involved nodes. This is achieved through the Connection Setup procedure, which is explained in detail in the following paragraphs. 
\subsection{Connection Setup in QuISP}
\label{sec:chap3_ConnSetupQuISP}
Despite quantum communication not being directional, establishing a quantum connection requires negotiating communication parameters, which are classical information and are therefore exchanged through classical channels. While in the following we describe the connection setup procedure in QuISP, it should be clear that this is only one possible implementation, and there is no standard protocol for how quantum connections are negotiated in general.

The first step is carried out by the Initiator, i.e.\@ the physical node tasked with starting the quantum application: when the Initiator's application submodule needs entanglement, it generates a message called the ConnectionSetupRequest (CSRq) and sends it to the node's ConnectionManager, which stores it in a queue. A CSRq contains the address of the prospect partner (called the Responder) and the number of requested entangled pairs. If the QNIC required to establish the connection is free, i.e.\@ it is not reserved by other connections, the CM reserves it and forwards the CSRq to the next hop towards the Responder.

On reception of a CSRq, a repeater's task is to relay it to the next hop on its route, which might be another repeater or the Responder itself. Before relaying the CSRq, the repeater checks whether the two QNICs required for the connection are free\paolofootnote{The repeater checks availability of the QNIC connected to the previous hop in the chain and of the one connected to the next hop.}, and if they are it reserves them and relays the CSRq. If one or both QNICs are already reserved by an earlier established connection, then the CSRq is not relayed to further nodes: the repeater creates a ConnectionSetupRejection message and sends it back to the initiator, which after a random back-off time will retry to initiate the rejected connection.

Once the Responder receives the CSRq, it checks availability of the required QNIC and reserves it, completing the first pass of the Connection Setup procedure.
At this point, the Responder generates the RuleSets and sends them to the other nodes, where they are translated into a low-level assembler-like language and executed by each node's RuleEngine. 
\subsection{The RuleSet}
The RuleSets are the core of QuISP's design, giving it its flexibility and providing endless extension potential: the RuleSets implement quantum routines such as tomography or entanglement swapping.
\begin{itemize}
    \item For the repeaters, the swapping RuleSets formalize the directive ``whenever entangled pairs are available, perform swapping and send correction messages to the new partners'';
    \item For the end nodes, the RuleSet depends on the specific application. As a simple example, a RuleSet for tomography would be ``whenever an entangled pair is available and entangled with partner end node, measure it in a random basis and log the result; when $N_\text{tomo}$ measurements have been carried out run statistics, reconstruct the quantum state and terminate execution.'';
\end{itemize}
From a concrete point of view, a RuleSet is a list of (\texttt{Condition}, \texttt{Action}) pairs called \textit{rules}. Conditions are composed of clauses such as ``There are $N$ entangled pairs with partner $i$ available'' and ``$K$ measurements have been performed'', combined through logical operators. The RuleSet is traversed from top to bottom, and when one of the Conditions is found to be true the corresponding Action is executed. Actions may include manipulations of the quantum hardware (such as measurements or other quantum operations on qubits) and classical communication operations (such as sending classical messages with control information inside). Rules are formulated in a custom and extendable assembler-like language that is interpreted by a Runtime, a container-like structure spawned by the RuleEngine on reception of a RuleSet. The Runtime can directly control the quantum hardware through the RealtimeController but also perform more abstract tasks through a callback interface to the RuleEngine. Users interested in extending the functionality of QuISP are required to formulate their quantum algorithms in the assembler language used by the Runtime.

While such an abstract architecture has numerous advantages, the most interesting one in this context is the decoupling between the quantum hardware, the classical communication components and the control modules. Since a quantum node acts as a self-contained processor for arbitrary RuleSets and the RuleSets themselves are shared through classical communication packets, it is possible to modify the physical layer implementation and create new interfaces for the nodes while leaving the quantum hardware and its control logic virtually untouched. In our case, this entails the possibility to interoperate modified satellite components with the original fiber ones in a large-scale hybrid network system. 

At this point, the connection is established and it runs until the termination condition of the RuleSets is reached. On RuleSet termination, a mechanism is required to terminate the connection, delete the RuleSets and release all QNICs that were reserved. This is known as a \emph{Connection Teardown} mechanism, and its implementation is one of the modifications to QuISP that make it applicable to scheduling problems.

In the following sections, we provide an account of all the modifications we performed to QuISP to make the work presented in this thesis possible. Readers interested in more detail concerning the unmodified QuISP code may consult \cite{quisp,quispsite}.
\section{Adapting QuISP to Dynamic Satellite Networks}
\label{sec:adapting_quisp_sat}
Since QuISP was designed with wired fiber networks in mind, several assumptions ingrained in the code base required modifications and extensions to adapt to the more dynamical nature
of satellite communications.
All our changes can be summed up as allowing the variation of parameters
which are fixed in fiber networks, and managing the consequences. The code changes corresponding to this work are available on GitHub \cite{PRCSVParser,PRFSChannels,PRFSChannelsTests,PRPointingSystem} and they are currently being merged into the main branch of QuISP.

The two main areas where QuISP's code base was extended to satellite communication were in the channel suite, where a new \texttt{FreeSpaceChannel} object was designed to account for intermittent visibility and varying transmission parameters across the free-space medium, and in the quantum node's classical communication interface, where new submodules such as the \texttt{PointingSystem} were introduced to regulate the flow of packets and prevent nodes from attempting to communicate when the receiving node is not in sight. 
\subsubsection{Free-Space Communication}
The first step towards satellite communication consisted in the definition of a \texttt{FreeSpaceChannel} object. Other than the obvious modification of parameters such as the speed of light, the peculiarity of a \texttt{FreeSpaceChannel} is that at any given moment the other end of the channel may not be in sight. Therefore, if a node attempts to send a message at a time when the recipient is not visible, the \texttt{FreeSpaceChannel} discards it. While a simple pass/fail model is enough for communication of classical control packets in free-space, the quantum version of the \texttt{FreeSpaceChannel} features a different transmission model from base QuISP.
To increase scalability, QuISP does not track the full quantum state of all qubits inside the simulation. Instead, each qubit is associated to an \textit{error vector}:
\begin{align*}
(P_\mathbf{I},P_X,P_Y,P_Z,P_0)
\end{align*}
where each component respectively represents the probability of being untouched, having undergone a Pauli X/Y/Z error or being lost.
Propagation through a channel is modeled by multiplying the error vector by the channel's \textit{propagation matrix}, which stores the probability for a qubit in state $i$ to be affected by error $j$ during propagation. The propagation matrix is calculated differently depending on the channel's physics and was the main object of our modifications.
Due to the exponential nature of losses and error in fiber channels, their transition matrix is first built in a per-kilometer form by taking loss and X/Y/Z error rates per kilometer and then suitably exponentiated depending on the length of the channel.

Since in free space the scaling of losses is quadratic, it is not possible to follow the same matrix exponentiation path. Moreover, while Pauli errors are a fundamental part of fiber propagation, they are not as prevalent in an isotropic medium such as free space. Due to these differences, the calculation of the propagation matrix for free space channels requires a new procedure.

To build a propagation matrix for free-space channels, the most important term is the loss one, which according to \cite{SatLosses,dFdP} is calculated as 
\begin{align}
    A(t) = \frac{L^2(t)(\theta^2_{diff} + \theta^2_{atm})}{D^2_R}A_{atm}(t).
    \label{eq:satellite_losses}
\end{align}
The terms included in this equation are described, together with the values we adopted, in tab.\@ \ref{tab:satellite_lossesparams}.
\begin{table}[]
    \centering
    \caption[Parameters from the free-space loss equation with definitions and typical values.]{Parameters from (\ref{eq:satellite_losses}) with definitions and typical values.}
    \begin{tabular}{|c|>{\centering\arraybackslash}m{.4\linewidth}|>{\centering\arraybackslash}m{.42\linewidth}|}
        \hline
        $L(t)$ & Satellite-ground distance & $500 \unit{km} \le L(t) \le1500 \unit{km}$ \\
        \hline
        $D_R$ & Diameter of the ground telescope & $\sim 1 \unit{m}$ \\
        \hline
        $\theta_\text{diff}$ & Diffraction-limited divergence angle of the transmitter telescope & $\frac{1.27\lambda}{D_T}$ \\
        \hline
        $\lambda$ & Wavelength of the photons & $\sim 800 \unit{nm} \le \lambda \le \sim 1500 \unit{nm}$ \\
        \hline
        $D_T$ & Diameter of the onboard telescope & $\sim 30 \unit{cm}$ \\
        \hline
        $\theta_\text{atm}$ & Atmospheric turbulence divergence angle of the transmitter telescope & $\frac{2.1\lambda}{r_0}$ \\
        \hline
        $r_0$ & Fried parameter (\cite{FriedParam}) & $5 \unit{cm} \le r_0 \le  20 \unit{cm}$ \\
        \hline
        $A_\text{atm}$ & Atmospheric Attenuation (dB) & $\sim 10 \unit{dB} \le A_{atm}  \le 25 \unit{dB}$ downlink, $\sim 35 \unit{dB} \le A_{atm}  \le 50 \unit{dB}$ uplink \\
        \hline
          \end{tabular}
    \label{tab:satellite_lossesparams}
\end{table}
For the Pauli error rates, due to the aforementioned optical isotropy of free-space, we employ flat, non distance-dependent user provided rates.

To avoid losing crucial messages, any module wishing to communicate across a \texttt{FreeSpaceChannel} needs to be equipped with a mechanism that checks whether the recipient is in sight before attempting communication. To this end, we implemented a \texttt{PointingSystem} submodule: before sending a classical message or a quantum signal, a node must poll its \texttt{PointingSystem} submodule for visibility of the recipient. If there is no visibility, classical control signals are buffered in a local queue and sent when visibility is restored whereas quantum packets (i.e.\@ photons) are discarded. This is because absence of visibility likely means the satellite has finished its passage and will not be visible for several hours, an interval of time that is too long to store qubits in current-generation quantum memories.

For each free-space link, the link distance, elevation and atmospheric transmission over time
are precomputed (see Fig.\@ \ref{fig:satellites_linkparams}) and stored in \texttt{.csv} files, which the \texttt{FreeSpaceChannel} takes as input 
to simulate the varying link over time. This \texttt{.csv} file based strategy
is intended to make the interfacing of our simulation to more elaborate orbital dynamics and/or
atmospheric simulation tools easy.
\subsubsection{Moving Nodes}
As discussed in Sec.\@ \ref{sec:satellite_TheorySingleLink}, one of the key physical challenges posed by a satellite system is its movement: varying the distance and elevation of the satellite changes a free-space channel's traversal time and transmission properties. In our implementation, the channel objects can read precalculated data on distance and atmospheric attenuation between a given ground station and a satellite. This solution allows for maximum interoperability with existing satellite software.

Another more impactful issue is the differential latency. In QuISP's base code, the timing for interferometric measurements is determined by the Transmitter node at boot, cached and sent to the Receiver node. In standard QuISP, the Transmitter schedules local emission at $t_0$ and instructs the Receiver to emit at $t_0 + t_{d}$, where $t_d$ is the channel's propagation delay. 
In our modified implementation, the Transmitter schedules local emission at time $t_0$, then calculates what the channel delay will be at time $t_0$ from the satellite's orbital data and instructs the Receiver to emit at $t_0 + t_\text{predicted}$. Instead of being cached at boot, the emission timing is re-negotiated at the beginning of each entanglement round. 
\section{Enabling Multiple Connections}
\label{sec:adapting_quisp_muxing}
At the time this work was started, network multiplexing, i.e.\@ the capability for a node to have multiple connections passing through it and to allocate resources in a way that warrants their simultaneous operation, was not available in QuISP. Since multiplexing is a crucial prerequisite for studying scheduling, its implementation is an active research topic. We report in the following the current state of our investigation, i.e.\@ a multiplexing implementation that works under a given set of assumptions and for specific, deterministic resource allocation policies.
\subsection{Terminating Communication: the Connection Teardown Mechanism}
\label{subsec:teardown-intro}
\begin{figure}
    \centering
    \subfloat[Incorrect behavior: RuleSets are executed as soon as they are received. Qubit allocation is asynchronous, swapping corrections are not guaranteed to arrive when the Initiator has booted its RuleSet and allocated the relevant qubit.]{\includegraphics[width=.5\linewidth]{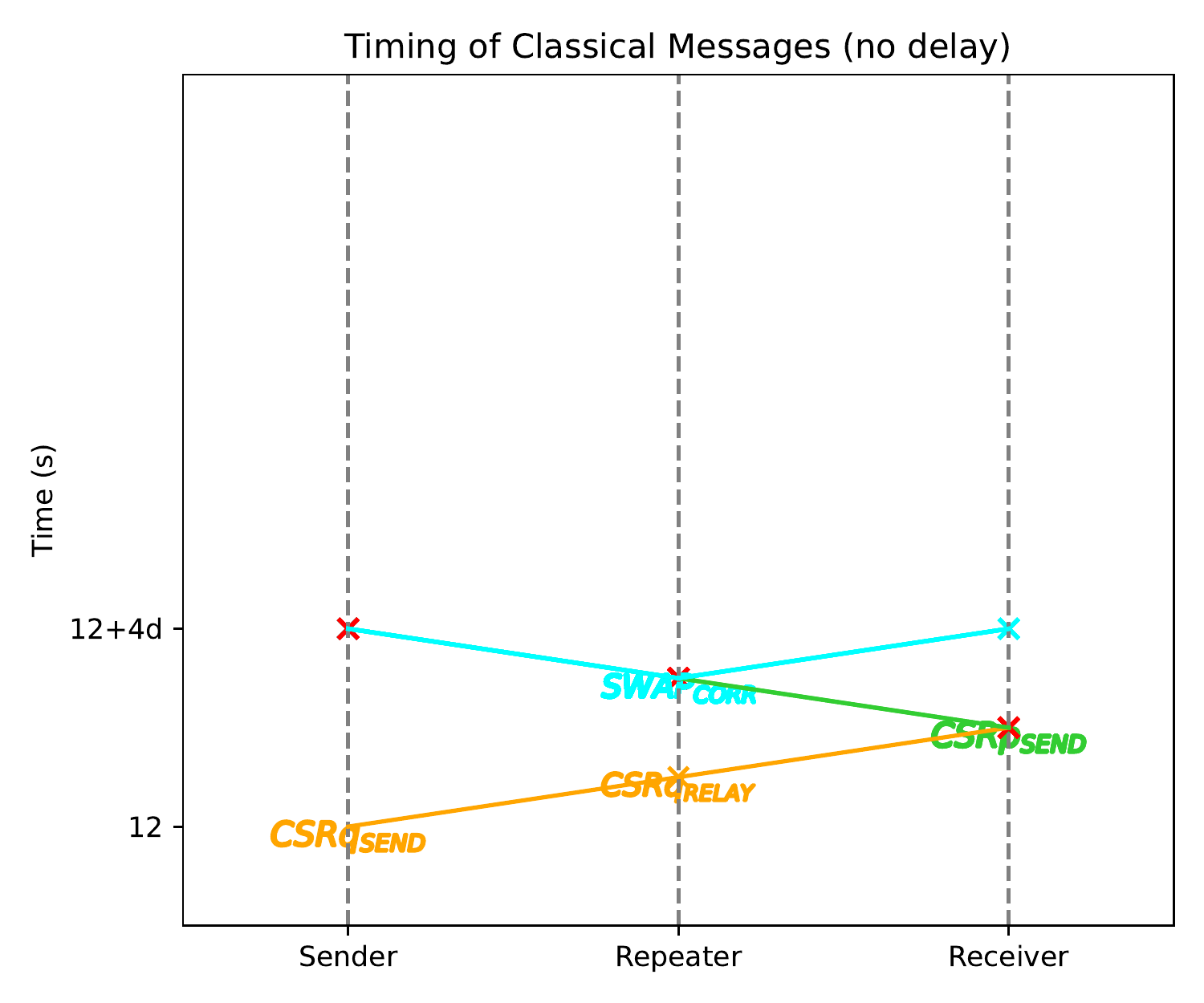}
    \label{fig:muxing_wrongswappingtiming}}
    \subfloat[Correct behavior: RuleSets execution starts after a programmable delay. Qubit allocation is synchronous. Correct selection of the start delay guarantees that swapping corrections arrive when the Initiator has booted its RuleSet and allocated the relevant qubit.]{\includegraphics[width=.5\linewidth]{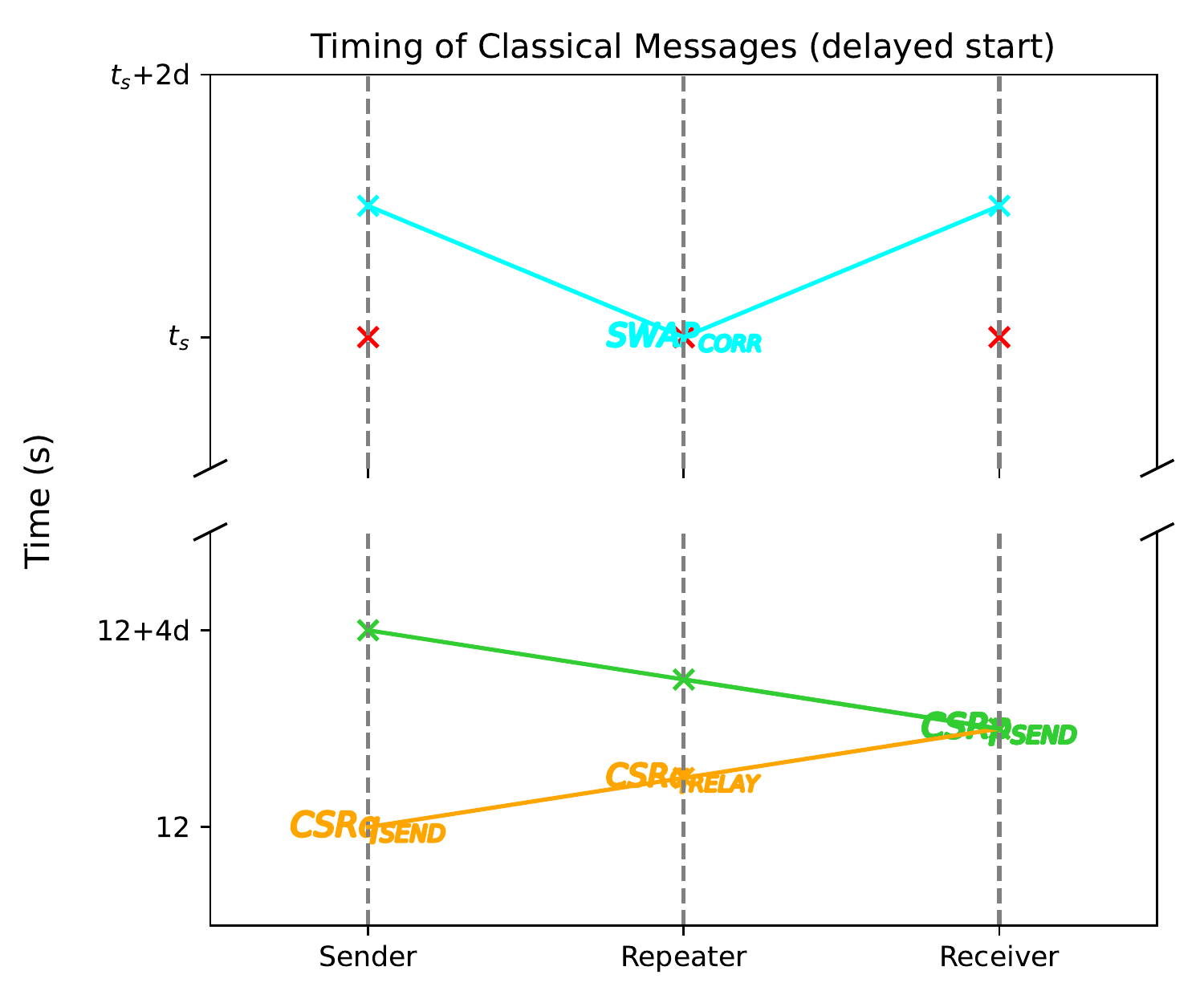}
    \label{fig:muxing_rightswappingtiming}}
    \caption[Timing scheme for QuISP's connection establishment procedure, with and without our corrections.]{A timing scheme showing our correction to QuISP's connection setup phase timing. $d$ represents the propagation delay across one fiber link, and $t_s$ is the programmable RuleSet execution start time. A red cross symbolizes allocation of one qubit, whereas other colored crosses represent the reception of the corresponding colored message. This plot was generated by parsing the output of custom logging routines we implemented in QuISP.}
\end{figure}

Tearing down a connection is a vital part of the operation of a (not necessarily quantum) network system: when an application terminates, the corresponding connection must be terminated and the reserved resources released. In QuISP, this means deleting the connection's RuleSets, freeing the reserved QNICs and resetting any leftover entangled qubits.\\
Notice that, whereas the Initiator and Responder's RuleSets have an explicit termination condition (stop once $N_\text{tomo}$ measurements have been performed), the repeaters' do not: without communication with other nodes, there is no immediate way for a quantum repeater to know by itself when enough entangled pairs have been distributed. Hence, the repeaters will perpetually allocate link-level pairs to the active RuleSet and swap them unless an external factor halts execution. The first step of a connection teardown procedure must therefore be a formal mechanism for the end nodes to clearly and promptly communicate termination to the repeaters.\\
In our implementation, we chose to let the Responder handle all the termination tasks: once its RuleSet's execution is terminated, the Responder's RuleEngine notifies its ConnectionManager, who sends out a list of ConnectionTeardown (CTD) messages requiring termination of swapping RuleSets, release of reserved QNICs and reset of all entangled qubits still allocated to the terminated RuleSet. Reception and compliance with the CTD message's instructions reverts the repeaters to the initial stage: no QNICs are reserved, no entangled pairs are allocated, and the nodes are ready to accept new reservations.\\
The implementation of teardown unmasked a timing problem in QuISP that was previously hidden by the restriction to a single connection: during normal operation, a link-level pair is generated, allocated to the active RuleSet, swapped until an end-to-end pair is created, and consumed by an application. However, during the connection setup stage in the unmodified QuISP code the repeaters start execution of the swapping RuleSet on reception (as shown in fig.\@ \ref{fig:muxing_wrongswappingtiming}), meaning that sometimes the first allocation and swapping are performed before the Initiator has actually received its RuleSet.
We remind the reader that in entanglement swapping the BSM is only part of the process: whenever an $AB$ and a $BC$ pair are swapped to form $AC$, $B$ must send classical messages to $A$ and $C$ to communicate the new partners for their qubits and the possible correction gates to apply. Until these messages have been received and the correction operations applied, the $AC$ entanglement cannot be used. This is precisely what created the timing issue: the correction messages can only be handled if a RuleSet is present, meaning that if they somehow arrive before the CSRp the concerned memory slot is never corrected nor recorded to have a new partner. When the CSRp is finally received, the slot looks like a link level entangled qubit to the newly established RuleSet and is therefore never consumed, effectively corrupting the memory slot in which it is stored and reducing the actual number of usable memory slots for entanglement by $1$.\\
This problem went undetected before our work because QuISP allowed a single connection: even if the buffers are reduced by a few qubits, the connection can still function and effectively distribute entanglement. However, when multiple connections are allowed, each subsequent connection corrupts some new qubits until the buffers are fully corrupted and no new entanglement can be established, effectively stalling the network.\\
To solve this problem, we added a \texttt{StartTime} field to the CSRp message, allowing the Responder to specify a time at which the RuleSet's execution should start. If the delay is correctly calculated (in our implementation, we set it to $10\times$ the time it took for the CSRq to reach the Responder), execution starts in a synchronous way and misallocations are prevented. The correct behavior is displayed in the timing plot in fig.\@ \ref{fig:muxing_rightswappingtiming}.
At this point, connection teardown is operational: it is possible to set a connection up, run it to termination, free its resources and accept a new one to restart the cycle. However, it is still only possible for each given QNIC to be reserved by a single RuleSet: true multiplexing requires multiple connections to be simultaneously operational, and in the following we will describe some additional prerequisites and design decisions.
\subsection{Multiple Concurrent Reservations}
\label{sec:sim_resregister}
In the unmodified QuISP code base, every QNIC has two possible reservation states, which are either busy or free. Whenever a connection requiring a given QNIC is established, said QNIC is set to the busy state and is unavailable for further connections until explicitly freed. To enable the establishment of multiple concurring connections, the reservation mechanism must be refined: not only need the ConnectionManager (CM) allow multiple reservations through the same QNIC, but it must also have a way to associate reserved QNIC to reserving connection. Such association enables the CM to delete reservations for terminated connections without perturbing the ones that are still running.\\
To satisfy these requirements, we introduced a new object called the \emph{ReservationRegister} (RR): the RR's purpose is to map a RuleSet ID to the QNICs it is reserving, and vice versa. When a connection is established, a new reservation is added to the register and when the termination message is received, the reservation is erased. The RR is handled by the CM, which uses it to store reservation information, and every CM has one RR.\\
Implementing the RR also uncovered a separate problem: in QuISP, QNIC reservations are made in the first pass of the connection setup procedure (CSRq from Initiator to Responder), but the RuleSet ID is only known in the second pass (CSRp from Responder to Initiator), once QNICs have already been reserved: there needs to be a way to associate QNICs that have been reserved during the first pass with the RuleSet ID received from the Responder during the second. To this end, we added a ConnectionSetupRequestID (CSRqID) field to the CSRq message. Reservations during the first pass are made and inserted in the RR under the provisional CSRqID, and later on updated to match the correct RuleSet ID. For this to be possible, the CSRqID needs to be communicated in the CSRp so that nodes that have a reservation under a matching CSRqID can modify it to be under the finalized RuleSet ID. In principle, there is no need to have two separate IDs: the provisional CSRqID generated by the Initiator could simply be promoted to a final RuleSet ID on acceptance of the connection. However, since QuISP's connection setup procedure was designed with the Responder generating the RuleSet ID, we opt for maximal compatibility with the base code and leave the generation of the final RuleSet ID to the Responder.\\
In practice, the RR is a wrapper object for two maps, one mapping a RuleSet ID to the set of QNICs reserved by it and one mapping each QNIC to the set of RuleSet IDs reserving it. Such a double map structure enables efficient lookup on both sides. Once the CM receives a connection termination message, the reservation matching the terminated connection's ID is looked up and erased.
\subsection{Simplifying Assumptions in the Absence of Link-Level Synchronization}
\label{subsec:mux_assumptions}
While the design and implementation work of the previous sections brings us closer to a working multiplexed simulation scenario, there are still some problems to solve before true multiplexing can be achieved. In the following, we outline these problems and lay down assumptions that enable us to extract preliminary scheduling results from QuISP. Link-Level Synchronization is the first issue that must be solved to develop a general-purpose multiplexing implementation: in QuISP, resource allocation is handled at the link level, which means entangled pairs are allocated to an end-to-end connection as soon as they are physical generated. This is a different approach from the one we took in chap.\@ \ref{ch:scheduling}, where the link-level pair was treated as a general resource that is not specifically allocated to any connection until consumption. While direct allocation is a less dynamic approach than the one we took in chap.\@ \ref{ch:scheduling}, there is no evidence in favor of either approach performing better than the other. We thus adhere to QuISP's design decision of allocation on creation, and keep the difference between the two allocation policies as an interesting potential investigation direction. When allocation on creation is employed, it is fundamental for both ends of a link to allocate their local qubit to the same end-to-end connection. The solution to this issue lies in link-level synchronization, and it has been detailed in the AQUA Team's design specifications for QuISP \cite[ConnectionSetupSpec.md, sec. 4.6 Barrier]{quispspecs}. In short, nodes need to negotiate a Link Allocation Policy (LAP), i.e.\@ an order of priority for allocation of resources to the active RuleSets. Nodes must agree on the LAP, and it is crucial they switch to a new LAP (e.g.\@ when a connection is established or torn down) synchronously. The design specification requires that this happen through special messages called Barrier Messages: whenever an event triggers a LAP change, nodes send a Barrier message to the affected neighbor. The message contains the new proposed LAP, the current Photonic Trial Number\paolofootnote{i.e.\@ the index of the photonic packet last processed by the node's BSA, only way to identify photons belonging to the same ebit across different nodes at the link level.} (PTN) and a random number. Each node compares the incoming message with its local variables, and the highest PTN is taken as starting point for the new LAP. In case the PTNs coincide, the LAP with the highest random number is adopted.\\
In the interest of exploring preliminary scheduling results, we investigate an alternative way of multiplexing inspired by time-division multiplexing (a multiplexing discipline briefly mentioned in \ref{subsec:Muxing_classical}), in which we divide the time axis in disjoint windows and allocate all generated ebits to a different connection in each window. This method can be implemented as a rudimentary resource allocation policy that allows the formulation of simple scheduling policies. The problem of link-level synchronization can be circumvented in this scenario by precalculating the allocation windows: nodes agree on a common start time (such as the delayed \texttt{StartTime} determined in our modified Connection Setup procedure as per sec. \ref{subsec:teardown-intro}) and then deterministically cycle active policies at prefixated time intervals. This approach has a major disadvantage in that it imposes that only deterministic precalculated scheduling policies be considered (a significant downgrade from the stochastic dynamic policies examined in chap.\@ \ref{ch:scheduling}), but its simplicity of implementation allows us to start exploring simple scheduling policies while the full, link-level synchronized allocation policy is under investigation.
\section{Simple Scheduling in Fiber with QuISP and Differences with our Framework}
\label{sec:DSSandQuISPSched}
\begin{figure}
    \centering
    \includegraphics[width=0.5\linewidth]{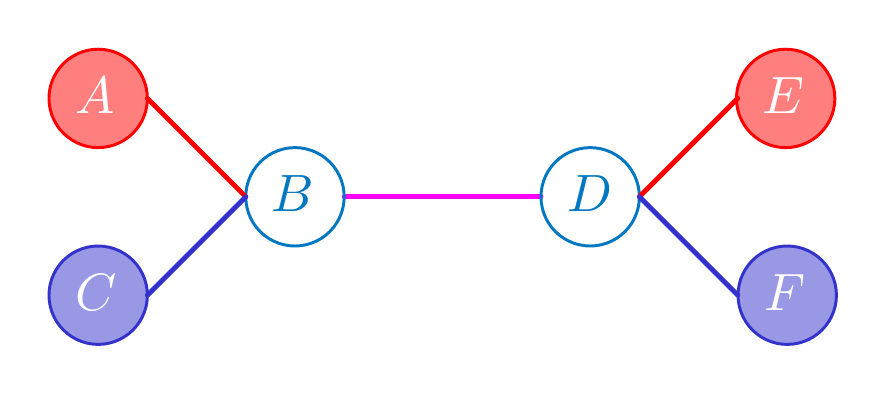}
    \caption[A simple dumbbell network topology.]{A simple dumbbell network topology, with two commodities $(A,E)$ and $(C,F)$ and a bottleneck across the shared link $BD$.}
    \label{fig:dumbbell_network}
\end{figure}

In this section, we examine a first example of scheduling problem with QuISP and cross-validate our modifications against the code developed in chap.\@ \ref{ch:scheduling}. To do so, we outline a simple scheduling policy that is compatible with both tools and examine the small differences between the two approaches.\\
Before going in more detail, there is a difference in language to clarify between the previous scheduling discussion and the same issue in QuISP. In this chapter, we presented QuISP as a simulator that classically negotiates a connection\paolofootnote{In classical network science, one would speak of \emph{connection based protocols}, which are protocols that negotiate a point-to-point connection before running communication. An example of such a protocol is TCP \cite{rfc9293TCP}, the most used transport layer protocol that needs to negotiate a connection between two nodes (e.g.\@ an end-user's computer and a remote server) before communication. A counter-example (also known as \emph{connectionless protocol}) would be the UDP \cite{rfc8304UDP} protocol, which is commonly used in situations where low latency is more important than reliability, such as taking real-time measurements from remote sensors.} requesting a fixed number of entangled pairs, runs quantum communication and then closes the connection. On the other hand, in our discussion in chap.\@ \ref{ch:scheduling} we examined a network in which in principle any pair of nodes may request entanglement, and the scheduler will attempt to serve it. In the context of scheduling problems, the two approaches are equivalent: a network with demand across several pairs might be seen as a network with several active connections, and whenever demand across a new pair is received, it is equivalent to a new connection being established. The only conceptual difference between our treatment of scheduling and QuISP's is in the traffic model: while in chap.\@ \ref{ch:scheduling} we adopted a Poissonian traffic model, a connection-based setup produces requests in a more impulsive way, placing a batch of demands on the network instead of the more distributed stream of a Poissonian model. As discussed during the presentation of our framework (sec.\@ \ref{sec:scheduling_queuingmodel}), the ebit and demand generation terms may act as open interfaces for alternative models. We therefore employ in this section a modified version of our simulator that generates demands in batches, with a user-defined number and period of generation.\\
As mentioned in the previous section, the current iteration of the multiplexing code in QUISP only supports deterministic scheduling policies. Therefore, we study a simple scheduling policy in the context of the bottleneck network shown in fig.\@ \ref{fig:dumbbell_network}.
The policy under exam supports two commodities, in this case $(A,E)$ and $(C,F)$), and is designed to alternate the complete allocation of the bottleneck link between them. It should be noted that this policy is not equivalent to the greedy one presented in sec.\@ \ref{sec_sub_Greedy}, which completely neglected demand: when the demand is zero across one of the two commodities, the corresponding route is ignored and the link is allocated completely to the other connection.
Therefore, we can summarize the policy under consideration as:
\begin{itemize}
    \item When there is no demand, nothing is scheduled. There are no swaps and no end-to-end ebits consumption. 
    \item When there is only demand across $(A,E)$ ($(C,F)$), the policy is equivalent to a greedy scheduler with a single $ABDE$ ($CBDF$) route.
    \item When both connections are active, the scheduler cycles between greedy scheduling across $ABDE$ or $CBDF$ exclusively depending on a user-specified switching period.
    \end{itemize}
\subsection{Validation: Comparing our Linear Algebraic Framework and QuISP}
\begin{sidewaysfigure}
\vspace{-2cm}
\hspace{-2cm}
    \subfloat[Performance metrics with our simple simulator, assuming instant classical communication and unlimited memory.]{\includegraphics[width=.59\linewidth,trim=8 5 60 5,clip]{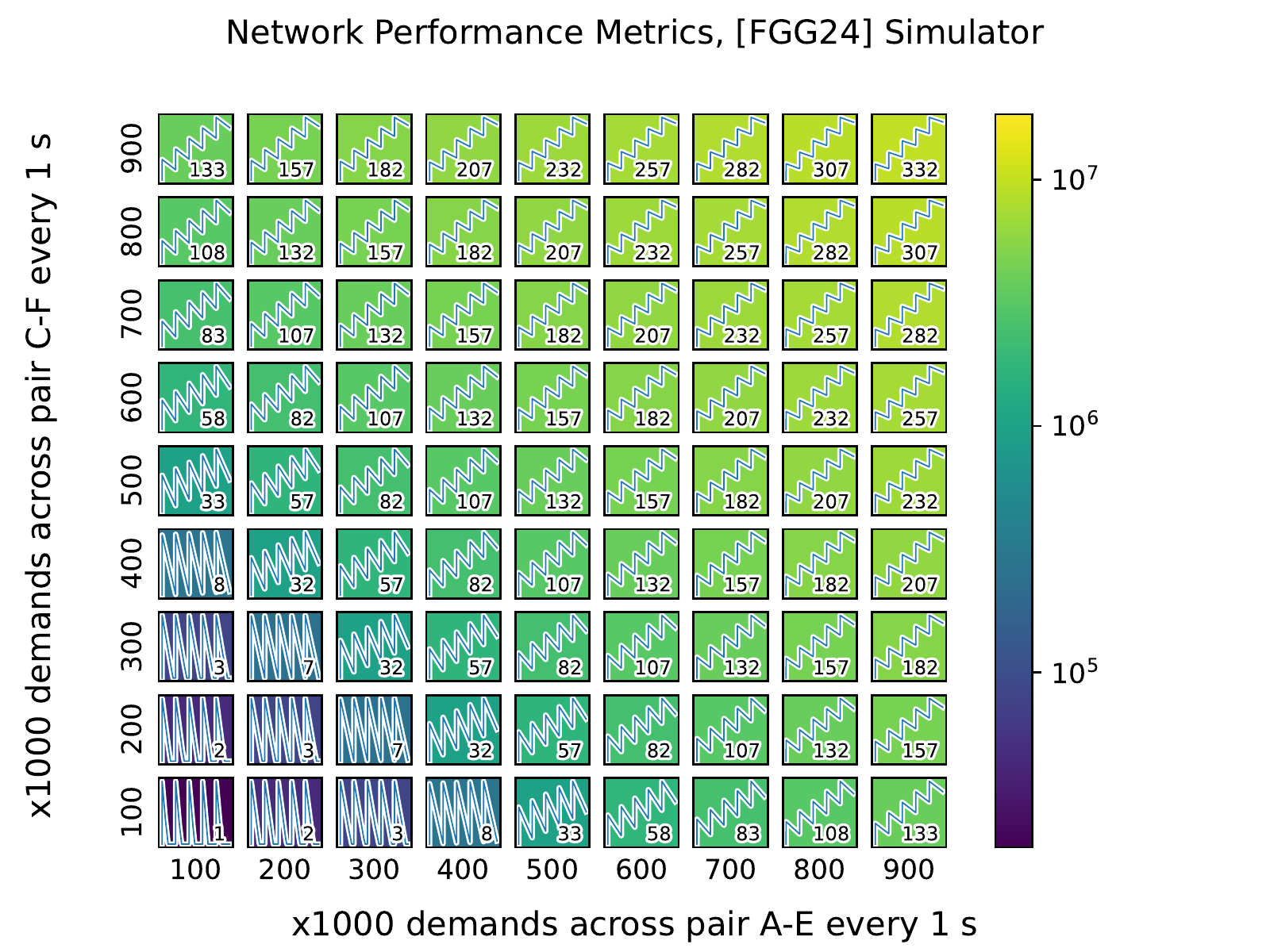}}{\hspace{.1em}\vrule\hspace{.1em}}
    \subfloat[Performance metrics extracted from QuISP's output, giving a slightly worse picture but still within the expected order of magnitude.]{\includegraphics[width=.59\linewidth,trim=8 5 60 5,clip]{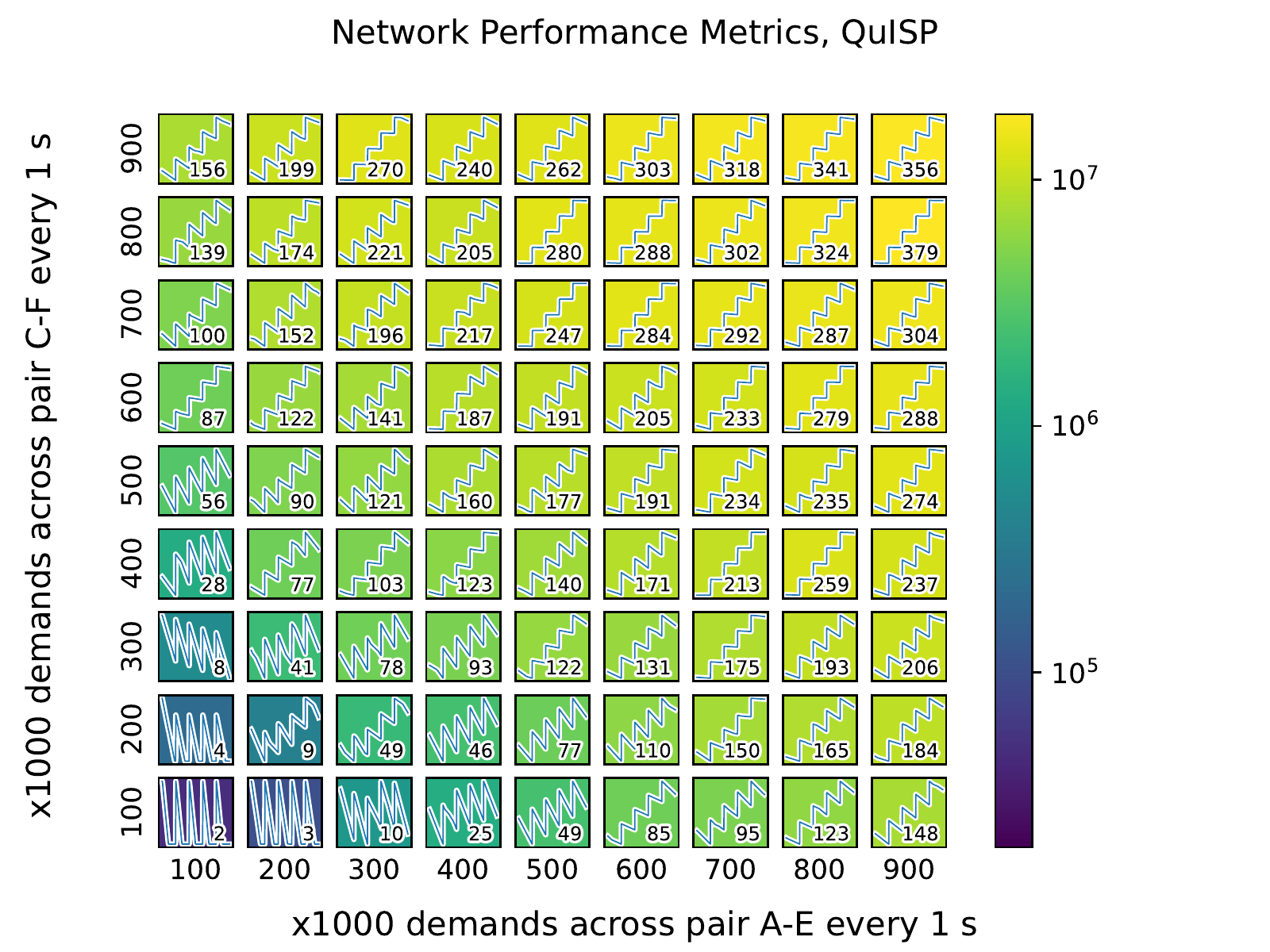}}
    \caption[Comparison of results between the simulator used in \cite{scheduling_journal} and QuISP.]{In this figure, we provide a comparison between our simple scheduling simulator and QuISP, as a mean of validation for both approaches and to investigate how much our optimistic physical assumptions are impacting the presented results. As in fig.\@ \ref{fig:scheduling_perfmetrics}, we plot in each cell the temporal trend of the net user demand in the network at a given load: a trend that reliably returns to zero shows a network that is likely stable (i.e.\@ able to serve all the requests in a finite time) while one that does not is most likely unstable. Looking at the trend of strictly increasing cells (e.g.\@ $(300,300)$ and above in both simulators) it is possible to notice how the less stable a configuration is, the more linear and steep its demand accumulation is. 
    Through these temporal plots, it is also possible to give a pictorial impression of the stability of the network under different loads by assigning a color to each cell representing the average demand backlog through the full simulation time. By calculating the average and thus plotting the stability region, we can appreciate how our simulator underestimates the demand backlog, making the network look more stable. Considering how relaxed the physical assumptions were on our simple model, we deem the two simulators to be in satisfactory qualitative agreement, giving reason to believe that the theoretical advantages observed in chap.\@ \ref{ch:scheduling} should carry over to a more realistic simulation environment and ideally to real hardware.
    } 
    \label{fig:sim_quispvsdss}
\end{sidewaysfigure}

To cross-validate the results of our simple Python simulator and QuISP, we need to unify the two tools under a single figure of merit. To this end, extensive use of OMNeT++'s built-in signal mechanism\paolofootnote{The signal mechanism is one of the most useful features in OMNeT++: inside the modules' C++ code, special instructions allow the simulation designer to record variables at any moment. The values of the recorded variables can be later analyzed through a Python interface. In our case, the ConnectionManager and RuleEngine were modified to respectively emit a signal when a ConnectionSetupRequest is created (i.e.\@ a batch of demands is received by the network) and when an end-to-end ebit is consumed.} allows us to extract stability regions from QuISP simulations and compare them with our simulator.\\
The obtained results are presented in fig.\@ \ref{fig:sim_quispvsdss}.
\subsubsection{Compatible Parameters: Unifying Inputs}
To properly compare the two simulators, it is important to unify their languages as much as possible. Other than the two simulators outputting the same figure of merit, they must also accept inputs that allow a fair comparison, especially considering the differences in physical assumptions. To this end, we implemented some additional features for our simulator:
\paragraph{Capped Physical Queues} The most radical difference between our simulator and QuISP is in the management of quantum memory slots, assumed infinite in our simulator. Since a direct comparison with QuISP would not be fair --- our simulator would accumulate infinite pairs in the memories and greatly overestimate performance --- but implementing finite memory would require a substantial redesign, we seek order-of-magnitude validation by capping the amount of ebits physical queues can generate. At every time step, physical queues can generate no more ebits than the difference between the cap and the number of currently held ebits. This solution is not perfectly accurate, as only physical queues are capped while virtual queues remain unaffected, but it is a reasonable modification to obtain order of magnitude results.
   \paragraph{Connection-based Traffic Model} While our simulator was developed with a continuous random stream of user demand in mind, QuISP works with a connection-based model, where users request batches of entangled pairs at fixed time intervals. To adapt our simulator, we implemented a second traffic model that allows users to specify a time interval and a number of ebits that will be requested by service pairs. The new traffic model can be exchanged with the old Poissonian one by toggling a variable in the simulator's inputs.
    \paragraph{Dumbbell Topology and Round-Robin Scheduler} The new topology and scheduler were added to the simulator by leveraging the custom topology and custom scheduler abstractions. This modification allowed us to practically gauge how open the simulator is to user extension, and it required a minor modification outside our abstractions: to implement a round-robin scheduler, we had to expose the simulation time as a parameter accessible to custom schedulers. This parameter is now available for custom policies and should prove useful for formulating future policies both in a possible extension to our work and in independent research projects by other users. We implemented the round-robin scheduling policy as a set of three precalculated scheduling vectors (zero load, nonzero load across either commodity alone) and applied the correct vector depending on the load conditions and on the current simulation time: when load is present across both commodities, the scheduler cycles between the two single-commodity decision vectors.
    \paragraph{Choice of the Time Step} We recall that our simulator is a DTS, as defined in sec.\@ \ref{sec:DTSvsDES}. DTSs are simulators that advance through the time axis uniformly, each time progressing by a fixed amount of time called a time step $\Delta t$. While $\Delta t$ carries no physical meaning, it is an important actor in the simulation and choosing its value improperly could invalidate results. In our case, we recall that scheduling decisions are taken at the end of every time step: this implies $\Delta t$ has an impact on how often entanglement is swapped and consumed. At every time step, a random number (either capped or uncapped) of qubits are generated, and then swapping and consumption decisions are taken for the entire network. Since this is coherent with QuISP's runtime loop (attempt generation of entanglement at the physical layer, perform swapping and consumption, repeat) it is sensible to choose our time step to be the duration of one entanglement generation round plus consumption. While QuISP has no fixed round duration (since it works with finite memory slots, the number of sequential generation attempts is not fixed), a good approximation is the ratio between the memory cap and the generation rate. With a generation rate of $1 \unit{\mega\hertz}$, this yields a time step of $0.1 \unit{\milli\second}$. Finally, link-level ebit generation in QuISP is handled through a Bell State Measurement with success probability $p_\text{BSM} = \sfrac{1}{2}$. Since our simulator does not account for such a measurement, we set its link-level ebit generation rate to half the value used in QuISP to account for the $\sfrac{1}{2}$ factor introduced by the BSM and properly match the generation rate of the two simulators.
\section{Summary of the Current State}
\label{sec:quisp_currentstate}
After discussing the modifications applied to QuISP throughout this thesis, we conclude this chapter with a brief summary of what has been done and which parts of the code still need development.
\paragraph{Satellite Communication} is completely implemented and functional: it can be used to simulate satellite-to-ground and ground-satellite-ground individual links, and it supports integration with the rest of QuISP's components to simulate larger satellite communication scenarios. The code is expected to also support multisatellite scenarios with satellite-to-satellite communication, and plans for future work include in-depth testing of such more refined scenarios. The code is available in the form of multiple pull requests to QuISP's GitHub repository \cite{PRCSVParser,PRFSChannels,PRFSChannelsTests,PRPointingSystem}: the first one was merged into the main code base and the remaining ones are scheduled for merging in the near future. These extensions to QuISP enabled us to simulate quantum networking in the new satellite regime and produce all simulation results presented in chap.\@ \ref{ch:satellites}.
\paragraph{Connection Teardown} and its timing corrections are functional in all cases tested in this thesis. However, the code has been in development until very recently, and as such test cases have yet to be written to accurately verify its functionality in all cases covered by QuISP. Other than in-depth testing and polishing, further development work will be required once a full multiplexing implementation with link-layer sync is available, as connection teardown is one of the triggering events for renegotiation of link allocation policies. Despite still requiring some work, the teardown mechanism is still useful because it enables QuISP to dynamically establish and close connections between different nodes, making the traffic model more realistic by lifting the single-connection restriction from the base code. The implementation of teardown allowed us to collect the data displayed in fig.\@ \ref{fig:sim_quispvsdss}.
\paragraph{Network Multiplexing} and scheduling in the general case are not at this moment possible in QuISP without major simplifying assumptions. A sizable amount of work needs to be invested in the implementation of a link-layer synchronization scheme to avoid misallocation of resources and enable dynamic management of link allocations by the nodes. Nevertheless, the application of some simplification assumptions such as deterministic resource allocation has allowed us to use QuISP to study a simple round-robin scheduling scenario and to validate our results against our Python-based simulator. As soon as link-layer syncing is implemented, general-case multiplexing will enable QuISP to simulate complex scheduling policies over large-scale fiber and satellite quantum networks.
\chapter{Entanglement Swapping in Orbit: A Satellite Quantum~Link Case Study}
\label{ch:satellites}
In this chapter, we model and simulate a satellite-mediated free-space quantum link. This chapter's results stem from a collaboration with Keio University's AQUA Team \cite{aquawebsite}, led by Rodney Van Meter. The results of this work were presented at the \textit{IEEE International Conference for Quantum Computing and Engineering 2024} \cite{fittipaldisatellites}.
\section{Introduction}
Designing a global-scale quantum network will most likely require the connection of metropolitan subnetworks through long-distance backbone links. Such an interconnection may prove to be extremely difficult to realize with quantum repeaters, leaving interesting research space in the exploration of alternative physical layer implementations that can coexist with them. An interesting candidate for long-distance backbone links is satellite communication: light attenuation in free-space is quadratic (as opposed to exponential in fiber) with distance, making satellite links very interesting candidates for long-distance backbone connections.

Satellite quantum links are an emerging trend in quantum network science, with active theoretical and experimental effort: in 2017, the Chinese Academy of Science launched Micius, the first ever quantum-enabled satellite \cite{Micius,YinMicius}. Micius was recently used to realize real-time Quantum Key Distribution \cite{MiciusQKD}, experimentally showing the feasibility of a satellite QKD link.
In \cite{VictorSatCVQKD}, the feasibility of a CVQKD LEO satellite-to-ground link is studied, with an account of the required experimental parameters to achieve useful performance through a statistical and simulated investigation of the quantum key rate. In the same subfield, \cite[Section VI]{PirandolaQKD} provides a wider scope overview of the current state of satellite QKD.\\
Much of the optimization revolving around satellite communication is devoted to the link layer, and in particular to minimizing losses. \cite{ValentinaSat1} and \cite{ValentinaSat2} describe how advanced adaptive optics techniques can be used at the communication terminals to boost respectively prepare-and-measure and entanglement-based QKD performance.

Beyond QKD, \cite{GundoSat} is a proposal for a quantum repeater protocol involving spaceborne nodes: both the cases in which satellites have onboard memory and that in which they only have photon pair sources (with swapping on the ground) are analyzed from the analytical and simulation point of view.  Moreover, in \cite{GundoSneakSat}, the authors further develop this scheme by studying a sneakernet-like approach mediated by a satellite node.

At the network level, \cite{RajaConnectingCities} is an effort to explicitly gauge the effectiveness of satellites as backbone interconnects between metropolitan fiber quantum networks using QKD as a performance benchmark and NetSquid as a simulation backend. Developing such an interconnected architecture would pave the way to the construction of a global quantum internet capable of distributing quantum entanglement to arbitrary groups of users, making worldwide distributed quantum applications \cite{protocolzoopaper,protocolzoo,AwschalomRoadmap} possible. 

In the context of interconnecting multiple subnetworks, this chapter focuses on studying the specificities of a point-to-point satellite link, with the long-term goal of integrating satellite links into our scheduling discussion and employing QuISP to study scheduling problems in large-scale fiber and satellite quantum networks. The main novelty in our work is that, since we predominantly focus on memory-endowed ground stations and satellites, we are required to keep into account the latency introduced by classical communication, which is related to the speed of light and puts significant constraints on the achievable performance, since no more photons can be in flight than there are available memory slots and every entanglement generation attempt also needs to wait for classical correction messages, effectively binding attainable rates to the round-trip time of the free-space channel between satellite and ground. The delays introduced by classical communication latency are often neglected in quantum network science, either because classical communication through the Internet is assumed to be ``easy'' with respect to quantum communication or because most research on quantum interconnects focuses on QKD as a benchmark, an application that does not generally require the long-term storage of ebits and is therefore unaffected by latency. Moreover, quantum communication is usually studied with optical fiber as a physical medium, which implies shorter links and much weaker latency-induced nonidealities. We show in this chapter that the instant communication assumption cannot apply to satellite networks, requiring a model that accounts for it and a simulator able to mirror classical message passing, core reason why QuISP was chosen as our simulation toolbox.

We show in this chapter a detailed 
study of a quantum link between Paris and Nice which could be created through
a single passage of a quantum-memory equipped satellite following the orbit of Micius \cite{Micius}.
We approach the problem through two complementary points of view: a simple 
analytical model and an event-based simulation with QuISP.

Given that the speed of light bounds the ground--satellite communication latencies to a millisecond time scale, during which the satellite does not move appreciably, a notion of rate of the link at a given time makes sense, allowing
us to derive a set of simple analytical formulas ((\ref{eq:theoretical-rate-corrected}), (\ref{eq:twinrate})) depending on known link parameters through which we can estimate the performance of a given satellite link over time. 
This model provides valuable insight during the feasibility study and network design phases, while also posing as a validation tool for results obtained through other means.

We analyze the performance of a single quantum satellite link in the satellite--ground and ground--satellite--ground scenarios:
\begin{itemize}
    \item In the first case, we study entanglement distribution between a quantum-memory-equipped satellite
following Micius' orbit \cite{Micius} and a ground station in Nice, in the south of France.
We assume a given number $m_S$ of quantum memory slots
(both onboard the satellite and at the ground station) and demonstrate the impact of memory 
size on the performance of our link.
    \item In the second case, we analyze a dual link between the satellite and two ground stations, 
one in Nice and one in Paris, with the satellite node acting as a quantum repeater. 
Once link-level entanglement is established between the satellite and each of the stations, 
the satellite creates a direct Nice--Paris connection through entanglement swapping. 
\end{itemize}
The choice to employ memory and entanglement swapping was mainly dictated by software compatibility: at the time this work was carried out, it was not possible to simulate midpoint-source entanglement generation in QuISP. Under some very minor adaptations, the discussion we present is also relevant for midpoint-source links (memories on the ground, twin photon source on the satellite), whose technological requirements are much lower than a swapping link.

Our goal is to provide a simple model and a simulation interface that are modular enough to be useful in researching quantum scheduling over large hybrid networks: our analytical model's simplicity makes it 
highly integrable in large network design calculations, while our simulation modules seamlessly work inside a realistic and highly scalable quantum network simulator, enabling simulation of large hybrid networks including both satellite and fiber links.


\section{System Description}
\label{sec:SysDesc}
\begin{figure}
    \centering
    \frame{\includegraphics[width=0.7\linewidth]{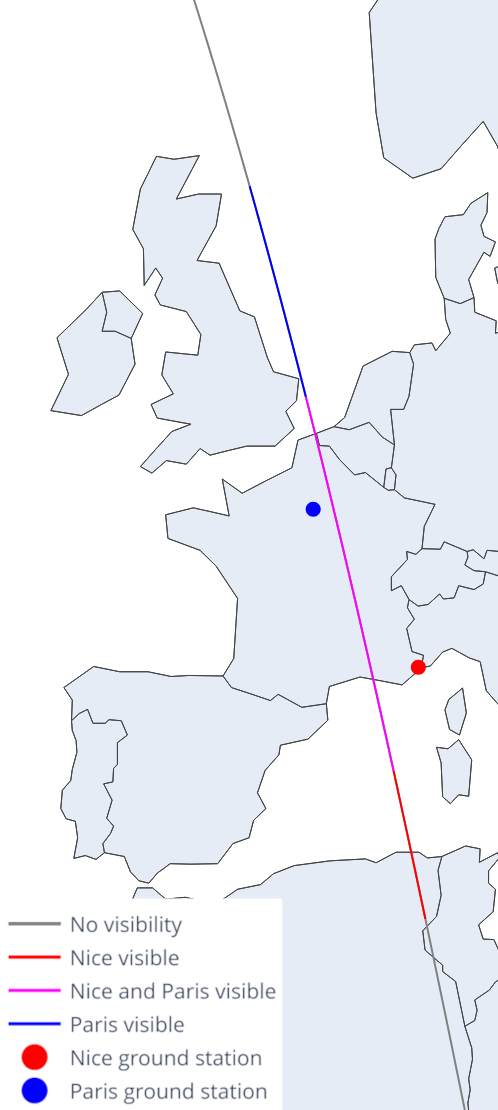}}
    \caption[Passage of the Micius Satellite over Nice and Paris.]{Passage of the Micius Satellite over the Nice and Paris ground stations.}
    \label{fig:sat-parisnicemap}
\end{figure}
\begin{figure}
    \centering
    \includegraphics[width=\linewidth]{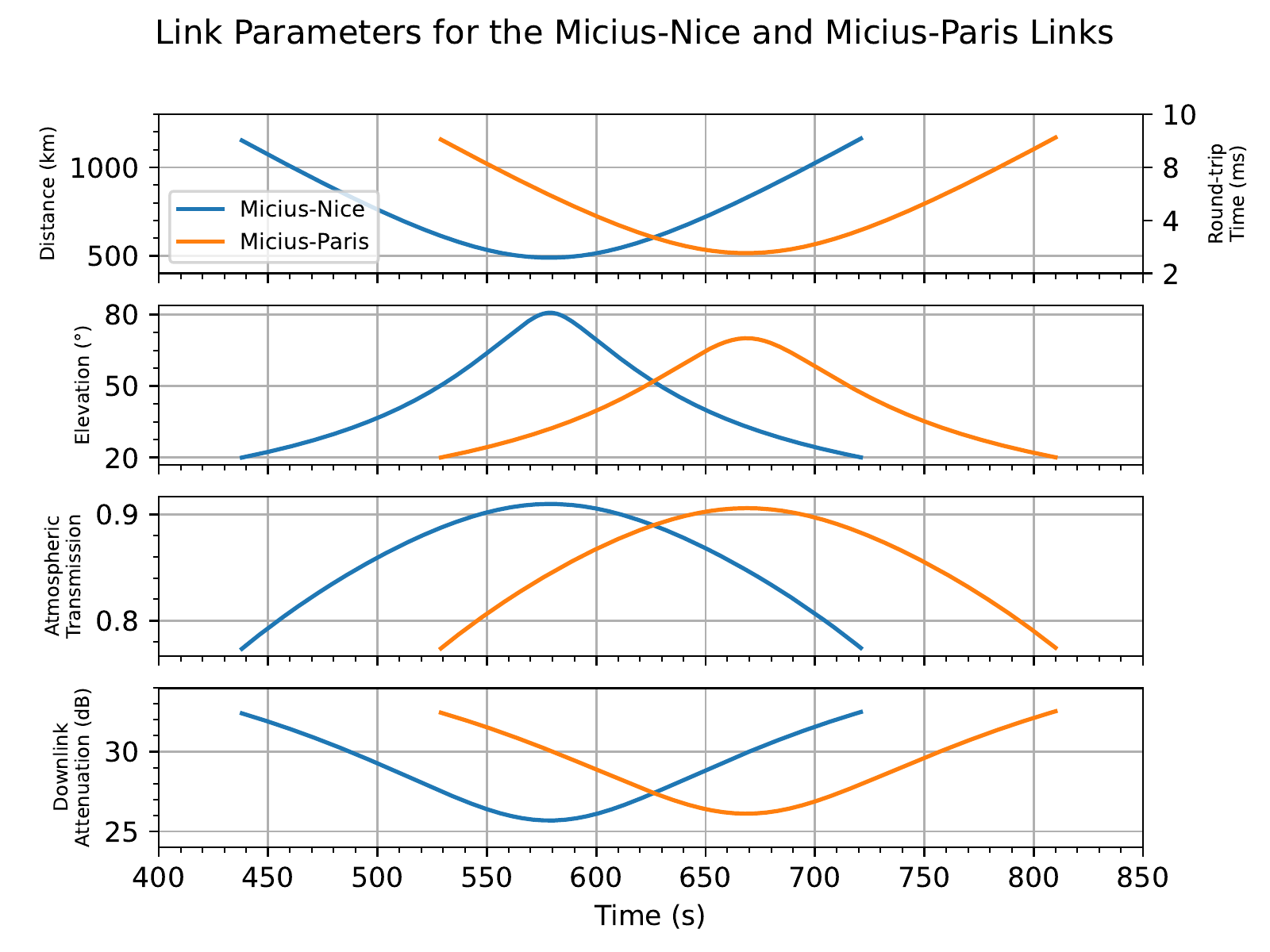}
    \caption[Parameters of the Micius--Nice and Micius--Paris satellite links during the passage of the Micius satellite
        starting July 27$^{\text{th}}$, 2021 at 10:40 AM UTC.]{Parameters of both the Micius--Nice and the Micius--Paris links during the passage of the Micius satellite
        starting July 27$^{\text{th}}$, 2021 at 10:40 AM UTC. 
        The distance and elevation are computed using \texttt{orekit} \cite{orekit} and the atmospheric attenuation with \texttt{LOWTRAN} \cite{lowtran}, in a rural setting with $23 \unit{km}$ visibility  and 
        no aerosol at a wavelength of $1550\,\unit{nm}$.
        The channel attenuation is computed using the previous parameters, using 
        (1) of \cite{SatLosses} for a satellite-to-ground (downlink) transmission
        and the parameters of \cite{dFdP}, 
        assuming telescope diameters of 10~cm on the satellite and a 1~m diameter on the ground. The link is taken to be available only for elevations above 20°.
    \label{fig:satellites_linkparams}
    }
\end{figure}
The link we study operates between two ground stations located in Nice and Paris, 
two French cities approximately 680 km apart (as shown in fig.\@ \ref{fig:sat-parisnicemap}, link parameters shown in fig. \ref{fig:satellites_linkparams}). To keep our discussion realistic, we employ orbital data from
a single passage of the Micius satellite \cite{Micius}. 
The large latitude difference between the two cities explains the $91$ seconds time difference in closest approach of the satellite from both locations, 
a nonnegligible difference for a 5 minute long satellite passage, while the longitude difference explains the $\ang{11}$
difference in their maximal elevation.
As a result, the link budget varies independently along the two legs of a ground-satellite-ground link.

This study is in the context of first generation quantum repeater networks \cite{1g2g3g}, 
where each node --- both on the ground and in space --- has access to a quantum memory, 
which it uses to synchronize entanglement swapping.
A key parameter of such memory is its lifetime $\tau$, the timescale at which
decoherence occurs. 
Given the unavoidable $\trt \ge 2\unit{ms}$ satellite--to--ground roundtrip communication time,
any quantum memory where $\tau \ll 1\unit{ms}$ would be essentially useless for 
our purpose. At the opposite end, quantum memories with very long
lifetimes --- $\tau\ge 100 \unit{s}$ --- would allow sneakernet-like
protocols where the satellite catches some qubits to carry them elsewhere, as in \cite{QNetShip}. 
To simplify our analysis, we set ourselves in an intermediate regime, 
where $\trt\ll\tau < 10 \unit{s}$: the decoherence
of the memory is negligible, but the satellite either uses or swaps the qubits within a few
seconds ---actually most often within a few tens of milliseconds--- a timescale practically
instantaneous compared to its movement.
The very relevant $\tau\sim 1\unit{ms}$ regime, where the decoherence of the quantum memory
induces noise, is kept for future work. A limitation of the quantum memory we take into account is its finite storage capacity
$m$. In practice, the limiting factor will be the number of qubits $m_S$ the satellite
can store, which we take to be $m_S\in \{10,50,100\}$ for numerical applications.
The satellite is able to emit a photon entangled with a qubit of its memory, as well as
to perform a perfect Bell measurement between two qubits stored in its memory.

At each ground station, we assume the same hardware as the satellite plus a Bell State Analyzer, which is required by the link-level entanglement generation protocol.
Link level entanglement is established according to the \texttt{SenderReceiver} protocol described in \cite{Jones}. In short, the protocol requires one of the nodes establishing entanglement (the Sender) to generate a stream of photons entangled with local memory slots. The stream of photons is to be relayed to the Receiver, which performs interference measurements in order to ``latch'' the incoming qubits in a local quantum memory. Entanglement generation through this protocol works in rounds of $N$ attempts, where $N$ is the minimum number of available memory slots at the two nodes. Once $N$ attempts have been made, the Receiver shares the measurement results with the Sender, failed entanglement attempts are cleared from memory and a new round begins. After link level entanglement has been generated, we assume the satellite to be capable of entanglement swapping. 
\section{Satellite--Ground Link Analysis}
\label{sec:satellite_TheorySingleLink}
It is already apparent from the high-level description of the protocol we adopt how classical communication latency, usually neglected in quantum literature, plays a crucial role in the calculation of the entanglement distribution rate: an entanglement generation round requires one round trip of the channel 
---milliseconds for a satellite link--- effectively stunting the attainable entanglement distribution rate. 

\subsection{Entanglement Rate}
Let $m_G$ and $m_S$ be the number of memory slots available respectively at the ground station and onboard the satellite. For technological reasons, we assume $m_S\le m_G$.
Let $\eta$ be the downlink transmission coefficient of the satellite to ground channel calculated in Fig.\@ \ref{fig:satellites_linkparams}. 
The generation of a satellite-ground entangled pair needs the storage of one
qubit in the satellite quantum memory for a roundtrip time $\trt$ --- until confirmation 
of the reception (or lack thereof) of the photon by the ground station is received ---
and is successful with a probability $\eta$. 
Thus, taking $N = m_S$ and letting $p_{\text{BSM}}$ be the success probability of the latching Bell State Measurement ($\sfrac{1}{2}$ in our case \cite{weinfurter1994experimental}), the best attainable entanglement distribution rate will be
\begin{align}
\label{eq:theoretical-rate-uncorrected}
    r \leq p_{\text{BSM}}\frac{\eta N}{\trt}.
\end{align}
Although (\ref{eq:theoretical-rate-uncorrected}) is already surprisingly accurate given its
simplicity (as shown later in Fig.\@ \ref{fig:validation-matchrate} and \ref{fig:validation-matchnumber}), 
some discrepancy appears for high $m_S=100$. 
As discussed below, this is explained by the differential latency induced by the satellite's speed.

\subsection{Differential Latency}
\label{sec:satellites_difflatency}
\begin{figure}
    \centering
    \includegraphics[width=\linewidth]{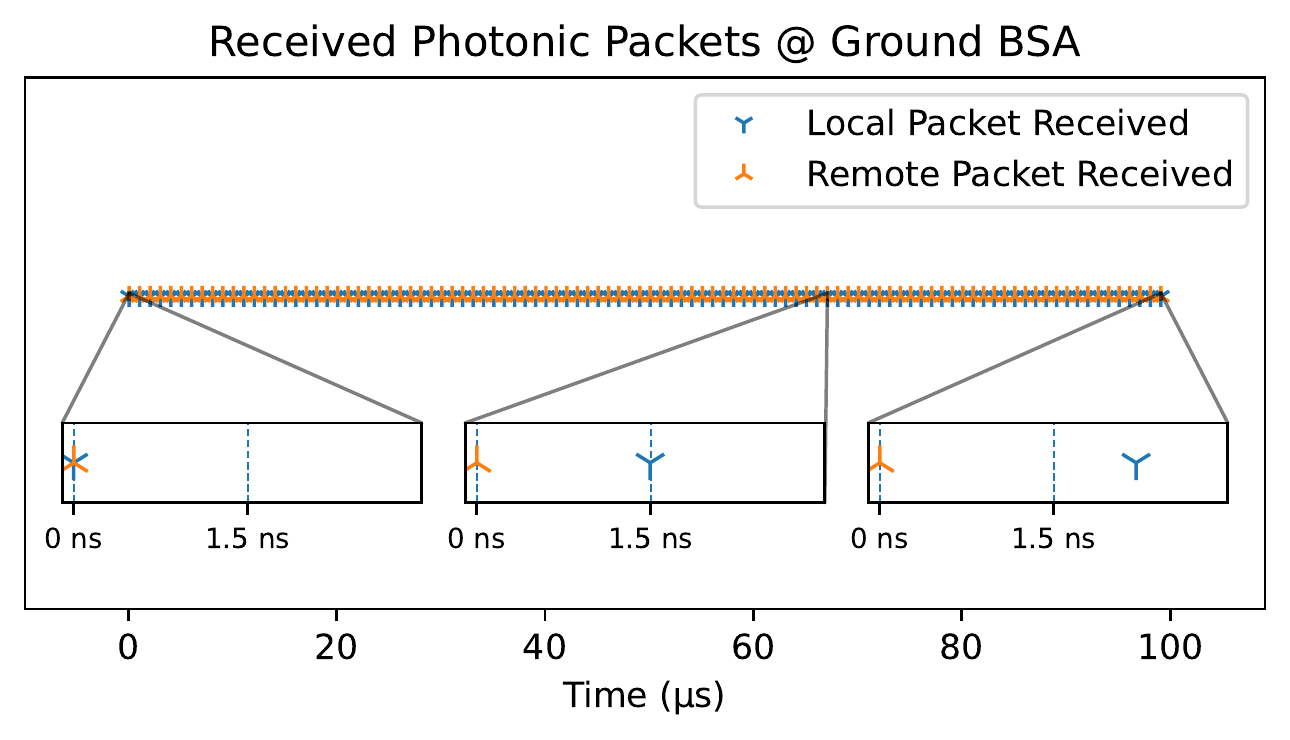}
    \caption[Timing of a satellite to ground photon train, showing the problem of differential latency.]{Timing of the first photon train, for a Micius--Nice
        link with $m_S=100$, extracted from our simulation. The reception of the first packet sets $t=0$, and the repetition rate is 1 \unit{MHz}. The local packet corresponds to
        the photon emitted by the ground quantum memory and the remote one to the photon received
        from the satellite. As shown by the inset, while initially in sync, they progressively 
        desynchronize, the 67$^\text{th}$ being the last one inside the acceptance window. A value of 1.5 \unit{ns} (default in QuISP) was employed for the acceptance window. More details on this choice are provided in the main text. To better display the timing of the incoming photons, this plot was generated from a lossless simulation.\label{fig:difflatency_demo}
    }
\end{figure}
Keeping in mind how crucial latency is to the efficient generation of entangled pairs, 
the high speed of satellites --- a few km/s --- introduces additional complexity in the form of \textit{differential latency},
that is a change of the timing of arrival of the photons due to the movement of the emitting satellite.
The \texttt{SenderReceiver} protocol requires the Receiver to perform interference measurements on the incoming photon stream, implying that the timing of incoming photons must be known throughout the procedure. 
In practice, this means that before entanglement distribution, a message with all the emission times is sent to the Receiver, which adjusts its local source to match incoming qubits. Whereas in optical fiber the arrival times are easily computed once and cached, moving nodes add a time discrepancy that must be countered.
Perfect intra-round matching is technologically challenging because it would require on-the-fly variation of the emission frequency of the photon pair source (often controlled by physical parameters like the emitting cavity length, which are difficult to adjust at runtime), but we demonstrate that good enough performance is attainable by simply syncing the first photon in each round. 
Of course, this implies that successive photons in the same train will eventually drift out of sync (as shown in fig. \ref{fig:difflatency_demo}), introducing a deep interplay between emission frequency, acceptance window for the interferometric measurement and satellite velocity. 
The net observed effect of differential latency consists of a time discrepancy that accumulates over successive photons until two corresponding photons do not interfere anymore and latching fails%
. Letting $\delta t_{DL}$ be the time shift introduced by the satellite's displacement, $v_r$ the radial component of the satellite's velocity, $T_{em}=1\unit{\micro s}$ the period of photon emission and $c=3\cdot10^8 \unit{m/s}$ the speed of light, we have
\begin{align}
    \delta t_{DL} = \frac{v_r T_{em}}{c}.
\end{align}
Since each successive photon undergoes cumulative desyncing, if we assume $v_r$ to be constant over the duration $m_ST_{em}$ of
one photon train we can place an upper bound on the useful photon train length $N$:
\begin{align}
\label{eq:maxN}
    N \leq \frac{w_i c}{|v_r| T_{em}},
\end{align}
where $w_i$ corresponds to the acceptance window of the Bell State Measurement at the ground station. The value of $w_i$ is technology dependent, with a typical value for it being a few nanoseconds: this parameter needs to be long compared to typical photodetector latencies and short compared to the photons' pulse length (which is in turn determined by the specific quantum memory in use). In our case, we opt for a value of 1.5 \unit{ns} (default in QuISP), as shown in fig.\@ \ref{fig:difflatency_demo}. Substituting (\ref{eq:maxN}) inside (\ref{eq:theoretical-rate-uncorrected}) yields a new expression for the theoretical rate, corrected for differential latency:
\begin{align}
\label{eq:theoretical-rate-corrected}
     r^* = p_{\text{BSM}}\frac{\eta}{\trt}\min\left(m_S,\frac{w_i c}{|v_r| T_{em}}\right).
\end{align}
This expression demonstrates how not only the classical communication latency drastically lowers entanglement distribution rates regardless of how optimized other parameters are, but the number of photons that is useful to exchange in a single experimental round is linked to the satellite's velocity, further impairing performance in the faraway orbit sections. 
\subsection{Validation of the Single Link Scenario}
\begin{figure}
    \centering
    \includegraphics[width=\linewidth]{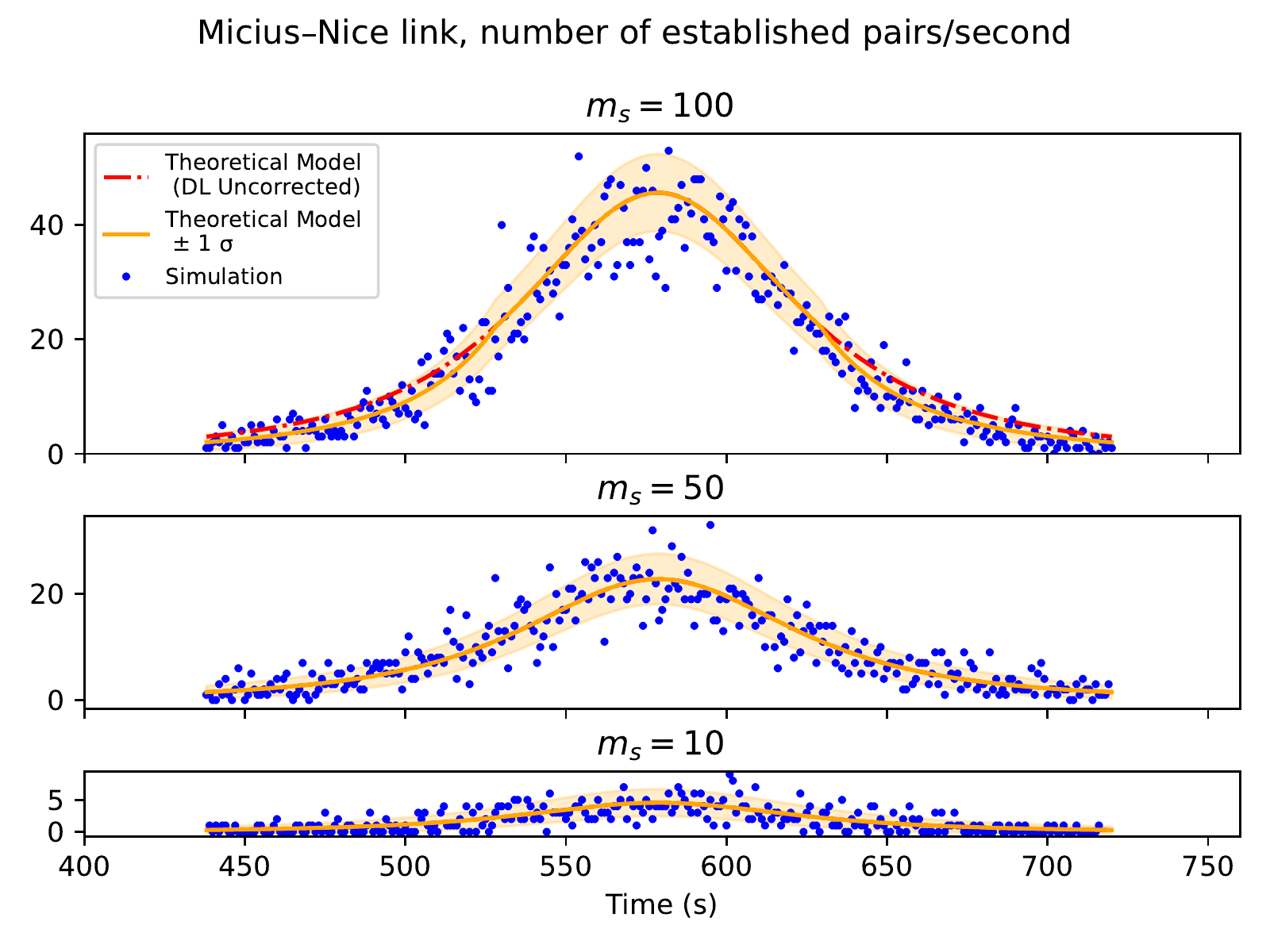}
    \caption[Rate of entanglement generation over a single Micius--Nice link.]{Rate of entanglement generation over a single Micius--Nice link for quantum memory sizes 
        $m_S\in\{10,50,100\}$.
        The orange shaded area corresponds to the expected $1\sigma = \sqrt{n}$ fluctuations.
        The simulation datapoints (blue points) for each second are within expected statistical fluctuations.}
    \label{fig:validation-matchrate}
\end{figure}
\begin{figure}
    \centering
    \includegraphics[width=\linewidth]{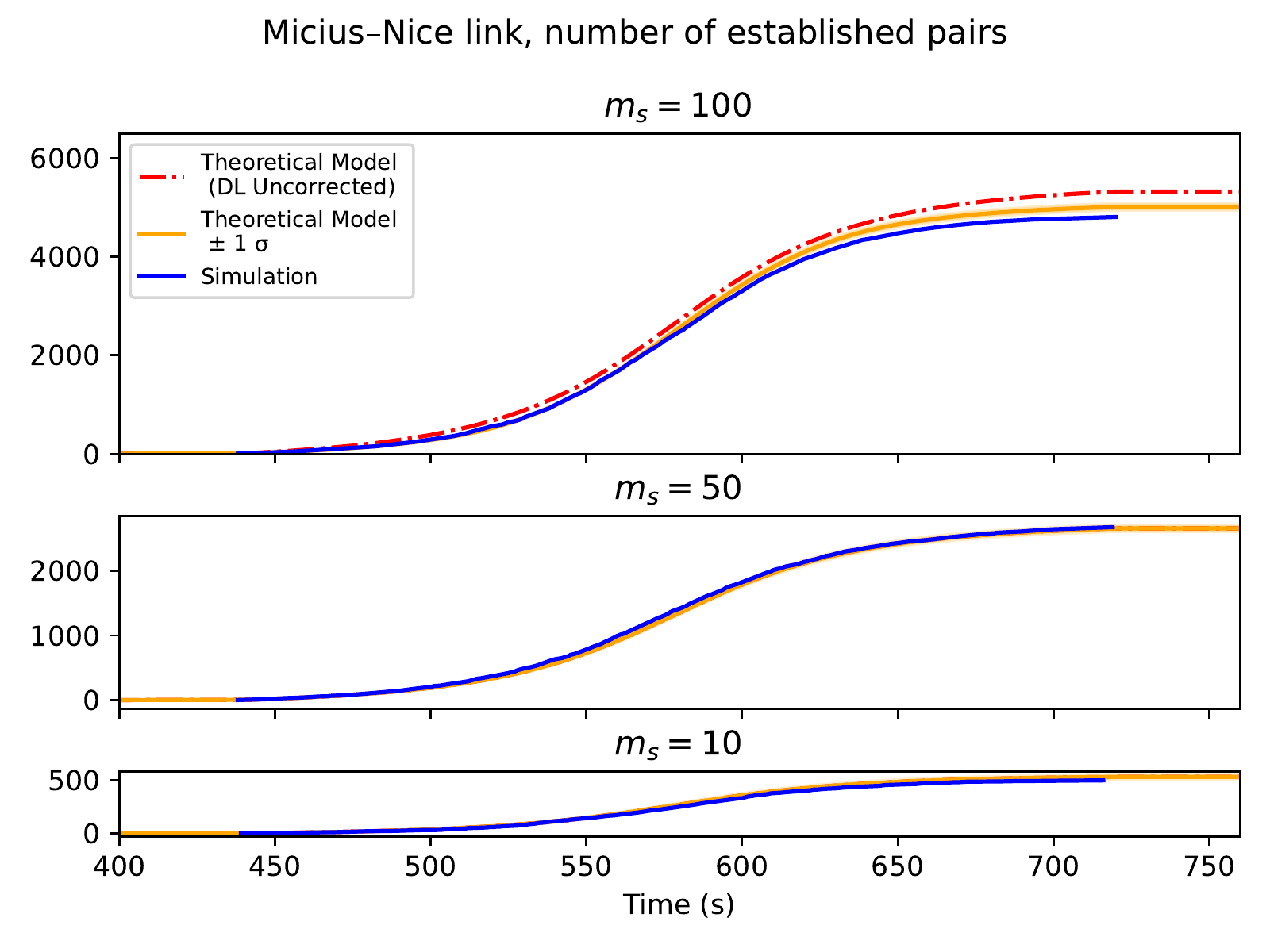}
    \caption[Number of generated entangled pairs over a single Micius--Nice link.]{Number of generated entangled pairs over a single Micius--Nice link for quantum memory $m_S\in\{10,50,100\}$.
    The orange shaded area corresponds to the expected $1\sigma$ statistical fluctuations.
    The simulation (blue curve) is within expected fluctuations.
    \label{fig:validation-matchnumber}}
\end{figure}
In this section, we provide results that cross-validate our model and simulation. We ran our simulations for a single passage 
of the Micius \cite{Micius} satellite, as seen by the Nice ground station for different entanglement round lengths and observed satisfactory agreement with the theoretical
calculations both in terms of entanglement distribution rate and total number of entangled pairs as detailed in 
fig.\@ \ref{fig:validation-matchrate} and \ref{fig:validation-matchnumber}.  
Additionally, we provide a plot of the predicted uncorrected rate of incoming photons $r$ (as per (\ref{eq:theoretical-rate-uncorrected})). 
For the maximal radial velocity of the satellite compared to Nice which is 6998~\unit{m/s}, 
(\ref{eq:maxN}) becomes $N\le 67$, which explains why the correction is only relevant for $m_S=100$, but not for $m_S\in \{10, 50\}$.
A close look at Fig. \ref{fig:validation-matchrate} further shows that the correction is most relevant at the beginning and end of the passage,
when the satellite is further away from the zenith
and its radial velocity higher.
\section{Ground--Satellite--Ground Link Analysis}
\label{sec:TheoryDualLink}
\begin{figure}
\centering
\subfloat[]{\includegraphics[height=6.3cm]{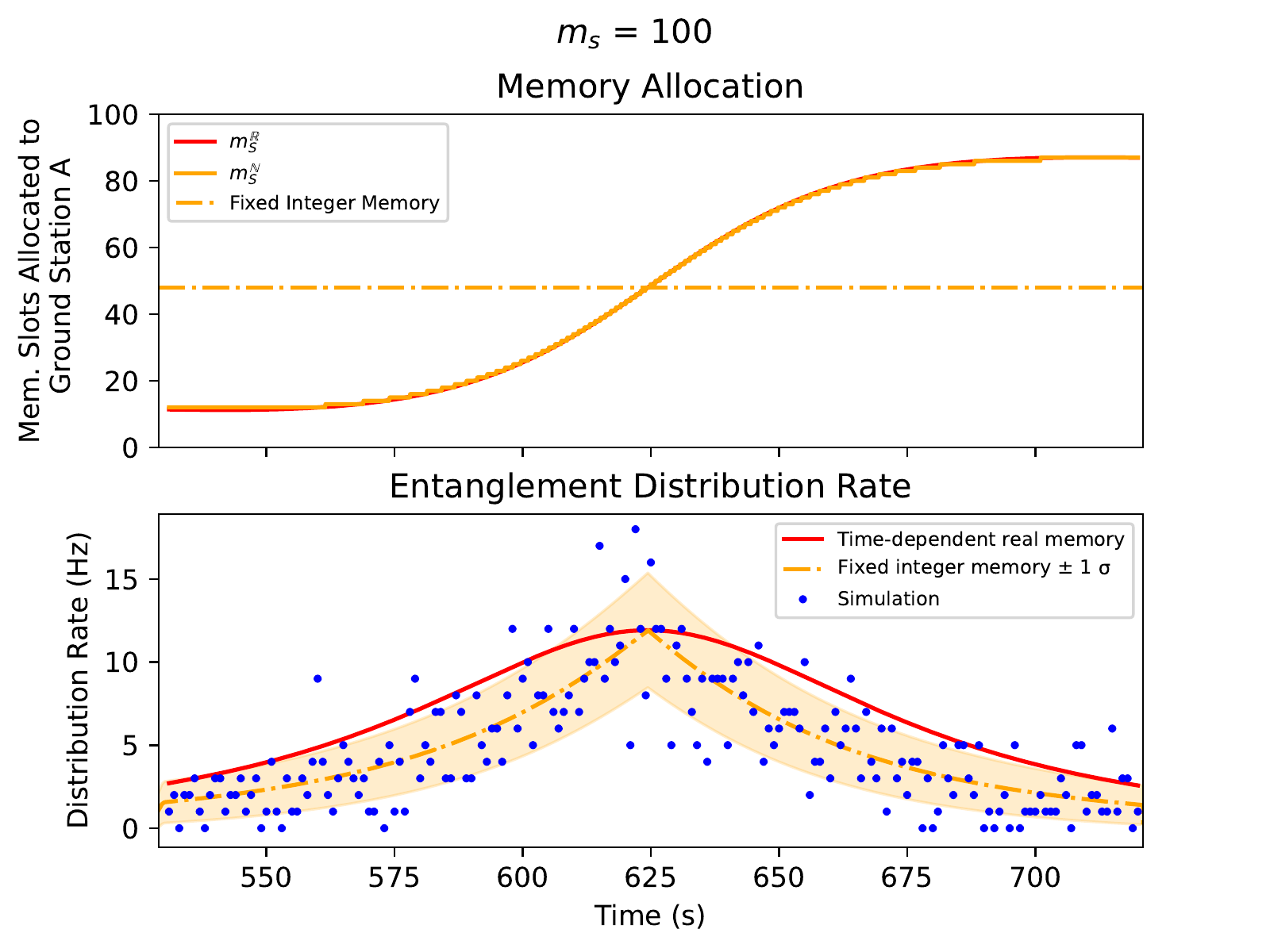}
\label{fig:satellite_malloc100}}
\subfloat[]{\includegraphics[height=6.3cm]{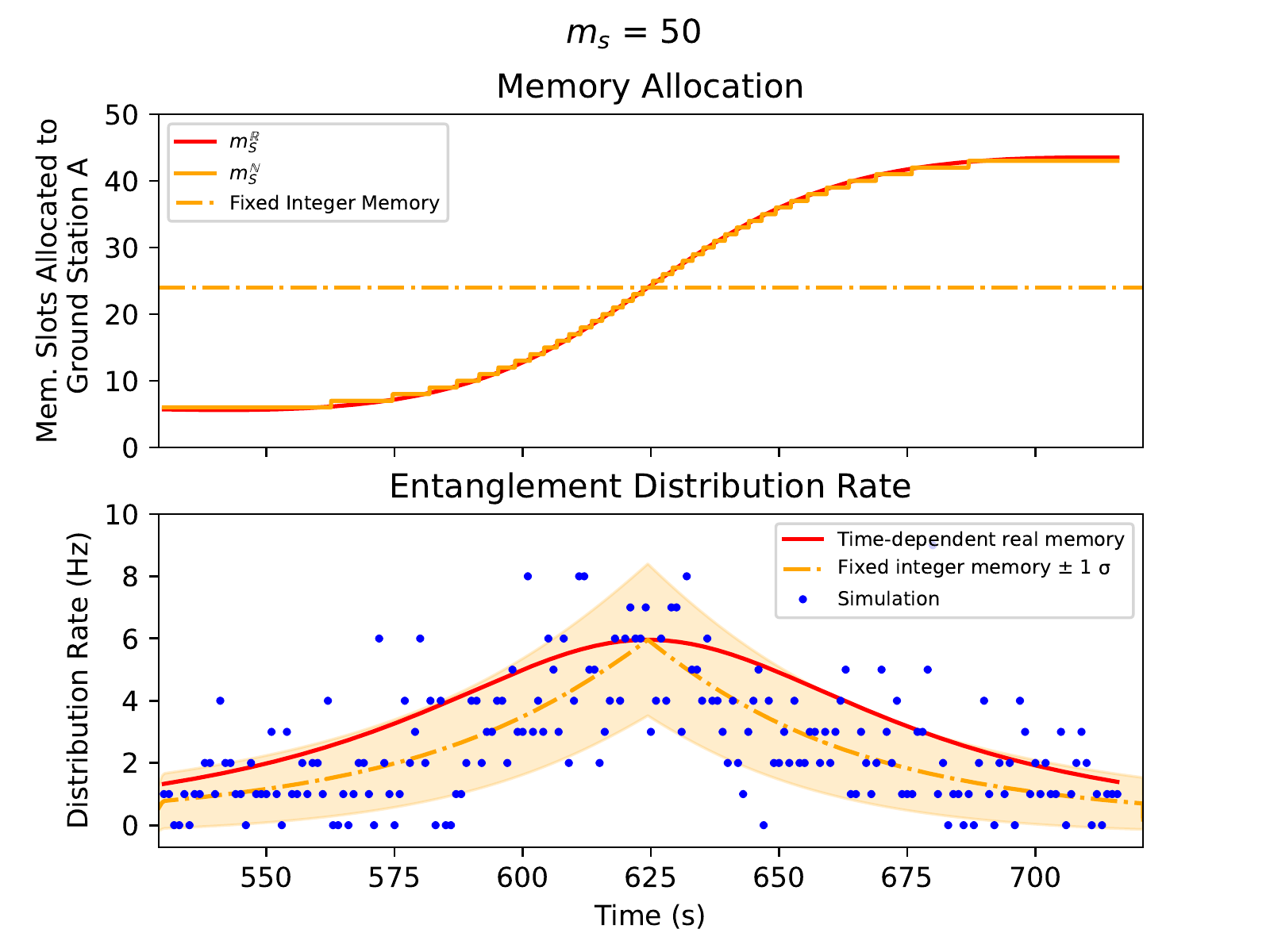}
\label{fig:satellite_malloc50}}

\subfloat[]{\includegraphics[height=6.3cm]{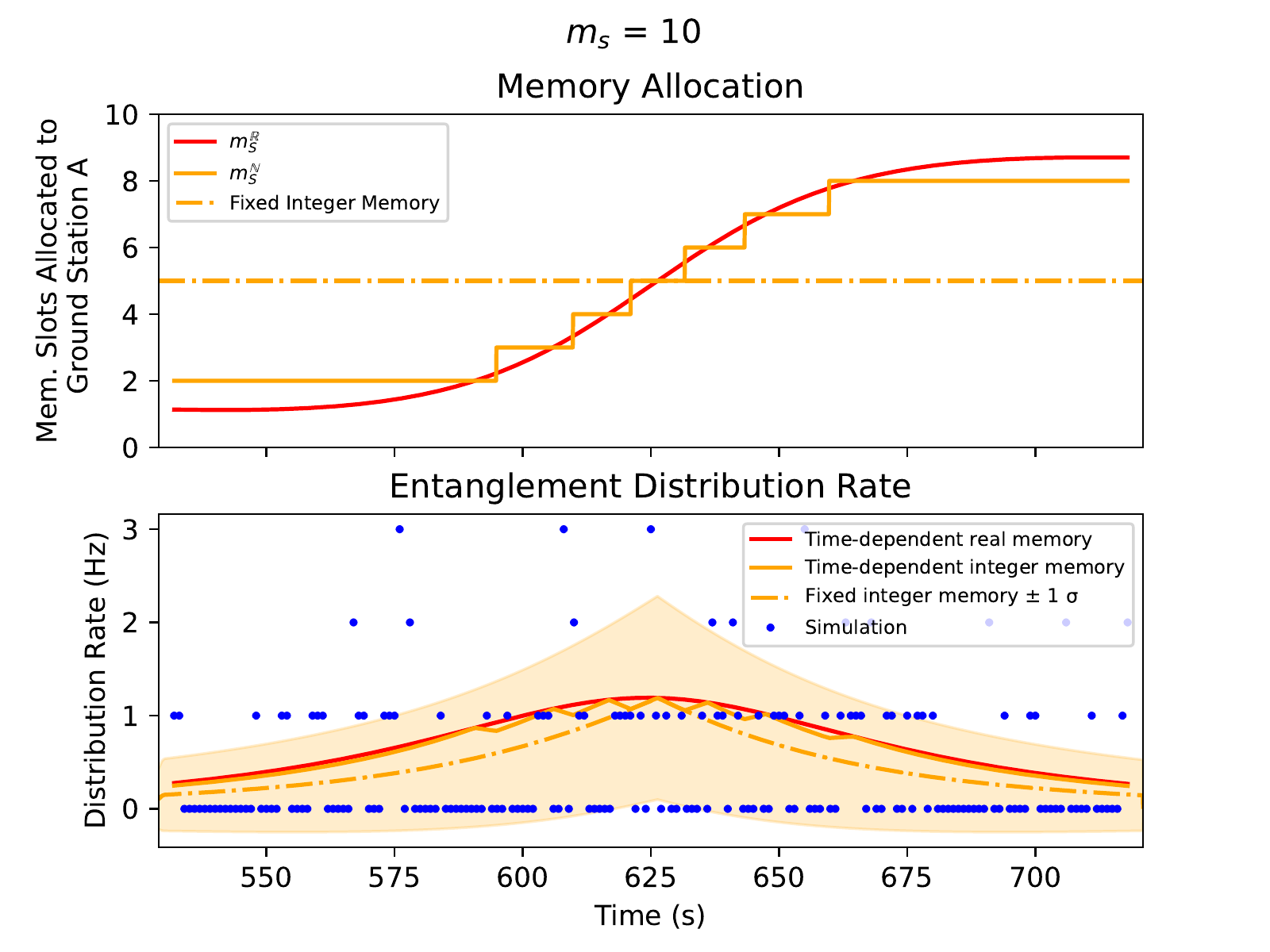}
\label{fig:satellite_malloc10}}
\caption[Theoretical entanglement distribution rate over a swapping Nice--Micius--Paris link with different quantum memory allocation policies.]{\label{fig:malloc}
Comparison of the theoretical entanglement distribution rate over a swapping Nice--Micius--Paris link with dynamic memory allocation against the theoretical and simulated 
    distribution rates with fixed allocation.
    The simulation datapoints (blue points) for each second are within expected statistical fluctuations.}
\end{figure}
\begin{figure}
    \centering
    \includegraphics[width=\linewidth]{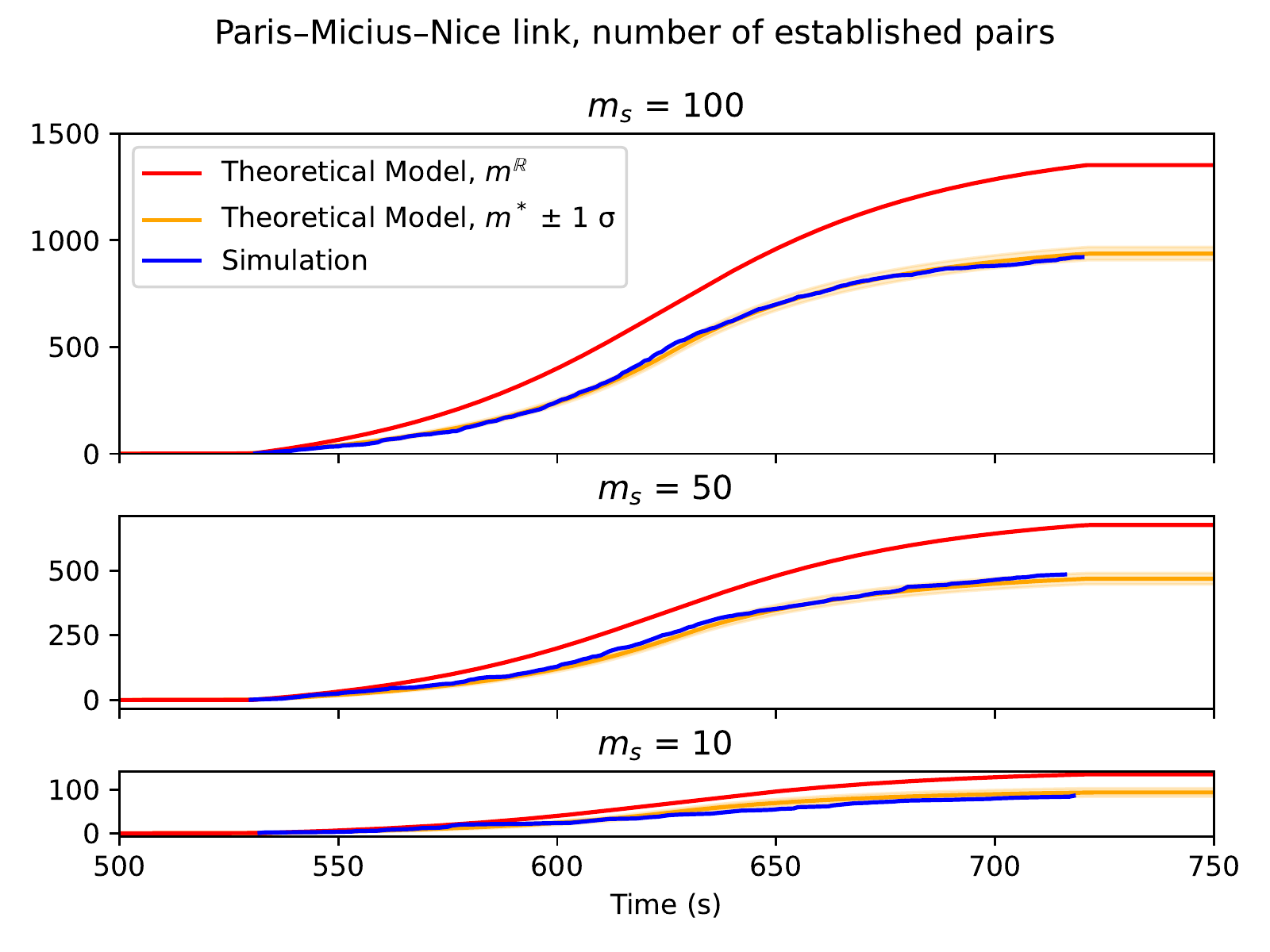}
    \caption[Number of generated entangled pairs over a swapping Nice--Micius--Paris link.]{Number of generated entangled pairs over a swapping Nice--Micius--Paris link for 
    quantum memory sizes $m_S\in\{10,50,100\}$.
     The performance difference between static and dynamic allocation is of 36\%, showing the interest of the latter for future work, even if both strategies do not differ by orders of magnitude.
     The simulation results are within expected statistical fluctuations.
     }
    \label{fig:validation-twinmatchrate}
\end{figure}
To analyze the dual link case, we adopt once again the \texttt{SenderReceiver} protocol to generate entanglement between the satellite and each of the ground nodes, and subsequently perform entanglement swapping aboard the satellite, thus merging the two satellite-to-ground links.
The dual entanglement distribution rate is simply the minimum between the entanglement distribution rates of each leg of the system,
computed as previously shown. However, a question remains: given a fixed amount $m_S$ of available memory slots aboard the satellite, 
how should they be assigned to each link?

Since the parameters of both downlinks evolve differently, there is no reason for a static $50:50$ split of the available slots to be the best solution. In the following, we start by studying optimal dynamic memory allocation. However, for various reasons, a constant split is simpler (for instance, when working with separate network interfaces connected to distinct quantum memories, as per \cite{QInternetArchitecture}). Thus, we work with a fixed value, extracted from the preceding dynamical analysis. We use the obtained values to validate the theoretical model against simulation and assess the performance degradation introduced by adopting static vs dynamic memory allocations. 

To simplify our theoretical analysis, let us first assume the number of allocated memory slots to each link, $m_A$ and $m_B$, can take real (non-integer) time-dependent values, where $A$ and $B$ label the two satellite--ground downlinks, and $m_A+m_B = m_S$.
The entanglement distribution rate $r_{AB}$ between the two ground stations is equal to the minimum of the two single-link rates.
Using (\ref{eq:theoretical-rate-uncorrected}), neglecting the differential latency correction for simplicity, we have
\begin{align}
\label{eq:twinrate}
    r_{AB} = \min\left(p_{\text{BSM}}^A\frac{m_S^A}{\trt^A}\eta^A,p_{\text{BSM}}^B\frac{ m_S^B}{\trt^B}\eta^B \right),
\end{align}
where $m_S^A$ and $m_S^B = m_S - m_S^A$ represent the number of memory slots associated to links $A$ and $B$, 
$\trt^\chi$ is the round-trip time across channel $\chi\in \{A,B\}$, $\eta^\chi$ the corresponding transmission coefficient and $p_{\text{BSM}}^\chi$ the success probability of the latching Bell State Measurement at the corresponding ground station.
This expression is maximized when both arguments of the $\min$ are equal, leading to
\begin{align}
\label{eq:memory_repartition}
    m^{\mathbb{R}}_A(t) &= \frac{\sfrac{\trt^A}{\eta^A}}{\sfrac{\trt^A}{\eta^A} + \sfrac{\trt^B}{\eta^B}}m_S,&
    m^{\mathbb{R}}_B(t) &= \frac{\sfrac{\trt^B}{\eta^B}}{\sfrac{\trt^A}{\eta^A} + \sfrac{\trt^B}{\eta^B}}m_S.
\end{align}
Note that all quantities except for $m_S$ are time-dependent. 

Substituting the relevant channel data, we obtain a time-dependent, real expression for the memory allocated to either link.
Of course, any actual memory partition can only accept integers value $m^{\mathbb{N}}_\chi(t)$,
optimally given by
\begin{align}
\label{eq:Nmemory_repartition}
    m^{\mathbb{N}}_A &= \left\lceil\frac{(m_S-1)\sfrac{\trt^A}{\eta^A}}{\sfrac{\trt^A}{\eta^A} + \sfrac{\trt^B}{\eta^B}}\right\rceil,&
    m^{\mathbb{N}}_B &=  \left\lfloor\frac{(m_S-1)\sfrac{\trt^B}{\eta^B}}{\sfrac{\trt^A}{\eta^A} + \sfrac{\trt^B}{\eta^B}} +1\right\rfloor.
\end{align}
As shown in Fig.\@ \ref{fig:validation-twinmatchrate}, even for $m_S=10$, this integer allocation
leads to rates close to the ones obtained using (\ref{eq:memory_repartition}).

At this point, technological considerations are in order: 
from the point of view of architecture and control, it is more convenient to design a system with two separate blocks of memory, each dedicated to one network interface. However, the situation may arise where one needs to design a satellite system where the more pressing constraint is that of memory availability. Such a system would benefit from a dynamical memory allocation policy that determines how much memory is allocated to each link in real time (akin to our simple analytical expression \ref{eq:memory_repartition}). 
Since QuISP was initially designed for fiber networks, every node is equipped with several independent quantum network interface cards (QNICs), each with its own set of memory slots. Therefore, we provide analytical results for the dynamic allocation case and both analytical and simulation results for the fixed memory one.
To choose the fixed memory values for our simulation, we adopt the following procedure:\begin{enumerate}
     \item We calculate the real, time-dependent value $m_A^\mathbb{R}(t)$ by substituting link parameters in (\ref{eq:memory_repartition});
    \item We round the obtained expression to obtain $m_A^\mathbb{N}(t)$;
    \item We 
            take as integer fixed value $m^*_A$ 
            the memory value $m_A^\mathbb{N}(t)$  for which  $r_{AB}(m_A^\mathbb{N}(t))$ attains its maximal value.
\end{enumerate}
This yields $(m^*_A,m^*_B)$, fixed integer memory allocation values that we can both substitute in (\ref{eq:twinrate}) to obtain a theoretical rate and input to our simulation as memory slot numbers.
From following the procedure above for the case of $m_S\in\{ 100,50,10\}$ with link $A$ associated to the Micius--Nice leg and link $B$ to the Micius--Paris one we obtain memory splits $(m^*_A,m^*_B) = (48,52), (24,26)$ and $(5,5)$ respectively. 
Fig.\@ \ref{fig:malloc} and \ref{fig:validation-twinmatchrate} show a comparison between the different allocation strategies, together
with the simulation results for a fixed memory split.
\section{Conclusions and Outlook}
We have presented a simple analytical model estimating the performance of a quantum satellite link equipped with quantum memory, and applied it to the entanglement distribution rate of a quantum satellite link to one and two ground stations, showing the key importance of quantum memory size and classical latency in the performance of the links.
As discussed in sec.\@ \ref{sec:adapting_quisp_sat}, we also developed an extension to QuISP to enable simulation of satellite links and cross-validated it with our theoretical model. Leveraging QuISP's scalability and versatility, the modifications developed in this chapter allow simulation of satellite quantum links accounting for all the quantum imperfections built in QuISP while simultaneously considering classical latency and memory allocation concerns.\\

In the immediate future, our work can be extended to apply to more complex cases. We considered the simplest possible iteration of our building block (one LEO satellite, one passage over two ground nodes), but extension is possible in multiple directions by exploring different orbit altitudes, multiple passages, or multiple ground terminals in a star-like arrangement. Moreover, the main use case of our extended QuISP version is the simulation of multiple satellite links interconnecting metropolitan-scale fiber networks.

On the long term, our analytical framework may be extended in several ways, such as including satellite--to--satellite links for intra- and inter-orbit communication, or a discussion of the impact of quantum memory noise and channel quality on entanglement distribution.\\
Simulation-wise, it could be interesting to further extend QuISP to allow for satellites with dynamic memory allocation: this would not only enable the validation of the performance margin we observed in sec. \ref{sec:TheoryDualLink}, but also pave the way for the study of more refined dynamic scheduling policies for memory allocation.
%
\chapter{Outlook and Conclusion}
\label{ch:Conclusion}
Throughout this thesis, we carried out an in-depth analysis of various issues related to scheduling in quantum networks. However, due to timing constraints and some delays during the software development phase, we were not able to fully explore dynamic scheduling in fiber and satellite quantum networks. As a goal for future research, we report our formulation of the problem and identify its requirements before moving to the final conclusions.

\section{Scheduling in Satellite-aided Quantum Networks}
In chap.\@ \ref{ch:scheduling}, we outlined a framework for designing and benchmarking complex scheduling policies over quantum networks. To extend the discussion to multiple interconnected subnetworks, we proposed in chap.\@ \ref{ch:satellites} a discussion of the peculiarities of satellite quantum links with respect to fiber ones. While it would be possible to simulate a fiber and free-space network with the simulator developed in chap. \ref{ch:scheduling}, one of the assumptions underneath that simulator was the absence of communication latency: while this can be a reasonable assumption over the individual links of a quantum repeater network in optical fiber, the insight gained in chap. \ref{ch:satellites} concerning the entanglement distribution rate over a free-space link shows that such an assumption is not viable over high-latency satellite links: when free-space links are considered, communication latency becomes a crucial limiting factor that must be taken into account.\\
Other than stunting performance, classical latencies also have other, non-obvious negative effects over network operation. Going back to chap. \ref{ch:scheduling}, the problem preventing the implementation of a Max Weight policy that has full information about the instantaneous network state was the fact that information needs finite time to propagate through the network. We argued that a centralized scheduler would take decisions based on outdated information, and developed new schedulers that circumvent the problem while providing acceptable performance. Translating our work to QuISP would give a way to gauge exactly how propagation latencies impair the performance of a fully informed scheduler. Furthermore, a classification we made for scheduling policies was that of centralized policies, where one global authority takes decisions for the whole network, versus distributed ones, where every node or cluster of nodes can autonomously take a scheduling decision for itself. Working with QuISP gives this classification a new depth, as it is possible to realistically study where the scheduler is deployed and how its position changes the service of the network's demands: a centralized scheduler that is closer to a subsection of the network will have more updated information about it and possibly take decisions that are biased to its needs.\\
Finally, a side benefit of working with an accurate network simulator is that it requires scheduling policies to be thought of in a distributed way: local information availability constraints are strictly enforced, and any remote information required must be passed through a message, better clarifying the communication overhead requirements of different policies.
\subsection{Formulating the Problem}
\begin{figure}
    \centering
    \includegraphics[width=.8\linewidth]{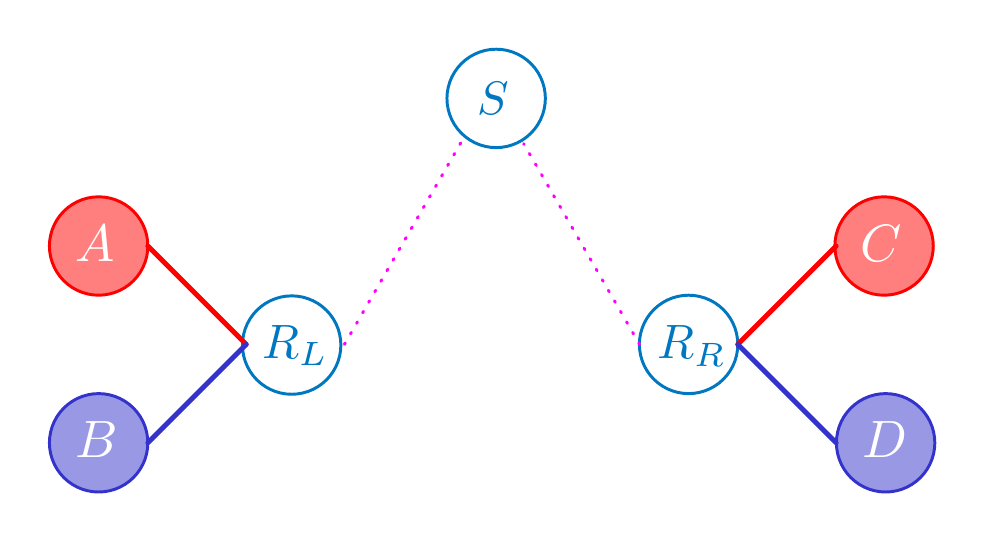}
    \caption[A simple satellite network topology.]{A simple satellite network topology: two elementary fiber subnetworks, each connected to its own repeater, which is in turn connected through a ground-satellite-ground link to the other subnetwork.}
    \label{fig:simple-sat-net}
\end{figure}

In its most basic form, a large-scale quantum network is a set of subnetworks interconnected by long-distance links. For our purposes, the simplest nontrivial case is that of two nontrivial networks interconnected by a satellite link (fig. \ref{fig:simple-sat-net}).
Users in network $A$ will initiate connections with network $B$, and the chosen scheduling policy will need to manage their concurrence.\\
Even though this is a simple case, there are already some interesting non-obvious issues: first of all, given the restriction to binary entanglement with connection-based transport layer, the most realistic and interesting version of the problem would be the one in which users in the system are allowed to randomly generate connections to others. This would not only create the problem of managing the satellite bottleneck, but also internal concurrence between connections in the same subnetwork: a scheduler would need to balance using pairs for local service versus swapping them for internetwork traffic. Before moving to fully dynamic scheduling problems, we note that an interesting direction would be precalculating deterministic, nonconstant schedules for a network with multiple satellite links. Despite the problem we discussed with the variability of satellite quantum links (sec.\@ \ref{sec:satellites_difflatency}), the trajectory of a satellite is usually available before the passage, and allows to predict the link quality. Therefore, when working with multiple satellite links (such as when interconnecting three or more metropolitan networks) it is possible to precalculate memory allocations depending on the expected quality of various links. Such a discussion would enable the study of slightly more dynamic networks, all while keeping to the deterministic time-division policies that are currently supported by our version of QuISP. To move to truly dynamic scheduling, we need to refine our multiplexing implementation to support link-level synchronization, enabling nodes to correctly allocate resources without the need for deterministic external synchronization.\\
\section{Additional Future Direction: Scheduling Multipartite Entanglement}
 As mentioned in chap. \ref{ch:simulation}, QuISP works in terms of connections, with an initiator negotiating a connection with a responder for a given number of Bell pairs and running an application. This is a common practice in classical networks, but not necessarily the best approach for quantum networking. As briefly hinted in sec. \ref{sec:intro_quantumentanglement}, quantum entanglement is not necessarily bipartite: it is possible --- and interesting for applications such as Quantum Electronic Voting \cite{CentroneEVoting}, Multiparty Quantum Computation \cite{CrepeauMPComp} and Conference Key Agreement \cite{MurtaCKA} --- to share entanglement between three or more parties, making the notion of initiator and responder more difficult to define: while it is possible to distribute multipartite states starting from bipartite ones \cite{MeignantMultipartite,PirkerMultiParty}, this is often suboptimal in terms of required entangled pairs, justifying the need for a truly multipartite network stack.\\
 Multipartite entanglement would also be an interesting nontrivial extension of our scheduling framework, as including multipartite queues and transitions would create scaling issues that require new solutions, such as reducing the amount of queues by greedily scheduling some sections of the network and only optimizing over relevant decisions.\\
\section{Summary and Conclusions}
\paragraph{Quantum Scheduling in Fiber Networks (chap.\@ \ref{ch:scheduling})}The first contribution we presented is a linear algebraic framework that allows formulation of the scheduling problem over 1G quantum networks. Our discussion is compatible with memory-endowed quantum networks in that it allows preparation of intermediate-range entangled pairs, enabling study of the design tradeoffs that arise when a scheduler needs to balance between preemptive swapping and responsiveness to instantaneous demand.\\
Our framework has open interfaces for the generation of ebits and for the traffic model that determines the arrival of user demand. We demonstrated our framework using Poissonian processes for both ebit and traffic generation, while the ebit losses were managed through a binomial model. We showed in the following chapters how the framework can flexibly adapt to a batch-based traffic model to examine systems in which users ask for ebits all at once, instead of requesting them in real time.\\
We leveraged our framework together with Lyapunov Drift Minimization to derive a family of scheduling policies for quantum networks based on quadratic optimization, and showed how these policies offer only minor advantages with respect to their computationally lighter linear counterparts.\\
Through a simple Python simulator based on our analytical discussion we analyzed the performance of the presented scheduling policies over a simple network topology (a chain of six quantum repeaters) and then generalized to larger topologies, both deterministic (square grid, possibly with some nodes removed) and procedurally generated (Erdős–Rényi, Watts-Strogatz). This allowed us to deeply explore the performance of the policies and their resiliency against traffic from a nontrivial number of users. Our simulator supports the analysis of the temporal trend of total accumulated demand in a network under a given scheduler, the average demand backlog (to approximate stability regions with high resolution) and of the total excursion of accumulated demand (to obtain a preliminary gauge of the order of magnitude of the memory requirements of a network).\\
Among our policies, we proposed an original localized one that blends global, non time-sensitive information such as generation rate and traffic averages with fresh, up-to-date information from the immediate neighborhood of each node. This policy was found to offer interesting performance margins over the greedy scheduler (common lower bound for performance) while retaining practical feasibility.
\paragraph{A Small Detour: Multiplexing in QuISP (chap.\@ \ref{ch:simulation})}
We performed some development work to refine QuISP's handling of quantum end-to-end connections. We implemented a mechanism for the end nodes of a quantum connection to terminate the execution of a quantum application (such as quantum tomography) and to release the reserved resources to the common pool. We achieved this by making the way QuISP handles reservations more granular, switching from a simple boolean variable for each quantum memory register to a full-fledged reservation register, complete with the ability for multiple connections to reserve the same resource. When an application returns, the node in charge of terminating the connection shares a connection teardown message with all the other connected nodes, who remove the connection from their local reservation registers.\\
Once teardown was implemented, several connections could reserve the same set of quantum memory registers. However, the default resource allocation mechanism at the nodes would still serve first all the demands of the first-established connection and then move to the second. To counter this and study true simultaneous service, we designed a preliminary implementation for network multiplexing, allowing quantum networks simulated in QuISP to manage multiple concurrent requests. To validate our findings, we also modified the small scheduling simulator developed in the first chapter to be compatible with the batched traffic model employed by QuISP, in which users request batches of $N$ entangled pairs instead of a continuous stream of random demands. This allowed us to examine the same figures of merit in QuISP and our simulator starting from compatible parameters: despite our simulation's simplistic assumptions, the results obtained from our code and from QuISP are in the same order of magnitude, which is a promising result concerning future attempts at using QuISP to analyze more refined scheduling policies over complex topologies.\\
Once link-level synchronization is implemented, our algebraic framework and simulation toolbox will enable users to provide a custom scheduling policy and obtain stability regions over realistic, noisy and lossy quantum internetworks at the metropolitan and continental scale.
\paragraph{Entanglement Swapping in Orbit (chap.\@ \ref{ch:satellites})}To better understand the problems of fiber and satellite quantum networks, we shifted the focus to the details of satellite quantum communication to ascertain how it differs from fiber systems. Given two elementary examples (satellite to one ground station, satellite to two ground stations) we provided an analytical account of the expected entanglement distribution rates. This analysis yielded interesting insight regarding the technical limitations of satellite quantum setups, which are direct consequences of the underlying physics, and highlighted the shortcomings in our previous approach that neglected classical communication latencies under the assumption that they would not appreciably impact communication rates.\\
Our analysis of the dual satellite to ground link included a discussion regarding quantum memory. Our work assumed the presence of a number of memory slots at the ground stations and onboard the satellite, and it was therefore relevant to explore how the limited amount of quantum memory slots should be assigned to the two links. We demonstrated that, even though it would be an improvement to have a dynamic allocation policy for quantum memory on the satellite, the performance of the link can attain the same order of magnitude with a simple static policy.\\
To complement our analytical study of satellite links, we presented some simulated results. Thanks to a collaboration with the AQUA Team in Keio University, we developed several extensions to the Quantum Internet Simulation Package (QuISP), enabling the simulation of satellite links inside quantum networks. We implemented a new channel model that takes data from standard orbital simulation toolboxes and exploits it to calculate losses along a free-space quantum channel using a model found in the relevant literature. Moreover, we created new submodules that verify whether the other end of a free-space link is currently in sight and buffer control messages if that is not the case. Once some additional software development work is carried out, leveraging the depth of features in QuISP and its scalability will allow in-depth investigation of large-scale fiber and satellite quantum networking systems.
%
{%
\setstretch{1.1}
\renewcommand{\bibfont}{\normalfont\small}
\setlength{\biblabelsep}{0pt}
\setlength{\bibitemsep}{0.5\baselineskip plus 0.5\baselineskip}
\printbibliography[nottype=online]
\newrefcontext[labelprefix={@}]
\printbibliography[heading=subbibliography,title={Webpages},type=online]
}
\cleardoublepage

\listoffigures
\cleardoublepage

\listoftables
\cleardoublepage





\end{document}